\documentclass[varenna]{cimento}

\usepackage{graphicx}
\usepackage{amsmath}
\usepackage{amssymb}
\usepackage{dsfont}
\usepackage{setspace}
\usepackage{verbatim}

\begin{document}
\catcode`\ä = \active \catcode`\ö = \active \catcode`\ü = \active
\catcode`\Ä = \active \catcode`\Ö = \active \catcode`\Ü = \active
\catcode`\ß = \active \catcode`\é = \active \catcode`\è = \active
\catcode`\ë = \active \catcode`\ô = \active \catcode`\ê = \active
\catcode`\ø = \active \catcode`\ò = \active \catcode`\í = \active
\defä{\"a} \defö{\"o} \defü{\"u} \defÄ{\"A} \defÖ{\"O} \defÜ{\"U} \defß{\ss} \defé{\'{e}}
\defè{\`{e}} \defë{\"{e}} \defô{\^{o}} \defê{\^{e}} \defø{\o} \defò{\`{o}} \defí{\'{i}}
\newcommand{\one}{$\left|1\right>$}
\newcommand{\two}{$\left|2\right>$}
\newcommand{\vect}[1]{\mathbf #1}
\newcommand{\li}{$^6$Li}
\newcommand{\na}{$^{23}$Na}
\newcommand{\Int}[1]{\int {\rm d}^3\mathbf #1}
\newcommand{\Intl}[2]{\int_{#2} {\rm d}^3\mathbf #1}
\newcommand{\Intp}[1]{\int \frac{{\rm d}^3\mathbf #1}{(2\pi\hbar)^3}}

\singlespacing
\title{Making, probing and understanding ultracold Fermi gases}

\author{Wolfgang Ketterle \atque Martin W. Zwierlein}

\institute{Department of Physics, MIT-Harvard Center for Ultracold Atoms, and Research Laboratory of Electronics,\\ Massachusetts Institute of Technology, Cambridge, Massachusetts, 02139, USA}

\maketitle
\vspace{3in}
W. Ketterle and M. W. Zwierlein, {\it Making, probing and understanding ultracold Fermi gases}, in Ultracold Fermi Gases, Proceedings of the International School of Physics "Enrico Fermi", Course CLXIV, Varenna, 20 - 30 June 2006, edited by M. Inguscio, W. Ketterle, and C. Salomon (IOS Press, Amsterdam) 2008

\pagestyle{plain}
{\bf Table of contents}\\
\tableofcontents

\section{Introduction}
\label{c:introduction}

\subsection{State of the field} This paper summarizes the
experimental frontier of ultracold fermionic gases.  It is based
on three lectures which one of the authors gave at the Varenna
summer school describing the experimental techniques used to
study ultracold fermionic gases, and some of the results obtained
so far. In many ways, the area of ultracold fermionic gases has
grown out of the study of Bose-Einstein condensates. After their
first experimental realizations in 1995~\cite{ande95,davi95bec},
the field of BEC has grown explosively. Most of the explored physics
was governed by mean-field interactions, conveniently described
by the Gross-Pitaevskii equation. One novel feature of trapped
inhomogeneous gases was the spatially varying density, that allowed
for the direct observation of the condensate, but also led
to new concepts of surface effects and collective excitations
which depended on the shape of the cloud.  The experimental and
theoretical explorations of these and other features have been a
frontier area for a whole decade!

A major goal had been to go beyond mean field physics, which is in
essence single particle physics, and to find manifestations of
strong interactions and correlations.  Three avenues have been
identified:  lower dimensions that enhance the role of
fluctuations and correlations, optical lattices that can suppress
the kinetic energy in the form of
tunnelling~\cite{grei01mott,bloc05lattice}, and Feshbach
resonances~\cite{stwa76,ties93,inou98,cour98fesh} that enhance
interactions by resonantly increasing the interparticle scattering
length. In bosonic systems, the tuning of interactions near
Feshbach resonances was of limited applicability due to rapid
losses. Feshbach resonances were used mainly to access molecular
states of dimers and trimers. In contrast, for fermions, losses
are heavily suppressed (see below), and so most of this review
focuses on strongly interacting fermions near Feshbach resonances.

By addressing the physics of strongly correlated matter, the field
of ultracold atoms is entering a new stage where we expect major
conceptional advances in, and challenges to many-body theory. We
regard it as fortunate that BEC turned out to be a less complex
target (both experimentally and theoretically), and over a decade,
important techniques and methods have been developed and validated,
including experimental techniques to confine and cool nanokelvin
atoms, the use of Feshbach resonances to modify their properties,
and many theoretical concepts and methods to describe trapped
ultracold gases and their interactions.  What we are currently
experiencing is the application of these powerful methods to
strongly correlated systems, and due to the maturity of the field,
the developments have been breath-taking, in particular with
bosons in optical lattices and fermions interacting via Feshbach
resonances. It is possible that the most important conceptional
advances triggered by the advent of Bose-Einstein condensation are
yet to be discovered.

It is amusing to note that in certain limits, strongly correlated
fermion pairs are again described by a mean-field theory. Their wave function is a product of identical pair wave
functions (albeit correctly anti-symmetrized), that for strong binding of the pairs turns into the
state described by the Gross-Pitaevskii equation.  This is the
simplest description of the BEC-BCS crossover. Still, the fact that
pairing has now become a many-body affair stands for the advent of a new
era in ultracold atom physics.

\subsection{Strongly correlated fermions - a gift of nature?}

It shows the dynamics of the field of ultracold atoms that the
area of strongly interacting fermions has not been expected or
predicted.  This may remind us of the pre-BEC era, when many
people considered BEC to be an elusive goal, made inaccessible by
inelastic interactions at the densities required~\cite{kett99var}.
When Feshbach resonances were explored in bosonic systems, strong
interactions were always accompanied by strong losses, preventing
the study of strongly interacting
condensates~\cite{inou98,sten98stro,corn00JILA_collapse}. The
reason is that a Feshbach resonance couples the atomic Hilbert
space to a resonant molecular state which is vibrationally highly
excited.  Collisions can couple this state to lower lying states
(vibrational relaxation).

What occurred in Fermi gases, however, seemed too
good to be true:  all relaxation mechanisms were dramatically
suppressed by the interplay of the Pauli exclusion principle and
the large size of the Feshbach molecules.  So what we have got is a
Hilbert space which consists of atomic levels plus one single
molecular level resonantly coupled to two colliding atoms. All
other molecular states couple only weakly. As a
result, pair condensation and fermionic superfluidity could be
realized by simply ramping down the laser power in an optical trap
containing $^6$Li in two hyperfine states at a specific magnetic
field, thereby evaporatively cooling the system to the superfluid state.
Even in our boldest moments we would not have dared to ask Nature
for such an ideal system.

Before the discovery of Feshbach resonances, suggestions to
realize fermionic superfluidity focused on lithium because of the
unusually large and negative triplet scattering
length~\cite{abra97,stoo96superfluid,houb98}. However, a major
concern was whether the gas would be stable against inelastic
collisions. The stability of the strongly interacting Fermi gas
was discovered in Paris in the spring of 2003, when long-lived
${\rm Li}_2$ molecules were observed despite their high
vibrational excitation~\cite{cubi03}\footnote{The observation of
long lifetimes of molecules outside a narrow Feshbach
resonance~\cite{stre03} is not yet understood and has not been
used to realize a strongly interacting gas.}. This and subsequent
observations~\cite{joch03lith,rega03lifetime} were soon explained
as a consequence of Pauli suppression~\cite{petr03three_body}.
Within the same year, this unexpected stability was exploited to
achieve condensation of fermion pairs.  This unique surprise has
changed the field completely.  Currently, more than half of the
research program of our group is dedicated to fermions interacting
near Feshbach resonances.

There is another aspect of Fermi gases, which turned out to be
more favorable than expected.  Early work on the BCS state in
ultracold gases suggested a competition between superfluidity and
collapse (for negative scattering length) or coexistence and phase
separation (for positive scattering length) when the density or
the absolute value of the scattering length $a$ exceeded a certain
value, given by $k_F |a| = \pi/2$, where $k_F$ is the Fermi wave
vector~\cite{stoo96superfluid,stoo99var,houb97}. This would have implied that the highest
transition temperatures to the superfluid state would be achieved
close to the limit of mechanical stability, and that the BCS-BEC
crossover would be interrupted by a window around the Feshbach
resonance, where phase separation occurs. Fortunately, unitarity
limits the maximum attractive energy to a fraction of the Fermi
energy ($\beta E_F$ with $\beta \approx - 0.58$), completely
eliminating the predicted mechanical instability.

Finally, a third aspect received a lot of attention, namely how
to detect the superfluid state.  Since no major change in the
spatial profile of the cloud was expected~\cite{houb97},
suggested detection schemes included a change in the decay rate
of the gas~\cite{houb97}, optical light scattering of Cooper
pairs~\cite{zhan99,weig00}, optical breakup of Cooper
pairs~\cite{torm00}, modification of collective
excitations~\cite{bara00,ming01}, or small changes in the spatial
shape~\cite{chio02}.  All these signatures are weak or
complicated to detect. Fortunately, much clearer and more easily
detectable signatures were discovered.  One is the  onset of pair
condensation, observed through a bimodal density distribution in
expanding clouds, observed either well below the Feshbach
resonance or after rapid sweeps of the magnetic field. Another
striking signature was the sudden change in the cloud shape when
fermion mixtures with population imbalance became superfluid, and
finally, the smoking gun for superfluidity was obtained by
observing superfluid flow in the form of quantized vortices.

Our ultimate goal is to control Nature and create and explore new
forms of matter.  But in the end, it is Nature who sets the rules, and in the case of ultracold fermions, she has
been very kind to us.

\subsection{Some remarks on the history of fermionic superfluidity}
\subsubsection{BCS superfluidity}

Many cold fermion clouds are cooled by sympathetic cooling with a
bosonic atom. Popular combinations are $^6$Li and $^{23}$Na, and
$^{40}$K and $^{87}$Rb.  It is remarkable that the first
fermionic superfluids were also cooled by a Bose-Einstein
condensate. Kamerlingh Onnes liquefied $^4$He in 1908, and
lowered its temperature below the superfluid transition point
(the $\lambda$-point) at $T_\lambda = 2.2$~K. In his
Nobel lecture in 1913, he notes ``that the density of the helium, which at
first quickly drops with the temperature, reaches a maximum at
2.2 K approximately, and if one goes down further even drops
again. Such an extreme could possibly be connected with the
quantum theory''~\cite{onne13}. But instead of studying, what we
know now was the first indication of superfluidity of bosons, he first focused on the behavior of metals at low
temperatures. In 1911, Onnes used $^4$He to cool down mercury,
finding that the resistivity of the metal suddenly dropped to
non-measurable values at $T_C = 4.2$ K, it became
``superconducting''. Tin (at $T_C = 3.8$ K) and lead (at $T_C = 6$
K) showed the same remarkable phenomenon.  This was the discovery
of superfluidity in an electron gas.

The fact that bosonic superfluidity and fermionic superfluidity
were first observed at very similar temperatures, is due to
purely technical reasons (because of the available cryogenic methods) and
rather obscures the very different physics behind these two
phenomena.

Bosonic superfluidity occurs at the degeneracy temperature, i.e. the temperature $T$ at which
the spacing between particles $n^{-1/3}$ at density $n$
becomes comparable to the thermal de Broglie wavelength $\lambda
= \sqrt{\frac{2\pi \hbar^2}{m k_B T}}$, where $m$ is the particle
mass.  The predicted transition temperature
of $T_{\rm BEC} \sim \frac{2\pi\hbar^2}{m}n^{2/3} \approx 3$ K for
liquid helium at a typical density of $n = 10^{22}\; \rm cm^{-3}$
coincides with the observed lambda point.

In contrast, the degeneracy temperature (equal to the Fermi
temperature $T_F \equiv E_F/k_B$) for conduction electrons is higher by the mass ratio
$m(^4 \rm He)/m_{e}$, bringing it up to several ten-thousand
degrees. It was only in 1957 when it became clear why in fermionic
systems, superfluidity occurs only at temperatures much smaller
than the degeneracy temperature.

Of course, the main difference to Bose gases is that electrons, being fermions,
cannot be in one and the same quantum state but instead
must arrange themselves in {\it different} states.  An obvious
scenario for superfluidity might be the formation of tightly bound pairs of electrons that can
act as bosons and could form a condensate. But apart from the problem
that the condensation temperature would still be on the order of
$E_F/k_B$, there is no known interaction which could be
sufficient to overcome the strong Coulomb repulsion and form
tightly bound electron pairs (Schafroth pairs~\cite{scha58}). The
idea itself of electrons forming pairs was indeed correct, but
the conceptual difficulties were so profound that it took several
decades from the discovery of superconductivity to the correct
physical theory.

In 1950, it became clear that there was indeed an effective
attractive interaction between electrons, mediated by the crystal
lattice vibrations (phonons), that was responsible for
superconductivity. The lattice vibrations left their mark in the
characteristic variation $T_C \propto 1/\sqrt{M}$ of the critical
temperature $T_C$ with the isotope mass $M$ of the crystal ions,
the isotope effect~\cite{reyn50,maxw50} predicted by H.
Fr\"ohlich~\cite{froh50}. Vibrational energies in the lattice are
a factor $\sqrt{m_{e}/M}$ smaller than the typical electronic
energy\footnote{The average distance between electrons $r_0$ is on
the order of atomic distances (several Bohr radii $a_0$), the
Fermi energy $E_F \sim \hbar^2/m_{e} r_0^2$ is thus on the scale
of typical Coulomb energies in an atom. Vibrational energies of
the lattice ions are then on the order $\hbar \omega_D \approx
\hbar \sqrt{\frac{\partial^2 U_{\rm Coulomb}}{\partial r^2} / M}
\sim \hbar \sqrt{E_F / M r_0^2} \sim \sqrt{m_{e}/M}\, E_F$.} $E_F$,
on the order of $k_B \times$ several 100 K (the Debye temperature
$T_D$ of the metal). While the isotope effect strongly argues for
$T_C $ being proportional to $T_D$, the Debye temperature is still
one or two orders of magnitude higher than the observed critical
temperature.

A breakthrough came in 1956, when L. Cooper realized that fermions
interacting via an arbitrarily weak attractive interaction on top
of a filled Fermi sea can form a bound pair~\cite{coop56}. In
other words, the Fermi sea is unstable towards pair formation.
However, unlike the tightly bound pairs considered before, the
``Cooper'' pair is very large, much larger than the
interparticle spacing. That is, a collection of these pairs
necessarily needs to overlap very strongly in space. In this
situation, it was far from obvious whether interactions
between different pairs could simply be neglected. But it was this simplifying idea that led to the final goal: Bardeen, Cooper and Schrieffer (BCS)
developed a full theory of superconductivity starting from a new,
stable ground state in which pair formation was included in a
self-consistent way~\cite{bard57}. Using the effective phonon-mediated
electron-electron interaction $V$, attractive for energies smaller
than $k_B T_D$ and assumed constant in this regime, the pair
binding energy was found to be $\Delta = 2 k_B T_D\,
e^{-1/\rho_F|V|}$, with $\rho_F = m_{e} k_F/2\pi^2 \hbar^2$ the
density of states at the Fermi energy and $\rho_F|V|$ assumed small
compared to 1. The bound state energy or the pairing gap depended in
the non-analytic fashion $e^{-1/\rho_F|V|}$ on the effective electron-electron
interaction $V$, explaining why earlier attempts using
perturbation theory had to fail. Also, this exponential factor can
now account for the small critical temperatures $T_C \simeq 5\,\rm K$:
Indeed, it is a result of BCS theory that $k_B T_C$ is simply
proportional to $\Delta_0$, the pair binding energy at zero
temperature: $k_B T_C \approx 0.57\, \Delta_0$. Hence, the
critical temperature $T_C \sim T_D\, e^{-1/\rho_F|V|}$ is
proportional to the Debye temperature $T_D$, in accord with the
isotope effect, but the exponential factor suppresses $T_C$ by a
factor that can easily be 100.

\subsubsection{The BEC-BCS crossover}

\begin{figure}
\centering
\includegraphics[width=4.5in]{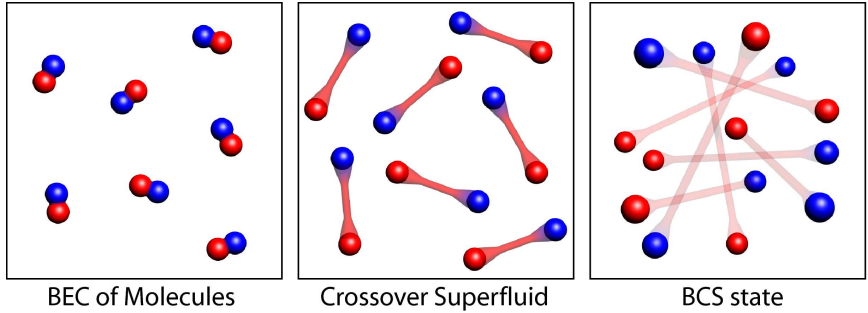}
  \caption[The BEC-BCS crossover]{The BEC-BCS crossover. By tuning
the interaction strength between the two fermionic spin states,
one can smoothly cross over from a regime of tightly bound
molecules to a regime of long-range Cooper pairs, whose
characteristic size is much larger than the interparticle spacing.
In between these two extremes, one encounters an intermediate
regime where the pair size is comparable to the interparticle
spacing.}\label{f:BECBCSCrossover}
\end{figure}

Early work on BCS theory emphasized the different nature of BEC
and BCS type superfluidity.  Already in 1950 Fritz London had
suspected that fermionic superfluidity can be understood as a pair
condensate in momentum space, in contrast to a BEC of tightly
bound pairs in real space~\cite{lond64}. The former will occur for
the slightest attraction between fermions, while the latter
appears to require a true two-body bound state to be available to
a fermion pair. Schrieffer points out that BCS superfluidity is
not Bose-Einstein condensation of fermion pairs, as these pairs do
not obey Bose-Einstein statistics~\cite{schr99super}. However, it
has become clear that BEC and BCS superfluidity are intimately
connected. A BEC is a special limit of the BCS state.

It was Popov~\cite{popo66}, Keldysh and
collaborators~\cite{keld68} and Eagles~\cite{eagl69} who realized
in different contexts that the BCS formalism and its ansatz for
the ground state wave function provides not only a good description
for a condensate of Cooper pairs, but also for a Bose-Einstein
condensate of a dilute gas of tightly bound pairs. For
superconductors, Eagles~\cite{eagl69} showed in 1969 that, in the
limit of very high density, the BCS state evolves into a
condensate of pairs that can become even smaller than the
interparticle distance and should be described by Bose-Einstein
statistics. In the language of Fermi gases, the scattering length
was held fixed, at positive and negative values, and the
interparticle spacing was varied. He also noted that pairing
without superconductivity can occur above the superfluid
transition temperature. Using a generic two-body potential,
Leggett showed in 1980 that the limits of tightly bound molecules
and long-range Cooper pairs are connected in a smooth
crossover~\cite{legg80}. Here it was the interparticle distance
that was fixed, while the scattering length was varied. The size
of the fermion pairs changes smoothly from being much larger than
the interparticle spacing in the BCS-limit to the small size of a
molecular bound state in the BEC limit (see
Fig.~\ref{f:BECBCSCrossover}). Accordingly, the pair binding
energy varies smoothly from its small BCS value (weak, fragile
pairing) to the large binding energy of a molecule in the BEC
limit (stable molecular pairing).

The presence of a paired state is in sharp contrast to the case of
two particles interacting in free (3D) space. 
Only at a critical interaction strength does a molecular state
become available and a bound pair can form. Leggett's result shows
that in the many-body system the physics changes smoothly with
interaction strength also at the point where the two-body bound
state disappears. Nozi\`eres and Schmitt-Rink extended Leggett's
model to finite temperatures and verified that the critical
temperature for superfluidity varies smoothly from the BCS limit,
where it is exponentially small, to the BEC-limit where one
recovers the value for Bose-Einstein condensation of tightly bound
molecules~\cite{nozi85}.

The interest in strongly interacting fermions and the BCS-BEC
crossover increased with the discovery of novel superconducting
materials. Up to 1986, BCS theory and its extensions and variations were largely successful in explaining the properties of superconductors. The record critical temperature increased only
slightly from 6 K in 1911 to 24 K in 1973~\cite{ginz91}. In 1986, however,
Bednorz and M\"uller~\cite{bedn86hightc} discovered
superconductivity at 35 K in the compound $\rm
La_{2-x}Ba_xCuO_4$, triggering a focused search for even higher
critical temperatures. Soon after, materials with transition
temperatures above 100 K were found. Due to the strong
interactions and quasi-2D structure, the exact mechanisms leading
to High-$\rm T_C$ superconductivity are still not fully
understood.

The physics of the BEC-BCS crossover in a gas of interacting
fermions does not directly relate to the complicated phenomena
observed in High-$T_C$ materials. However, the two problems share several features: In the crossover regime, the pair size is comparable to the interparticle distance. This relates to
High-$T_C$ materials where the correlation length (``pair size'')
is also not large compared to the average distance between
electrons. Therefore, we are dealing here with a strongly correlated ``soup''
of particles, where interactions between different pairs of
fermions can no longer be neglected. In both systems the normal
state above the phase transition temperature is far from being an
ordinary Fermi gas. Correlations are still strong enough to form
uncondensed pairs at finite momentum. In High-$T_C$ materials,
this region in the phase diagram is referred to as the ``Nernst
regime'', part of a larger region called the
``Pseudo-gap''~\cite{lee06hightc}.

One point in the BEC-BCS crossover is of special interest: When
the interparticle potential is just about strong enough to bind
two particles in free space, the bond length of this molecule
tends to infinity (unitarity regime). In the medium, this bond
length will not play any role anymore in the description of the
many-body state. The only length scale of importance is then the
interparticle distance $n^{-1/3}$, the corresponding energy scale
is the Fermi energy $E_F$. In this case, physics is said to be
universal~\cite{ho04uni}. The average energy content of the gas,
the binding energy of a pair, and ($k_B$ times) the critical
temperature must be related to the Fermi energy by universal
numerical constants. The size of a fermion pair must be given by a
universal constant times the interparticle distance.

It is at the unitarity point that fermionic interactions are at
their strongest.  Further increase of attractive interactions will
lead to the appearance of a bound state and turn fermion pairs
into bosons.  As a result, the highest transition temperatures for
fermionic superfluidity are obtained around unitarity and are on
the order of the degeneracy temperature. Finally, almost 100 years
after Kamerlingh Onnes, it is not just an accidental coincidence anymore that
bosonic and fermionic superfluidity occur at similar temperatures!

\subsubsection{Experiments on fermionic gases}

After the accomplishment of quantum degeneracy in bosons, one
important goal was the study of quantum degenerate fermions.
Actually, already in 1993, one of us (W.K.) started to set up dye
lasers to cool fermionic lithium as a complement to the existing
experiment on bosons (sodium).  However, in 1994 this experiment
was shut down to concentrate all resources on the pursuit of
Bose-Einstein condensation, and it was only in early 2000 that a
new effort was launched at MIT to pursue research on fermions.
Already around 1997, new fermion experiments were being built in
Boulder (using $^{40}$K, by Debbie Jin) and in Paris (using
$^6$Li, by Christophe Salomon, together with Marc-Oliver Mewes, a
former MIT graduate student who had worked on the sodium BEC
project).

All techniques relevant to the study of fermionic gases had
already been developed in the context of BEC, including magnetic
trapping, evaporative cooling, sympathetic
cooling~\cite{myat97,bloc01symp}, optical trapping~\cite{stam98odt} and Feshbach
resonances~\cite{inou98,cour98fesh}.  The first degenerate Fermi
gas of atoms was created in 1999 by B. DeMarco and D. Jin at JILA
using fermionic $^{40}$K~\cite{dema99}.  They exploited the rather
unusual hyperfine structure in potassium that allows magnetic
trapping of two hyperfine states without spin relaxation, thus
providing an experimental ``shortcut'' to sympathetic cooling. All
other schemes for sympathetic cooling required laser cooling of
two species or \emph{optical} trapping of two hyperfine states of
the fermionic atom.  Until the end of 2003, six more groups had
succeeded in producing ultracold degenerate Fermi gases, one more
using $^{40}$K (M. Inguscio's group in Florence,
2002~\cite{roat02}) and five using fermionic \li\ (R. Hulet's
group at Rice~\cite{trus01}, C. Salomon's group at the ENS in
Paris~\cite{schr01}, J. Thomas' group at Duke~\cite{gran02}, our
group at MIT~\cite{hadz02} in 2001 and R. Grimm's group in
Innsbruck in 2003~\cite{joch03bec}).

Between 1999 and 2001, the ideal Fermi gas and some collisional
properties were studied.  2002 (and late 2001) was the year of
Feshbach resonances when several groups  managed to optically
confine a two-component mixture and tune an external magnetic
field to a Feshbach
resonance~\cite{diec02fesh,loft02,ohar02,ohar02,joch02}. Feshbach resonances were observed by
enhanced elastic collisions~\cite{loft02}, via an increase in loss
rates~\cite{diec02fesh}, and by hydrodynamic expansion, the
signature of a strongly interacting gas~\cite{ohar02science}. The
following year, 2003, became the year of Feshbach molecules.  By
sweeping the magnetic field across the Feshbach resonance, the
energy of the Feshbach molecular state was tuned below that of two
free atoms (``molecular'' or ``BEC'' side of the Feshbach
resonance) and molecules could be produced~\cite{rega03mol}. These
sweep experiments were very soon implemented in Bose gases and
resulted in the observation of ${\rm Cs}_2$~\cite{herb03cs_mol},
${\rm Na}_2$~\cite{xu03na_mol} and ${\rm Rb}_2$~\cite{durr04mol}
molecules. Pure molecular gases made of bosonic atoms were created
close to~\cite{herb03cs_mol} or clearly in~\cite{xu03na_mol} the
quantum-degenerate regime. Although quantum degenerate molecules
were first generated with bosonic atoms, they were not called
Bose-Einstein condensates, because their lifetime was too short to
reach full thermal equilibrium.

Molecules consisting of fermionic atoms were much more
long-lived~\cite{cubi03,joch03lith,stre03,rega03lifetime} and were
soon cooled into a Bose-Einstein condensate.  In November 2003,
three groups reported the realization of Bose-Einstein
condensation of
molecules~\cite{grei03molbec,zwie03molBEC,joch03bec}. All three
experiments had some shortcomings, which were soon remedied in
subsequent publications.  In the $^{40}$K experiment the effective
lifetime of 5 to 10 ms was sufficient to reach equilibrium in only
two dimensions and to form a quasi- or nonequilibrium
condensate~\cite{grei03molbec}. In the original Innsbruck
experiment~\cite{joch03bec}, evidence for a long-lived condensate
of lithium molecules was obtained indirectly, from the number of
particles in a shallow trap and the magnetic field dependence of
the loss rate consistent with mean-field effects.
A direct observation followed soon after~\cite{bart04}.
The condensate observed at MIT was distorted by an anharmonic
trapping potential.

To be precise, these experiments realized already crossover
condensates (see section~\ref{c:expobservation}) consisting of large, extended
molecules or fermion pairs. They all operated in the strongly
interacting regime with $k_F a>1$, where the size of the pairs is
not small compared to the interparticle spacing.  When the
interparticle spacing $\sim 1/k_F$ becomes smaller than the
scattering length $\sim a$, the two-body molecular state is not
relevant anymore and pairing is a many-body affair. In fact, due
to the increase of collisional losses on the ``BEC'' side,
experiments have so far explored pair condensates only down to
$k_F a \approx 0.2$~\cite{zwie05vort}. Soon after these first experiments
on fermion pair condensates, their observation was extended throughout the whole
BEC-BCS crossover region by employing a rapid ramp to the
``BEC''-side of the Feshbach
resonance~\cite{rega04,zwie04rescond}.

During the following years, properties of this new crossover
superfluid  were studied in thermodynamic
measurements~\cite{bour04coll,kina05heat}, experiments on
collective excitations~\cite{kina04sfluid,bart04coll}, RF
spectroscopy revealing the formation of pairs~\cite{chin04gap},
and an analysis of the two-body part of the pair wave function was
carried out~\cite{part05}.  Although all these studies were
consistent with superfluid behavior, they did not address
properties unique to superfluids, i.e. hydrodynamic excitations
can reflect superfluid or classical hydrodynamics, and the RF
spectrum shows no difference between the superfluid and normal
state~\cite{schu07pair}. Finally, in April 2005, fermionic
superfluidity and phase coherence was directly demonstrated at MIT
through the observation of vortices~\cite{zwie05vort}. More recent
highlights (in 2006 and 2007) include the study of fermionic
mixtures with population
imbalance~\cite{zwie05imbalance,part06phase,shin06phase,part06deform,shin07phasediagram},
the (indirect) observation of superfluidity of fermions in an
optical lattice~\cite{chin06}, the measurement of the speed of
sound~\cite{jose07sound} and the measurement of critical velocities~\cite{mill07critical}.
Other experiments focused on two-body physics including
the formation of $p$-wave molecules~\cite{gaeb07pwave} and the
observation of fermion antibunching~\cite{rom06}.

\subsubsection{High-temperature superfluidity}

\begin{table}
  \centering
  \begin{tabular}{l|r|r|r}
    System & $T_C$ & $T_F$ & $T_C/T_F$  \\
    \hline
    Metallic lithium at ambient pressure~\cite{tuor07lithiumambient}
                                                & 0.4 mK & 55 000 K         & $10^{-8}$       \\
    Metallic superconductors (typical)         & 1--10 K    & 50 000 -- 150 000 K         & $10^{-4\dots-5}$       \\
    $^3$He                                     & 2.6 mK & 5 K              & $5\cdot 10^{-4}$    \\
    MgB$_2$                                    & 39 K   & 6 000 K &  $10^{-2}$       \\
    High-$T_C$ superconductors                 & 35--140 K  & 2000 -- 5000 K & $1\dots 5\cdot 10^{-2}$     \\
    Neutron stars                              & $10^{10}$ K & $10^{11}$ K & $10^{-1}$ \\
    Strongly interacting atomic Fermi gases    & 200 nK & 1 $\mu$K         & 0.2            \\
  \end{tabular}
  \caption{Transition temperatures, Fermi temperatures and their ratio $T_C/T_F$ for a variety of fermionic superfluids or superconductors.}\label{t:hightc}
\end{table}

The crossover condensates realized in the experiments on ultracold Fermi gases are a new type of fermionic superfluid.
This superfluid differs from $^3$He, conventional and even High-$T_C$ superconductors in its high
critical temperature $T_C$ when compared to the Fermi temperature
$T_F$. Indeed, while $T_C/T_F$ is about $10^{-5}\dots 10^{-4}$
for conventional superconductors, $5 \,10^{-4}$ for $^3$He
and $10^{-2}$ for High-$T_C$ superconductors, the strong
interactions induced by the Feshbach resonance allow atomic
Fermi gases to enter the superfluid state already at $T_C/T_F
\approx 0.2$, as summarized in table~\ref{t:hightc}. It is this large value which allows us to call this
phenomenon {\it ``high-temperature superfluidity''}. Scaled to the
density of electrons in a metal, this form of superfluidity would
already occur far above room temperature (actually, even above
the melting temperature).

\subsection{Realizing model systems with ultracold atoms}

Systems of ultracold atoms are ideal model systems for a host of
phenomena.  Their diluteness implies the absence of complicated or
not well understood interactions.  It also implies that they can
be controlled, manipulated and probed with the precision of atomic
physics.

Fermions with strong, unitarity limited interactions are such a
model system. One encounters strongly interacting fermions in a
large variety of physical systems: inside a neutron star, in the
quark-gluon plasma of the early Universe, in atomic nuclei, in
strongly correlated electron systems.  Some of the phenomena in
such systems are captured by assuming point-like fermions with
very strong short range interactions.  The unitarity limit in the
interaction strength is realized when the scattering length
characterizing these interactions becomes longer than the
interparticle spacing.  For instance, in a neutron star, the
neutron-neutron scattering length of about -18.8 fm is large
compared to the few fm distance between neutrons at densities of
$10^{38} \rm\,cm^{-3}$. Thus, there are analogies between results
obtained in an ultracold gas at unitarity, at densities of
$10^{12} \rm\,cm^{-3}$, and the physics inside a neutron star.
Several communities are interested in the equation of state, in
the value of the total energy and of the superfluid transition
temperature of simple models of strongly interacting
fermions~\cite{scha07b}.

Strongly interacting fermions can realize flow deep in the
hydrodynamic regime, i.e. with vanishing viscosity.  As discussed
in chapter~\ref{c:expobservation}, the viscosity can be so small
that no change in the flow behavior is observed when the
superfluid phase transition is crossed.  This kind of
dissipationless hydrodynamic flow allows to establish connections
with other areas.  For instance, the anisotropic expansion of an
elongated Fermi gas shares features with the elliptical (also
called radial) flow of particles observed in heavy ion collisions,
which create strongly interacting quark
matter~\cite{back05Phobos}.

The very low viscosity observed in strongly interacting Fermi
gases~\cite{kina04sfluid,kina05damping,bart04coll} has attracted
interest from the high energy physics community.  Using methods from string
theory, it has been predicted that the ratio of the shear
viscosity to the entropy density can not be smaller than
$\frac{1}{4 \pi}$~\cite{kovt05visc}. The two liquids that come
closest to this lower bound are strongly interacting ultracold
fermions and the quark gluon plasma~\cite{scha07visc}.

Another idealization is the pairing of fermions with different
chemical potentials. This problem emerged from superconductivity
in external fields, but also from superfluidity of quarks, where
the heavy mass of the strange quark leads to ``stressed pairing''
due to a shift of the strange quark Fermi energy~\cite{casa04,
alfo01}. One of the authors (W.K.) still remembers vividly how
an MIT particle physics colleague, Krishna Rajagopal, asked him
about the possibility of realizing pairing between fermions with
different Fermi energies (see~\cite{bowe02}), even before
condensation and superfluidity in balanced mixtures had become possible.
At this point, any realization seemed far away. With
some satisfaction, we have included in these Varenna notes our
recently observed phase diagram for population imbalanced
ultracold fermions~\cite{shin07phasediagram}.

This overlap with other areas illustrates a special role of cold
atom experiments:  They can perform ``quantum simulations'' of
simple models, the results of which may then influence research in
other areas.  Of course, those simulations cannot replace
experiments with real quarks, nuclei and condensed matter systems.

\subsection{Overview over the chapters}

With these notes we want to give a comprehensive introduction into
experimental studies of ultracold fermions.  The first focus of
this review is on the description of the experimental techniques
to prepare and manipulate fermionic gases
(chapter~\ref{c:exptechniques}), and the methods to diagnose the
system including image analysis (chapter~\ref{c:analysis}). For
those techniques which are identical to the ones used for bosons
we refer to our review paper on bosons in the 1998 Varenna
proceedings. The second focus is on the comprehensive description
of the physics of the BEC-BCS crossover
(chapter~\ref{c:BECBCStheory}) and of Feshbach resonances
(chapter~\ref{c:feshbach}), and a summary of the experimental
studies performed so far (chapters~\ref{c:expobservation} and
\ref{c:otherstudies}). Concerning the presentation of the material
we took a bimodal approach, sometimes presenting an in-depth
discussion, when we felt that a similar description could not be
found elsewhere, sometimes giving only a short summary with
references to relevant literature.  Of course, the selection of
topics which are covered in more detail reflects also the
contributions of the MIT group over the last six years. The theory
chapter on the BCS-BEC crossover emphasizes physical concepts over
formal rigor and is presented in a style that should be suitable
for teaching an advanced graduate course in AMO physics. We
resisted the temptation to include recent experimental work on
optical lattices and a detailed discussion of population
imbalanced Fermi mixtures, because these areas are still in rapid
development, and the value of a review chapter would be rather
short lived.

These notes include a lot of new material not presented elsewhere.
Chapter~\ref{c:analysis} on various regimes for trapped and
expanding clouds summarizes many results that have not been
presented together and can serve as a reference for how to fit
density profiles of fermions in all relevant limits.
Chapter~\ref{c:BECBCStheory} on BCS pairing emphasizes the density
of states and the relation of Cooper pairs in three dimensions to
a two-particle bound state in two dimensions.  Many results of BCS
theory are derived in a rigorous way without relying on
complicated theoretical tools. In chapter~\ref{c:feshbach}, many
non-trivial aspects of Feshbach resonances are obtained from a
simple model. Chapter~\ref{c:expobservation} presents density
profiles, not published elsewhere, of a resonantly interacting
Fermi gas after expansion, showing a direct signature of
condensation. In chapter~\ref{c:expobservation}, we have included
several unpublished figures related to the observation of
vortices.

\section{Experimental techniques}
\label{c:exptechniques}

The ``window'' in density and temperature for achieving fermionic
degeneracy is similar to the BEC window.  At densities below
$10^{11}$ cm$^{-3}$, thermalization is extremely slow, and
evaporative cooling can no longer compete with (technical) sources
of heating and loss.  At densities above $10^{15}$ cm$^{-3}$, three
body losses usually become dominant.  In this density window,
degeneracy is achieved at temperatures between 100 nK and 50 $\mu$K.

The cooling and trapping techniques to reach such low temperatures
are the same as those that have been developed for Bose-Einstein
condensates. We refer to our Varenna paper on BEC~\cite{kett99var}
for a description of these techniques. Table~\ref{t:cooldown}
summarizes the different cooling stages used at MIT to reach
fermionic superfluidity in dilute gases, starting with a hot
atomic beam at $450\,^\circ$C and ending with a superfluid cloud
of 10 million fermion pairs at 50 nK.

Although no major new technique has been developed for fermionic
atoms, the nature of fermionic gases emphasizes various aspects of
the experimental methods:

\begin{itemize}
    \item Different atomic species.  The most popular atoms for BEC, Rb and Na, do not have any stable fermionic isotopes.  The workhorses in the field of ultracold fermions are $^{40}$K and $^6$Li.
    \item Sympathetic cooling with a different species (Na, Rb, $^7$Li).  This
     requires techniques to load and laser cool two different kinds of atoms
     simultaneously, and raises the question of collisional
     stability.
     \item All optical cooling.  When cooling $^6$Li, the need for a different
     species can be avoided by all optical cooling using two
     different hyperfine states.  This required further development of optical traps with large trap depth.
    \item Two-component fermionic systems.  Pairing and superfluidity is observed in a two-component fermionic system equivalent to spin up and spin down.  This raises issues of
    preparation using radiofrequency (RF) techniques, collisional stability, and detection
    of different species.  All these challenges were already encountered in spinor
    BECs, but their solutions have now been further developed.
    \item Extensive use of Feshbach resonances.  Feshbach resonances
    were first observed and used in BECs.  For Fermi gases,  resonantly enhanced interactions were crucial to achieve
    superfluidity.  This triggered developments in rapid switching
    and sweeping of magnetic fields across Feshbach resonances, and in generating homogeneous fields for
    ballistic expansion at high magnetic fields.
    \item Lower temperatures.  On the BCS side of the phase
    diagram, the critical temperature decreases exponentially with the interaction strength between
    the particles. This provides additional motivation to cool far
    below the degeneracy temperature.
\end{itemize}

In this chapter, we discuss most of these points in detail.

\begin{table}
  \centering
        \begin{tabular}{p{0.1\linewidth}c|c|c|cp{0.4\linewidth}}
&Stage                 & Temperature   & Density & $T / T_F$&\\
\hline
&  & & &   &\\
&Two-species oven      &   720 K       &  $10^{14} \,\rm cm^{-3}$    &  $10^8$&\\
&  & & &   &\\
&Laser cooling         & 1 mK          &  $10^{10} \,\rm cm^{-3}$    & $10^4$&\\
&({\it Zeeman slower {\rm \&} MOT})  & & &   &\\
&Sympathetic cooling   & 1 $\mu$K     & $ 10^{13} \,\rm cm^{-3}$ & $0.3$&\\
&({\it Magnetic trap})     & & &  &\\
&Evaporative cooling   & 50 nK        & $ 5\cdot 10^{12} \,\rm cm^{-3}$ & $0.05$&\\
&({\it Optical trap})   & & & &\\
        \end{tabular}
  \caption{The various preparatory stages towards a superfluid Fermi gas in the MIT experiment. Through a combination of laser cooling, sympathetic cooling with sodium atoms, and evaporative cooling, the temperature is reduced by 10 orders of magnitude. The first steps involve a spin-polarized gas. In the last step, strong attractive interactions are induced in a two-state Fermi mixture via a Feshbach resonance. This brings the critical temperature for superfluidity up to about $0.3\, T_F$ - the ultracold Fermi gas becomes superfluid.}\label{t:cooldown}
\end{table}

\subsection{The atoms}

At very low temperatures, all elements turn into solids, with the
exception of helium which remains a liquid even at zero
temperature. For this reason, $^3$He had been the only known
neutral fermionic superfluid before the advent of laser cooling.
Laser cooling and evaporative cooling prepare atomic clouds at
very low densities, which are in a metastable gaseous phase for a
time long enough to allow the formation of superfluids.

Neutral fermionic atoms have an odd number of neutrons. Since
nuclei with an even number of neutrons are more stable, they are
more abundant.  With the exception of beryllium each element has
at least one isotope, which as a neutral atom is a boson.
However, there are still many choices for fermionic atoms
throughout the periodic table. Because alkali atoms have a simple
electronic structure and low lying excited states, they are ideal
systems for laser cooling.  Among the alkali metals, there are two
stable fermionic isotopes, $^6$Li and $^{40}$K, and they
have become the main workhorses in the field. Recently, degenerate
Fermi gases have been produced in metastable
$^3$He$^*$~\cite{mcna06helium} and
Ytterbium~\cite{fuku07ytterbium}, and experiments are underway in
Innsbruck to reach degeneracy in strontium.

\subsubsection{Hyperfine structure}
\label{s:hyperfinestructure}

\begin{figure}[t]
\begin{center}
\includegraphics[width=5in]{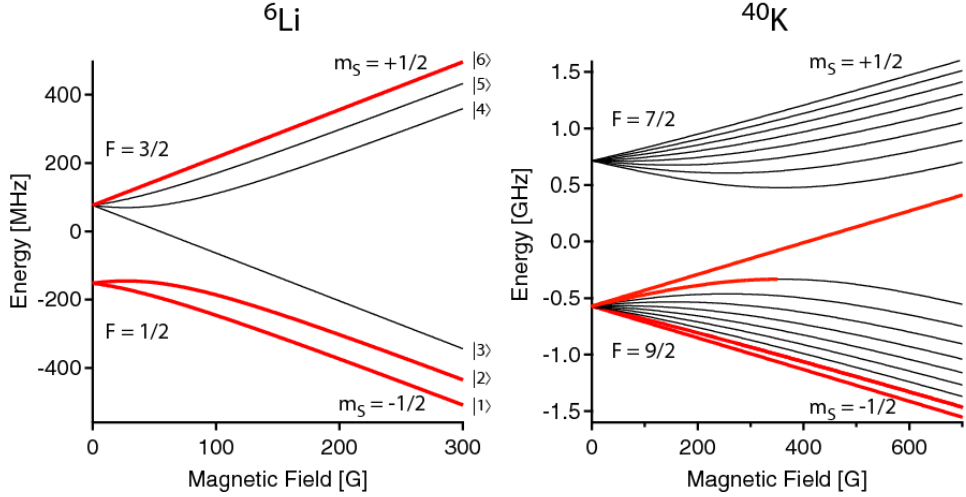}
\caption[Hyperfine states of $^6$Li and $^{40}$K]{Hyperfine states
of $^6$Li and $^{40}$K. Energies are relative to the atomic ground
state without hyperfine interaction. $^6$Li has nuclear spin $I =
1$, for $^{40}$K it is $I=4$. The $^6$Li hyperfine splitting is
$\Delta \nu_{\rm hf}^{\rm ^6Li} = 228$ MHz, for $^{40}$K it is
$\Delta \nu_{\rm hf}^{\rm ^{40}K} = -1.286$ GHz. The minus sign
indicates that the hyperfine structure is reversed for $^{40}$K,
with $F = 9/2$ being lower in energy than $F = 7/2$. Thick lines
mark hyperfine states used during cooling to degeneracy.}\label{f:breitrabi}
\end{center}
\end{figure}

Pairing in fermions involves two hyperfine states, and the choice
of states determines the collisional stability of the gas, e.g.
whether there is a possible pathway for inelastic decay to
lower-lying hyperfine states.  Therefore, we briefly introduce the
hyperfine structure of $^6$Li and $^{40}$K.

The electronic ground state of atoms is split by the hyperfine
interaction. The electrons create a magnetic field that interacts
with the nuclear spin $\vect{I}$. As a result, the total electron
angular momentum, sum of angular momentum and spin, $\vect{J} =
\vect{L} + \vect{S}$, is coupled to the nuclear spin to form the
total angular momentum of the entire atom, $\vect{F} = \vect{J} +
\vect{I}$. Alkali atoms have a single valence electron, so $S
= 1/2$, and in the electron's orbital ground state, $\vect{L} = 0$. Hence the
ground state splits into two hyperfine manifolds with total
angular momentum quantum numbers $F = I + 1/2$ and $F = I - 1/2$.
In a magnetic field $\vect{B}$, these hyperfine states split again
into a total of $(2S + 1)(2I+1) = 4I + 2$ states. The hamiltonian
describing the various hyperfine states is
\begin{equation}
    H_{\rm hf} = a_{\rm hf} \vect{I}\cdot\vect{S} + g_s \mu_B \vect{B}\cdot\vect{S} - g_i \mu_N \vect{B} \cdot \vect{I}
\end{equation}
Here, $a_{\rm hf}$ is the hyperfine constant of the alkali atom,
$g_s \approx 2$ and $g_i$ are the electron and nuclear $g$-factors,
$\mu_B \approx 1.4 \, \rm MHz/G$ is the Bohr magneton and $\mu_N$
the nuclear magneton.

The hyperfine states of $^6$Li and $^{40}$K are shown in
Fig.~\ref{f:breitrabi}. Good quantum numbers at low field are the
total spin $F$ and its $z$-projection, $m_F$. At high fields $B
\gg a_{\rm hf}/\mu_B$, they are the electronic and nuclear spin
projections $m_S$ and $m_I$.

\subsubsection{Collisional Properties}
\label{s:collprop}

The Pauli exclusion principle strongly suppresses collisions
between two fermions in the same hyperfine state at low
temperatures.  Because of the antisymmetry of the total
wave function for the two fermions, $s$-wave  collisions are forbidden.  Atoms in the same hyperfine state can
collide only in odd partial waves with $p$-wave as the lowest
angular momentum channel.

For p-wave collisions, with the relative angular momentum of
$\hbar$, atomic mass $m$ and a thermal velocity of $v_T$, the
impact parameter of a collision is $\hbar/m v_T$, which is equal
to the thermal de Broglie wavelength $\lambda_T=
\sqrt{\frac{2\pi\hbar^2}{m k_B T}}$. When the range of the
interaction potential $r_0$ is smaller than $\lambda_T$, the atoms
``fly by'' each other without interaction. For a van-der-Waals
potential, the range is $r_0 \approx (m C_6/\hbar^2)^{1/4}$. Below
the temperature $k_B T_p = \hbar^2/m r_0^2$, $p$-wave scattering
freezes out, and the Fermi gas becomes cohesionless, a truly ideal
gas!  For $^6$Li, $T_p \approx 6 \,\rm mK$, much larger than the
temperature in the magneto-optical trap (MOT). For $^{40}$K, $T_p
= 300 \,\rm \mu K$. Since these values for $T_p$ are much higher
than the window for quantum degeneracy, a second species or second
hyperfine state is needed for thermalization and evaporative
cooling. We now discuss some general rules for inelastic two-body
collisions.

\begin{itemize}
    \item Energy.  Inelastic collisions require the internal
    energy of the final states to be lower than that of the initial
    states.  Therefore, a gas (of bosons or fermions) in the
    lowest hyperfine state is always stable with respect to two-body collisions.  Since the lowest
    hyperfine state is a strong magnetic field seeking state, optical traps, or generally traps using ac magnetic or electric fields are required for confinement.
    \item Angular momentum.  The $z$-component $M$ of the total angular
momentum of the two colliding atoms (1 and 2) is conserved. Here,
$M=M_{int} + M_{rot}$, where the internal angular momentum
$M_{int}=m_{F,1}+m_{F,2}$ at low fields and  $M_{int}=
m_{I,1}+m_{I,2} + m_{S,1}+m_{S,2}$ at high fields, and $M_{rot}$
is the $z$-component of the angular momentum of the atom's
relative motion.
     \item Spin relaxation.  Spin relaxation occurs when an inelastic collision
     is possible by exchanging angular momentum between electrons and nuclei, without affecting  the motional angular momentum.
      Usually, the rate constant for this process is on the order of $10^{-11}\, \rm{cm}^3 \rm{s}^{-1}$ which implies
       rapid decay on a ms scale for typical densities.  As a general rule, mixtures of hyperfine states with allowed spin relaxation
       have to be avoided. An important exception is $^{87}$Rb where spin relaxation is suppressed by about three orders of magnitude by quantum interference~\cite{myat97}).  Spin relaxation is
       suppressed if there is no pair of states with lower internal energy
       with the same total $M_{int}$.  Therefore, degenerate gases in a state with  maximum $M_{int}$ cannot undergo
       spin relaxation.
    \item Dipolar relaxation.  In dipolar relaxation, angular momentum is transferred from the electrons and/or nuclei to the atoms' relative motion.
     Usually, the rate constant for this process is on the order
of $10^{-15}\, \rm{cm}^3 \rm{s}^{-1}$ and is sufficiently slow
(seconds) to allow the study of systems undergoing dipolar
relaxation. For instance, all magnetically trapped Bose-Einstein
condensates can decay by dipolar relaxation, when the spin flips
to a lower lying state.

 \item Feshbach resonances.
Near Feshbach resonances, all inelastic processes are usually
strongly enhanced.  A Feshbach resonance enhances the wave function
of the two colliding atoms at short distances, where inelastic
processes occur (see ~\ref{c:feshbach}). In addition, the coupling
to the Feshbach molecule may induce losses that are entirely due
to the closed channel. It is possible that the two enhanced
amplitudes for the same loss process interfere destructively.

\item (Anti-)Symmetry.
    At low temperature, we usually have to consider only atoms colliding in the $s$-wave incoming channel. Colliding fermions then have to be in two different
    hyperfine states, to form an antisymmetric total wave function. Spin relaxation is not changing the
    relative motion.  Therefore, for fermions, spin relaxation into a pair of
    identical states is not possible, as this would lead to a symmetric wave function.  Two identical final states are also forbidden for ultracold fermions undergoing dipolar relaxation, since
    dipolar relaxation obeys the selection rule $\Delta L=0, 2$ for
    the motional angular momentum and can therefore only connect
    even to even and odd to odd partial waves.
\end{itemize}

We can now apply these rules to the hyperfine states of alkalis.
For magnetic trapping, we search for a stable pair of magnetically
trappable states (weak field seekers, i.e. states with a positive
slope in Fig.~\ref{f:breitrabi}). For atoms with $J = 1/2$ and
nuclear spin $I = 1/2$, 1 or 3/2 that have a normal hyperfine
structure (i.e. the upper manifold has the larger $F$), there is
only one such state available in the lower hyperfine manifold. The
partner state thus has to be in the upper manifold. However, a
two-state mixture is not stable against spin relaxation when it
involves a state in the upper hyperfine manifold, and there is a
state leading to the same $M_{int}$ in the lower manifold.
Therefore, $^6$Li (see Fig.~\ref{f:breitrabi}) and also $^{23}$Na
and $^{87}$Rb do not have a stable pair of magnetically trappable
states. However, $^{40}$K has an inverted hyperfine structure and
also a nuclear spin of 4. It thus offers several combinations of
weak-field seeking states that are stable against spin relaxation.
Therefore, $^{40}$K has the unique property that evaporative
cooling of a two-state mixture is possible in a magnetic trap,
which historically was the fastest route to achieve fermionic
quantum degeneracy~\cite{dema99}.

An optical trap can confine both weak and strong field seekers.
Mixtures of the two lowest states are always stable against spin
relaxation, and in the case of fermions, also against dipolar
relaxation since the only allowed output channel has both atoms in
the same state.  Very recently, the MIT group has realized
superfluidity in $^6$Li using mixtures of the first and third or
the second and third state~\cite{schu07tbp}.  For the
combination of the first and third state, spin relaxation into the
second state is Pauli suppressed.  These two combinations can
decay only by dipolar relaxation, and surprisingly, even near
Feshbach resonances, the relaxation rate remained small. This might be caused by the small hyperfine energy, the small mass and the small van der Waals coefficient $C_6$ of $^6$Li, which lead to a small
release energy and a large centrifugal barrier in the $d$-wave
exit channel.

For Bose-Einstein condensates at typical densities of
$10^{14}$ cm$^{-3}$ or larger, the dominant decay is three-body
recombination. Fortunately, this process is Pauli suppressed for
any two-component mixture of fermions, since the probability to encounter three fermions in a small volume, of the size of the molecular state formed by recombination, is very small.  In
contrast, three-body relaxation is not suppressed if the molecular
state has a size comparable to the Fermi wavelength.  This has
been used to produce molecular clouds (see section~\ref{s:makingmolecules}).

After those general considerations, we turn back to the
experimentally most relevant hyperfine states, which are marked
with thick (red) lines in Fig.~\ref{f:breitrabi}.  In the MIT experiment, sympathetic
cooling of lithium with sodium atoms in the magnetic trap is
performed in the upper, stretched state $\left|6\right> \equiv
\left|F = 3/2,m_F = 3/2\right>$. In the final stage of the
experiment, the gas is transferred into an optical trap and
prepared in the two lowest hyperfine states of $^6$Li, labelled
$\left|1\right>$ and $\left|2\right>$, to form a strongly
interacting Fermi mixture around the Feshbach resonance at 834 G.
The same two states have been used in all $^6$Li experiments
except for the very recent MIT experiments on mixtures between
atoms in $\left|1\right>$ and $\left|3\right>$, as well as in
$\left|2\right>$ and $\left|3\right>$ states.  In experiments on
$^{40}$K at JILA, mutual sympathetic cooling of the $\left|F =
9/2,m_F = 9/2\right>$ and $\left|F = 9/2, m_F = 7/2\right>$ states
is performed in the magnetic trap. The strongly interacting Fermi
mixture is formed using the lowest two hyperfine states
$\left|F=9/2,m_F=-9/2\right>$ and $\left|F=9/2,m_F=-7/2\right>$
close to a Feshbach resonance at 202 G.

As we discussed above, evaporative cooling requires collisions
with an atom in a different hyperfine state or with a different
species. For the latter approach, favorable properties for
interspecies collisions are required.  Here we briefly summarize
the approaches realized thus far.

The stability of mixtures of two hyperfine states has been
discussed above.  Evaporation in such a system was done for
$^{40}$K in a magnetic trap~\cite{dema99} using RF-induced,
simultaneous evaporation of both spin states.  In the case of
$^6$Li, laser cooled atoms were directly loaded into optical traps
at Duke~\cite{gran02} and Innsbruck~\cite{joch03lith} in which a
mixture of the lowest two hyperfine states was evaporatively
cooled by lowering the laser intensity.  Other experiments used
two species. At the ENS~\cite{schr01} and at Rice~\cite{trus01},
spin-polarized $^6$Li is sympathetically cooled with the bosonic
isotope of lithium, $^7$Li, in a magnetic trap. At MIT, a
different element is used as a coolant, $^{23}$Na. This approach
is more complex, requiring a special double-species oven and two
laser systems operating in two different spectral regions (yellow
and red).  However, the $^6$Li-$^{23}$Na interspecies collisional
properties have turned out to be so favorable that this experiment
has led to the largest degenerate Fermi mixtures to date with up
to 50 million degenerate fermions~\cite{hadz03big_fermi}. Forced
evaporation is selectively done on $^{23}$Na alone, by using a
hyperfine state changing transition around the $^{23}$Na hyperfine
splitting of $1.77 \;\rm GHz$. The number of $^6$Li atoms is
practically constant during sympathetic cooling with sodium. Other
experiments on sympathetic cooling employ $^{87}$Rb as a coolant
for $^{40}$K ~\cite{inou04, kohl05fermilattice, ospe06hetero,
rom06} or for $^6$Li~\cite{silb05,tagl08fermifermi}.

Another crucial aspect of collisions is the possibility to enhance
elastic interactions via Feshbach resonances. Fortunately, for all
atomic gases studied so far, Feshbach resonances of a reasonable
width have been found at magnetic fields around or below one
kilogauss, rather straightforward to produce in experiments. Since
Feshbach  resonances are of central importance for fermionic
superfluidity, we discuss them in a separate chapter (\ref{c:feshbach}).

\subsection{Cooling and trapping techniques}
The techniques of laser cooling and magnetic trapping are
identical to those used for bosonic atoms.  We refer to the
comprehensive discussion and references in our earlier Varenna
notes~\cite{kett99var} and comment only on recent advances.

One development are experiments with two atomic species in order
to perform sympathetic cooling in a magnetic trap.  An important
technical innovation are two-species ovens which create atomic
beams of two different species.  The flux of each species can be
separately controlled using a two-chamber oven design~\cite{stan05oven}.  When
magneto-optical traps (MOTs) are operated simultaneously with two
species, some attention has to be given to light-induced interspecies collisions leading to trap loss.  Usually, the number of trapped atoms for each species after full loading is smaller than if the MOT is operated with
only one species.  These losses can be mitigated by using
sequential loading processes, quickly loading the second species,
or by deliberately applying an intensity imbalance between
counter-propagating beams in order to displace the two trapped
clouds~\cite{hadz03big_fermi}.

Another development is the so-called all-optical cooling, where
laser cooled atoms are directly transferred into an optical trap
for further evaporative cooling.  This is done by ramping down the
laser intensity in one or several of the beams forming the optical
trap.  All-optical cooling was introduced for bosonic atoms
(rubidium~\cite{barr01}, cesium~\cite{webe03},
sodium~\cite{dumk06na_all_optical}, ytterbium~\cite{taka03}) and
is especially popular for fermionic lithium, where evaporative
cooling in a magnetic trap is possible only by sympathetic cooling
with a second species.

In the following two sections, we address in more detail issues of
sympathetic cooling and new variants of optical traps, both of
relevance for cooling and confining fermions.

\subsubsection{Sympathetic cooling}

\emph{Overlap  between the two clouds.}  One limit to sympathetic
cooling is the loss of overlap of the coolant with the cloud of
fermions. Due to different masses, the sag due to gravity is
different for the two species. This is most severe in experiments
that employ $^{87}$Rb to cool
$^{6}$Li~\cite{silb05,tagl08fermifermi}. For harmonic traps, the
sag is given by $\Delta x_{1,2} = g/\omega_{1,2}^2$ for species 1
and 2, with $g$ the earth's gravitational acceleration, and
$\omega$ the trapping frequency along the vertical direction. The
spring constant $k = m \omega^2\approx \mu_B B''$ is essentially the same for all
alkali atoms, when spin-polarized in their stretched state and
confined in magnetic traps with magnetic field curvature $B''$. It is of the same order for alkalis confined in optical traps, $k = \alpha I''$, where the polarizability $\alpha$ is similar for
the alkalis and lasers far detuned from atomic resonances, and $I''$ is the curvature of the
electric field's intensity.  The thermal cloud size, given by
$\sqrt{k_B T/k}$, is thus species-independent, while the sag
$\Delta x_{1,2} \approx g m_{1,2} / k$ is proportional to the
mass. The coolant separates from the cloud of fermions once $g
(m_2 - m_1)/k \approx \sqrt{k_B T/k}$, or $k_B T \approx g^2 (m_2
- m_1)^2/k$. For trapping frequencies of 100 Hz for $^6$Li, and
for $^{87}$Rb as the coolant, this would make sympathetic cooling
inefficient at temperatures below 30 $\mu$K, more than an order of
magnitude higher than the Fermi temperature for 10 million
fermions. For $^{23}$Na as the coolant, the degenerate regime is
within reach for this confinement. Using the bosonic isotope
$^7$Li as the coolant, gravitational sag evidently does not play a
role. To avoid the problem of sag, one should provide strong
confinement along the axis of gravity. A tight overall confinement
is not desirable since it would enhance trap loss due to
three-body collisions.

\emph{Role of Fermi statistics.} When fermions become degenerate,
the collision rate slows down.  The reason is that scattering into
a low-lying momentum state requires this state to be empty, which
has a probability $1 - f$, with $f$ the Fermi-Dirac occupation
number. As the occupation of states below the Fermi energy gets
close to unity at temperatures $T \ll T_F$, the collision rate is
reduced. Initially, this effect was assumed to severely limit
cooling well below the Fermi temperature~\cite{dema99}. However,
it was soon realized that although the onset of Fermi degeneracy
changes the kinetics of evaporative cooling, it does not impede
cooling well below the Fermi temperature~\cite{holl00_evap_fermi,
geis02}.  The lowest temperature in evaporative cooling is always
determined by heating and losses.  For degenerate Fermi systems,
particle losses (e.g. by background gas collisions) are more
detrimental than for Bose gases, since they can create hole
excitations deep in the Fermi sea~\cite{timm01fermi-heat, carr04,
idzi05}.

\emph{Role of Bose statistics.} If the coolant is a boson, the
onset of Bose-Einstein condensation changes the kinetics of
evaporation.
It has been proposed that sympathetic cooling becomes highly
inefficient when the specific heat of the coolant becomes equal or
smaller than that of the Fermi system~\cite{trus01, wout02}. However,
although an almost pure Bose-Einstein condensate has almost zero
specific heat, its capacity to remove energy by evaporating out of
a trap with a given depth is even larger than that of a Boltzmann
gas, since the initial energy of the Bose gas is lower. On the
other hand, the rate of evaporation is lower for the Bose
condensed gas, since the number of thermal particles is greatly
reduced.  In the presence of heating, a minimum rate of
evaporation is required~\cite{wout02}. This might call for
additional flexibility to independently control the confinement
for bosons and fermions, which is possible via the use of a
two-color trap~\cite{onof02}. In particular, on can then expand
the bosonic coolant and suppress the onset of Bose-Einstein
condensation.

Other work discussed phenomena related to the interacting
condensate.  When the Fermi velocity becomes smaller than the
critical velocity of a superfluid Bose-Einstein condensate, then
the collisional transfer of energy between the fermions and bosons
becomes inefficient~\cite{timm98super}. Another phenomenon for
sufficiently high boson density is mean-field attraction or
repulsion of the fermions, depending on the relative sign of the
intraspecies scattering length~\cite{molm98}. Attractive
interactions can even lead to a collapse of the condensate as too
many fermions rush into the Bose cloud and cause three-body
collisions, leading to losses and heating~\cite{modu02, ospe05}.

Given all these considerations, it is remarkable that the simplest
scheme of evaporating bosons in a magnetic trap in the presence of
fermions has worked very well.  In the MIT experiment, we are
currently limited by the number of bosons used to cool the Fermi
gas. Without payload (the fermions), we can create a sodium
Bose-Einstein condensate of 10 million atoms. When the
fermions outnumber the bosons, the cooling becomes less efficient,
and we observe a trade-off between final number of fermions and
their temperature.  We can achieve a deeply degenerate Fermi gas
of $T/T_F = 0.05$ with up to 30 million fermions~\cite{hadz03big_fermi}, or aim for even larger atom numbers at the cost of
degeneracy. On a day-to-day basis, we achieve 50 million fermions
at $T/T_F = 0.3$. This degenerate, spin-polarized Fermi gas can
subsequently be loaded into an optical trap for further
evaporative cooling as a two-component Fermi mixture.

The preparation of a two-component mixture by an RF pulse and
decoherence (see section~\ref{s:decoherence}) lowers the maximum occupation number to 1/2 and
increases the effective $T/T_F$ to about 0.6.  Therefore, there is no benefit of cooling the
spin polarized Fermi cloud to higher degeneracy.

\subsubsection{Optical trapping}
\label{s:opticaltrap}
Optical traps provide the confinement for almost all experiments
on ultracold fermions.  The reason is that most of the current
interest is on interacting two-component systems.  Optical traps
confine both strong- and weak-field seeking states.  Trapping
atoms in the lowest lying hyperfine states (which are always
strong-field seeking) suppresses or avoids inelastic collisions,
as discussed in section~\ref{s:collprop}.

Most importantly, using electric fields for trapping frees the
magnetic field to be tuned to Feshbach resonances and thereby to
enhance elastic interactions.  There is only one experiment on
ultracold atoms that studied Feshbach resonances in a magnetic
trap (in $^{85}$Rb~\cite{corn00JILA_collapse}), all others have
been performed in optical traps.

The important case of $^6$Li has led to advances in optical traps
with large volume and trap depth. All-optical cooling to BEC has
been convenient in some experiments with rubidium and sodium
Bose-Einstein condensates eliminating the need for magnetic
traps~\cite{barr01, dumk06na_all_optical}. However, standard
magnetic traps are not considerably increasing the complexity of
the experiment. One could take the position that a magnetic trap
is easier to operate and to maintain than a high power laser or an
enhancement cavity. However, bypassing the magnetic trap for
$^6$Li also bypasses the need for another species ($^7$Li, Rb, Na)
and therefore an additional laser system\footnote{In other cases
magnetic trapping has not been an option due to inelastic
collisions~\cite{webe03} or vanishing magnetic
moment~\cite{taka03}.}.

In the following section, we discuss some optical trapping
geometries used in ultracold Fermi experiments. For a more
detailed discussion on optical trapping, we refer the reader to~\cite{kett99var} and~\cite{grim00}.

So far, all optical traps for fermions have used red detuned laser
beams where the atoms are confined in the intensity maximum of the
laser beam(s).  The trapping potential is given by the AC Stark shift
\begin{equation}
    U(\vect{r}) = - \frac{\hbar \omega_R^2(\vect{r})}{4} \left(\frac{1}{\omega_0 - \omega_L} + \frac{1}{\omega_0 + \omega_L}\right) \simeq \frac{\hbar \omega_R^2(\vect{r})}{4\Delta}
\label{e:dipolepotential}
\end{equation}
where $\omega_0$ is the atomic resonance frequency, $\omega_L$ is
the frequency of the laser light, and $\Delta = \omega_L -
\omega_0$ the laser's detuning from resonance. The approximation
on the right hand side holds for $|\Delta| \ll \omega_0$.
$\omega_R$ is the position-dependent Rabi frequency describing the
strength of the atom-field coupling. In terms of the intensity
$I(\vect{r})$ of the laser light and atomic parameters, it is
defined by $2\omega_R^2(\vect{r})/\Gamma^2 = I(\vect{r})/I_{\rm
SAT}$, where $\Gamma$ is the natural decay rate of the atom's
excited state, and $I_{\rm SAT} = \hbar \omega_0^3 \Gamma / 12 \pi
c^2$ is the saturation intensity.
For $^6$Li, $\Gamma = 2\pi\cdot 6 \,\rm MHz$ and $I_{\rm SAT} = 3
\,\rm mW/cm^2$, for $^{40}$K, $\Gamma = 2 \pi\cdot 6 \,\rm MHz$
and $I_{\rm SAT} = 2 \,\rm mW / cm^2$.

\paragraph{Single-beam optical trap}

The simplest trap consists of a single,
red-detuned, focused gaussian laser beam, with intensity profile
\begin{equation}
I(\rho,z) = \frac{2P}{\pi {\rm w}^2 \left(1 + z^2/z_R^2\right)}
\exp\left(- \frac{2\, \rho^2}{{\rm w}^2 \left(1 +
z^2/z_R^2\right)}\right).
\end{equation}
The beam parameters are the laser power $P$, the $1/e^2$ beam
waist radius w, and the Rayleigh range $z_R$.  $\rho$ and $z$ are
the distances from the beam focus along the radial and axial
directions, respectively. The Rayleigh range is related to the
beam waist and the wavelength of the laser via $z_R = \pi {\rm
w}^2/\lambda$. The bottom of the potential well formed by the
laser beam can be approximated as a harmonic oscillator with
trapping frequencies $\omega_\rho/2\pi = \sqrt{2P/\pi^3 m {\rm
w}^4}$ and $\omega_z/2\pi = \sqrt{P/\pi^3 m {\rm w}^2 z_R^2}$. For
$^6$Li, a laser beam operating at a wavelength $\lambda = 1064$ nm
with 100 mW of power, focused down to a waist of ${\rm w} = 25
\,\mu \rm m$, provides a trap depth $U \simeq 6\,\mu \rm K$, a
radial frequency $\omega_\rho/2\pi = 1.2 \,{\rm kHz}$ and an
aspect ratio $\omega_\rho/\omega_z = \sqrt{2}\pi {\rm w}/\lambda
\simeq 100$. This is sufficient for loading atoms that were
evaporatively or sympathetically pre-cooled in a magnetic trap.

Loading atoms directly from a millimeter-sized MOT, at
temperatures of several $100\,\mu\rm K$, into a single beam
optical trap requires larger trap depths, a larger waist and
ideally a smaller aspect ratio to enhance the overlap with the
rather spherical MOT region. One solution is the use of
Quasi-Electrostatic Traps (QUEST)~\cite{take95cs} formed by a
focused $\rm CO_2$ laser at $\lambda = 10.6 \,\mu$m. Due to the
large detuning from atomic resonance the trap operates in the
quasi-electrostatic regime where $\omega_L/\omega_0 \rightarrow 0$
and the dipole potential $U = \frac{\hbar \Gamma^2}{4\omega_0}
\frac{I}{I_{\rm SAT}}$ no longer depends on the frequency of the
laser light. The longer wavelength allows for a larger waist at
still moderate aspect ratios. In the group at Duke
University~\cite{gran02}, 65 W of power was focused to a
waist of
47 $\mu\rm m$ (Rayleigh range 660 $\mu\rm m$), providing a
trap
depth for $^6$Li atoms of 690 $\mu$K. The resulting
radial and
axial frequencies were 6.6 kHz and 340 Hz,
respectively. This deep
trap allowed to capture $1.5 \times 10^6$
atoms from the MOT at
Doppler-limited temperatures of $150\, \mu$K. The tight
confinement ensured good starting conditions for
evaporative
cooling.

\paragraph{Hybrid trap}

A large beam waist is preferable for several purposes, e.g. for
creating a large trap volume or for controlling any aberrations
which would cause a deviation from cylindrical symmetry --- this
was crucial for the creation of vortices~\cite{zwie05vort}.  To
avoid the large aspect ratio of the optical trap, a confining
magnetic curvature can be added along the axial direction by using
two coils with a separation larger (smaller) than the Helmholtz
configuration (distance equals radius) for low field (high field) seekers.  Maxwell's equations then
require an anti-confining curvature along the radial direction,
which, however, is negligible compared to the tight optical
confinement.  As a result, this hybrid trap features optical
radial confinement and axial magnetic confinement. In addition,
high bias fields are needed to tune across the Feshbach resonance.
Such a setup has been used in many experiments in Innsbruck, at
Rice, and at MIT.   Details of the MIT magnetic field
configuration are discussed in section~\ref{s:highfields}. In our
experiments, the axial confinement is almost purely magnetic
($\omega_z/2\pi \simeq 23\,\rm Hz$). The optical trap provides radial
confinement with $\omega_r/2\pi$ in the range of 50 to 300 Hz, which
varies the aspect ratio of the cloud between about 2 and 12.

We will now discuss two other important aspects of optical traps.
One is the compensation of gravity that is crucial for creating
traps with cylindrical symmetry, the other one is the issue of
the trap depth that controls evaporative cooling.

In the MIT experiment~\cite{zwie05vort}, the hybrid trap has a typical aspect
ratio of $\omega_r/\omega_z = 6$. The optical trapping beam and
the magnetic field coils are horizontally aligned. Compensation
for gravity is ensured by ``sitting on one side of the saddle''.

Along the vertical ($x$-)axis, the combined potential of
gravity+magnetic fields is $-\frac{1}{4} m \omega_z^2 x^2 - m g
x$, where we used $i \omega_z/\sqrt{2}$ as the anti-confining curvature.   Thus, gravity
shifts the saddle potential by an amount $2 g /
\omega_z^2 \approx
1 \,\rm mm$. The ``sweet spot'' in the radial
plane to which the
optical trap needs to be aligned is thus not
the center of the
magnetic field coils, but about 1 mm above it. In this position no
gradients act
on the atoms. If the optical trap is round in the
radial plane,
the combined potential experienced by the atoms is
round as well.
Round traps are crucial for the observation of
vortices, and also
for the study of collective excitations with
radial symmetry.

\begin{figure}[t]
\begin{center}
\includegraphics[width=4in]{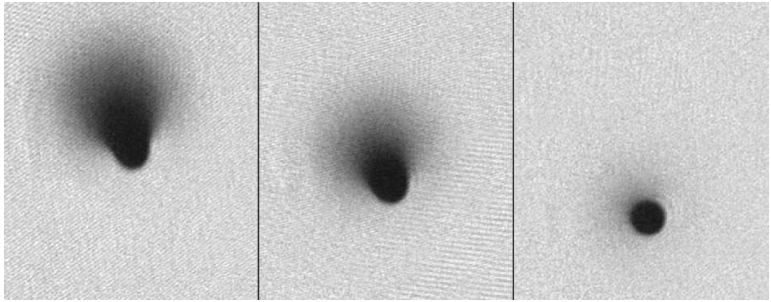}
\caption[Alignment of the optical trap to achieve a radially
symmetric potential]{Alignment of the optical trap to achieve a
radially symmetric potential. In the left image, the trap is still
far from the ``sweet spot''. In the right image, stray gradients are
almost completely cancelled. The absorption images are of a
lithium pair condensate after 10~ms of expansion. The field of
view for each image is 1.1 $\times$ 1.3 mm.}\label{f:odtalign}
\end{center}
\end{figure}

The alignment procedure of the optical trap is shown in
Fig.~\ref{f:odtalign}. At the end of evaporation of the lithium
condensate, the trap depth is reduced in about 30 ms to a very
shallow depth which is not sufficient to hold the atoms if they
are not in the ``sweet spot''. After 10 ms of expansion from the
optical trap one clearly observes in which direction the atoms
spill out, and one can counteract by moving the optical trap.

A low intensity tightly focused beam and a larger intensity beam
with a softer focus provide the same radial confinement.  However,
the trap depth is very different.  This is important if the cloud
needs to be cooled by evaporation, e.g. during the nucleation of a
vortex lattice after stirring up the cloud.  Cooling of the cloud
will be efficient if the trap depth $U$ is not much higher than
the Fermi energy $E_F$. This condition sets a stringent constraint
for the beam waist.

We illustrate this by discussing the situation in the MIT vortex
experiment, where we wanted to have a rather small aspect ratio $a
= \omega_r / \omega_z$.  The axial trapping frequency $\omega_z$
was fixed
by the magnetic field curvature.
The relation between
$U$ and the waist w is
\begin{equation}
    U = \frac{1}{4} m \omega_r^2 {\rm w}^2 = \frac{1}{4} m \omega_z^2 a^2 {\rm w}^2
\end{equation}
The Fermi energy per spin state for a total number of atoms $N$ is
given by (using the harmonic approximation for the radially
gaussian potential):
\begin{equation}
    E_F = \hbar (\omega_r^2 \omega_z)^{1/3} (3 N)^{1/3} = \hbar \omega_z a^{2/3} (3 N)^{1/3}
\end{equation}
Requiring $U \gtrsim E_F$ results in
\begin{equation}
    \label{e:waist}
    {\rm w} \gtrsim 2 \sqrt{\frac{\hbar}{m \omega_z}} a^{-2/3} (3N)^{1/6}
\end{equation}
If we want to trap $N = 1 \times 10^7$ atoms with an aspect ratio
$a=5$ and an axial trapping frequency $\omega_z/2\pi = 20\,\rm Hz$
(typical values), we need the waist to be larger than 100 $\mu$m.
Note that this requirement is quite stringent. Changing $\omega_z$
is limited: Increasing the current in the curvature coils by a
factor of two (which increases power dissipation in the curvature
coils by four) only reduces the required waist by 20\%. Allowing
for an aspect ratio of 10 would give another reduction by only
35\%.  A longer aspect ratio would have had adverse effects for
the alignment of the stirring beam and the observation of vortex
cores after expansion.

For our choice of ${\rm w} = 120 \,\mu$m, the Rayleigh range is
$z_0 = \pi {\rm w}^2 / \lambda > 4 \,\rm cm$ while a typical axial
cloud size is 1 mm.   The maximum power in the laser beam is $4$
W, which limits the trap depth to about 10 $\mu$K. This is still
deep enough to load about $3 \times 10^7$ degenerate fermions from
the magnetic trap after the sympathetic cooling stage with sodium
(The Fermi temperature in the combined magnetic and optical trap
during this loading is 5 $\mu$K, and the degenerate cloud at
$T/T_F \approx 0.3$ is not much larger than a zero-temperature
Fermi sea).  These numbers illustrate that optical traps for
fermions need much more power than for a Bose-Einstein condensate
because of the combined need for a deeper and larger trap.

\paragraph{Crossed dipole trap}
Another option for loading atoms from a MOT into an optical
potential is the use of crossed laser beams. This geometry
provides a roughly spherical trapping volume, and offers a good
trade-off between trap depth and volume. This configuration
allowed the first demonstration of Bose-Einstein condensation of
atoms by all-optical means~\cite{barr01}. Fermionic atoms were
loaded into a crossed dipole trap  by the Paris
group~\cite{bour03} after pre-cooling in a magnetic trap. When
magnetic fields are applied, e.g. for tuning near a Feshbach
resonance, the tight optical confinement in all three dimensions
makes the trap more robust against potential magnetic field
gradients which could drag atoms out of the trap. Crossed dipole
traps have also been used to prepare fermionic clouds for loading
into optical lattices~\cite{chin06,rom06}.

\paragraph{Resonator-enhanced standing wave trap}
The Innsbruck group enhanced the laser intensity and thus the trap
depth by forming a standing-wave optical resonator~\cite{mosk01res}.
The power of a 2 W Nd:YAG laser at $\lambda = 1064\,\rm nm$ was
resonantly enhanced by a factor of $\sim$ 120, resulting in a trap
depth of $\sim 1$ mK in the focus with $115\,\rm\mu m$ waist. This
was deep enough to capture atoms directly from the MOT. The
standing wave presented a 1D lattice potential to the atoms, that
were thus tightly confined in several pockets along the direction
of the standing wave. The high density in each pocket provided
good starting conditions for evaporative cooling. After some
initial cooling, the atoms were transferred into a single-beam
optical trap~\cite{joch03bec}.

\subsection{RF spectroscopy}
\label{e:chap2RFspectroscopy}

A single-component ultracold Fermi gas, with all atoms occupying
the same spin states, is an almost perfect realization of an ideal
non-interacting gas. $s$-wave collisions are forbidden due to the
Pauli principle, and $p$-wave collisions are frozen out. In the
absence of $p$-wave or higher partial wave scattering resonances,
no phase transition occurs down to exponentially small
temperatures~\footnote{For attractive $p$-wave interactions with scattering length $a$, the critical temperature is
$T_C \sim (E_F/k_B) \exp[-\pi/2(k_F |a|)^3]$~\cite{houb97}.}.

Physics becomes interesting only in the presence of
interactions, and the obvious way to introduce interactions into a
Fermi gas is by forming a two-component system, such as a mixture
of the two lowest hyperfine states of $^6$Li. $s$-wave scattering is
then allowed between fermions of opposite spin. More accurately
speaking, as the spin-part of the two-particle wave function can
now be antisymmetric, symmetric $s$-wave scattering is now allowed.
Such a two-state mixture can be created via optical pumping after
the MOT phase, or via RF spectroscopy, starting from a pure
single-component gas. Since RF spectroscopy is an invaluable tool
to prepare, manipulate and probe ultracold gases, we review it
here in more detail. First, we summarize basic aspects of RF
spectroscopy, and then focus on clock shifts and mean field
energies.

\subsubsection{Basics}

Let us note some important properties of RF spectroscopy: a)
The RF field has a very long wavelength ($\approx$ 3m), so there
is negligible momentum transfer. The coupling takes place only
between internal states of each individual atom. b) The RF field
(typically from a $\sim$ cm-large antenna) is essentially constant
over the size of the sample ($\sim 100 \,\mu\rm m$). Thus, the
entire cloud is simultaneously addressed by the same coupling. c)
The RF pulse generally creates a superposition of the two coupled states. Such coherences can be
long-lived in the absence of decay mechanisms.

In many cases, one can approximate a system of two coupled states
$\left|1\right>$ and $\left|2\right>$ with energies $E_1$ and
$E_2$ as an isolated two level system driven by a field $V =
V(t)\,\left(\left|2\right>\left<1\right|+e^{i\phi}\left|1\right>\left<2\right|\right)$
oscillating close to the resonant frequency $\omega_0 = (E_2 -
E_1)/\hbar$. Such a two-level system is conveniently described as
a pseudo spin-1/2, for which $\left|\uparrow\right> \equiv
\left|1\right>$ and $\left|\downarrow\right> \equiv
\left|2\right>$. Keeping only the part of the interaction that
resonantly drives the transition (``rotating wave
approximation''), the Hamiltonian is written as
\begin{eqnarray}
    H = H_0 + V; \quad H_0 \equiv -\frac{\hbar\omega_0}{2} \sigma_z; \quad V \equiv -\frac{\hbar \omega_R}{2}\left(\sigma_x \cos\omega t + \sigma_y \sin \omega t\right)
\end{eqnarray}
where $\sigma_i$ are the Pauli spin matrices and $\omega_R$ is the
Rabi frequency, giving the strength of the coupling. $\omega_R$ depends
on the drive field (in our experiments a magnetic field generated
by an antenna) and the coupling matrix element between the two
hyperfine states. A typical value for $\omega_R/2\pi$ is several
kHz. The free Hamiltonian $H_0$ has its natural interpretation as
a constant magnetic field in the $z$-direction of
pseudospin-space. In the same way, the interaction $V$ represents
a (real or fictitious) rotating magnetic field in the transverse ($x$-$y$) plane.
Transforming into the frame rotating at frequency $\omega$, the
direction of the transverse field is constant, while the $z$-field
(and thus the energy splitting between the two states in the
rotating frame) is reduced to $-\hbar \delta =
\hbar(\omega_0-\omega)$. For a resonant drive with $\delta = 0$,
only the constant transverse field is left, and - borrowing from
the classical picture - the spin (or pseudospin) precesses around
it at frequency $\omega_R$. A complete inversion of the
spin-direction - and thus a complete transfer from state
$\left|1\right>$ into state $\left|2\right>$ - is achieved for a
RF pulse length $t = \pi/\omega_R$ (so-called $\pi$-pulse). An
equal superposition $\frac{1}{\sqrt{2}}\left(\left|1\right> +
\left|2\right>\right)$ is achieved for $t = \pi/2\omega_R$
($\pi/2$-pulse).

\subsubsection{Adiabatic rapid passage}

For general detuning $\delta$, the (fictitious) magnetic field in
the rotating frame is $\vect{B}_{\rm rot} = \frac{\hbar\delta}{2}
\hat{\vect{z}} -\frac{\hbar\omega_R}{2} \hat{\vect{x}}$. At large
positive detuning $\delta \gg \omega_R$, it is predominantly
pointing in the $+z$-direction, and the pseudospin precesses around
it. If the initial state is either state $\left|1\right>$ or state $\left|2\right>$, then the
pseudospin is pointing up or down, and the angle between it and
the fictitious magnetic field is small. If the detuning is now
{\it slowly} swept from $\delta \gg \omega_R$ through resonance
($\delta = 0$) and towards large and negative values, the
pseudospin will adiabatically follow the direction of the changing
magnetic field and thus end up, for $\delta \ll -\omega_R$,
aligned opposite to its original direction. One has thus
adiabatically transferred the atom from state $\left|1\right>$ to
state $\left|2\right>$ (or vice versa). The condition of
adiabaticity requires that the pseudospin's precession frequency
is always fast compared to the change of the magnetic field's
direction, given by the azimuthal angle $\theta =
\arctan\frac{\delta}{\omega_R}$. This condition is most stringent
on resonance, where it reads $\omega_R \gg \dot{\theta} =
\frac{\dot{\delta}}{\omega_R}$ or $\dot{\delta} \ll \omega_R^2$.
For a non-adiabatic transfer, the probability for a successful
transfer is given by the formula due to Landau and Zener:
\begin{equation}
    P_{\left|1\right> \rightarrow \left|2\right>} = 1 - \exp\left(-2\pi \frac{\omega_R^2}{\dot{\delta}}\right)
\end{equation}

It is important to realize that both a short RF-pulse as well as a
non-adiabatic Landau-Zener transfer will leave the atom in a
superposition state $\cos\alpha \left|1\right> + \sin\alpha \, e^{i
\phi} \left|2\right>$. If these processes are applied to a Fermi
gas initially polarized in a single spin state $\left|1\right>$,
it will still be a fully polarized Fermi gas after the RF-pulse,
with the only difference that now all the atoms are in one and the
same superposition state of $\left|1\right>$ and $\left|2\right>$.

It is only by decoherence discussed below
(section~\ref{s:decoherence}) that the superposition state
transforms into an incoherent mixture of atoms in states
$\left|1\right>$ and $\left|2\right>$.

\subsubsection{Clock shifts}
\label{s:clockshifts}

Clock shifts are density dependent shifts of transition
frequencies due to interactions between the atoms. The name derives
from their presence in atomic clocks. Indeed, they are one of the dominant sources of systematic error in current cold atom fountain clocks~\cite{sort00}.
The absence of
clock shifts in two-state Fermi mixtures
facilitates the use of RF transitions to create spin mixtures and
allows to accurately
calibrate magnetic fields. The emergence of
clock shifts in three
component Fermi systems provides an important tool
to probe the many-body
physics underlying such shifts.

\paragraph{Absence of clock shifts in a fully polarized Fermi gas}

It is tempting - but incorrect - to calculate clock shifts by
considering the energy shift due to interactions of the two atomic
states involved and then associating the resulting energy
difference with a shifted transition frequency. Let us take for
example a Fermi gas fully polarized in state $\left|1\right>$, and
let $a_{12}$ be the scattering length for collisions between
$\left|1\right>$ and $\left|2\right>$. An atom in state
$\left|2\right>$ would  experience a mean-field energy shift
$\delta E_2 = \frac{4\pi\hbar^2}{m} a_{12} n_1$ due to
interactions with the cloud of $\left|1\right>$-atoms at density
$n_1$. As an RF pulse transfers $\left|1\right>$ atoms into state
$\left|2\right>$, one might incorrectly conclude that the RF
resonance is shifted by an amount $\delta E_2$. However, the RF
pulse does not incoherently transfer some atoms into state
$\left|2\right>$, where they would experience the shift $\delta
E_2$. Such a process would increase entropy, while RF radiation is
a unitary transformation which conserves entropy. Rather, the RF
pulse transfers all atoms simultaneously into a new superposition
state $\left|\alpha\right> \equiv \cos\alpha \left|1\right> +
\sin\alpha \,e^{i \phi} \left|2\right>$. The fermions are still
fully spin-polarized, they all occupy the same (rotated) quantum
state. Therefore, each pair of fermions has to be in an
antisymmetric motional state which excludes $s$-wave collisions.
As a consequence, there is  no interaction energy in the final
state and the clock shift is zero. {\it Clock shifts are absent in
an ultracold spin-polarized Fermi gas}. This was directly
demonstrated in the MIT experiment~\cite{gupt03rf}.

A similar argument shows that in the case of thermal bosons (with
intrastate scattering lengths $a_{11} = a_{22} = 0$), there {\it
is} a clock shift, but it is twice the energy shift $\delta E_2$
for an infinitesimal RF transfer~\cite{zwie03}. The factor of two
results from correlations in a spin-polarized thermal
Bose gas, which are preserved during the RF pulse: In a coherent
collision, two indistinguishable thermal bosons either preserve
their momenta or exchange them. If, in contrast, RF spectroscopy is performed on
a pure Bose-Einstein condensate, the colliding bosons have the
same momenta, hence exchange collisions are absent and the
mean-field shift is indeed $\delta E_2$~\cite{okte99,okte02}.

\paragraph{Absence of clock shifts in a two-state mixture of fermions}
\begin{figure}[t]
\begin{center}
\includegraphics[width=4in]{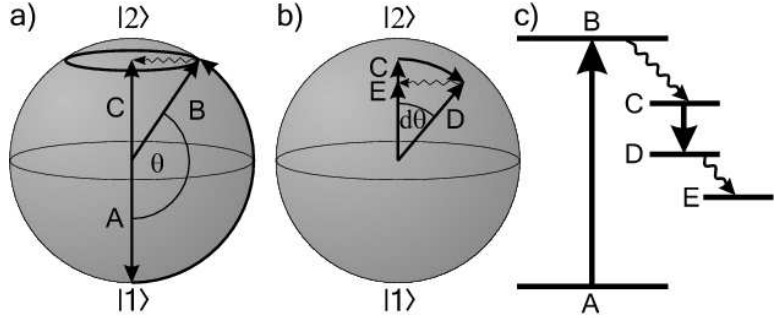}
\caption[Bloch sphere representation of RF transitions]{Bloch
sphere representation of RF transitions. a) An RF pulse
rotates a pure state A into B. The superposition state decoheres into a ``ring'' distribution which has ``lost'' its definite phase and is represented by the vertical vector C.
b) A second RF pulse transforms the fully decohered state C into a partially coherent
state D. The final state E is reached only after further
decoherence. c) Transfers A$\rightarrow$B and C$\rightarrow$D are
coherent and reversible. B$\rightarrow$C and D$\rightarrow$E are
irreversible. From~\cite{zwie03}.}\label{f:bloch}
\end{center}
\end{figure}
Switching back to fermions, one may ask whether there is a clock
shift if the initial state is {\it not} spin-polarized, but a
decohered mixture of populations in state $\left|1\right>$ and
$\left|2\right>$. Interactions are now clearly present, the energy
levels of atoms in states $\left|1\right>$ and $\left|2\right>$ are now truly shifted by
$\delta E_1 = \frac{4\pi\hbar^2}{m} a_{12} n_2$ and $\delta E_2$, and one might
(incorrectly) expect a clock shift $\delta E_2 - \delta E_1$ for
transitions from state $\left|1\right>$ to $\left|2\right>$.
However, similar to the case of the spin-polarized sample, one has
to distinguish between the state which can be accessed by the
transition, and some other incoherent state which can be reached
only by an entropy increasing decoherence process (see
Fig.~\ref{f:bloch}). Even though the
initial mixture had no
coherence (i.e. the off-diagonal elements of the density matrix
were zero), the RF pulse reintroduces
coherence into the system.
The final state after the RF pulse is
not an incoherent mixture
with different particle numbers, but a
state in which each
$\left|1\right>$ atom has been transferred
into the superposition
state $\left|\alpha\right>$, and each
$\left|2\right>$ atom into
the orthogonal superposition state. One
can show that for
fermions, interaction energies are invariant
under such a coherent
transfer. The outcome is that in spite of
possibly strong
interactions between atoms in state
$\left|1\right>$ and
$\left|2\right>$, {\it there is no interstate
clock-shift in a
two-state mixture of fermionic
atoms}~\cite{zwie03}.

\paragraph{Clock shifts in transitions to a third state}

While RF transitions between two populated fermionic states do not
reveal energy shifts, transitions from e.g. state $\left|2\right>$
into a third, empty state reveal the presence of interactions. A
priori, such transitions require knowledge of three scattering
lengths, $a_{12}$, $a_{13}$ and $a_{23}$. However, the preceding
discussion shows that the coherent transfer from $\left|2\right>$
to state $\left|3\right>$ is not affected by interactions between
these two states. To first approximation, the clock shift will be
given by the differential mean-field shift experienced by an atom
in state $\left|3\right>$ compared to that experienced by an atom
in state $\left|2\right>$. The clock shift should thus read
$\Delta \omega = \frac{4\pi\hbar}{m} \left(a_{13} -
a_{12}\right) n_1$. This dependence was used to observe the change
of the scattering length near a Feshbach resonance~\cite{rega03fesh,gupt03rf}.
However, this expression is valid only for small
scattering lengths, much smaller than the range of the potential
$r_0$ and $1/k_F$. Its extension will be discussed in the next
section.

\paragraph{Sum rule expression for the average clock shift}

It is possible to derive a general expression for the average
clock shift of an RF transition, valid for any many-body state of
bosons and fermions~\cite{baym07}. Of course, knowledge of the
average shift may be of limited use in cases where the RF spectrum
has a complex line shape, e.g. is asymmetric, has high-energy
tails or shows  several peaks, but it still provides an important
consistency check. The starting point is an initial many-body
state $\left|12\right>$ with energy $E_{12}$ that contains atoms
in state $\left|1\right>$ and $\left|2\right>$. The RF pulse
resonantly drives transitions between state $\left|2\right>$ and
$\left|3\right>$. As before in the discussion of RF transitions in
a two-state mixture, it is incorrect to calculate the clock shift
via the energy difference between two fully decohered states with
different particle numbers in state $\left|2\right>$ and
$\left|3\right>$. The expression for the differential mean-field
shift $\Delta\omega = \frac{4\pi\hbar}{m} (a_{13}-a_{12}) n_1$
thus cannot be generally true. Rather, to obtain the average clock
shift, one must calculate the average energy cost per atom for
rotating all atoms in state $\left|2\right>$ into a coherent
superposition of $\left|2\right>$ and $\left|3\right>$, a state
$\left|\vartheta\right> = \cos\vartheta\left|2\right> +
\sin\vartheta\left|3\right>$. The many-body state is then changed
into $\left|1\vartheta\right>$, which has the same spatial
many-body wave function as the original state. The total number of
transferred atoms is $N_2 \vartheta^2$ for small $\vartheta$, with
$N_2$ the number of atoms originally in $\left|2\right>$. The
average energy cost per atom for this rotation is thus
\begin{equation}
\label{e:clockshift1} \hbar\bar{\omega} = \hbar\omega_{23} +
\hbar\Delta \omega = \lim_{\vartheta\rightarrow
0}\frac{\left<1\vartheta\right|H\left|1\vartheta\right> -
\left<12\right|H\left|12\right>}{N_2 \vartheta^2}
\end{equation}
where $\hbar\omega_{23}$ is the hyperfine+Zeeman energy difference
between $\left|2\right>$ and $\left|3\right>$, and $H$ is the
total Hamiltonian of the interacting mixture in three hyperfine
states. The latter can be split into the internal hyperfine+Zeeman
Hamiltonian and the external Hamiltonian $H_{\rm ext}$ describing
the kinetic and interaction energy. The rotated state
$\left|1\vartheta\right>$ is generated by the many-body analogue
of the transverse (fictitious) magnetic field above, $S_x =
\frac{1}{2} \int {\rm d}^3 r (\varPsi^\dagger_3 \varPsi_2 +
\varPsi^\dagger_2 \varPsi_3)$. So we have $\left|1\vartheta\right>
= e^{-i\vartheta S_x} \left|12\right> \approx (1- i \vartheta S_x)
\left|12\right>$, and Eq.~\ref{e:clockshift1} gives
\begin{equation}
\hbar\Delta\omega = \frac{1}{2 N_2}
\left<12\right|\left[S_x,[H_{\rm ext},S_x]\right]\left|12\right>
\end{equation}
An identical expression for the clock shift can be calculated
using Fermi's Golden Rule~\cite{baym07}. This general sum rule for
the spectral response can be applied to strongly interacting
fermions~\cite{baym07,punk07rf}. For weak interactions with the
scattering lengths small compared to the characteristic size $r_0$
of the interatomic potential, one indeed obtains the mean field
expression of the previous section. For scattering lengths larger
than $r_0$ (but still smaller than $1/k_F$) this expression is
modified to
\begin{equation}
\Delta
\omega = \frac{4\pi\hbar}{m} \frac{a_{12}}{a_{13}} (a_{13}-a_{12})
n_1. \end{equation} The general result, valid for all scattering
lengths large than $r_0$, is \begin{equation} \Delta \omega =
\frac{1}{\hbar}\left(\frac{1}{k_F a_{13}} - \frac{1}{k_F
a_{12}}\right)
\frac{\partial{(E_{12}/N_2)}}{\partial{\left(\frac{1}{k_F
a_{12}}\right)}}
\end{equation}
where $\partial{(E_{12}/N_2)}/\partial{\left(\frac{1}{k_F
a_{12}}\right)}$ is the change in the energy of the original state
$\left|12\right>$ under a change of the interaction strength
$1/k_F a_{12}$. This change is varying smoothly as a function of
$1/k_F a_{12}$ and is well-behaved even for resonant interactions,
$1/k_F a_{12} = 0$. This expression shows that for strong
interactions, the clock shift is expected to approach zero. This
explains, at least qualitatively, the observation of vanishing
clock shifts in a strongly interacting, unpaired Fermi
gas~\cite{gupt03rf}.

\subsubsection{The special case of $^6$Li}

The usefulness of RF spectroscopy strongly depends on the spectral
resolution one can achieve in the laboratory. The characterization
of interaction effects on the order of a tenth of the Fermi energy
requires a resolution on the kilohertz level. At high magnetic
fields around the Feshbach resonance in $^6$Li, typical
magnetic field stabilities are about 10 mG, day-to-day
fluctuations can be ten times larger. It is one of the many
fortuitous facts about the $^6$Li atom that due to its small
hyperfine interactions, magnetic fields of several hundred Gauss
completely decouple the electron from the nuclear spin. Therefore
there are several RF transitions which flip only the nuclear spin
and thus have only a very weak field sensitivity.  For
example, the $\left|1\right>$-$\left|2\right>$ atomic resonance
has a field dependence smaller than 2.7 kHz/G above 600 G, which
makes it easy to have sub-kHz resolution without any special field
stabilization.

This unique property of $^6$Li has allowed numerous RF
experiments on unitarity limited interactions~\cite{gupt03rf}, on
strong interaction effects in resonantly interacting
gases~\cite{chin04gap,shin07rf}, precision spectroscopy of atoms
and molecules~\cite{bart04fesh} and on imbalanced Fermi
gases~\cite{schu07pair}. In contrast, $^{40}$K has a field dependence of 170 kHz/G for
transitions between states $\left|2\right>$ and $\left|3\right>$
near the Feshbach resonance at 202 G. This was still sufficient
for RF-dissociation of molecules~\cite{rega03mol} and the
characterization of a Feshbach resonance~\cite{rega03fesh}.

\subsubsection{Preparation of a two-component system}
\begin{figure}[t]
\begin{center}
\includegraphics[width=3in]{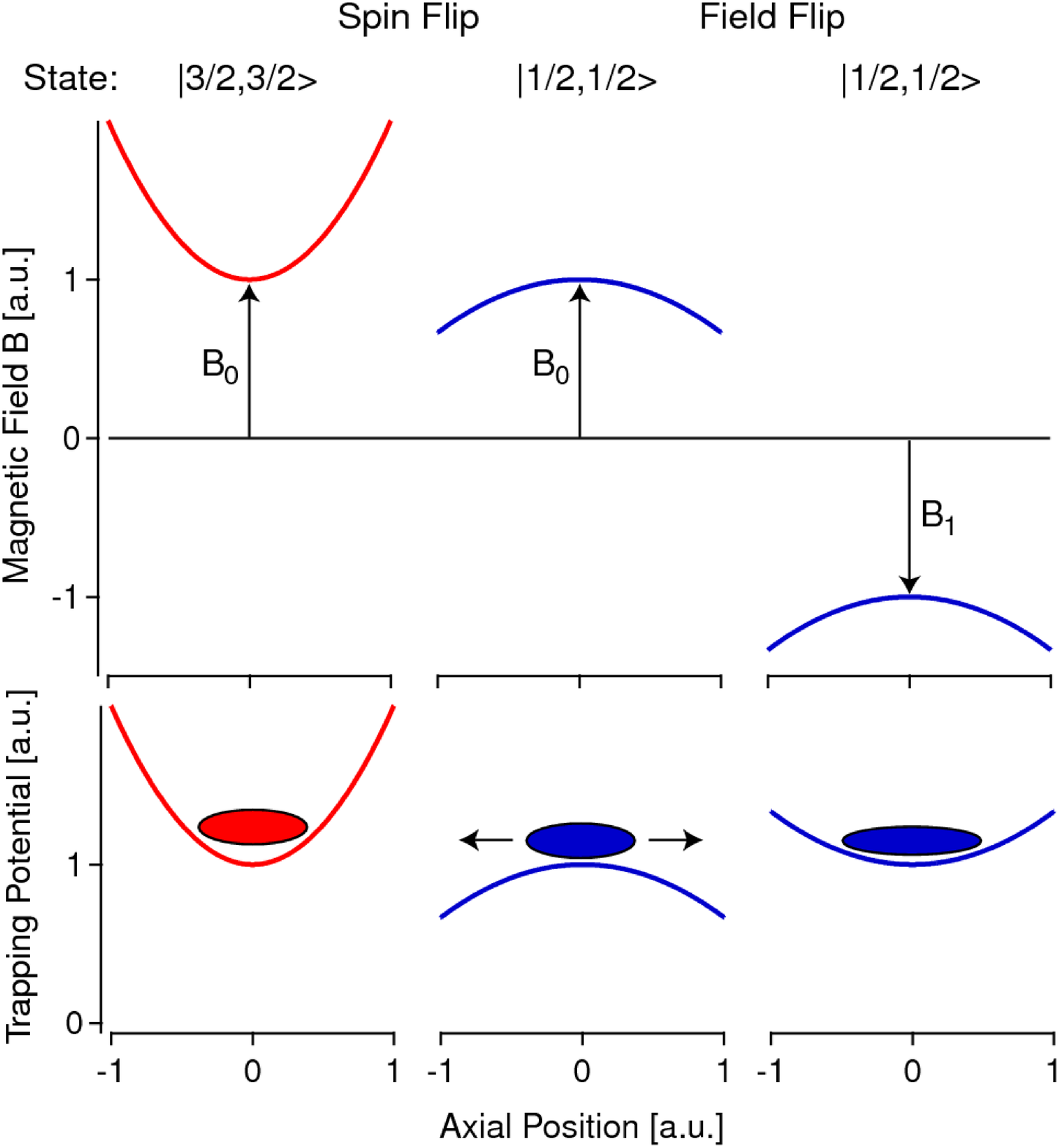}
\caption[Hyperfine transfer]{Hyperfine transfer of the cloud in a magnetic field
curvature. The atoms are initially trapped in state
$\left|3/2,3/2\right>$. After the spin transfer into state
$\left|1/2,1/2\right>$, atoms are no longer trapped. A quick
adiabatic reversal of the sign of the magnetic field retraps the
atoms.\label{f:hyperfinetransfer}}
\end{center}
\end{figure}

Here we discuss how we use RF pulses and magnetic fields to
transform a single-component Fermi cloud at low magnetic fields into
a strongly interacting two-component mixture near a high-field
Feshbach resonance.
\paragraph{Experimental Procedure}
In the MIT experiment, a spin-polarized Fermi gas is produced by
sympathetic cooling in a magnetic trap (see
section~\ref{s:highfields} and Fig.~\ref{f:magtrap} for details).
Loading into the optical trap is performed in several steps.
First, the radial confinement of the magnetic trap is removed by
reducing the current in the cloverleaf coils to zero. This is a
delicate process, as the center of the magnetic trap needs to
remain aligned with the optical trap at all times during the
current ramp-down. The atoms are still polarized in the stretched
state $\left|F,m_F\right> = \left|3/2,3/2\right>$. They experience
the radial confinement of the optical trap plus the axial magnetic
curvature. After the transfer into state $\left|1\right> \equiv
\left|F=1/2,m_F = 1/2\right>$ (an adiabatic Landau-Zener
RF-transfer close to the zero-field hyperfine splitting of
$228\,\rm MHz$), the atoms are in a high-field seeking state and
thus experience an {\it anti-trapping} axial curvature. By quickly
reversing the sign of the axial magnetic bias field the atoms are
trapped again (see Fig.~\ref{f:hyperfinetransfer}).
At this stage, the magnetic field is increased to values near the
Feshbach resonance between state \one\ and \two, located at $B =
834\, \rm G$ (see chapter~\ref{c:feshbach}). Starting with the
fully polarized gas in state \one, a non-adiabatic Landau-Zener RF
sweep (around the hyperfine splitting of $\sim 76 \,\rm MHz$ on
resonance) transfers atoms into a superposition of states \one\
and \two. The admixture of state \two\ can be freely controlled via the Landau-Zener sweep
rate.

\paragraph{Decoherence}
\label{s:decoherence} RF spectroscopy will not produce a decohered two-state
mixture, but a coherent superposition state, by applying
a suitable RF pulse or via a non-adiabatic Landau-Zener sweep.  A
decoherence mechanism is required for the gas to develop into a
mixture of two states, i.e. to incoherently populate two distinct
quantum states described by a diagonal density matrix.  Only such
a mixture will interact via $s$-wave collisions and possibly show
pairing and superfluidity at achievable temperatures.

We found experimentally that an efficient decoherence mechanism
for the trapped gas is provided by the magnetic field curvature of
the optical/magnetic hybrid trap.  Atoms that follow different
trajectories in the inhomogeneous field will acquire different
phases. After some time, the relative phases of atoms are
scrambled and one is left with an incoherent mixture. Being no
longer in identical states, $s$-wave interactions between atoms are
allowed. To demonstrate that decoherence has occurred, the
emergence of clock shifts in transitions to a third, empty state
has been recorded in~\cite{gupt03rf}. The timescale for
decoherence was found to be tens of milliseconds.

We can estimate the decoherence time from the spread of magnetic
fields across the sample.  Since the axial potential is mainly
magnetic, the atoms experience a spread of Zeeman energies equal
to the Fermi energy.  At high magnetic fields, the magnetic
moment of the two lowest states differs only by the nuclear
magnetic moment, which is three orders of magnitude smaller than
the electron's magnetic moment. We thus estimate the decoherence rate
to be a thousand times smaller than the Fermi energy divided by
$\hbar$. For a typical Fermi energy of $\hbar \times 100$ kHz, we thus
expect a decoherence time of 10 ms, in agreement with
observations.

\subsection{Using and characterizing Feshbach resonances}
Feshbach resonances are crucial for realizing strongly interacting
Fermi systems.  In this section, we present the Feshbach resonance
as an experimental tool to prepare and analyze such systems. This
section assumes a basic understanding of the physics of a Feshbach
resonance.  A detailed discussion of the underlying concepts and a
theoretical description of Feshbach resonances can be found in
chapter~\ref{c:feshbach}.

\subsubsection{High magnetic fields}
\label{s:highfields}

In $^6$Li, the broad Feshbach resonance between the lowest two
hyperfine states lies at 834 G~\cite{bart04fesh}. To access the
BEC-BCS crossover, magnetic fields of about 1000 G or more are
necessary. To generate these fields with sufficient homogeneity
while maintaining good optical access requires a careful design,
usually with some compromises.

\begin{figure}[t]
\begin{center}
\includegraphics[width=4in]{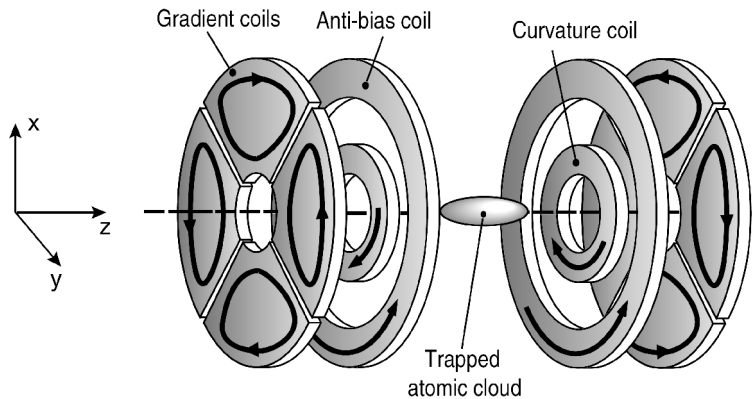}
\caption[Magnetic trap]{Magnetic trap in the MIT experiment, used
for sympathetic cooling of \li\ with \na. The trap consists of a
``curvature coil'' that produces an axially confining potential. Its
offset magnetic field is cancelled to about 0.5 G by the
``anti-bias coil''. Radial confinement is provided by the gradient
coils which are wound in ``cloverleaf'' configuration. They replace
the four Ioffe bars in a standard Ioffe-Pritchard trap. After the
fermions are loaded into a single-beam optical trap, the anti-bias
coils access the wide Feshbach resonance between the two lowest
hyperfine states of \li\ at 834 G.} \label{f:magtrap}
\end{center}
\end{figure}

If magnetic field coils with $N$ windings are placed in Helmholtz
configuration outside the vacuum chamber or glass cell of typical
diameter d = 3 cm, a current of about $I \sim B d/ \mu_0 N \sim
3000\,\rm A/N$ is required. For such current densities, water
cooling is essential. For a given total coil cross section $A$,
the coil resistance is $R = \rho N^2 (2\pi d)/A$, with $\rho = 1.7
\cdot 10^{-8}\,\Omega$m
 the resistivity of copper. The required electric power is $P = R
I^2 = E_B / \tau$, where $E_B = B^2 d^3/2\mu_0$ is the magnetic
energy of a homogeneous field $B$ stored in a volume $d^3$, and
$\tau \sim \mu_0 A/\rho$ is the $1/e$ decay time of the field
energy if the coils were shorted ($\tau$ = coil
inductance/resistance). Both $\tau$ and $P$ are independent of the
number of windings $N$.  The division of the designated volume of
copper into wires determines the voltage and current of the power
supply at constant power.

As the required magnetic field and the dimension $d$ are
determined by design constraints, the only variable here is the
total cross section of the coils $A$ which is often chosen to be a
fraction of $d^2$. For $A = (1\,\rm cm)^2$, a power of $P \sim 300
\,\rm W$ is dissipated in each coil and the time constant $\tau \sim
2\,\rm ms$. The time constant gives the fastest possible magnetic field ramp-up time, unless the power supply has a higher maximum
power than the power $P$ for steady operation.  Fortunately, the
field decay time can be reduced by using a ``ring-down'' resistance
in parallel with the coil. A diode ensures that this ring-down
path is opened once the power supply is switched off.

Fig.~\ref{f:magtrap} shows the magnetic field configuration used
in the MIT experiment. It allows for independent control of the bias
field, the magnetic field curvature, and the radial gradient
through the use of independent coils. The ``cloverleaf'' coils are
needed for tight radial confinement during the sympathetic
cooling stage of \li\ with \na\ in the magnetic trap. In order to
tune the interatomic interactions across the Feshbach resonance,
the bias field should be an independent parameter.
This is accomplished by arranging a pair of coils as close as
possible to the Helmholtz configuration.  Our ``Feshbach'' coils
(which also serve as ``anti-bias'' coils~\cite{kett99var} during magnetic
trapping) generate a residual magnetic field curvature that
corresponds to an axial trapping frequency of 11.0 Hz for $^6$Li at 834 G
(resonance).  If necessary we can compensate for this curvature by
using the ``curvature'' coils.  In practice, these two pairs of
coils contribute both to curvature and bias field, and controlling the two currents independently allows a wide range of possible values.  In most of our
experiments, the ``curvature'' coils provide the bulk of the
axial confinement.  Thus, varying the bias field with the ``Feshbach'' coils between
$B_0 = 700$ G to $B_0 = 1000$ G changes the axial trap frequency
by only 0.5 Hz around the value at the Feshbach resonance ($\omega_z/2\pi
= 22.8\,\rm Hz$).

\subsubsection{Methods for making molecules}
\label{s:makingmolecules}
\begin{figure}[ht]
\begin{center}
\includegraphics[width=4in]{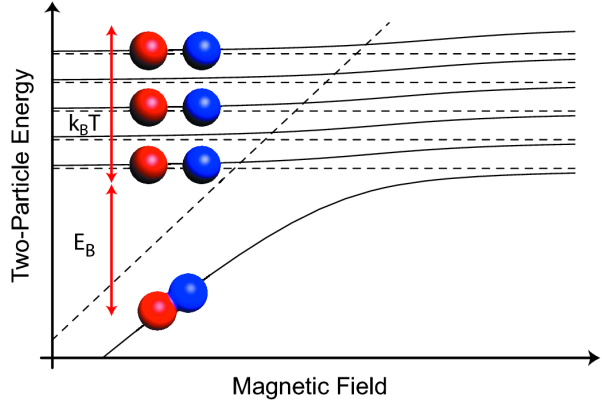}
\caption[Creating molecules via three-body collisions]{Creating
molecules via three-body collisions. A molecular state is coupled
to the continuum.  As the gas is cooled on the molecular side, the
Feshbach molecular state is populated via
three-body collisions.
If the binding energy is not much larger
than $k_B$ times the
temperature, the energy carried away by the
third body does not
substantially heat the sample. For fermionic
particles, further
decay into lower lying vibrational
states is strongly suppressed
due to Pauli
blocking.}\label{f:threebodycartoon}
\end{center}
\end{figure}

\begin{figure}[ht]
\begin{center}
\includegraphics[width=3.5in]{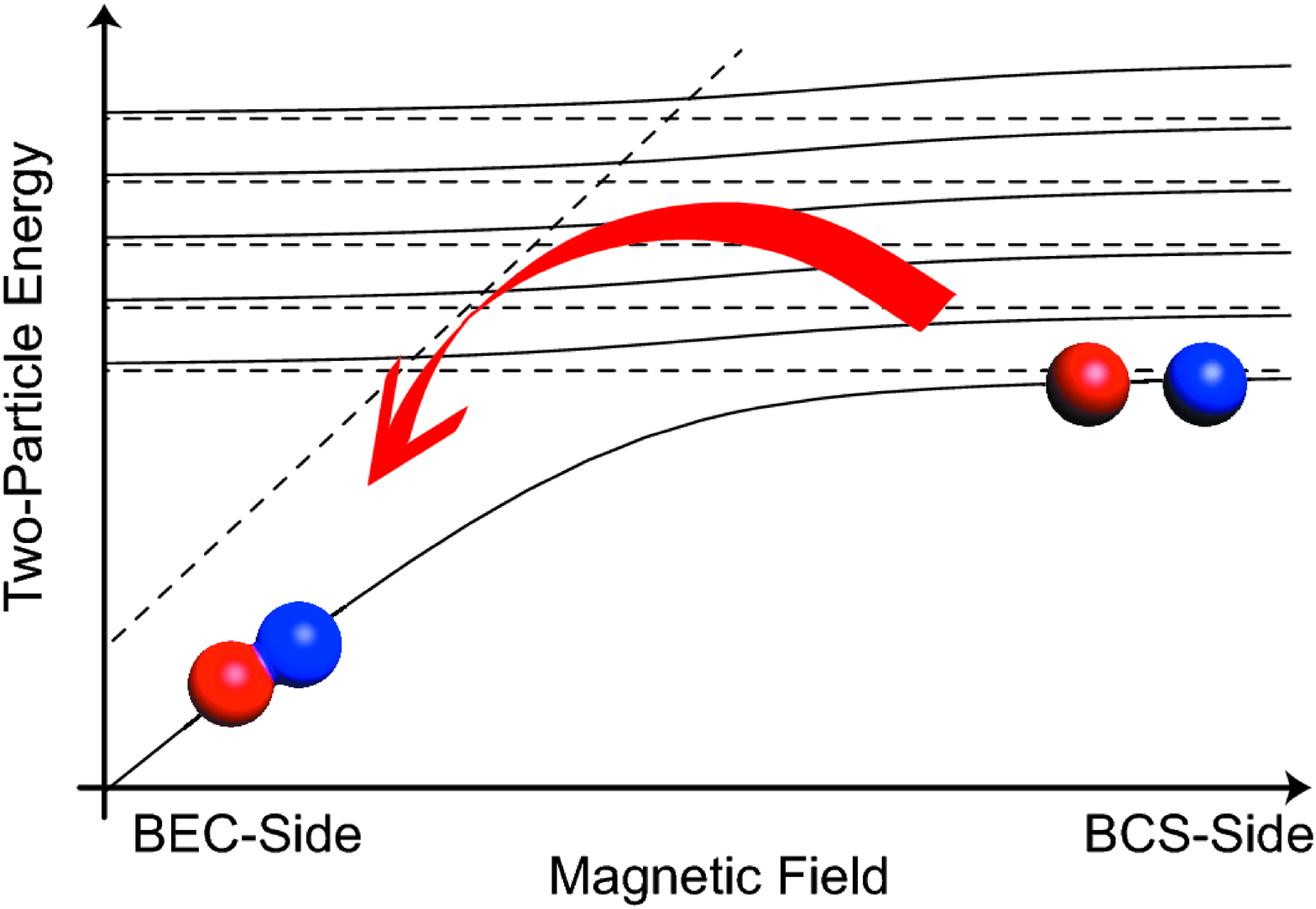}
\caption[Creating molecules via magnetic field ramps]{Creating
molecules via magnetic field ramps. A magnetic field sweep can
transfer unbound atoms adiabatically into the molecular state,
much like a two-level Landau-Zener
transition.}\label{f:feshbachcartoon}
\end{center}
\end{figure}

Molecules are one form of pairing, and therefore play a major role
in studying pairing between fermions.   Many of our experiments
use a purely molecular cloud as an intermediate step.   Several
methods have been demonstrated to create molecules from ultracold
atoms:
\begin{itemize}

\item  Photoassociation.  In photoassociation two colliding atoms
are optically excited to a bound state, which is electronically
excited. By using a second step or a Raman transition, the
electronic ground state can be accessed, usually with high
vibrational excitation. This method is discussed in other recent
reviews  \cite{jone06,huts06}. So far, the phase space density of
    molecular clouds has been limited by heating from near resonant
    light and collisions involving the electronically excited
    intermediate state, or the vibrationally excited final state.
    \item Three-body recombination near a Feshbach resonance.
   \item Coherent two-body transfer near a Feshbach resonance via (1) a magnetic field sweep,  (2) RF association, and (3) magnetic field modulation.
Note that many theoretical descriptions of photoassociation are directly applicable to Feshbach resonances, as they can be regarded as photoassociation resonances with zero frequency photons. Sweeps of the magnetic field across the Feshbach resonance are equivalent to frequency sweeps across the photoassociation resonance.
\end{itemize}

In a three-body recombination, two of the colliding atoms form a
molecule, the third particle (atom or molecule) carrying away the
leftover energy and momentum. This process preferentially
populates the most weakly bound molecular states. Their binding
energies lie between zero and $\approx -160 \,\hbar^2/m r_0^2$
(for an asymptotic van-der-Waals potential $V(r) \sim - C_6/r^6$),
depending on rotational quantum numbers and boundary conditions at
the inner turning point of the potential~\cite{lero69,gao00}. With
the van-der-Waals range $r_0 = (m C_6/\hbar^2)^{1/4} \approx 60\,
a_0$ for $^6$Li, these binding energies can be up to $k_B$ times 1
K. The released energy in such a collision heats up the cloud,
leading to trap loss (hence the name ``bad collisions''). However,
in the case of the very weakly bound molecular state on the
molecular side of a Feshbach resonance (scattering length $a>0$),
the binding energy can be on the order of the temperature, and
molecules can efficiently form without severe heating and trap
loss (Fig.~\ref{f:threebodycartoon}). Subsequently, leftover atoms
can be evaporated from the optical trap. This process is very
efficient, since weakly bound molecules have twice the atomic
polarizability, hence the optical trap is twice as deep for
molecules than it is for single atoms.

The Feshbach molecules are in the highest vibrational state of the
interatomic potential (see chapter~\ref{c:feshbach}). They are only stable
if the decay into lower lying vibrational levels is slow.  It
turns out that for fermions this decay is suppressed by the Pauli
principle (see section~\ref{s:history}).

Producing molecules coherently by a magnetic field sweep is
reversible and does not generate heat.  It exploits the tunability
of the Feshbach molecular state: Starting with unbound atoms in the
continuum, one can sweep the magnetic field across the resonance
and form a bound molecule (Fig.~\ref{f:feshbachcartoon}). Some
aspects of  this sweep can be described as a two-level
Landau-Zener sweep through an avoided crossing. For a coupling
matrix element $V$ between two ``bare'' states, $\left|a\right>$ and
$\left|b\right>$, and an energy sweep rate $\dot{E}$, one
finds~\cite{mies00}
\begin{equation}
\label{e:LandauZenersweep}
    P_{\left|a\right> \rightarrow \left|b\right>} = 1 - e^{-c \frac{\left|V\right|^2}{\hbar \dot{E}}}
\end{equation}
for the probability $P_{\left|a\right> \rightarrow
\left|b\right>}$ to make a transition from $\left|a\right>$ to
$\left|b\right>$ as the bare state energies are swept through
resonance. Here $c$ is a numerical constant on the order of 1.

In the case of Feshbach resonances, the two ``levels'' are the
molecular state and the state of two unbound atoms. $V$ is the
coupling matrix element between these states, $V = \sqrt{N/\Omega}
\, g_0$, an expression that we will discuss in
chapter~\ref{c:feshbach}. The number $N$ of atom pairs that
appears in $|V|^2$ accounts for the fact that each spin up atom
has $N$ chances to form a molecule with a spin down atom per
quantization volume $\Omega$. Alternatively, one can consider
two-body physics in a box of volume $\Omega/N$, which emphasizes
the local picture of two atoms forming a molecule.

If we take the simple Feshbach model of
chapter~\ref{c:feshbach}, we can replace $g_0^2 =
\frac{4\pi\hbar^2}{m} a_{\rm bg} \, \Delta\mu \Delta B$, with $a_{\rm bg}$ the background scattering length, $\Delta\mu$ the difference in magnetic moments between the molecular state and two free atoms, and $\Delta B$ describing the width of the Feshbach resonance. The bare
state energies are tuned via the magnetic field, so $\dot{E} =
\Delta\mu\,  \dot{B}$. We then have
\begin{equation}
    P_{\rm atoms \rightarrow molecules} \simeq 1 - \exp\left(- A \frac{n}{\dot{B}}\right)
    \label{e:LandauZener}
\end{equation}
with $A = c \frac{4\pi\hbar a_{\rm bg} \Delta B}{m}$. The higher
the density and the slower the magnetic field ramp across
resonance, the more efficient is the production of
molecules~\cite{hodb05mol,kohl06feshbachreview}.

The schematic figure of the Feshbach resonance
(Fig.~\ref{f:feshbachcartoon}) suggests that the simple two-state
picture applies only to the lowest state of relative motion
between the two atoms.  Excited states of relative motion on the
BCS side are adiabatically connected to the next lower lying state
of relative motion on the BEC side. Therefore, the  Landau-Zener
probability discussed above should have a prefactor which is the
probability for two atoms to be in the same phase space cell,
proportional to the phase space density.  Indeed, it has been
experimentally verified in~\cite{hodb05mol} that the  efficiency
of forming molecules during a slow adiabatic sweep increases
monotonously with the phase space density and that it can exceed
50 \% for both bosonic and fermionic thermal clouds (up to 90\%
transfer was achieved for $^{40}$K).

The coherent conversion of two atoms into molecules can be
accomplished not only by sweeping the Feshbach field, but also by
modulating the magnetic field close to the Feshbach resonance, at
a frequency corresponding to the molecular binding
energy~\cite{thom05mol,gaeb07pwave}. Yet another method is to
drive a free-bound RF transition~\cite{ospe06hetero}, where initially
one of the free atoms occupies a different hyperfine state.

\subsubsection{Observation of Feshbach resonances}

A Feshbach resonance is an ``intimate'' encounter between two atoms,
which collide and temporarily form a molecule before they separate
again.  Many collisional processes are enhanced and have been used
to locate the magnetic field position of these resonances.

\begin{itemize}
    \item Trap loss by enhanced inelastic collisions.
    The first observations of Feshbach resonances were made by monitoring loss due to three-body recombination~\cite{inou98} and due to an enhanced photoassociation rate~\cite{cour98fesh}.  The broad Feshbach resonance in $^6$Li was mapped
    out using trap loss~\cite{diec02fesh}.  However, since the molecules
    formed in three-body recombination are long-lived close to resonance, the center
    of the loss feature was found at fields well below the
    Feshbach resonance.  In addition, an unpredicted narrow Feshbach
    resonance at 543 Gauss was found~\cite{diec02fesh,stre03,schu04fesh}.  Trap loss spectroscopy is
    usually applied to find new resonances and has been used, for
    example, to discover $p$-wave Feshbach resonances in $^{40}$K~\cite{rega03pwave} and
    $^6$Li~\cite{zhan04pwave,schu04fesh} and interspecies Feshbach resonances~\cite{stan04,inou04}.
    \item Rapid thermalization.
    The increased scattering length leads to rapid thermalization
    of the gas.  This method was used to study the resonance in $^{85}$Rb~\cite{robe98}, and in $^{40}$K~\cite{loft02}. The absence of thermalization was used to locate the position of the zero-crossing of the scattering length in $^6$Li at 528 G~\cite{ohar02,joch02}.
     \item Change of interaction energy.
        For Bose-Einstein condensates, this is observed by the
        change in mean field energy and therefore the size of the
        cloud, either in trap or in ballistic
        expansion~\cite{inou98,corn00JILA_collapse}.  For
        fermions, the change of the interaction energy has been
        monitored via clock shifts (see section~\ref{s:rfspectroscopy}).  The size of
        the fermionic cloud varies smoothly and monotonously
        across resonance, a direct consequence of the smooth change of the cloud's energy in the BEC-BCS crossover (see chapter~\ref{c:BECBCStheory}).
    \item RF spectroscopy of Feshbach molecules. Using RF spectroscopy, one can determine the onset of molecular
dissociation, and then, by extrapolation, find the value of the
magnetic field at which the molecular binding energy
vanishes~\cite{bart04fesh,rega03mol} . The most precise value for
the broad $^6$Li Feshbach resonance was derived from RF
spectroscopy between weakly bound molecular states using a
multi-channel scattering model~\cite{bart04fesh}.
    \item Threshold for molecule formation.
    When the magnetic field is swept across the Feshbach
    resonance, molecules will appear with a sharp onset at the
    resonance.
    \item Threshold for molecule dissociation.
    Feshbach molecules start to dissociate when the magnetic field
    is raised to a value above the Feshbach resonance.
\end{itemize}

Since the last two methods are directly related to the formation
and detection of molecules, we discuss them in more detail.

\begin{figure}[ht]
\begin{center}
\includegraphics[width=4.5in]{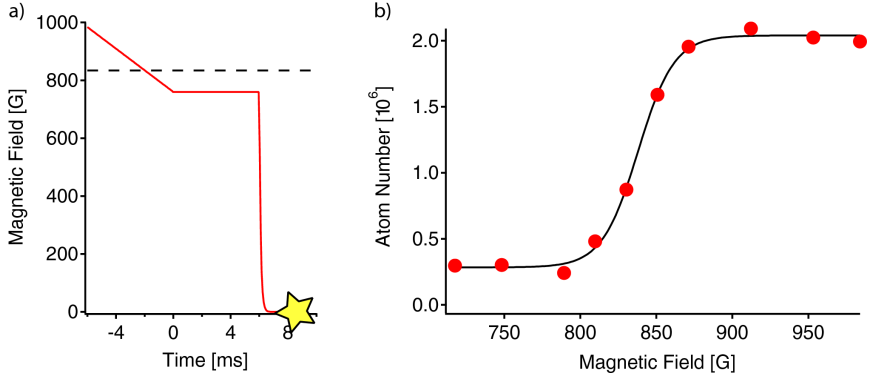}
\caption[Molecule formation by magnetic field sweep across the
Feshbach resonance]{Molecule formation by magnetic field sweep
across the Feshbach resonance. a) Experimental procedure. A Fermi
mixture prepared on the BCS-side of the Feshbach resonance is
swept across resonance (shown as the dashed line) to form
molecules. The gas is released from the trap at the end point of
the ramp at time $t= 0$ ms. Zero-field imaging, indicated by the
star, detects the leftover atoms. b) Atomic signal vs end point of
the magnetic field sweep. The line is a fit to an error function,
whose center is determined to be $838 \pm 27$ G, with an
uncertainty given by the 10\%-90\% width (54 G).}\label{f:moleculeramp}
\end{center}
\end{figure}

To observe the onset of molecule formation, one prepares  a Fermi
mixture on the ``BCS''-side of the Feshbach resonance, where no
two-body molecular bound state exists in vacuum (see
chapter~\ref{c:feshbach}).   As the magnetic field is swept across
the resonance, molecules will form and, accordingly, the signal
from unbound atoms will diminish
(Fig.~\ref{f:moleculeramp})~\cite{rega03mol,herb03cs_mol,stan04}.
From Fig.~\ref{f:moleculeramp} we determine a value of $B_0 = 838
\pm 27$ G for the position of the resonance.  The loss of atomic
signal is  reversible: Ramping back across the resonance will
dissociate the molecules, and reestablish all or most of the
atomic
signal~\cite{rega03mol,cubi03,stre03,joch03lith,herb03cs_mol,xu03na_mol,durr04mol}.

\begin{figure}[ht]
\begin{center}
\includegraphics[width=5.2in]{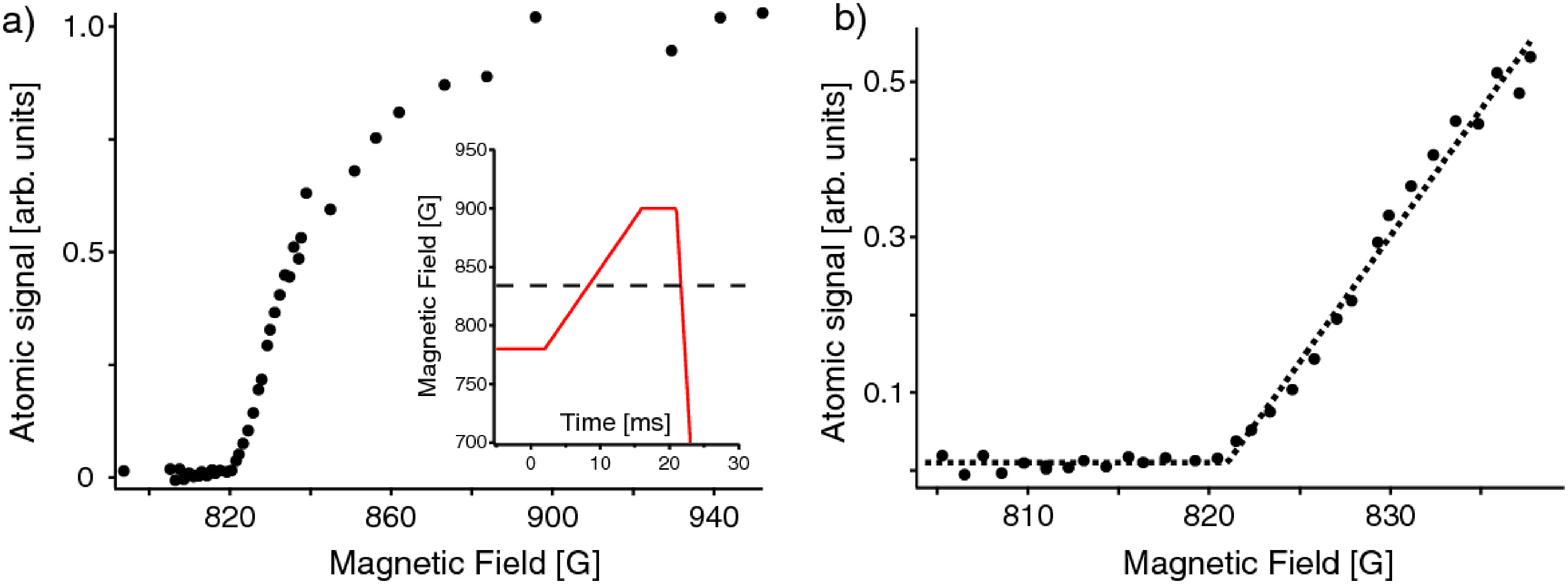}
\caption[Locating the Feshbach resonance by molecule
dissociation]{Locating the Feshbach resonance by molecule
dissociation. The experimental procedure is shown in the inset. A
molecular cloud is prepared on the BEC-side of the Feshbach
resonance, at 780 G, and released from the trap at $t = 0$ ms.
After some expansion, the field is ramped to a test value around
resonance (shown as the dashed line), held constant and is finally
brought to zero field, where only unbound atoms are imaged. a) The
atomic signal as a function of the test field shows a sharp
threshold behavior at $821\pm 1$ G, where the uncertainty is the
statistical error of a threshold fit, shown in
b).}\label{f:feshbachlocation}
\end{center}
\end{figure}

In fact, the dissociation method gives a more accurate
determination of the location of the Feshbach
resonance~\cite{rega04,zwie04rescond}. To avoid effects due to the
high density in the trap (i.e. many-body physics), in~\cite{zwie04rescond} the molecular
cloud is expanded to a 1000 times lower density, about
$10^{10}\,\rm cm^{-3}$. Then the magnetic field is ramped to a
value $B_{\rm test}$. If $B_{\rm test}$ lies above the Feshbach
resonance, the molecules will dissociate into unbound atoms, which
are subsequently detected at low field.

The very sharp onset of the atomic signal at $B_{\rm test} = 821
\pm 1$ G is striking (see Fig.~\ref{f:feshbachlocation}) and
suggests this magnetic field value as the position of the Feshbach
resonance. However, via molecular RF spectroscopy the location of
the Feshbach resonance has been determined to lie at $834.1 \pm
1.5$ G~\cite{bart04fesh}. The reason for this discrepancy is
probably that molecules at threshold are extremely fragile and
might break apart before the resonance is reached, thus shifting
the observed threshold to lower values.  See
Ref.~\cite{schu04fesh} for a discussion and characterization of
such shifts.  RF spectroscopy addresses more tightly bound
molecules and identifies the resonance by extrapolation, thus
avoiding stability issues very close to resonance.

From Fig.~\ref{f:feshbachlocation} we can directly see that before
dissociation, more than 99\% of the gas exists in form of
molecules. The reason is that this molecular cloud was formed via
the three-body process, by simply cooling the gas at the fixed
field of 780 G (the BEC-side of the resonance). The lifetime of
the weakly bound molecules is so long, and the binding energy is
so small, that losses and heating are negligible, and, after
evaporation of leftover unbound atoms, essentially all particles
are bound into molecules.

\subsubsection{Determination of the coupling strength of Feshbach resonances}
\label{s:coupling}
\begin{figure}[t]
\begin{center}
\includegraphics[width=5.2in]{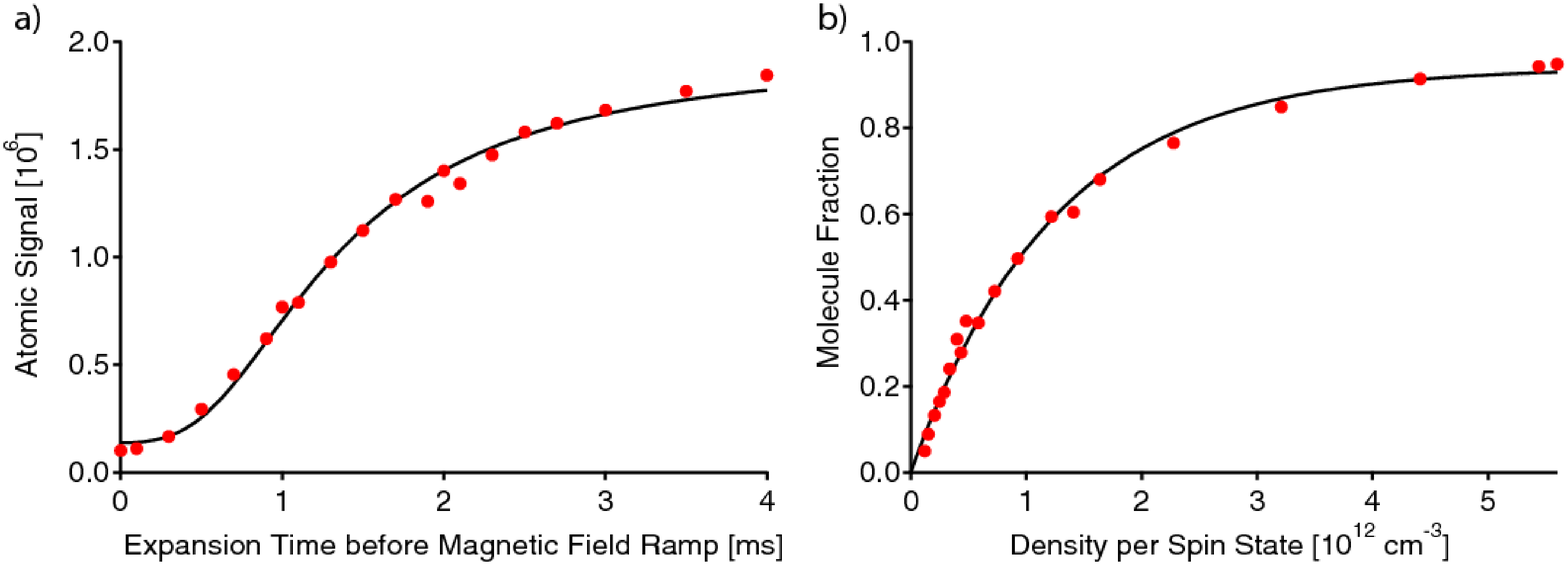}
\caption[Revival of the atomic signal during expansion and
strength of Feshbach coupling]{Revival of the atomic signal during
expansion and strength of Feshbach coupling. a) The magnetic field
is switched off after varying expansion times for a cloud released
at $840\,\rm G$. The field ramp creates molecules more efficiently
at the high densities of the trap than at low densities after long
expansion. In b), all of the atomic signal
loss is interpreted as molecular conversion and plotted as a
function of density. The density was calibrated by imaging the
cloud at high field for varying expansion times. All fits are for
the simple Landau-Zener-model described in the text.}\label{f:atomrevival}
\end{center}
\end{figure}

The ``strength'' of a Feshbach resonance is determined by the square
of the matrix element which couples the closed and open channels,
proportional to $g_0^2 = \frac{4\pi\hbar^2}{m} a_{\rm bg}
\, \Delta\mu \Delta B$ (see section~\ref{s:makingmolecules} and chapter~\ref{c:feshbach}). This expression depends
on the background scattering length only because of the definition of $\Delta B$ in the formula for the scattering length $a(B) = a_{\rm bg} \left(1 - \frac{\Delta B}{B-B_0}\right)$. A Feshbach resonance with the same strength but on top of a larger background scattering length then has a narrower width $\Delta B$. So one way to determine the strength
of a Feshbach resonance is by measuring or knowing $a_{\rm bg}$,
$\Delta\mu$ and $\Delta B$.

The matrix element can be measured more directly from the dynamics
of molecule dissociation and formation.  When Feshbach molecules in $^{23}$Na
were dissociated with variable field ramp, the kinetic energy of
the fragments was shown to increase with the ramp
speed~\cite{muka04}. This reflects the finite lifetime of the
Feshbach molecules, which are ``ramped up'' in energy for about one
lifetime, before they decay through their coupling to the open
channel.  This method was also applied to $^{87}$Rb~\cite{durr04diss}.

Here we describe experiments using the reverse process, i.e. the
formation of molecules by a variable field ramp, as introduced in
section~\ref{s:makingmolecules} above. Fig.~\ref{f:atomrevival}
demonstrates the extremely strong coupling strength of the \li\
Feshbach resonance. In this experiment, a Fermi mixture is
released from the trap at $B = 840$ G, slightly above the Feshbach
resonance.  When the magnetic field is suddenly switched off at
the same time (at an initial slew rate of $\dot{B} = 30 \,\rm
G/\mu s$), almost the entire atomic signal vanishes, i.e. the
conversion into molecules is almost 95\%. The $^6$Li Feshbach
resonance is so strong that the quantitative conversion from
trapped fermions into molecules during a Feshbach sweep can only
be avoided by using small magnetic field coils with low
self-inductance and correspondingly fast switch-off
time~\cite{cubi03}.

However, when we allow the gas to expand and lower its density
before the sweep, then the conversion to molecules is only
partial, and we can determine the strength of the Feshbach
coupling.  In Fig.~\ref{f:atomrevival} b) we  interpret all the
disappeared atomic signal as conversion into the weakly-bound
molecular state neglecting other loss-channels like unobserved
molecular states.  The conversion efficiency decreases with
decreasing density and can be fit using the simple the
Landau-Zener formula Eq.~\ref{e:LandauZener}.  We find that the constant $A$ in Eq.~\ref{e:LandauZenersweep} is $A \approx 24\, {\rm G}/({10^{12}\,\rm cm^{-3} \, \mu s})$ with a relative
error of $50\%$ due to the uncertainty in the atom number.

The theoretical prediction is Eq.~\ref{e:LandauZener} with $c = 2
\pi$~\cite{chwe04mol}. With the parameters for $^6$Li we find
\begin{equation}
    \label{e:landauzenerparam}
    A = \frac{8\pi^2\hbar a_{\rm bg} \Delta B}{m} = 19 \, \frac{\rm G}{10^{12}\,\rm cm^{-3} \, \mu s}
\end{equation}
For comparison, for the $^{40}$K Feshbach resonance at $B=224.2$ G
used in some experiments the prediction is $A = 0.011 \, \frac{\rm
G}{10^{12} \,\rm cm^{-3}\, \mu s}$ ($a_{\rm bg} = 174\,a_0, \Delta
B = 9.7\,\rm G$~\cite{rega03fesh}). This is not far from the value
$A \approx 0.004 \frac{\rm G}{10^{12} \,\rm cm^{-3}\, \mu s}$ one
extracts from the measurement in ~\cite{rega03mol} ($0.05\,\rm
G/\mu s$ was the ramp speed that resulted in a $1/e$ transfer of
molecules, at a peak density of $n_{\rm pk} = 1.4 \times 10^{13}
\,\rm cm^{-3}$). The broad resonance in $^6$Li can efficiently
convert atoms into molecules at 2,000 times faster sweep rates.

The good agreement with the simple Landau-Zener model might be
fortuitous.  Ref. \cite{hodb05mol} points out that the conversion
efficiency must depend on the phase space density and presents
data which, in the case of $^{85}$Rb, disagree with simple
theoretical predictions by a factor of eight.

\subsubsection{The rapid ramp technique}
\label{s:rapidramp}
\begin{figure}[ht]
\begin{center}
\includegraphics[width=3.5in]{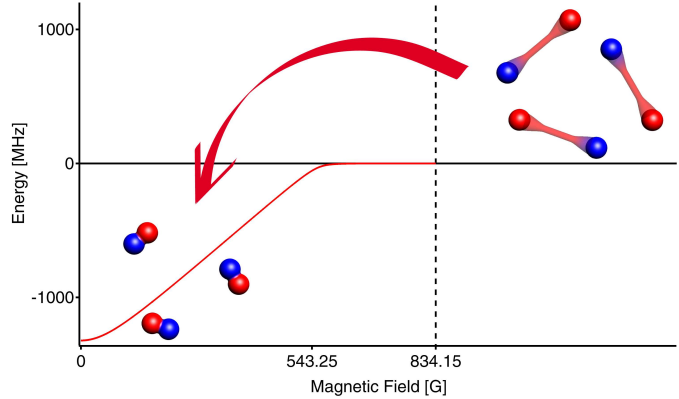}
\caption[Rapid ramp to the molecular side to observe pair
condensation]{Rapid ramp to the molecular side to observe pair
condensation. Immediately after switching off the trapping beam,
the magnetic field is ramped to zero field. This converts
long-range pairs into stable, tightly bound molecules, preserving the momentum
distribution of the original pairs.}\label{f:rampingfields}
\end{center}
\end{figure}

So far, we have discussed the time scale for two-body physics,
namely the association of two atoms into a molecule.  For isolated
atom pairs, this process is independent of the total momentum of
the pair, which is preserved due to Galilean invariance. In a
many-body system, fermion pairs interact and collide, and their
momentum changes.  If the two-body time scale is faster than the
many-body time scale, there is an interesting window of ramp rates for the sweep across the Feshbach resonance: One can be slow enough to quantitatively convert atom pairs into molecules, but also fast enough such that the momentum distribution of the final molecules reflects the momentum distribution of the
fermion pairs prior to the sweep (see Fig.~\ref{f:rampingfields}).

This method was introduced by the JILA group~\cite{rega04}, and
later adapted to \li\ by our group~\cite{zwie04rescond}. It made
it possible to analyze the momentum distribution of fermion pairs
across the whole BEC-BCS crossover and detect the pair condensate
(see chapter~\ref{c:expobservation}).

The problem with the rapid ramp technique is that it is not clear
what the many-body time scale is. In particular, one wants to rule
out that the pair momentum distribution changes during the sweep
or that a condensate is formed while ramping.  We address this
question by listing several time scales of the system
(table~\ref{t:timescales}).
For theoretical modelling of the ramp process, see refs.
~\cite{ho04proj,pera05,altm05,yuzb05noneq}.  The final
demonstration that the rapid ramp does preserve the absence or
presence of a pair condensate before the sweep has to come from
experiment (see section~\ref{s:formation}).

\begin{table}[h]
\centering
\begin{tabular}{p{0.02\linewidth}l|c|rp{0.2\linewidth}}
&  Time scale & Formula & Value &\\ \hline
&  Two-body physics & $\hbar/g_0 \sqrt{2 \pi n}$ & $20\,\rm ns $ &\\
&  Magnetic field ramp through anti-crossing & $g_0 \sqrt{2 \pi n}/\Delta \mu \dot{B}$ & $80\,\rm ns$ &\\
&  Inverse Fermi energy & $\hbar/ E_F$ & $3 \,\rm \mu s$    &\\
&  Time required to leave strongly interacting region & $\delta B/\dot{B}$ & $3\,\rm \mu s$&\\
&  Evolution of the gap at $k_F |a| = 2$ & $\Delta/\dot{\Delta}$ & $10\,\rm \mu s$&\\
&  Gap at $k_F |a| = 2$ &$\hbar / \Delta$ & 15 $\mu s$ &\\
&  Inverse collision rate at unitarity and $T/T_F = 0.1$ & $\approx 0.23\, \hbar E_F/(k_B T)^2$ \cite{gehm03coll} & $70 \,\rm \mu s$&\\
&  Growth time of a pair condensate at $k_F |a| = 2$ & $\approx \hbar E_F/\Delta^2$ \cite{Bara04Coex} & $ 75\,\rm \mu s$&\\
&  Radial trapping period & $2 \pi/\omega_r$ & 2 ms& \\
\end{tabular}
\caption{Time scales involved in the rapid ramp technique.  The
given values are typical for the MIT experiment and assume a
density of $1.5\cdot 10^{13} \,\rm cm^{-3}$.} \label{t:timescales}
\end{table}

The fastest timescale, given by $\hbar$ divided by the Feshbach coupling strength,
governs the two-body physics (in the Landau-Zener picture, this is
the inverse of the anti-crossing gap).  The timescale at which the
magnetic field sweeps through the anti-crossing is given in the
second line of Table~\ref{t:timescales}.  As pointed out above, in
the MIT experiment, even switching off the current through the
Feshbach coil is still slower than the in-trap two-body time
scale,  resulting in conversion efficiencies into molecules of
larger than 90\%.

The next fastest time scale is set by the Fermi energy, which in
the unitarity regime on resonance would set the timescale for
collisions in the normal Fermi gas, were it not for Pauli
blocking. Indeed, if we multiply the local density $n$ with the
rms velocity in the Fermi-Dirac distribution $\propto v_F$ and
with the unitarity limited cross section for elastic collisions
$\sim 4\pi/k_F^2$, we obtain a ``classical'' collision rate of $\sim
E_F/\hbar$. Also, the Fermi energy should set the time scale at
which local fluctuations of the gas density can ``heal'', as the
local chemical potential on resonance is given by $\mu \approx 0.5
E_F$. As the gas is brought into the weakly interacting regime on
the BEC-side, where $k_F a < 1$, this many-body relaxation rate
$\mu/\hbar$ should decrease to the smaller mean-field rate of a
molecular BEC. It is thus interesting to know whether the ramp is
adiabatic with respect to this local ``healing'' or relaxation,
averaged over the sweep. If we use $\hbar/E_F$ as an upper bound
for the relaxation rate around resonance, and neglect relaxation
outside this region, the relevant scale is the time it takes to
leave the strongly interacting regime. For typical parameters in
our experiment, $k_F a \approx 1$ around $750$ G, $\delta B \sim
85$ G away from resonance, and the time scale is $\sim \delta
B/\dot{B}$.

The time it takes to leave the unitarity limited region in our gas
is on the order of the Fermi time scale, and should be smaller
than the inverse collision rate. This would mean that the original
momentum distribution of fermion pairs is  ``frozen in'' during the
ramp, and the momentum distribution of the molecules at the end of
the sweep reflects that of the fermion pairs on the BCS-side.
However, since a collisional model for a weakly interacting gas
should not be taken too seriously to estimate the relaxation time,
experimental tests were required, which will be discussed in
section~\ref{s:formation}.

The ramp is non-adiabatic with respect to the time scale of the
gap, which is forced to evolve faster than it can adiabatically
respond to the change in interaction strength,
$\dot{\Delta}/\Delta \gtrsim \Delta / \hbar$. On the BCS-side of
the resonance,  the average binding energy of pairs is $\frac{3}{4}\Delta^2/E_F$.
The last condition implies that
pairs cannot adiabatically adjust their size during the fast ramp.
On the BEC-side, the pair binding energy evolves into the
molecular binding energy, $E_B = -\hbar^2/m a^2$. If one ramps far
enough on the molecular side, $a$ becomes so small and $E_B$
becomes so large that the molecular state can follow the ramp
adiabatically. This observation is used in~\cite{altm05} to split
the discussion of the field ramp into a ``sudden'' and an
``adiabatic'' part, connected at the scattering length $a^*$ for
which $\dot{E_B}/E_B = E_B/\hbar$. The ``sudden'' part is then
modelled as a projection of the initial to the final pair wave
function. One finds $a^* = (A/16\pi^2\dot{B})^{1/3}$, with $A$
given by Eq.~\ref{e:landauzenerparam}, and $k_F a^* = (3 A n /16
\dot{B})^{1/3}$, which is just the third root of the Landau-Zener
parameter entering Eq.~\ref{e:LandauZener}. The latter is $\gtrsim
1$ if the molecule conversion is efficient, as it is in our case,
directly implying that the ``sudden'' to ``adiabatic'' transition
still occurs in the strongly interacting regime, $k_F a^* \gtrsim
1$. The ramp time needed to enter the adiabatic regime is thus
smaller or about equal to the time required to leave the unitarity
region.

Finally, there is the relaxation time scale of
the gas in response to a change in the particle distribution. In a
normal Fermi gas of $N$ particles at temperatures $T \ll T_F$,
relaxation occurs via collisions of particles close to the Fermi
surface, of number $N T/T_F$. Pauli blocking reduces the available
final states for collisions by another factor of $T/T_F$, giving a
relaxation time $\tau_R \approx \hbar E_F / (k_B T)^2$. In
general, if the Fermi surface is smeared over an energy width
$\Delta E$, the relaxation time is $\tau \approx \hbar E_F /
\Delta E^2$. In the case of a (BCS-type) superfluid, $\Delta E =
\Delta$, and the relaxation time thus scales as $\tau_R = \hbar
E_F / \Delta^2$~\cite{bara04Rabi}.

\subsection{Techniques to observe cold atoms and molecules}
\label{s:techniques}
\subsubsection{Basics}
The basic techniques of imaging ultracold fermions are identical
to those for imaging bosons, which were described in great detail
in the 1999 Varenna lecture notes~\cite{kett99var}. The two main
techniques are absorption and dispersive imaging. In absorption
imaging, a laser beam tuned to the atomic resonance is absorbed by
the atoms, whose shadow image is recorded on a CCD-camera. It is
often applied after expansion of the cloud from the atom trap, as
the optical density of the trapped cloud is so high that the
absorption is strongly saturated. Detuning the laser frequency to
avoid strong absorption often results in image distortions due to
dispersive effects. Dispersive imaging relies on the phase shift
that atoms impart on the laser light and is usually implemented
with a sufficiently large detuning $\delta$ so that the phase shift
is on the order of unity.

Both types of imaging heat the sample by the recoil of Rayleigh
scattered photons. However, in dispersive imaging, the signal is
due to forward scattering which is enhanced similarly to
superradiance.  As a result, for the same amount of heating, the
number of signal photons is larger than in absorption imaging by
the resonant optical density divided by four~\cite{kett99var}.
This factor can be big (on the order of
one hundred) for large trapped clouds and has made it possible to take several dispersive
images of the same sample without noticeable smearing (so-called
non-destructive imaging).

With regard to imaging, the main difference between experiments on
ultracold fermions and bosons is that typically, in the boson case
one deals with a single spin state (an exception are experiments
on spinor-BEC~\cite{sten98spin,chan04,schm04spinor,higb05spinBEC}), while
in Fermi gases
at least two hyperfine states are involved.
Especially for the
study of imbalanced Fermi systems where the
spatial profile is
different for the two components, double-shot
imaging techniques
are essential.  In such techniques, an image of
one spin state is
rapidly succeeded by an image of the second
state. In case of
residual off-resonant absorption of the first
imaging light pulse,
the second image has to be taken after less
then a few tens of
$\mu$s, to avoid blurring as atoms move due to
photon recoil.
Current CCD cameras allow rapid successive exposure
by shifting
each pixel row of the first image underneath a mask,
where it is
safely stored during the second exposure.  Both
absorption imaging~\cite{part06phase} and dispersive imaging~\cite{shin06phase} have
been used in this way.  Another
technique that has been employed
for RF
spectroscopy~\cite{gupt03rf} was to use several independent
laser
beams, each resonant with a different atomic transition,
that were
simultaneously recorded on different parts of the CCD
chip.

\begin{figure}[t]
\begin{center}
\includegraphics[width=4in]{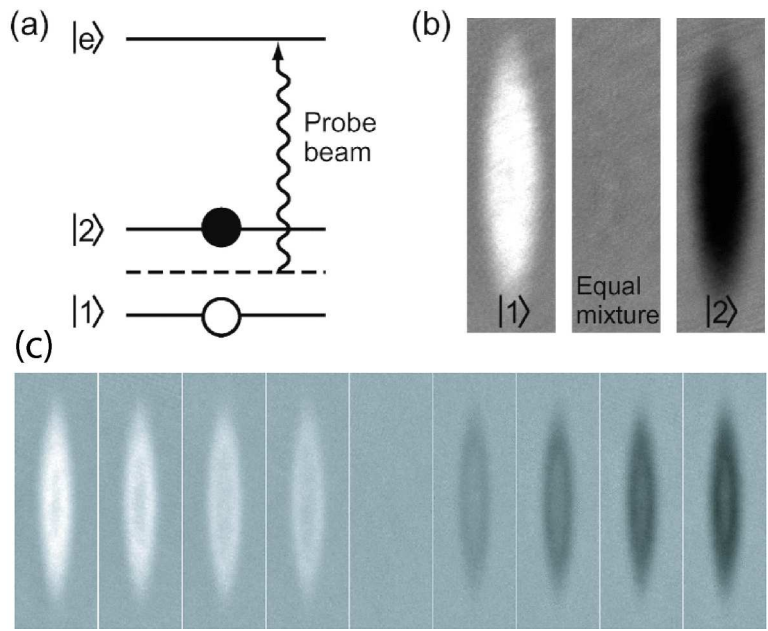}
\caption[Phase contrast technique]{In-situ phase-contrast imaging
of the density difference of two spin states of $^6$Li at the
834
G Feshbach resonance. (a) The probe beam is tuned to the red of
the resonance for state 1, and to the blue for state 2. The
resulting optical signal in the phase-contrast image is
proportional to the density difference of the atoms in the two
states. (b) Phase-contrast images of trapped atomic clouds in
state \one (left) and state \two (right) and of an equal mixture
of the two states (middle).  (c) The imbalance in the populations
$N_1, N_2$ of the two
states, defined as $(N_1-N_2)/(N_1+N_2)$,
was chosen to be
-50, -37, -30, -24, 0, 20, 30, 40 and 50\%.  The
observation of a
distinctive core shows the shell structure of the
cloud caused by
phase separation. The height of each image is
about 1 mm.  See
Ref.~\cite{shin06phase} for further details.}
\label{f:phasecontrast}
\end{center}
\end{figure}

The probe frequency in dispersive imaging can be chosen to record
a weighted sum of the column densities of the two components. In
particular, by adjusting the laser detuning to lie in between the
two resonance frequencies, phase-contrast imaging~\cite{kett99var}
then records directly the density difference~\cite{shin06phase}
without the need of subtracting two large signals.  Since spin
polarization is proportional to the density difference, this
technique was crucial in the study of imbalanced Fermi systems~\cite{shin06phase} (see Fig.~\ref{f:phasecontrast}).
\subsubsection{Tomographic techniques}
\label{s:tomographic}
Both absorption and dispersive imaging
integrate along the
line-of-sight and provide information about
the column densities.
However, by taking such projections along an
infinite number of
angles, one can reconstruct the
three-dimensional density
distribution tomographically using the
so-called inverse Radon
transformation.

If the sample has cylindrical symmetry, then one line-of-sight
integrated image $n(x,z)$ is sufficient for the reconstruction of
the atomic density $n(r,z)$ using the inverse Abel transform~\cite{drib02abel}
\begin{equation}
   n(r,z) = -\frac{1}{\pi}\int_{r}^\infty {\rm d}x\,\frac{1}{\sqrt{x^2 - r^2}}\frac{{\rm d}n(x,z)}{{\rm d}x}
   \label{e:abeltransform}
\end{equation}

This transformation takes the derivative of the column density
image, and so-to-speak inverts the line-of-sight integration.  In
particular, discontinuous jumps in the three dimensional density
appear as kinks (jumps in the derivative) of the column density.

The derivative is sensitive to noise. Due to the derivative in
Eq.~\ref{e:abeltransform}, this technique requires a very good
signal-to-noise ratio. Reducing noise by averaging (blur filter)
is not an option if high spatial resolution has to be maintained,
i.e. for the reconstruction of sharp phase boundaries.

The inverse Abel transformation has been used to reconstruct the
propagation of phonons in a Bose-Einstein
condensate~\cite{ozer02}, and to reconstruct $s$- and $d$-wave
scattering halos in the collision of two Bose-Einstein
condensates~\cite{bugg04scatt,thom04scatt}.  Tomographic
reconstruction was essential in distinguishing fermionic
superfluids with equal densities of the two components from
polarized superfluids (which have a density
imbalance)~\cite{shin06phase}.  The signal-to-noise was high
enough to identify sharp phase boundaries between the superfluid
and the normal phase in the reconstructed images, characteristic
for a first-order phase transition~\cite{shin07phasediagram}.

Tomographic reconstruction has been extended to RF spectroscopy.
RF spectra have usually been recorded for the whole cloud.
However, using sufficiently short RF pulses followed immediately
by spatial imaging of the cloud, it has been possible to record
the spatial distribution of the RF induced changes in the
population of the initial or final state~\cite{gupt03rf,shin07rf}.
When many such images are recorded for different RF frequencies,
and processed with an inverse Abel transformation, one obtains
local RF spectra~\cite{shin07rf}.  These spectra are no longer
affected by inhomogeneous broadening due to the spatially varying
density, and one even obtains a series of spectra each at a
different local density.  This method was developed to reveal a
gap in the RF spectrum of a fermionic superfluid and to observe
its homogeneous width and line shape~\cite{shin07rf}.

\subsubsection{Distinguishing atoms from molecules}

On the BEC side of the Feshbach resonance, molecules are stable,
and in order to verify the presence of molecules or to
quantitatively characterize the system, it became necessary to
distinguish atomic from molecular densities.

The following properties of these Feshbach molecules are important
for their detection. (1) The molecules are stable on the BEC side,
not too far away from the Feshbach resonance, and are dissociated
by sweeping the magnetic field across it.  (2) Close to the
Feshbach resonance, the size of the molecules is approximately
given by the atomic scattering length $a$ and can become very
large. Here, the molecular wave function has ``open channel''
character, i.e. the molecule is formed out of the same continuum
of states in which two free atoms collide in.  (3) As a
consequence, the Feshbach molecule can be regarded as ``two atoms
on a stick'', and the frequency for resonant transitions is very
close to the atomic frequencies.  More precisely, the molecules
are expected to absorb most strongly near the outer turning point
R. The excited state potential is split by the resonant van der
Waals interaction $\zeta \hbar \Gamma (\lambda /2\pi R)^3$ where
$\lambda$ is the resonant wavelength and $\zeta$ is $\pm (3/4)$ or
$\pm (3/2)$ for different excited molecular states. Therefore, as
long as $a \gtrsim \lambda/2\pi$, the Feshbach molecules resonate
at the atomic transition frequencies. For $a \lesssim
\lambda/2\pi$, the molecules should be almost transparent at the
atomic resonances.  (4) Sufficiently far away from the resonance,
the Feshbach molecules assume more and more closed channel
character, and due to the different hyperfine interaction in the
closed channel, have a magnetic moment different from the free
atoms.

Various methods use these properties to obtain the molecular
populations:

\begin{itemize}
    \item  Subtract from the signal of (atoms+molecules)
    the signal of (atoms only).  The atoms only signal is
    obtained by sweeping to low or zero magnetic field, where the
    molecules no longer absorb at the atomic resonance. The field ramp needs to be slow compared
to two-body timescales (i.e. should not ``rip'' the molecule apart),
but fast compared to losses. The
    atoms+molecules signal is obtained either at a magnetic field
close to resonance~\cite{zwie03molBEC} or after a sweep across the resonance, which
dissociates the molecules~\cite{rega03mol,cubi03,herb03cs_mol,xu03na_mol,durr04mol,joch03lith,zwie03molBEC}.
\item
Distinguish molecules by Stern-Gerlach separation.  In
ballistic expansion at magnetic fields sufficiently far away from
the Feshbach resonance, molecules can be spatially separated from
the atoms and distinguished on absorption images.  This technique
was used to detect molecules formed of bosonic atoms~\cite{herb03cs_mol,xu03na_mol,durr04mol}. For $^6$Li, it was used to measure the magnetic moment
and hence the contribution of the closed channel to the Feshbach
molecule~\cite{joch03lith}.
\item Distinguish molecules by RF spectroscopy.  The molecular RF
spectrum is shifted from the atomic line by the molecular binding
energy.  Therefore, an RF pulse populating an unoccupied state~\cite{rega03mol}
can be tuned to either spin flip atoms or
dissociate molecules.  Imaging light in resonance with this
initially unoccupied state can record the molecular population.
\end{itemize}

In most of our studies at MIT in the BEC-regime of the Feshbach resonance, the temperature of the cloud was so
low that it consisted purely of molecules, i.e. we then did not discern any atomic population using the first of the methods listed above~\cite{zwie03molBEC}. Therefore, we routinely image the
whole cloud at fields slightly below the Feshbach resonance
knowing that this (atoms+molecules) signal is purely molecular.

\section{Quantitative analysis of density distributions}
\label{c:analysis}

The purpose of imaging and image processing is to record density
distributions of the atomic cloud, either trapped or during
ballistic expansion.  All our knowledge about the properties of
cold atom systems  comes from the analysis of such images.  They
are usually compared to the results of models of the atomic gas.
Some models are exact (for the ideal gas), others are
phenomenological or approximations.  Many important models for
bosonic atoms have been presented in our 1999 Varenna notes. Here
we discuss important models for fermions, which allow us to infer
properties of the system from recorded (column) density
distributions.

\subsection{Trapped atomic gases}

\subsubsection{Ideal Bose and Fermi gases in a harmonic trap}

The particles in an atom trap are isolated from the surroundings,
thus the atom number $N$ and total energy content $E_{\rm tot}$ of
the atomic cloud is fixed. However, it is convenient to consider
the system to be in contact with a reservoir, with which it can
exchange particles and energy (grand canonical ensemble). For
non-interacting particles with single-particle energies $E_i$, the
average occupation of state $i$ is
\begin{eqnarray}
    \left<n_i\right> =  \frac{1}{e^{(E_i - \mu)/k_B T} \mp 1}
    \label{e:BoseFermidist}
\end{eqnarray}
with the upper sign for bosons, the lower sign for fermions. These
are the Bose-Einstein and Fermi-Dirac distributions, respectively.
For a fixed number of particles $N$ one chooses the chemical potential $\mu$ such that $N
= \left<N\right> = \sum_i \left<n_i\right>$.

Let us now apply these distributions to particles confined in a
harmonic trap, with trapping potential
\begin{equation}
V(\vect{r}) = \frac{1}{2} m (\omega_x^2 x^2 + \omega_y^2 y^2 +
\omega_z^2 z^2) \label{e:potential}
\end{equation}
We assume that the thermal energy $k_B T \equiv 1/\beta$ is much
larger than the quantum mechanical level spacings $\hbar
\omega_{x,y,z}$ (Thomas-Fermi approximation). In this case, the
occupation of a phase space cell
$\left\{\vect{r},\vect{p}\right\}$ (which is the phase-space
density times $h^3$) is given by Eq.~\ref{e:BoseFermidist}
\begin{eqnarray}
f(\vect{r},\vect{p}) = \frac{1}{e^{(\frac{\vect{p}^2}{2m} +
V(\vect{r}) - \mu)/k_B T} \mp 1} \label{e:FermiBose}
\end{eqnarray}
The density distribution of the thermal gas is
\begin{eqnarray}
n_{th}(\vect{r}) &=& \Intp{p}\; f(\vect{r},\vect{p})\nonumber\\
&=& \pm \frac{1}{\lambda_{ dB}^3}\, {\rm Li}_{3/2}\left(\pm
e^{\beta\left(\mu - V(\vect{r})\right)}\right) \label{e:density}
\end{eqnarray}
where $\sqrt{\frac{2\pi \hbar^2}{m k_B T}}$ is the de Broglie
wavelength. ${\rm Li}_n(z)$ is the $n^{th}$-order
Polylogarithm, defined as
\begin{eqnarray}
    {\rm Li}_n(z)\; \equiv\; \frac{1}{\pi^n} \int {\rm d}^{2n}r \frac{1}{e^{\vect{r}^2}/z - 1}\; \stackrel{n\ne 0}{=}\; \frac{1}{\Gamma(n)}\int_0^\infty {\rm d}q \frac{q^{n-1}}{e^q/z - 1}
\label{e:polylog}
\end{eqnarray}
where the first integral is over $2n$ dimensions, $\vect{r}$ is
the radius vector in $2n$ dimensions, $n$ is any positive
half-integer or zero and $\Gamma(n)$ is the Gamma-function
\footnote{The Polylogarithm appears naturally in integrals over
Bose-Einstein or Fermi-Dirac distributions.  Some
authors~\cite{huan87} use different functions for bosons
$g_n(z)={\rm Li}_n(z)$ and for fermions
$f_n(z)=-{\rm Li}_n(-z)$.  The Polylogarithm can be
expressed as a sum ${\rm Li}_n(z) = \sum_{k=1}^\infty
\frac{z^k}{k^n}$ which is often used as the definition of
the Polylogarithm.  This expression is valid for all complex
numbers $n$ and $z$ where $|z|\le 1$. The definition
given in the text is valid for all $z\le l$.

Special cases: ${\rm Li}_0(z) = \frac{1}{1/z - 1}$,
${\rm Li_1}(z) = -\ln(1-z)$. $f(\vect{r},\vect{p})$ can
be written as $\pm{\rm Li}_0(\pm
\exp[\beta(\mu-\frac{\vect{p}^2}{2m} - V(\vect{r}))])$. When integrating density distributions to obtain column densities, a useful formula is:
\begin{equation}
    \int_{-\infty}^\infty dx \;{\rm Li}_n(z\,e^{- x^2}) = \sqrt{\pi}\; {\rm Li}_{n+1/2}(z)
    \label{e:polyintegral}.
\end{equation}
 Limiting values: ${\rm Li}_n(z) \stackrel{z \ll 1}{\rightarrow} z$ and $-{\rm Li}_n(-z) \stackrel{z\rightarrow\infty}{\rightarrow} \frac{1}{\Gamma(n+1)}\; \ln^n(z)$.}.
Note that expression~\ref{e:density} is correct for any potential
$V(\vect{r})$. The constraint on the number of thermal particles
is
\begin{equation}
N_{th} = \Int{r} \; n_{th}(\vect{r})
\end{equation}
For a harmonic potential (~\ref{e:potential}), we obtain
\begin{equation}
N_{th} = \pm \left(\frac{k_B T}{\hbar \bar{\omega}}\right)^3 {\rm
Li}_3(\pm\,e^{\beta\mu}) \label{e:numberofatoms}
\end{equation}
with $\bar{\omega} = (\omega_x \omega_y \omega_z)^{1/3}$ the
geometric mean of the trapping frequencies.

In the classical limit at high temperature, we recover the
Maxwell-Boltzmann result of a gaussian distribution,
\begin{equation}
    n_{cl}(\vect{r}) = \frac{N}{\pi^{3/2} \sigma_x \sigma_y \sigma_z} e^{- \sum_i x_i^2/\sigma_{x_i}^2} \qquad {\rm with} \; \sigma_{x,y,z}^2 = \frac{2 k_B T}{m \omega_{x,y,z}^2}
\end{equation}

The regime of quantum degeneracy is reached when $\lambda_{dB}
\approx n ^{-1/3}$, or when the temperature $T \approx T_{\rm
deg}$.  The  degeneracy temperature $T_{\rm deg} =
\frac{\hbar^2}{2m k_B} n^{2/3}$ is around or below one $\mu \rm K$
for typical experimental conditions.

For {\bf bosons}, it is at this point that the ground state
becomes macroscopically occupied and the condensate forms. The
density profile of the ideal gas condensate is given by the square
of the harmonic oscillator ground state wave function:
\begin{equation}
n_c(\vect{r}) = \frac{N_0}{\pi^{3/2} d_x d_y d_z} e^{-\sum_i
x_i^2/d_{x_i}^2}
\end{equation}
where $d_{x_i} = \sqrt{\frac{\hbar}{m \omega_{x_i}}}$ are the
harmonic oscillator lengths. The density profile of the thermal,
non-condensed component can be obtained from Eq.~\ref{e:density}
if the chemical potential $\mu$ is known. As the number of
condensed bosons $N_0$ grows to be significantly larger than 1,
the chemical potential $\mu \approx - \frac{k_B T}{N_0}$ (from
Eq.~\ref{e:BoseFermidist} for $E_0 = 0$) will be much closer to
the ground state energy than the first excited harmonic oscillator
state. Thus we set $\mu = 0$ in the expression for the
non-condensed density $n_{th}$ and number $N_{th}$ and obtain
\begin{eqnarray}
n_{th}(\vect{r}) &=& \frac{1}{\lambda_{dB}^3} {\rm Li}_{3/2}(e^{-V(\vect{r})/k_B T})\\
N_{th} &=& N (T/T_C)^3\qquad \mbox{for $T<T_C$}
\end{eqnarray}
with the critical temperature for Bose-Einstein condensation in a
harmonic trap
\begin{equation}
T_C \equiv \hbar \bar{\omega}\; (N / \zeta(3))^{1/3} = 0.94 \;
\hbar \bar{\omega} N^{1/3}
\end{equation}
where $\zeta(3) = {\rm Li}_3(1) \approx 1.202$. At $T=T_C$, the
condition for Bose condensation is fulfilled in the center of the
trap, $n = {\rm Li}_{3/2}(1)/\lambda_{dB}^3 =
2.612/\lambda_{dB}^3$. For lower temperatures, the maximum density
of the thermal cloud is ``quantum saturated'' at the critical value
$n_{th} = 2.612/\lambda_{dB}^3 \propto T^{3/2}$. The condensate
fraction in a harmonic trap is given by
\begin{equation}
N_0/N = 1 - (T/T_C)^3
\end{equation}
For $T/T_C = 0.5$ the condensate fraction is already about 90\%.

For {\bf fermions}, the occupation of available phase space cells
smoothly approaches unity without any sudden transition:
\begin{equation}
  f(\vect{r},\vect{p}) = \frac{1}{e^{(\frac{\vect{p}^2}{2m} + V(\vect{r}) - \mu)/k_B T} + 1} \stackrel{T \rightarrow 0} \rightarrow \left\{%
\begin{array}{ll}
    1, & \hbox{$\frac{\vect{p}^2}{2m} + V(\vect{r}) < \mu$} \\
    0, & \hbox{$\frac{\vect{p}^2}{2m} + V(\vect{r}) > \mu$} \\
\end{array}%
\right. \label{e:fermiphasespace}
\end{equation}
Accordingly, also the density profile changes smoothly from its
gaussian form at high temperatures to its zero temperature shape:
\begin{eqnarray}
n_F(\vect{r}) &=& \Intp{p} \, f(\vect{r},\vect{p}) \stackrel{T\rightarrow 0}{\rightarrow} \int_{\left|\vect{p}\right|< \sqrt{2m(\mu-V(\vect{r}))}} \frac{{\rm d}^3\vect{p}}{(2\pi\hbar)^3}\nonumber\\
&=& \frac{1}{6\pi^2} \left(\frac{2m}{\hbar^2}\right)^{3/2}
\left(\mu - V(\vect{r})\right)^{3/2}.
\end{eqnarray}

From Eq.~\ref{e:fermiphasespace} we observe that at zero
temperature, $\mu$ is the energy of the highest occupied state of
the non-interacting Fermi gas, also called the Fermi energy $E_F$.
The (globally) largest momentum is $p_F \equiv \hbar k_F \equiv
\sqrt{2 m E_F}$, the Fermi momentum. {\it Locally}, at position
$\vect{r}$ in the trap, it is $p_F(\vect{r}) \equiv \hbar
k_F(\vect{r}) \equiv \sqrt{2 m \epsilon_F(\vect{r})} \equiv \hbar
(6\pi^2 n_F(\vect{r}))^{1/3}$ with the local Fermi energy
$\epsilon_F(\vect{r})$ which equals $\mu(\vect{r},T=0) = E_F -
V(\vect{r})$. The value of $E_F$ is fixed by the number of
fermions $N$, occupying the $N$ lowest energy states of the trap.
For a harmonic trap we obtain
\begin{eqnarray}
    N &=& \Int{r} \; n_F(\vect{r}) = \frac{1}{6} \left(\frac{E_F}{\hbar \bar{\omega}}\right)^3\nonumber\\
\Rightarrow    E_F &=& \hbar \bar{\omega} (6 N)^{1/3}
    \label{e:Ferminumber}
\end{eqnarray}
and for the zero-temperature profile
\begin{eqnarray}
n_F(\vect{r}) &=& \frac{8}{\pi^2} \frac{N}{R_{Fx} R_{Fy} R_{Fz}}
\; \left[\max \left(1 - \sum_i
\frac{x_i^2}{R_{Fi}^2},0\right)\right]^{3/2}
\label{e:Fermidensity}
\end{eqnarray}
with the Fermi radii $R_{F{x,y,z}} = \sqrt{\frac{2 E_F}{m
\omega_{x,y,z}^2}}$. The profile of the degenerate Fermi gas has a
rather flat top compared to the gaussian profile of a thermal
cloud, as the occupancy of available phase space cells saturates
at unity.

At finite $T \lesssim T_F$, we can understand the shape of the
cloud by comparing $k_B T$ with the local Fermi energy
$\epsilon_F(\vect{r})$. For the outer regions in the trap where $k_B
T \gg \epsilon_F(\vect{r})$, the gas shows a classical (Boltzmann)
density distribution $n(\vect{r}) \propto e^{-\beta V(\vect{r})}$. In
the inner part of the cloud where $k_B T \ll
\epsilon_F(\vect{r})$, the density is of the zero-temperature form $n(\vect{r})
\propto (E_F - V(\vect{r}))^{3/2}$. The Polylogarithm smoothly
interpolates between the two regimes. We notice here the
difficulty of thermometry for very cold Fermi clouds: Temperature
only affects the far wings of the density distribution. While for
thermal clouds above $T_F$, the size of the cloud is a direct
measure of temperature, for cold Fermi clouds one needs to extract
the temperature from the shape of the distribution's wings.

Note that the validity of the above derivation required the Fermi
energy $E_F$ to be much larger than the level spacing $\hbar
\omega_{x,y,z}$. For example, in very elongated traps and for low
atom numbers one can have a situation where this condition is
violated in the tightly confining radial dimensions.

\subsubsection{Trapped, interacting Fermi mixtures at zero temperature}
\label{s:trappedmixtures}
\begin{figure}
    \centering
    \includegraphics[width=4in]{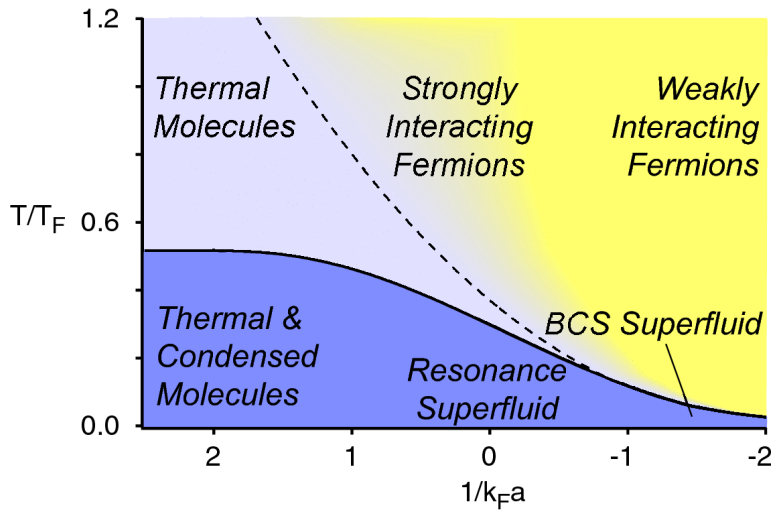}
    \caption{Phase diagram of interacting Fermi mixtures in a
    harmonic trap, as a function of temperature and interaction
    strength $1/k_F a$. Shown is the critical temperature $T_C$
    for the formation of a superfluid as a function of $1/k_F a$
    (full line) as well as the characteristic temperature $T^*$ at
    which fermion pairs start to form (dashed line),
    after~\cite{pera04temp}. The shading indicates that pair
    formation is a smooth process, not a phase transition.}
    \label{f3:phasediagram}
\end{figure}

We now consider the case of $N$ fermionic atoms equally populating
two hyperfine states (``spin up'' and ``spin down''). Atoms in
different spin states interact via $s$-wave collisions
characterized by the scattering length $a$. A dimensionless
parameter measuring the strength and sign of the interaction
strength is $1/k_F a$, essentially the ratio of the interparticle
spacing to the scattering length. For weak attractive
interactions, $1/k_F a \rightarrow -\infty$, the ground state of
the system is a BCS superfluid (see chapter~\ref{c:BECBCStheory}).
As the magnitude of the scattering length increases to a point
where $a \rightarrow \mp \infty$ diverges (thus $1/k_F a
\rightarrow 0$), a  two-body molecular bound state enters the
interparticle potential. For weak repulsive interactions, $1/k_F a
\rightarrow +\infty$, the ground state of the system is then a
Bose-Einstein condensate of weakly-interacting molecules of mass
$M = 2m$, in which two fermions of opposite spin are tightly
bound.

Fig.~\ref{f3:phasediagram} summarizes the different regimes within
this BEC-BCS crossover. We see that the character of the
Fermi mixture drastically changes as a function of temperature and
interaction strength. For temperatures $T \gg T^*$ fermions are
unpaired, and a free Fermi mixture exists on the BEC- and the
BCS-side of the phase diagram. On resonance, the mixture might
still be strongly interacting even at high temperatures, thus
possibly requiring an effective mass description of the
interacting gas. The density distribution will have the same shape
as a free Fermi gas at all interaction strengths. Below $T^*$,
fermion pairs start to form. On the BEC-side, where fermions are
tightly bound, the thermal distribution should now be that of a
gas of bosons with mass $M = 2m$. As a consequence, the cloud will
shrink. Below $T_C$, we will finally observe a superfluid,
condensed core, surrounded by a thermal cloud of molecules in the
BEC-limit, or of unpaired fermions in the BCS-limit.

In general, the calculation of density
distributions in the strongly interacting regime is a difficult
affair. Simple expressions for the densities can be derived for
superfluid gases at zero temperature, for molecular gases on the
``BEC''-side at large and positive $1/k_F a$, for weakly interacting
Fermi gases on the ``BCS''-side for large and negative $1/k_F a$,
and in the classical limit at high temperatures.

\paragraph{BEC limit}

The molecular Bose-Einstein condensate is described by a many-body
wave function $\psi(\vect{r})$ which obeys the {\rm
Gross-Pitaevskii equation}~\cite{peth02bec}
\begin{equation}
\left(-\frac{\hbar^2 \nabla^2}{2M} + V_M(\vect{r}) + g
\left|\psi(\vect{r},t)\right|^2\right) \psi(\vect{r},t) = i \hbar
\frac{\partial}{\partial t} \psi(\vect{r},t)
\label{e:grosspitaevskii}
\end{equation}
where $V_M(\vect{r})$ is the trapping potential experienced by the
molecules, and $g = \frac{4\pi \hbar^2 a_M}{M}$ describes the
intermolecular interactions. We can identify $\left|\psi\right|^2$
with the condensate density $n_c$, which for weak interactions and
at zero temperature equals the density of molecules $n_M$. The
validity of Eq.~\ref{e:grosspitaevskii} is limited to {\it weakly
interacting} gases of molecules, for which the gas parameter $n_M
a_M^3 \approx (\frac{k_F a}{6.5})^3 \ll 1$. In typical experiments
on BECs of bosonic atoms, the corresponding condition is very well
fulfilled. For a sodium BEC with $n \approx 10^{14} \, \rm
cm^{-3}$ and $a = 3.3\,\rm nm$, we have $n a^3 \approx 4 \cdot
10^{-6}$. However, for molecular condensates near a Feshbach
resonance,  this condition can be easily violated (see
chapter~\ref{c:expobservation}).

In equilibrium, the ground-state wave function is $\psi(\vect{r},t)
=  e^{-i \mu_M t/\hbar} \psi(\vect{r})$, where $\mu_M$ is the
ground state energy and is identified with the molecular chemical
potential, and $\psi(\vect{r})$ is a solution of the stationary
Gross-Pitaevskii equation
\begin{equation}
\left(-\frac{\hbar^2 \nabla^2}{2M} + V_M(\vect{r}) + g
\left|\psi(\vect{r})\right|^2\right) \psi(\vect{r}) = \mu_M
\psi(\vect{r}) \label{e:statgrosspitaevskii}
\end{equation}
In the {\it ideal gas limit}, $g n_c \ll \hbar \omega_{x, y, z}$,
we recover the harmonic oscillator result for the condensate's
density distribution $n_c(\vect{r})$. In the {\it Thomas-Fermi
limit}, on the other hand, interactions dominate over the kinetic
energy of the condensate wave function, $g n_c \gg \hbar
\omega_{x,y,z}$. Already for weakly interacting alkali gases, this
condition is very well fulfilled, with typical interaction
energies of $g n_c \sim k_B \times 150 \,\rm nK$ and
$\hbar\omega_r \approx k_B \times 5 \,\rm nK$. In this
approximation we obtain the condensate density $n_c(\vect{r}) =
\left|\psi(\vect{r})\right|^2$:
\begin{equation}
    n_c(\vect{r}) = \max \left(\frac{\mu_M - V_M(\vect{r})}{g},0\right)
\end{equation}
Thus, a condensate in the Thomas-Fermi approximation ``fills in''
the bottom of the trapping potential up to an energy $\mu_M$,
which is determined by the total number of molecules, $N_M = N/2 =
\Int{r} \; n_c(\vect{r})$. Taking $V_M(\vect{r}) = 2 V(\vect{r})$
with the harmonic trapping potential for single atoms in
Eq.~\ref{e:potential}, one obtains a parabolic density profile,
\begin{equation}
    n_c(\vect{r}) = \frac{15}{8\pi} \frac{N_M}{R_x R_y R_z} \max \left(1 - \sum_i \frac{x_i^2}{R_i^2},0\right)
    \label{e:BECparabola}
\end{equation}
where the {\it Thomas-Fermi radii} $R_i = \sqrt{\frac{2 \mu_M}{M
\omega_i^2}}$ give the half-lengths of the trapped condensate
where the density vanishes. The chemical potential is given by
\begin{equation}
    \mu_M = \frac{1}{2} \hbar \bar{\omega} \left(\frac{15 N_M a_M}{\bar{d}_{h.o.}}\right)^{2/5}
    \label{e:chempot}
\end{equation}
where $\bar{d}_{h.o.} = (d_x d_y d_z)^{1/3} = \sqrt{\hbar/M
\bar{\omega}}$ is the geometric mean of the harmonic oscillator
lengths for molecules.

Interactions thus have a major effect on the shape of the
Bose-Einstein condensate, changing the density profile from the
gaussian harmonic oscillator ground state wave function to a broad
parabola, as a result of the interparticle repulsion. The
characteristic size of the condensate is no longer given by the
harmonic oscillator length but by the generally much larger
Thomas-Fermi radius $R_{x,y,z} = d_{x,y,z}
\sqrt{\frac{\bar{\omega}}{\omega_{x,y,z}}} \left(\frac{15 N_M
a_M}{\bar{d}_{h.o.}}\right)^{1/5}$. Also the aspect ratio changes,
for example in the $x$-$y$ plane from $\frac{d_x}{d_y} =
\sqrt{\frac{\omega_y}{\omega_x}}$ to $\frac{R_x}{R_y} =
\frac{\omega_y}{\omega_x}$. Nevertheless, weakly interacting
condensates are still considerably smaller in size than a thermal
cloud at $k_B T > \mu_M$, and more dense. This leads to the clear
separation between the dense condensate in the center of the cloud
and the large surrounding thermal cloud, the ``smoking gun'' for
Bose-Einstein condensation (both in the trapped and in the
expanding cloud, see section~\ref{s:expansion} below). In the case
of strong interactions, when the chemical potential $\mu_M$
becomes comparable to $k_B T_C$, this direct signature of
condensation will be considerably weaker. In this regime we also
have to account for the mutual repulsion between the thermal cloud
and the condensate (see section~\ref{s:molecularclouds} below).

\paragraph{BCS limit}

In the weakly interacting BCS limit ($1/k_F a \rightarrow
-\infty$), pairing of fermions and superfluidity have very small
effects on the density profile of the gas. The sharp Fermi surface
in $k$-space at $k_F$ is modified only in an exponentially narrow
region of width $\sim k_F \exp(-\frac{\pi}{2 k_F |a|})$. The
density, i.e. the integral over occupied $k$-states, is thus
essentially identical to that of a non-interacting Fermi gas. The
result is Eq.~\ref{e:Fermidensity} with the number of spin-up
(spin-down) atoms $N_{\uparrow,\downarrow} = N/2$ and Fermi energy
$E_{F} = \hbar \bar{\omega} (6
N_{\uparrow,\downarrow})^{1/3} = \hbar \bar{\omega} (3 N)^{1/3}$.
As one approaches the strongly interacting regime $1/k_F a \approx
-1$, it is conceivable that the formation of the superfluid leaves
a distinct trace in the density profile of the gas, as this is the
situation in the BEC-limit, and the crossover between the two
regimes is smooth. Indeed, several theoretical studies have
predicted kinks in the density profiles signalling the onset of
superfluidity~\cite{holl01,pera04temp,ho04uni,staj05dens}. We were able to observe
such a direct signature of condensation on resonance ($1/k_F a =
0$) and on the BCS-side ($1/k_F a < 0$) in unequal Fermi mixtures
(see section~\ref{s:imbalance}). In equal mixtures, we detected a
faint but distinct deviation from the Thomas-Fermi profile on
resonance (see~\ref{s:directunitarity}).

\paragraph{Unitarity}
\label{s:unitarity}

The regime on resonance ($1/k_F a = 0$) deserves special
attention. The scattering length diverges and leaves the
interparticle distance $n^{-1/3} \sim 1/k_F$ as the only relevant
length scale. Correspondingly, the only relevant energy scale is
the Fermi energy $\epsilon_F = \hbar^2 k_F^2 / 2 m$. The regime is
thus said to be universal. The chemical potential $\mu$ can then
be written as a universal constant times the Fermi energy: $\mu =
\xi \epsilon_F$. In the trapped case, we can use this relation
locally  (local density approximation) and relate the local
chemical potential $\mu(\vect{r}) = \mu - V(\vect{r})$ to the
local Fermi energy $\epsilon_F(\vect{r}) \equiv \hbar^2
k_F(\vect{r})^2 / 2 m \equiv \frac{\hbar^2}{2 m} (6\pi^2
n_{\uparrow}(\vect{r}))^{2/3}$, where $n_{\uparrow}(\vect{r})$ is the density of atoms in one spin state. We then directly obtain a relation for the density profile $n_{\uparrow U}(\vect{r})$ of the unitary Fermi gas:
$$n_{\uparrow U}(\vect{r}) = \frac{1}{6\pi^2} \left(\frac{2m}{\xi \hbar^2}\right)^{3/2} \left(\mu - V(\vect{r})\right)^{3/2}.$$
The constraint from the number of particles in spin up, $N_\uparrow = N/2$, determines $\mu$:
\begin{eqnarray}
    N_\uparrow &=& \Int{r} \; n_{\uparrow U}(\vect{r}) = \frac{1}{6} \left(\frac{\mu}{\sqrt{\xi} \hbar \bar{\omega}}\right)^3\nonumber\\
\Rightarrow    \mu &=& \sqrt{\xi} E_F.
\end{eqnarray}
The density profile becomes
\begin{eqnarray}
n_{\uparrow U}(\vect{r}) &=& \frac{8}{\pi^2} \frac{N_\uparrow}{R_{Ux} R_{Uy} R_{Uz}}
\; \left[\max \left(1 - \sum_i
\frac{x_i^2}{R_{Ui}^2},0\right)\right]^{3/2}
\label{e:Unitaritydensity}
\end{eqnarray}
with the radii $R_{U{x,y,z}} = \xi^{1/4} R_{F{x,y,z}}$.
Table~\ref{tab:zerotemptrappeddensity} summarizes the various
density profiles of interacting Fermi mixtures.

Remarkably, the functional form $n_{\uparrow U}(\vect{r}) \propto
(\mu-V(\vect{r}))^{3/2}$ is identical to that of a non-interacting
Fermi gas.  The underlying reason is that the equation of state
$\mu \propto n^{2/3}$ has the same power-law form as for
non-interacting fermions. The universal constant $\xi$ simply
rescales the radii (by a factor $\xi^{1/4}$) and the central
density (by a factor $\xi^{-3/4}$).  One thus has direct
experimental access to the universal constant $\xi$ by measuring
the size of the cloud at unitarity (see section~\ref{s:energymeasurements}).

\begin{table}
  \centering
  \begin{tabular}{p{0.02\linewidth}c|c|c|cp{0.3\linewidth}}
&     &  BEC-limit  &   Unitarity  & BCS-limit \\
&$\frac{1}{k_F a}$ & $\infty$ & 0& $-\infty$\\ \hline
&    $\gamma$ (in $\mu \propto n^\gamma$) & 1 & 2/3 & 2/3 \\
&    $n_\uparrow(\vect{r})/n_\uparrow(\vect{0})$ & $1 - \sum_i \frac{x_i^2}{R_{i}^2}$ & $(1 - \sum_i \frac{x_i^2}{R_{Ui}^2})^{3/2}$ & $(1 - \sum_i \frac{x_i^2}{R_{Fi}^2})^{3/2}$ \\
&$n_\uparrow(\vect{0})$ & $\frac{15}{8\pi} \frac{N_\uparrow}{R_x R_y R_z}$ &
$\frac{8}{\pi^2}  \frac{N_\uparrow}{R_{Ux} R_{Uy} R_{Uz}}$ &
$\frac{8}{\pi^2}  \frac{N_\uparrow}{R_{Fx} R_{Fy} R_{Fz}}$ \\
&Radii & $R_i = \sqrt{\frac{2 \mu_M}{M \omega_i^2}}$ & $R_{Ui} =
\xi^{1/4} R_{Fi}$ & $R_{Fi} = \sqrt{\frac{2 E_F}{m \omega_i^2}}$
\end{tabular}
  \caption{Zero-temperature density profiles of a trapped, interacting Fermi mixture in the BEC-BCS crossover. The density is zero when the expressions are not positive. For definitions see the text.}
  \label{tab:zerotemptrappeddensity}
\end{table}

\subsection{Expansion of strongly interacting Fermi mixtures}
\label{s:expansion}

Intriguingly rich physics can be uncovered by the simple release
of ultracold gases from their confining trap. From the size of the
expanded cloud and the known time-of-flight one directly obtains
the energy content of the gas: the temperature in the case of
thermal clouds, the Fermi energy for non-interacting degenerate
Fermi gases, the mean-field energy for Bose-Einstein condensates.
In the case of free ballistic expansion, where no collisions occur
during expansion, the density distribution of the expanded cloud
directly reveals the original momentum distribution of the
particles in the trap. Thermal clouds will become spherical after
ballistic expansion, reflecting their isotropic momentum
distribution in the trap. The expansion of Bose-Einstein
condensates is not ballistic but mean-field driven, leading to
superfluid hydrodynamic expansion. As mean-field energy is
preferentially released in the direction(s) of tight confinement,
this allows for the famous ``smoking gun'' signature of
Bose-Einstein condensation: inversion of the condensate's aspect
ratio after expansion out of an anisotropic trap. In strongly
interacting gases the normal, uncondensed cloud can be
collisionally dense, and will expand according to classical
hydrodynamics. As particles will preferentially leave the cloud
along the narrower dimensions, where they undergo fewer
collisions, this also leads to an inversion of the cloud's initial
aspect ratio. It is thus no longer a ``smoking gun'' for
condensation, but merely for strong interactions. Expansion is
also useful to measure correlations in momentum
space~\cite{grei05corr}. Finally, in the case of harmonic
trapping, expansion of a superfluid cloud can often be described
as a ``magnifying glass'', a mere scaling of the density
distribution in the trap. This allows for example to observe
quantized vortices~\cite{zwie05vort}, which are too small to be
observable in the trap.  In this section, we show how quantitative
information can be derived from images of expanding clouds.

\subsubsection{Free ballistic expansion}
\label{s:ballistic}

Let us consider the expansion of a non-condensed thermal cloud. If
the mean free path $\lambda_c$ between collisions is longer than
the size of the trapped cloud $R$, we can neglect collisions
during expansion, which is hence ballistic. The collision rate is
$\Gamma = n \sigma v$, with density $n$, collisional cross section
$\sigma$ and thermal (root mean square) velocity $v$, which gives
$\lambda_c = v/\Gamma = 1/n \sigma$. As $R = v/\omega$ for a
harmonic trap, the condition $\lambda_c \gg R$ is equivalent to
having $\Gamma \ll \omega$, that is, the mean time interval
between collisions should be larger than a period of oscillation
in the trap.

This condition can be fulfilled for the cloud of uncondensed
molecules in the BEC limit where $1/k_F a \gg 1$ and collisions
are negligible (this has been the case also for atomic BECs with
the exception of very large thermal clouds,
see~\cite{stam98coll,shva03hydro}), and for the cloud of unpaired
fermions in the BEC- and in the BCS-limit for $k_F |a| \ll 1$ (the
exact criterion is still $\Gamma \ll \omega$). For molecules with
mass $M$, we need to replace $m \rightarrow M$ in the following
discussion.

In the ballistic case, a particle initially at point $\vect{r}_0$
in the trap, will reach point $\vect{r} = \vect{r}_0 +
\frac{\vect{p}_0}{m} t$ after expansion time $t$. We obtain the
density at point $\vect{r}$ at time $t$ by adding the
contributions from particles at all points $\vect{r}_0$ that had
the correct initial momentum $\vect{p}_0 = m(\vect{r} -
\vect{r}_0)/t$. In terms of the semi-classical distribution
$f(\vect{r},\vect{p})$, Eq.~\ref{e:FermiBose}, this is
\begin{eqnarray}
n(\vect{r},t) &=& \Int{r}_0 \Intp{p_0} \; f(\vect{r_0},\vect{p_0})\; \delta\left(\vect{r} - \vect{r_0} - \frac{\vect{p}_0}{m}t\right) \nonumber\\
&=& \Intp{p_0}\; f\left(\vect{r}-\frac{\vect{p}_0}{m}t,\vect{p}_0\right) \nonumber\\
&=& \Intp{p_0}\; \left\{\exp\left[\beta\frac{\vect{p}_0^2}{2m} +
\beta V\left(\vect{r}-\frac{\vect{p}_0}{m}t\right) - \beta\mu\right] \mp
1\right\}^{-1} \label{e:expandeddensity}
\end{eqnarray}
The integral can be carried out analytically in the case of a
harmonic potential (Eq.~\ref{e:potential}):
\begin{eqnarray}
n(\vect{r},t) &=& \Intp{p_0}\; \left\{\exp\left[\beta\sum_i \left(1+\omega_i^2 t^2\right) \frac{p_{0i}^2}{2m} +\beta\sum_i\frac{1}{2}m\frac{\omega_i^2 x_i^2}{1+\omega_i^2 t^2} - \beta\mu\right] \mp 1\right\}^{-1} \nonumber \\
&=& \pm \frac{1}{\lambda_{dB}^3}\prod_i
\frac{1}{\sqrt{1+\omega_i^2 t^2}}\; {\rm Li}_{3/2}\left[\pm
\exp\left(\beta\mu - \beta\sum_i\frac{1}{2}m\frac{\omega_i^2
x_i^2}{1+\omega_i^2 t^2}\right)\right] \nonumber
\end{eqnarray}
Note that this has the same form as the density distribution in
the trap, but with spatial dimension $i={x,y,z}$ rescaled by the
factor $b_i(t) = \sqrt{1 + \omega_i^2 t^2}$. Ballistic expansion
of a thermal (bosonic or fermionic) cloud from a harmonic trap is
thus a scaling transformation:
\begin{eqnarray}
    n(\vect{r},t) = \frac{1}{\mathcal{V}(t)}\;n\left(\frac{x}{b_x(t)},\frac{y}{b_y(t)},\frac{z}{b_z(t)},t =0\right)
\label{e:scalingtrafo}
\end{eqnarray}
where the unit volume scales as $\mathcal{V}(t) = b_x b_y b_z$.
 After an expansion time long compared to the trapping periods ($t \gg 1/\omega_i$), we have
\begin{eqnarray}
n(\vect{r},t \gg 1/\omega_i) &=& \pm \frac{1}{\lambda_{
dB}^3}\frac{1}{(\bar{\omega} t)^3}\, {\rm Li}_{3/2}\left[\pm
\exp\left(\beta\mu -
\beta\frac{1}{2}m\frac{\vect{r}^2}{t^2}\right)\right]
\label{e:expdensitylongtimes}
\end{eqnarray}
As expected, we obtain an isotropic density profile, reflecting
the original isotropic momentum distribution of the trapped gas.
Importantly, the {\it shape} of the density profile, i.e. its
variation with $\vect{r}$, becomes insensitive to the trapping
potential. Eq.~\ref{e:expdensitylongtimes} thus holds for a
general trapping geometry, for expansion times long compared to
the longest trapping period. Even if the trapping potential is not
known in detail, one can still determine the cloud's temperature
and even decide whether the gas is degenerate. Note that the {\it
momentum distribution} at point $\vect{r}$ after long expansion
times $t \gg 1/\omega_i$ has become {\it anisotropic}:
\begin{eqnarray}
f(\vect{r},\vect{p_0},t) &=& \Int{r}_0 \; f(\vect{r_0},\vect{p_0})\; \delta\left(\vect{r} - \vect{r_0} - \frac{\vect{p}_0}{m}t\right) \nonumber\\
&=&  f\left(\vect{r}-\frac{\vect{p}_0}{m}t,\vect{p}_0\right) \nonumber\\
\stackrel{t \gg 1/\omega_i}{\rightarrow} &=&
\left(\exp\left[\beta\left(\sum_i \omega_i^2 t^2
\frac{\left(p_{0i} - m \frac{x_i}{t}\right)^2}{2m}
+\frac{1}{2}\,m\frac{\vect{r}^2}{t^2} - \mu\right)\right] \mp
1\right)^{-1} \label{e:ballisticmomentum}
\end{eqnarray}
The momentum distribution at point $\vect{r}$ is ellipsoidal,
centered at $\bar{\vect{p}} = m\frac{\vect{r}}{t}$, and with
characteristic widths $\Delta p_i \propto m\frac{\Delta x_i}{t}
\propto \frac{1}{\omega_i}$ directly mirroring the ellipsoidal
atomic distribution in the trap.

\paragraph{Ballistic expansion into a saddle potential}
\label{s:saddleballistic}

In many experiments, atoms are released from an optical trap, but
magnetic fields (Feshbach fields) are still left on.  In general,
these magnetic fields are inhomogeneous, either due to technical
limitations, or deliberately, e.g. in case of the optical-magnetic
hybrid trap discussed in section~\ref{s:opticaltrap}.  We focus here on the
important case of a magnetic field created by pair of coils which
generates a saddle point potential.

So we assume that at $t>0$, the gas is not released into free
space, but into a new potential.  We define $V(\vect{r},t>0) =
\frac{1}{2} m \left(\omega_{Sx}^2 x^2 + \omega_{Sz}^2 y^2 +
\omega_{Sz}^2 z^2\right)$, and can describe expansion into
anticonfining potentials with imaginary frequencies. For
example, for the magnetic saddle potentials relevant for the MIT
experiments, the radial dimension is anticonfining and
$\omega_{Sx,y} = {\rm i} \frac{1}{\sqrt{2}}\omega_{Sz}$. In the
potential $V(\vect{r},t>0)$, particles with initial position
$\vect{r_0}$ and momentum $\vect{p_0}$ will reach the point
$\vect{r}$ with $x_i = \cos(\omega_{Si} t) x_{0i} +
\frac{1}{\omega_{Si}} \sin(\omega_{Si} t) \frac{p_{0i}}{m}$ after
expansion time $t$. The calculation of the density profile is
fully analogous to the case of free expansion, after the change of
variables $x_{0i} \rightarrow \tilde{x}_{0i}/\cos(\omega_{Si}t)$
and the substitution $t \rightarrow
\sin(\omega_{Si}t)/\omega_{Si}$. We again obtain a scaling
transformation, Eq.~\ref{e:scalingtrafo}, but for this case with
scaling parameters $b_i(t) = \sqrt{\cos^2(\omega_{Si}t) +
\frac{\omega_i^2}{\omega_{Si}^2} \sin^2(\omega_{Si}t)}$. For
expansion into the magnetic saddle potential, this gives
$b_\perp(t) = \sqrt{\cosh^2(\frac{1}{\sqrt{2}}\omega_{Sz}t) +
\frac{2\omega_\perp^2}{\omega_{Sz}^2}
\sinh^2(\frac{1}{\sqrt{2}}\omega_{Sz}t)}$ and $b_z(t) =
\sqrt{\cos^2(\omega_{Sz}t) + \frac{\omega_z^2}{\omega_{Sz}^2}
\sin^2(\omega_{Sz}t)}$. For the MIT trap, the initial axial
trapping potential is dominated by the magnetic field curvature,
while the initial radial potential is almost entirely due to the
optical trap. After switching off the optical trap, we have
$\omega_{Sz} = \omega_z$ and $\omega_{Sx} = {\rm
i}\frac{1}{\sqrt{2}}\omega_z$. In this case, $b_z(t) = 1$ and the
cloud expands only into the radial direction.

\subsubsection{Collisionally hydrodynamic expansion}
\label{s:collhydro}

If the mean free path $\lambda_c$ is short compared to the cloud
size, the gas is in the hydrodynamic regime, and collisions during
expansion can no longer be neglected.  Collisions will tend to
reestablish local thermal equilibrium, in particular an isotropic
momentum distribution. For anisotropic traps, this directly leads
to anisotropic expansion, in strong contrast to the ballistic
case: Particles trying to escape in one direction suffer
collisions that redistribute their momenta equally in all
directions. The escape is hindered more for the weakly confined
directions where the cloud is long initially and particles can
undergo more collisions. For cylindrically symmetric clouds, this
leads to an inversion of the aspect ratio of the cloud during
expansion.

Hydrodynamic expansion can take place for $1/k_F |a| < 1$, which
includes (for $a>0$) strongly interacting clouds of
uncondensed molecules, and (for $a<0$) a strongly interacting,
normal Fermi mixture.  There is no sharp boundary between
molecular hydrodynamics and fermionic hydrodynamics, since $1/k_F
|a| < 1$ is the strongly interacting regime where many-body
physics dominates and the single-particle description (molecules
in one limit, unbound fermions in the other) is no longer valid.

In the hydrodynamic regime, the evolution of the gas is governed
by the continuity equation for the density $n(\vect{r},t)$ and,
neglecting friction (viscosity), the Euler equation for the
velocity field $\vect{v}(\vect{r},t)$:
\begin{eqnarray}
\frac{\partial n}{\partial t} + \nabla \cdot (n \vect{v}) &=& 0 \\
m \frac{{\rm d}\vect{v}}{{\rm d}t} = m \frac{\partial
\vect{v}}{\partial t} + m (\vect{v} \cdot \nabla) \vect{v} &=& -
\nabla V(\vect{r},t) - \frac{1}{n} \nabla P(\vect{r},t)
\end{eqnarray}
where $P$ is the pressure. Friction is
negligible deep in the hydrodynamic regime, when the mean free
path approaches zero.  The Euler equation is simply Newton's law
for the collection of gas particles at $\vect{r}$. In steady
state, we recover the equilibrium solution
\begin{equation}
\nabla P_0(\vect{r}) = n_0(\vect{r}) \nabla \mu_0(\vect{r}) = -
n_0(\vect{r}) \nabla V(\vect{r},0) \label{e:initialpressure}
\end{equation}
where we have used the expression for the local chemical potential
$\mu_0(\vect{r}) = \mu - V(\vect{r})$.
\paragraph{Scaling solution for harmonic potentials}
In the case of free expansion, the potential $V(\vect{r},t)$ is
the initial harmonic trapping potential for $t<0$, with radial and
axial trapping frequencies $\omega_{\perp}(0)$ and
$\omega_{z}(0)$, and zero for $t>0$. We can more generally
consider here an arbitrary time variation $\omega_\perp(t)$ and
$\omega_z(t)$ of the trapping frequencies. For this case, the
Euler equation allows a simple scaling solution for the
coordinates and velocities~\cite{kaga97bose}
\begin{eqnarray}
x_i(t) &=& b_i(t)\, x_{0i} \nonumber\\
v_i(t) &=& \frac{\dot{b}_i}{b_i}\, x_i(t)
\end{eqnarray}
with initial conditions $b_i(0) = 1$ and $\dot{b}_i(0) = 0$. The
unit volume scales as $\mathcal{V}(t) = b_x\,b_y\,b_z$, the
density varies as $n(\vect{r},t) = n_0(\vect{r}_0)/\mathcal{V}$,
where the fluid element at initial position $\vect{r}_0$ has
propagated to $\vect{r}$ at time $t$.

\paragraph{Pressure}
The thermodynamic properties of a simple fluid or gas only depend
on three variables, that are, in the grand canonical description,
the temperature $T$, the chemical potential $\mu$ and the volume
$V$. From the grand canonical partition function Z, one obtains in
this case the pressure $P = k_B T \frac{\ln Z}{V}$. For a
non-interacting, ideal gas of bosons or fermions, the average
energy is $E = \frac{3}{2} k_B T \ln Z$, leading to the relation
$PV = \frac{2}{3} E$. This equation is no longer true for an
interacting gas, for example the van der Waals gas. It is very
remarkable, then, that this relation nevertheless holds also for
the strongly interacting, unitary gas on resonance, for all
temperatures~\cite{ho04uni,thom05virial}~\footnote{On resonance,
universality requires that the energy $E = N \epsilon_F f(T/T_F)$
with a universal function $f$. Entropy can only be a function of
$T/T_F$, so adiabaticity requires this ratio to be constant. The
pressure is then $P = -\partial E/\partial V|_{S,N} = -N f(T/T_F)
\partial \epsilon_F/\partial V = \frac{2}{3} E/V$.}. Under an
adiabatic expansion, the energy $E$ changes according to ${\rm d}E
= - P {\rm d}V$. Hence $\frac{3}{2}{\rm d} (PV) = \frac{3}{2}
(V{\rm d} P + P{\rm d} V) = - P {\rm d}V$, which leads to the law
$P V^{5/3} = {\rm const}$ for adiabatic expansion. The pressure
thus scales as $\mathcal{V}^{-5/3}$, and the force, using
Eq.~\ref{e:initialpressure},
\begin{eqnarray}
    -\frac{1}{n} \frac{\partial}{\partial x_i} P(\vect{r},t) &=& -\frac{\mathcal{V}}{n_0} \frac{1}{b_i} \frac{\partial}{\partial x_{0i}} \frac{P_0(\vect{r}_0)}{\mathcal{V}^{5/3}} = \frac{1}{b_i\mathcal{V}^{2/3}} \frac{\partial}{\partial x_{i0}} V(\vect{r}_0,0)\nonumber \\
&=& \frac{1}{b_i\mathcal{V}^{2/3}} m \omega_{i}^2(0) x_{i0}
\end{eqnarray}

The Euler equations then reduce to equations for the scaling
parameters $b_i(t)$, which can be solved numerically:
\begin{eqnarray}
    \label{e:eulerclass}
    \ddot{b}_i = -\omega_i^2(t)\, b_i + \frac{\omega_{i}^2(0)}{b_i\mathcal{V}^{2/3}}
\end{eqnarray}

In the following section we will see that superfluid hydrodynamics
leads to very similar scaling equations, with the exponent $2/3$
for the volume scaling parameter $\mathcal{V}$ replaced by the
parameter $\gamma$ in the equation of state of the superfluid
$\mu(n) = n^\gamma$. The discussion of free expansion, the
long-time behavior, inversion of the aspect ratio etc. will be
identical for superfluid hydrodynamics, so we defer the topic until
the next section.

\paragraph{From ballistic to hydrodynamic expansion}

The regime in between ballistic, collisionless expansion and pure
hydrodynamic, collisional expansion can be treated approximately.
For the effects of interactions on a classical gas,
see~\cite{guer99osc,pedr03}, for the case of Fermi gases with
attractive interactions, see~\cite{meno02}.

\subsubsection{Superfluid hydrodynamic expansion}

In the simplest (scalar) case, a superfluid is described by a
macroscopic, complex order parameter $\psi(\vect{r},t) =
\sqrt{n(\vect{r},t)} e^{{\rm i} \phi(\vect{r},t)}$ parameterized by
the {\it superfluid density} $n(\vect{r},t)$ and a phase
$\phi(\vect{r},t)$. The dynamics of the order parameter are well
described by a time-dependent Schr\"odinger equation of the type
\begin{equation}
    {\rm i} \hbar \frac{\partial}{\partial t} \psi(\vect{r},t) = \left(-\frac{\hbar^2}{2m}\nabla^2 + V(\vect{r},t) + \mu(n(\vect{r},t))\right)\psi(\vect{r},t)
\label{e:GPequation}
\end{equation}
where $\mu(n)$ is the chemical potential given by the equation of
state of the superfluid. In the case of weakly interacting BECs,
this is the Gross-Pitaevskii equation for the condensate
wave function from section~\ref{s:trappedmixtures}. For fermionic superfluids, a formally similar equation is the Ginzburg-Landau equation, which is however valid only close to $T_C$.
Rewriting Eq.~\ref{e:GPequation}
in terms of the superfluid density $n$ and velocity $\vect{v}$,
neglecting the curvature $\nabla^2 \sqrt{n}$ of the magnitude of
$\psi$ and using the fact that the superfluid is
irrotational $\nabla \times \vect{r} = 0$, we arrive again at the
continuity equation and the Euler equation for classical inviscous
flow:
\begin{eqnarray}
\frac{\partial n}{\partial t} + \nabla \cdot (n \vect{v}) &=& 0\\
m \frac{\partial \vect{v}}{\partial t} + m (\vect{v} \cdot \nabla)
\vect{v} &=& - \nabla \left(V+ \mu(n)\right)
\end{eqnarray}
The validity of these hydrodynamic equations is restricted to superfluids whose
healing length is much smaller than the sample size and thus, for fermionic superfluids in
a harmonic trap, for a superfluid gap larger than
the harmonic oscillator energies $\hbar
\omega_{x,y,z}$~\cite{meno02}.
For a power-law equation of state $\mu(n) \propto n^\gamma$, the
equations allow a scaling solution for (possibly time-varying)
harmonic potentials. The scaling parameters $b_i(t)$ are
given by the differential equations~\cite{cast96,kaga97bose,meno02,hu04coll}
\begin{eqnarray}
    \ddot{b}_i = -\omega_i^2(t)\, b_i + \frac{\omega_{i}^2(0)}{b_i\mathcal{V}^{\gamma}}
    \label{e:euler}
\end{eqnarray}
Important limiting cases in the BEC-BCS crossover are:
\begin{itemize}
\item[-] BEC-limit ($1/k_F a \gg 1$): Here, the mean-field repulsion between molecules leads to a chemical potential per fermion $\mu(n) = \frac{\pi\hbar^2 a_M n}{m}$, so $\gamma = 1$.
\item[-] BCS-limit ($1/k_F a \ll -1$): In the BCS-limit, the dominant contribution to the chemical potential comes from the kinetic energy of the constituent fermions, given by the Fermi energy. So here $\mu(n) = \epsilon_F \propto n^{2/3}$ and $\gamma = 2/3$.
\item[-] Unitarity limit ($1/k_F a = 0$): In the unitarity limit, the only remaining energy scale is the Fermi energy. One necessarily has $\mu(n) \propto \epsilon_F \propto n^{2/3}$ and $\gamma = 2/3$, just as in the BCS-limit.
\end{itemize}

Note that the scaling laws for the BCS- and the unitarity limit~\cite{cast04scal}
are identical to those found for a collisionally hydrodynamic gas
in section~\ref{s:collhydro}.
For a derivation of superfluid hydrodynamics in the BCS-limit, we refer the reader to the contribution of Y. Castin to these lecture notes.

\begin{figure}
    \centering
    \includegraphics[width=3in]{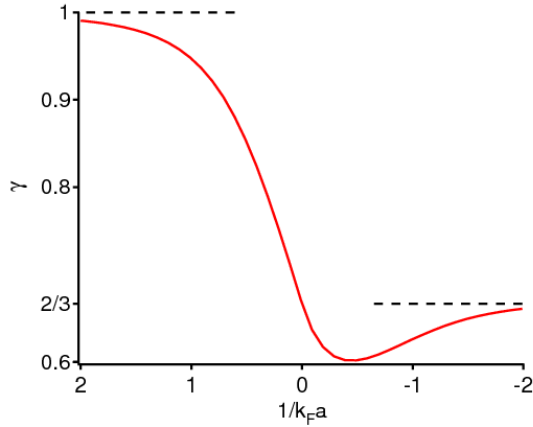}
    \caption{The exponent $\gamma$ as a function of the interaction parameter $1/k_F a$. $\gamma$ approximately describes the superfluid equation of state $\mu(n) \sim n^\gamma$ in the BEC-BCS crossover. A similar figure can be found in~\cite{hu04coll}.}
    \label{f:gamma}
\end{figure}

The Leggett ansatz (see section~\ref{s:crossoverwavefunction}) allows to interpolate
between the BEC- and the BCS-regime and gives a chemical potential
$\mu(n)$ that correctly captures the physics in the two limits.
With its help, we can define an effective exponent $\gamma =
\frac{n}{\mu}\frac{\partial \mu}{\partial n}$ and write $\mu(n)
\simeq n^\gamma$, assuming that $\gamma$ varies slowly with the
interaction parameter $1/k_F a$. This exponent, shown in
Fig.~\ref{f:gamma}, attains the correct limiting values in the
BEC- and the BCS-limit, as well as on resonance, so we may use it
for the present purpose as an approximate description of the gas'
equation of state throughout the crossover.

\paragraph{In-trap density profile}
The in-trap density profile of the superfluid at zero temperature can be deduced from
the Euler equations in steady state. Neglecting kinetic energy
$\frac{1}{2}\, m\, \vect{v}^2$ (Thomas-Fermi approximation), the
equation simply reads $V(\vect{r}) + \mu(n(\vect{r})) = {\rm
const.} = \mu(n(\vect{0}))$. For the power-law equation of state
$\mu(n)\propto n^\gamma$, we directly obtain
\begin{equation}
    n(\vect{r}) \propto \left(\mu(n(\vect{0})) - V(\vect{r})\right)^{1/\gamma}
\end{equation}
for $\mu(n(\vect{0})) > V(\vect{r})$ and zero otherwise. For a BEC
and harmonic trapping, we recover the inverted parabola,
Eq.~\ref{e:BECparabola}, for a BCS superfluid in the limit of weak
interactions the density distribution of an ideal Fermi gas,
Eq.~\ref{e:Fermidensity}. Note that in the crossover $1/k_F |a|
\lesssim 1$, the correct calculation of the density profile is
less straightforward, as the parameter $1/k_F(\vect{r}) a$ depends
on position, and the equation of state varies across the cloud.
The power-law approach, using a fixed $\gamma =
\gamma(1/k_F(\vect{0}) a)$, will only provide an approximate
description. Fortunately, on resonance evidently $1/k_F(\vect{r})a
= 0$ across the entire cloud, and the power-law equation of state
becomes exact at $T=0$.

\paragraph{Free expansion out of a cylindrically symmetric trap}

In this case $\omega_i(t>0) = 0$, and $\omega_x(0) = \omega_y(0)
\equiv \omega_\perp$. We have
\begin{eqnarray}
\label{e:cylindricscaling1}
\ddot{b}_\perp &=& \frac{\omega_\perp^2}{b_\perp^{2\gamma+1}
b_z^{\gamma}}\\ \ddot{b}_z &=&
\frac{\omega_z^2}{b_\perp^{2\gamma} b_z^{\gamma+1}}
\label{e:cylindricscaling2}
\end{eqnarray}
The MIT trap is cigar-shaped, with an aspect ratio of short to
long axes $\epsilon = \omega_z/\omega_\perp \ll 1$. In such a
case, expansion is fast in the radial, initially tightly confined
dimensions, whereas it is slow in the $z$-direction. For times
short compared to $\tau_\epsilon = \frac{1}{\omega_z \epsilon}$,
many axial trapping periods, we can set $b_z \approx 1$ on the
right side of Eqs.~\ref{e:cylindricscaling1} and \ref{e:cylindricscaling2}, decoupling the
transverse from the axial expansion. For $\gamma = 1$, the case of
a Bose-Einstein condensate of tightly bound molecules, the
simplified equations for $t\ll\tau_\epsilon$ have an analytic
solution~\cite{cast96,kaga97bose}: $b_\perp(t) = \sqrt{1 +
\omega_\perp^2 t^2}$ and $b_z(t) = 1 + \epsilon^2
\left(\omega_\perp t \arctan(\omega_\perp t) - \ln\sqrt{1 +
\omega_\perp^2 t^2}\right)$.  For long times $t$, the
expansion is linear in time: $b_\perp(t) = \omega_\perp t$ for $t \gg 1/\omega_\perp$ and
$b_z(t) = (\pi/2) \epsilon^2 \omega_\perp t$ for $t \gg \tau_\epsilon$. Note that the radial
expansion accidentally follows the same scaling law as that of a
ballistically expanding normal cloud.

\begin{figure}
    \centering
    \includegraphics[width=3in]{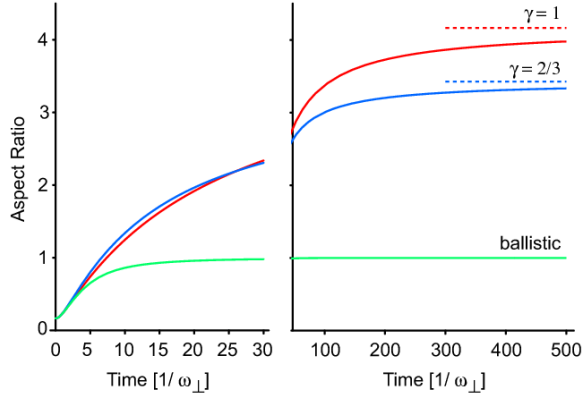}
    \caption{Aspect ratio $\epsilon(t) = R_x(t)/R_z(t)$ as a function of time for the MIT trap ($\epsilon = 1/6$) in ballistic, collisional or superfluid hydrodynamic expansion ($\gamma = 2/3$) and superfluid hydrodynamic expansion of a molecular BEC ($\gamma = 1$). }
    \label{f:aspect}
\end{figure}

\begin{figure}
    \centering
    \includegraphics[width=3in]{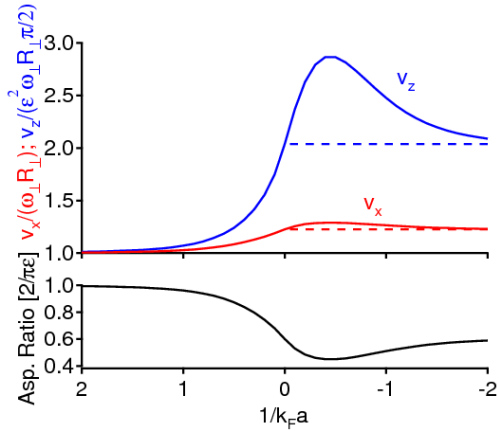}
    \caption{Asymptotic velocities and aspect ratio for hydrodynamic expansion out of a very elongated cigar-shaped trap ($\epsilon = \omega_z/\omega_x \ll 1$), as a function of the interaction parameter $1/k_F a$. The dashed lines show the asymptotic values in the BCS-limit. A similar figure can be found in~\cite{hu04coll}.}
    \label{f:velocityplot}
\end{figure}

The general behavior of the expanding gas is the same for all
relevant $\gamma$. Driven either by repulsive interactions
(BEC-case) or by kinetic energy (BCS-case), the gas first expands
radially at constant acceleration $\ddot{R}_\perp(t\ll
1/\omega_\perp) = R_\perp(0) \omega_\perp^2$, and over a radial
trapping period reaches a final expansion velocity $\dot{R}(t\gg
1/\omega_\perp) \approx \omega_\perp R_\perp(0)$. The axial size
grows as $b_z(t) -1 \approx \epsilon^2 \omega_\perp t$, leading to
an inversion of the cloud's aspect ratio from initially $\epsilon$
to $\sim 1/\epsilon$. This inversion is in contrast to the
isotropic aspect ratio of a ballistically expanding gas, and is
thus  characteristic for hydrodynamic expansion, which can be of
collisional {\it or} of superfluid origin. Fig.~\ref{f:aspect} and
table~\ref{t:expansionsummary} summarize the time evolution of the
cloud's radii and aspect ratios for $\gamma = 1$ (BEC) and $\gamma
= 2/3$ (BCS and Unitarity), while Fig.~\ref{f:velocityplot}
compares the long-time behavior of the velocities and aspect
ratios across the BEC-BCS crossover. For expansion out of an
elongated cigar-shaped trap and $\gamma = 2/3$, which holds in the
BCS-limit, at unitarity, but also for a collisionally hydrodynamic
gas, the asymptotic expansion velocity is $v_\perp =
\sqrt{\frac{3}{2}} \omega_\perp R_\perp(0) \approx 1.22\;
\omega_\perp R_\perp(0)$. This can be understood by noting that
the cloud's kinetic energy, initially distributed isotropically,
is released only into the radial direction during hydrodynamic
expansion, so $\frac{1}{2} m v_\perp^2 = \frac{3}{2} \mu =
\frac{3}{4} m \omega_\perp^2 R_\perp(0)^2$.

\begin{table}
  \centering
  \begin{tabular}{p{0.03\linewidth}r|c|c|cp{0.25\linewidth}}
&   & Ballistic & Hydrodynamic (BEC) & (BCS, unitarity, collisional) \\ \hline
&  $b_\perp(t)$ & $\sqrt{1+\omega_\perp^2 t^2}$ & $\sqrt{1+\omega_\perp^2 t^2}$ & $\sim 1.22\; \omega_\perp t$ \\
&  $b_z(t)$ & $\sqrt{1+\omega_z^2 t^2}$ & $1 + \epsilon^2 (\omega_\perp t \arctan(\omega_\perp t)$ & $\sim 2.05 \;\frac{\pi}{2}\epsilon^2 \omega_\perp t$ \\
&& & $\qquad \quad- \ln\sqrt{1 + \omega_\perp^2 t^2})$ & \\
&  AR & 1 & $\frac{2}{\pi} \frac{1}{\epsilon}$ & $\sim 0.60\; \frac{2}{\pi} \frac{1}{\epsilon}$
\end{tabular}
  \caption{Comparison between ballistic and hydrodynamic expansion.
Formulas for hydrodynamic expansion assume a long cigar-shaped
trap ($\epsilon = \omega_z/\omega_x \ll 1$), formulas for the
aspect ratio (AR) and for the BCS, Unitarity, collisional limit
give the asymptotic behavior. The formula for BEC-expansion is valid at short times $t \ll \omega_\perp/\omega_z^2$, but also captures the correct long time limit.}\label{t:expansionsummary}
\end{table}

\paragraph{Hydrodynamic expansion into a saddle potential}

As discussed in section~\ref{s:saddleballistic}, expansion may not occur into free
space, but into an inhomogeneous magnetic field which is often
described by a saddle potential.

The Euler equations (\ref{e:euler}) now read for $t>0$
\begin{equation}
    \ddot{b}_i = -\omega_{S,i}^2\, b_i + \frac{\omega_{i}^2(0)}{b_i\mathcal{V}^{\gamma}}
\end{equation}
Here, $\omega_{S,i}$ are the real or imaginary frequencies
characterizing the saddle point potential. These equations
typically need to be solved numerically. For a Bose-Einstein
condensate of molecules expanding from long cigar-shaped traps
($\epsilon \ll 1$), the {\it radial} equation again allows for an
analytic solution identical to that for a ballistically expanding,
non-interacting gas. One obtains
\begin{equation}
b_\perp(t) = \sqrt{\cosh^2\left(\frac{1}{\sqrt{2}}\omega_{Sz}t\right) +
\frac{2\omega_\perp^2}{\omega_{Sz}^2}
\sinh^2\left(\frac{1}{\sqrt{2}}\omega_{Sz}t\right)}.
\end{equation}
However, the {\it axial} cloud size behaves drastically different
from a non-interacting cloud. For $\omega_{Sz} = \omega_z$, the
axial cloud size of a non-interacting gas would never change
($b_z(t) = 1$), whereas a hydrodynamic gas, released into the
radial dimensions, will start to {\it shrink} axially under the
influence of the confining axial potential. The cloud's energy
(interaction energy for a BEC, kinetic energy for a BCS
superfluid) escapes radially, hence there is not sufficient
pressure to maintain the axial cloud size.

Further discussions of superfluid hydrodynamics and scaling transformations can be found in the contributions of Y. Castin and S. Stringari to these proceedings.

\subsection{Fitting functions for trapped and expanded Fermi gases}

In the preceding sections we derived the 3D density distribution
of a Fermi mixture in various regimes. However, all imaging
techniques record column densities, density profiles integrated
along the line of sight (the $z$-axis in the following).

For condensed gases, where $n(\vect{r}) \approx n(\vect{0})\, (1 -
V(\vect{r})/\mu)^{1/\gamma}$, one obtains the column density
\begin{equation}
    n_{\rm 2D, c}(x,y) =  n_c\left(1 - \frac{x^2}{R_x^2} - \frac{y^2}{R_y^2}\right)^{\frac{1}{\gamma}+\frac{1}{2}}
\end{equation}
For thermal Bose (molecular) and Fermi clouds, we have
\begin{equation}
\label{e:coldens}
    n_{\rm 2D}(x,y) = n_{\rm 2D,0}\, {\rm Li}_2\left(\pm \exp\left[\beta\mu - \beta\frac{1}{2}m(\omega_x^2 x^2 + \omega_y^2 y^2)  \right]\right)/\,{\rm Li}_2\left(\pm\, e^{\beta\mu}\right)
\end{equation}

In the following, we will discuss the fitting functions valid in the different regimes of interaction, and the derived quantities.

\subsubsection{Non-interacting Fermi gases}
\paragraph{Cloud size}
In the classical regime at $T/T_F \gg 1$, the characteristic cloud
size is given by the gaussian radius $\sigma_i = \sqrt{\frac{2 k_B
T}{m\omega_i^2}}$. In the degenerate regime, however, the cloud
size saturates at the Fermi radius $R_{Fi} = \sqrt{\frac{2
E_F}{m\omega_i^2}}$. It is thus convenient to define a fit
parameter that interpolates between the two limits:
\begin{eqnarray}
    R_i^2 &=& \frac{2 k_B T}{m\omega_i^2} \; f(e^{\mu\beta}) \rightarrow \left\{%
\begin{array}{ll}
    \sigma_i, & T/T_F \gg 1 \\
    R_{Fi}, & T/T_F \ll 1 \\
\end{array}%
\right.    \nonumber \\
    \mbox{where } f(x) &=& \frac{{\rm Li}_1(-x)}{{\rm Li}_0(-x)} = \frac{1 + x}{x} \ln(1+x)
\end{eqnarray}
For all temperatures, $R_i$ is thus directly related to the
physical size of the cloud, and thus a better choice as a fit
parameter than $\sigma_i$, which goes to zero at $T=0$, or
$\sqrt{\frac{2\mu}{m\omega_i^2}}$, which goes to zero around
$T/T_F = 0.57$. Numerically, using $R_i$ is easier to implement
than using the root mean square radius of the cloud
\begin{equation}
    \left<x_i^2\right> = \frac{k_B T}{m \omega_i^2}\; \frac{{\rm Li}_4(-e^{\mu\beta})}{{\rm Li}_3(-e^{\mu\beta})}
\end{equation}

\paragraph{Fitting function}

The fit function used for the density profiles of Fermi clouds is
then in 2D
\begin{equation}
    n_{\rm 2D}(x,y) = n_{\rm 2D,0}\, \frac{{\rm Li}_2\left(\pm \exp\left[q - \left(\frac{x^2}{R_x^2} + \frac{y^2}{R_y^2}\right)\,f(e^q)\right]\right) }{{\rm Li}_2\left(\pm e^q\right)}
\end{equation} and for 1D
\begin{equation}
    n_{\rm 1D}(x) = n_{\rm 1D,0}\, \frac{{\rm Li}_{5/2}\left(\pm \exp\left[q -\frac{x^2}{R_x^2} \,f(e^q)\right]\right) }{{\rm Li}_{5/2}\left(\pm e^q\right)}.
\end{equation}
The parameter $q = \mu\beta$, the logarithm of the fugacity,
determines the {\it shape} of the cloud. For a small fugacity
(large and negative $q$), the above functions reduce to the simple
gaussian distribution of thermal clouds. For high fugacity (large
and positive $q$), they tend to the
zero-temperature distribution $n_{\rm 2D,0}(1 -
\frac{x^2}{R_{Fx}^2})^2$ (in 2D) and $n_{\rm 1D,0}(1 -
\frac{x^2}{R_{Fx}^2})^{5/2}$ (in 1D).

\paragraph{Derived quantities}

{\it Degeneracy}
The degeneracy parameter $T/T_F$ can be calculated by combining
Eq.~\ref{e:numberofatoms} with Eq.~\ref{e:Ferminumber}:
\begin{equation}
    \frac{T}{T_F} = \left[-6\, {\rm Li}_3(-e^{q})\right]^{-1/3}
    \label{e:Degeneracy}
\end{equation}
This parameter depends only on the {\it shape} of the cloud. A
characteristic point where shape deviations due to quantum
statistics start to play a role is the point where $\mu$ changes
sign, and we see from Eq.~\ref{e:Degeneracy} that this occurs at
$T/T_F \approx 0.57$. Many non-ideal aspects of imaging, such as
finite resolution, out of focus imaging, saturation, heating of
the cloud by the probe pulse etc., tend to wash out the
non-gaussian features of a highly degenerate Fermi cloud and hence
lead to a larger value of $T/T_F$. However, dispersive effects due
to non-resonant imaging light can potentially mimic sharp edges of
the cloud, which the fitting routine would then falsely interpret
to result from a very low $T/T_F$. It is clear that care has to be
taken when determining the degeneracy parameter from the shape of
the cloud alone.

\paragraph{Temperature}
\label{s:temperature}
The size of the cloud and the shape parameter $q$ give the
temperature as
\begin{equation}
    k_B T = \frac{1}{2} m \omega_i^2 \frac{R_i^2}{b_i(t)^2} \frac{1}{f(e^q)}
    \label{e:thermometry}
\end{equation}
where we have used the expansion factor $b_i(t)$ from
section~\ref{s:expansion}. We recall that $b_i(t) =
\sqrt{1+\omega_i^2 t^2}$ for the free expansion of a
non-interacting Fermi gas. For low temperatures $T \ll T_F$,
$f(e^{\mu\beta}) \rightarrow \mu\beta = \mu/k_B T$ and $R_i =
b_i(t)\,R_{Fi}$. In this case, temperature only affects the wings
of the density distribution, where the {\it local}
$T/T_F(\vect{r})$ is still large. In fact,
\begin{equation}
    n_{\rm 1D}(x) \propto \left\{%
\begin{array}{ll}
    (1 - \frac{x^2}{R_{Fx}^2})^{5/2} & \hbox{for } x \ll R_{Fx} \\
    e^{-\frac{x^2}{\sigma_x^2}} & \hbox{for } x \gg R_{Fx}, \\
\end{array}%
\right.
\end{equation}
and we see that temperature only affects the cloud's wings beyond
the zero-temperature Fermi radius. Thermometry of very low
temperature Fermi clouds is thus difficult, limited by the
signal-to-noise ratio in the low-density wings of the
distribution. This is different from thermometry of thermal clouds
at high temperature $T \gg T_F$, where the entire size of the
cloud $\sigma_i$ directly gives the temperature.

Because of the sensitivity to the cloud's wings, thermometry is
more robust when the full 2D distribution is used for the fit.
Alternatively, one can rely on the known trapping geometry plus
the local density approximation and perform an average over the
elliptical equipotential lines in the $x$-$y$ plane (line of sight
integration necessarily mixes points at different values of the
potential energy.) As the number of points included in the average
grows with the distance from the cloud's center, the
signal-to-noise will actually be best in the wings. Such an
average is superior to a simple integration along the $x$-axis,
for example, as this will more strongly mix regions that have
different local $T/T_F$.

The ideal gases (Fermi, Bose, Boltzmann) are the only systems for
which we have an exact description.  Therefore, they are
attractive as a thermometer, when brought in contact with strongly
interacting systems.  This concept has been recently carried out
by determining absolute temperatures for imbalanced Fermi gases at
unitarity~\cite{shin07phasediagram}.  In these systems, for sufficiently
high imbalance, the majority cloud extends beyond the minority
cloud, and is (locally) an ideal gas.  Therefore, in Ref.~\cite{shin07phasediagram} the spatial
wings of these clouds could be fitted with the functions for the
ideal Fermi gas discussed in this section, and absolute
temperatures for the superfluid phase transition could be
determined. The fitting of the majority wings had to be done with in-trap
profiles, which required to address the effect of anharmonicities of the
optical trapping potential.  Usually, for thermometry, ballistic
expansion is preferable since velocity distributions are
independent of the shape of the trapping potential.  However, in
the case of imbalanced Fermi gases, the atoms in the wings
can collide with the strongly interacting core during expansion,
modifying their velocity distribution.

Another way to perform ideal gas thermometry is done by converting
the sample to a non-interacting system by sweeping sufficiently
far away from the Feshbach resonance.  If such magnetic field
sweeps are adiabatic, they conserve entropy (but not temperature).
By fitting the spatial profiles of the non-interacting cloud, the
entropy $S$ of the strongly interacting system can be determined.
If it is possible to vary the energy of the strongly interacting
system in a controlled way (e.g. by using the virial theorem at
unitarity~\cite{ho04uni,thom05virial} or by providing controlled heating~\cite{kina05heat}), one
can determine the derivative $dS/dE$ which is equal to the inverse
absolute temperature.  So far this method could be implemented
only for a balanced Fermi system at unitarity~\cite{luo07entropy} and, due to the need
of determining a derivative, could only provide temperatures
averaged over a range of energies.

\paragraph{Number of atoms and Fermi energy}

The number of atoms in the observed spin state can be obtained
from the total absorption recorded in the cloud's CCD image. The
transmission of resonant light at pixel $(x,y)$ is given by
$\tilde{T}(x,y) = e^{-\sigma_0 \int n_{\rm 3D}(x,y,z)\,{\rm d}z}$,
where $\sigma_0$ is the resonant atom-photon cross section for
light absorption. Thus, the number of atoms is
\begin{equation}
    N_\uparrow = \frac{A}{M \sigma_0} \sum_{pixels} - \ln(\tilde{T}(x,y))
\end{equation}
where $A$ is the area per pixel and $M$ the optical magnification.

Typically, the fitting functions are applied to the optical
density $\sigma_0 \int_z n_{\rm 3D}(x,y,z) = -
\ln\left(\tilde{T}(x,y)\right)$. The fit parameter $n_{\rm 2D,0}$
thus measures the peak optical density of the cloud, while the
radii $\tilde{R}_x$ and $\tilde{R}_y$ have units of camera pixels.
The number of atoms described by the fitting function is thus
given by
\begin{eqnarray}
    N_{\rm fit} &=& \frac{A}{M \sigma_0}\pi\, n_{\rm 2D,0}\, \tilde{R}_x \tilde{R}_y \frac{{\rm Li}_3(-e^{q})}{{\rm Li}_2(-e^{q})}\frac{{\rm Li}_0(-e^{q})}{{\rm Li}_1-e^{q})} \nonumber \\
&\rightarrow& \frac{A}{\sigma_0}\left\{%
\begin{array}{ll}
    \frac{\pi}{3} \, n_{\rm 2D,0} \, \tilde{R}_{Fx} \tilde{R}_{Fy}, & T \ll T_F \\
    \pi\, n_{\rm 2D,0} \, \tilde{\sigma}_{x} \tilde{\sigma}_{y}, & T \gg T_F. \\
\end{array}%
\right.
\end{eqnarray}

From the number of atoms and the trapping frequencies, one can
calculate the Fermi energy $k_B T_F$:
\begin{equation}
k_B T_F = \hbar \bar{\omega} (6 N_{\rm fit})^{1/3}
\end{equation}

An independent determination of the Fermi energy is provided
by the measured (physical) size of the cloud $R_i$ for highly
degenerate clouds. For $T\rightarrow 0$, $R_i \approx b_i(t)
R_{Fi}$ and thus $ k_B T_F = \frac{1}{2} m \omega_i^2
\frac{R_i^2}{b_i(t)^2} $.  As only the trapping frequencies and
the magnification of the imaging system enter into this equation,
this relation allows a calibration of the light absorption cross
section which may be reduced from the resonant cross section by
detuning, non-ideal polarization of the probe light, and
saturation.

For arbitrary temperature, the shape parameter $q$ enters the
relation for the Fermi energy:
\begin{equation}
k_B T_F = k_B T \frac{T_F}{T} = \frac{1}{2} m \omega_i^2
\frac{R_i^2}{b_i(t)^2} \frac{\left(-6\, {\rm
Li}_3(-e^{q})\right)^{1/3}}{f(e^q)}
\end{equation}

\subsubsection{Resonantly interacting Fermi gases}
\label{s:resonantgases}

The calculation of density profiles of interacting gases is
delicate. Already above the superfluid transition temperature,
attractive interactions lead to a shrinking of the cloud. Since
interactions (parameterized by the local $k_F a$) vary across the
cloud, there is a priori no simple analytical function describing
interacting Fermi gases. Experimentally, it turns out that the
difference in the {\it shape} of a (balanced) interacting and a
non-interacting Fermi mixture is minute around resonance and on
the BCS-side. Especially for the resonant case ($1/k_F a = 0$),
this has led to the common practice of using the shape of the
non-interacting Fermi gas as  fitting function, and quote an
effective temperature $\tilde{T}$ and effective degeneracy
$\frac{\tilde{T}}{T_F}$ of resonantly interacting
clouds~\cite{ohar02science,kina05heat}. In fact, universality on
resonance implies that the gas' chemical potential must be
$\mu(\vect{r}) = \xi(T/T_F) \epsilon_F(\vect{r})$, with a
universal function $\xi(T/T_F)$ which only depends on the reduced
temperature $T/T_F$~\cite{ho04uni}. The zero-temperature limit of
$\xi \equiv \xi_0$ has been subject of extensive experimental and
theoretical studies (see section~\ref{s:energymeasurements}), and its
value is $\xi(0) \approx 0.42$. At $T=0$, we have for a trapped
gas $\mu(\vect{r}) = \mu_0 - V(\vect{r}) = \xi_0
\epsilon_F(\vect{r}) \propto n^{2/3}(\vect{r})$. The density
profile will then have the exact same shape as a non-interacting
Fermi gas, with a renormalized Fermi temperature. However, for
finite temperature, $\xi(T/T_F)$ differs from the temperature
dependence of a non-interacting gas~\cite{bulg06TC}, and there is
no a priori reason that the shape of the cloud at unitarity should
be similar to that of a non-interacting Fermi gas. It turns out
that the difference is very small.

\begin{figure}[h]
\begin{center}
\includegraphics[width=5.3in]{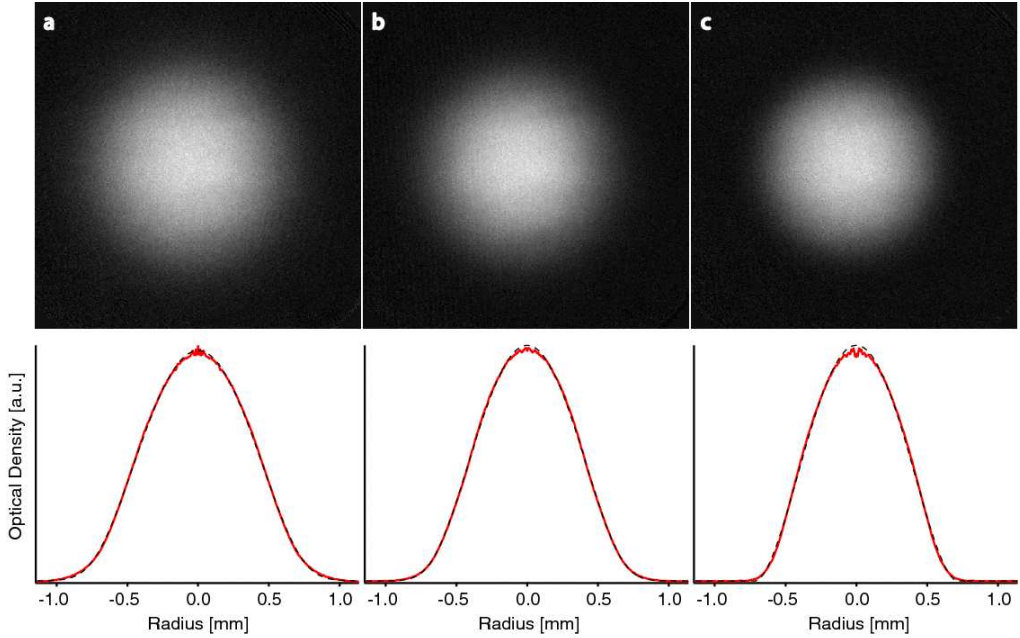}
\caption[The absence of a clear signature of condensation in the
spatial profile of strongly interacting Fermi gases]{The absence
of a signature of condensation in the spatial profile of strongly
interacting Fermi gases.  Shown are high-resolution images of spin
up atoms in a resonantly interacting, equal mixture of spin up and
spin down for different temperatures. The lower graphs show
azimuthally averaged radial profiles (noise level well below 1\%
of the maximum optical density). All three clouds are very well
fit using a finite-temperature Thomas-Fermi distribution (with
fugacity $e^{\mu / k_B \widetilde{T}}$, central density $n_0$ and
mean square radius $<r^2>$ as free parameters, see
Eq.~\ref{e:coldens}).   However, the
empirical temperatures of $\widetilde{T}/T_F = 0.22$ (a), $0.13$
(b) and $0.075$ (c) determined from the profiles' wings indicate
that at least clouds $b$ and $c$ should be in the superfluid
regime. Trap parameters $\nu_r = 162$ Hz, $\nu_z = 22.8$ Hz, 10 ms
time of flight, expansion factor 13.9,  atom numbers $N$ per spin
state were $10.2$ (a), 9.5 (b) and 7.5 $\times
10^6$.}\label{f:dilemma}
\end{center}
\end{figure}

The shape similarity was an important issue in the quest for
superfluidity in Fermi gases. In the case of weakly interacting
BECs, condensation is apparent from the sudden appearance of a
dense, central core in midst of a large thermal cloud. In contrast
to that, Fermi gases do not show such a  signature, at least at
first sight (see Fig.~\ref{f:dilemma}), and different detection
methods for superfluidity were explored.

The only loophole that may allow seeing a signature of superfluidity in the
spatial profile of balanced Fermi gases would be a rapid variation of $\xi(T/T_F)$ around the critical temperature $T_C$. This would
translate into a sudden variation of the density at the interface
between the normal and superfluid region, e.g. where the gas is
locally critical, $T = T_C(\vect{r})$.

We have indeed found a faint signature of condensation in density
profiles of the unitary gas on resonance after expansion. These
results will be presented in section~\ref{s:directunitarity}. Note
that the observation of such a feature in the density profiles
draws into question the common practice of determining an
``effective temperature'' from density profiles at unitarity using
the ideal gas fitting function. In contrast to balanced Fermi
mixtures, a striking signature of condensation can be observed in
the density profiles of mixtures with {\it imbalanced} populations
of spin up and spin down fermions. This will be discussed in
section~\ref{s:directimbalance}.

\subsubsection{Molecular clouds}
\label{s:molecularclouds}
In partially condensed molecular gases that are weakly
interacting, one can neglect the mutual repulsion between the
condensate and the surrounding thermal cloud of molecules. The
density distribution is typically well-fit by a bimodal sum of an
inverted parabola for the condensate
\begin{equation}
    n_c(x,y) = n_{c0}\left(1 - \frac{x^2}{R_{cx}^2} - \frac{y^2}{R_{cy}^2}\right)
    \label{e:condensatefit}
\end{equation}
and a Bose-function for the thermal cloud, as in
Eq.~\ref{e:coldens}, with the parameter $q = \mu\beta$ often left as an adjustable parameter (instead of fixing it via the condensate's chemical potential $\mu = g n_{c0}$):
\begin{equation}
    n_{th}(x,y) = n_{th0} \, {\rm Li}_2\left(\exp\left[q-\frac{x^2}{R_{th,x}^2} - \frac{y^2}{R_{th,y}^2}\right]\right)/{\rm Li}_2(\exp\left[q\right])
\label{e:bimodalfit}
\end{equation}
For practical purposes, this is often simplified by a gaussian, as if $q \ll 0$. Then $n_{th} \approx n_{th0} \, \exp\left(-\frac{x^2}{R_{th,x}^2} - \frac{y^2}{R_{th,y}^2}\right)$.

Once the condensate mean-field $2 g n_c$ (with $g =
\frac{4\pi\hbar^2 a_M}{M}$) experienced by thermal molecules is no
longer small compared to $k_B T$, the mutual repulsion can no
longer be neglected. The thermal molecules will then experience a
``Mexican-hat'' potential, the sum of the confining harmonic
potential $V_M(\vect{r})$, and the repulsion from the condensate $2
g n_c(\vect{r})$ and from the surrounding thermal cloud, $2 g
n_{th}(\vect{r})$. The thermal molecules themselves will in turn
repel the condensate. The situation can be captured by two coupled
equations for the condensate (in Thomas-Fermi approximation) and
the thermal cloud:
\begin{eqnarray}
    g n_c(\vect{r}) &=& {\rm Max}\left(\mu - V_M(\vect{r}) - 2 g n_{th}(\vect{r}),0\right) \nonumber \\
    n_{th}(\vect{r}) &=& \frac{1}{\lambda_M^3}\,{\rm Li}_{3/2}(e^{\beta\left(\mu - V_M(\vect{r}) - 2 g n_c(\vect{r}) - 2 g n_{th}(\vect{r})\right)}) \nonumber \\
&=& \frac{1}{\lambda_M^3}\,{\rm
Li}_{3/2}(e^{-\beta\left|\mu - V_M(\vect{r}) - 2 g
n_{th}(\vect{r})\right|})
\end{eqnarray}
where $\lambda_M$ is the thermal de Broglie wavelength for molecules. In the case of weakly interacting Bose gases, one can neglect the mean field term $2 g n_{th}(\vect{r})$~\cite{nara98semi}.

Note that these coupled equations are only an approximative way to
describe the strongly correlated gas. The mean field approximation
for the thermal molecules neglects phonons and other collective
excitations.  The above equations can be solved numerically. In
the limit of strong interactions, the condensate almost fully
expels the thermal molecules from the trap center, so that the
thermal cloud forms a shell around the condensate.

The practical implication of this discussion is that there is no
simple analytic expression for the density distribution of partially condensed
clouds in the strongly interacting regime. For fitting purposes
one may still choose for example the bimodal fit of
Eq.~\ref{e:bimodalfit}, but one must be aware that quantities like
the ``condensate fraction'' thus obtained depend on the model
assumed in the fit.   For tests of many-body calculations, the
full density distributions should be compared to those predicted
by theory.

\paragraph{Derived quantities}
{\it Temperature.}

For weakly interacting Bose gases, Eq.~\ref{e:bimodalfit} holds
and the temperature is given by
\begin{equation}
    k_B T = \frac{1}{2} m \omega_i^2 \frac{R_{th,i}^2}{b_i(t)^2}
\end{equation}
where $b_i(t) = \sqrt{1 + \omega_i^2 t^2}$ is the expansion factor
of the thermal gas. To ensure model-independent results, only the
thermal gas should be included in the fit, not the condensed core.
For strongly interacting clouds, temperature can in principle
still be obtained from the thermal wings of the {\it trapped}
molecular distribution, which is gaussian at distances $\vect{r}$
for which $V_M(\vect{r}) \gg \mu$.  However, a
possible systematic correction can occur in expansion due to
interactions of molecules in the wings with the core, which may be
either condensed or strongly interacting.

Note that unless the whole cloud is deep in the hydrodynamic
regime, there is no simple scaling law for the expansion of such
strongly interacting molecular gases.  Absolute thermometry of
strongly interacting, balanced gases is still a challenging
problem.

{\it Chemical potential}

In a confining potential, and at zero temperature, the chemical
potential is given by the size of the condensate, as $V(\vect{r})
= \mu$. It can be expressed by the fit parameters according to
Eq.~\ref{e:condensatefit} as
\begin{equation}
    \mu = \frac{1}{2} m \omega_i^2 \frac{R_{c,i}^2}{b_i(t)^2}
\end{equation}
with $b_i(t)$ the expansion factor for superfluid hydrodynamic
expansion into direction $i$. At finite temperatures and for
strong interactions, the thermal cloud will mostly reside outside
the condensate and can affect the actual or fitted condensate
size.

{\it Condensate fraction}

In the field of dilute atomic gases, the condensate fraction is a
key quantity to characterize the superfluid regime.  In contrast
to superfluid helium and superconductors, gaseous condensates can
be directly observed in a dramatic way. However, unless
interactions are negligible,  the determination of the condensate
fraction is  model dependent. For weakly interacting gases (or
those obtained after a rapid ramp into the weakly interacting
regime), the density distribution can typically be well fit with
the bimodal distribution of Eqs.~\ref{e:condensatefit} and
\ref{e:bimodalfit}. A robust way to define a ``condensate
fraction'' is then to ascribe the total number of molecules in the
narrower distribution to the condensate. For strong interactions
however, the mean-field repulsion of thermal and condensed
molecules (see above) will lead to the expulsion of a large part
of thermal molecules from the condensate. In addition, low-energy
excitations such as phonons, as well as quantum depletion will
modify the non-condensed fraction at the position of the
condensate, and the fitted condensate fraction depends on the form
of the fitting function for the bimodal fit.  In these cases, it
is better to directly compare density distributions with
theoretical predictions.

\section{Theory of the BEC-BCS crossover}
\label{c:BECBCStheory}

\subsection{Elastic collisions}
\label{s:elasticcollisions}

Due to their diluteness, most properties of systems of ultracold
atoms are related to two-body collisions. If we neglect the weak
magnetic dipole interaction between the spins, the interatomic
interaction is described by a central potential $V(r)$. At large
distances from each other, atoms interact with the van der
Waals-potential $-C_6/r^6$ as they experience each other's
fluctuating electric dipole\footnote{For distances on the order of
or larger than the characteristic wavelength of radiation of the
atom, $\lambda \gg r_0$, retardation effects change the potential
to a $-1/r^7$ law.}. At short distances on the order of a few Bohr
radii $a_0$, the two electron clouds strongly repel each other,
leading to ``hard-core'' repulsion. If the spins of the two valence
electrons (we are considering alkali atoms) are in a triplet
configuration, there is an additional repulsion due to Pauli's
exclusion principle. Hence, the triplet potential $V_T(r)$ is
shallower than the singlet one $V_S(r)$. The exact inclusion of
the interatomic potential in the description of the gas would be
extremely complicated. However, the gases we are dealing with are
ultracold and ultradilute, which implies that both the de Broglie
wavelength $\lambda_{dB}$ and the interparticle distance $n^{-1/3} \sim
5\,000-10\,000\, a_0$ are much larger than the range of the
interatomic potential $r_0$ (on the order of the van der Waals
length $r_0 \sim \left(\mu C_6 / \hbar^2\right) \sim 50\, a_0$ for
\li). As a result, scattering processes never explore the fine
details of the short-range scattering potential. The entire
collision process can thus be described by a single quantity, the
{\it scattering length}.

Since the description of Feshbach resonances and of the BCS-BEC
crossover require the concept of the effective range and
renormalization of the scattering length, we quickly summarize
some important results of scattering theory.

The Schr\"odinger equation for the reduced one-particle problem in
the center-of-mass frame of the colliding atoms (with reduced mass
$m/2$, distance vector $r$, and initial relative wave vector $\vect{k}$) is
\begin{equation}
   (\nabla^2 + k^2)\Psi_{\vect{k}}(\vect{r}) = v(r)\Psi_{\vect{k}}(\vect{r}) \quad\mbox{with } k^2 = \frac{m E}{\hbar^2} \quad \mbox{and } v(r) = \frac{m V(r)}{\hbar^2}
\label{e:schrodinger}
\end{equation}
Far away from the scattering potential, the wave function
$\Psi_{\vect{k}}(\vect{r})$ is given by the sum of the incident plane wave
$e^{i \vect{k} \cdot \vect{r}}$ and an outgoing scattered wave:
\begin{equation}
    \Psi_{\vect{k}}(\vect{r}) \sim e^{i \vect{k} \cdot \vect{r}} + f(\vect{k}',\vect{k}) \frac{e^{i k r}}{r}
    \label{e:psiasymptotic}
\end{equation}
$f(\vect{k}',\vect{k})$ is the scattering amplitude for scattering
an incident plane wave with wave vector $\vect{k}$ into the
direction $\vect{k}' = k\, \vect{r}/r$ (energy conservation implies $k' = k$).

Since we assume a central potential, the scattered wave must be
axially symmetric with respect to the incident wave vector
$\vect{k}$, and we can perform the usual expansion into partial
waves with angular momentum $l$~\cite{land77qm}.  For ultracold
collisions, we are interested in describing the scattering process
at {\it low momenta} $k \ll 1/r_0$, where $r_0$ is the range of
the interatomic potential.  In the absence of resonance phenomena
for $l \ne 0$, {\it $s$-wave scattering} $\,l=0$ is dominant over all
other partial waves (if allowed by the Pauli principle):
\begin{equation}
f \approx f_s = \frac{1}{2ik}(e^{2i\delta_s}-1) = \frac{1}{k \cot
\delta_s - i k}
\label{e:scattamp}
\end{equation}
where $f_s$ and $\delta_s$ are the $s$-wave scattering amplitude and phase shift, resp.~\cite{land77qm}. Time-reversal symmetry implies that $k\cot\delta_s$ is
an even function of $k$. For low momenta $k \ll 1/r_0$, we may expand it to order $k^2$:
\begin{equation}
k \cot \delta_s \approx -\frac{1}{a} + r_{\rm eff} \frac{k^2}{2}
\end{equation}
which defines the {\it scattering length}
\begin{equation}
a = -\lim_{k \ll 1/r_0} \frac{\tan \delta_s}{k},
\end{equation}
and the effective range $r_{\rm eff}$ of the scattering
potential. For example, for a spherical well potential of depth $V
\equiv \hbar^2 K^2/m$ and radius $R$, $r_{\rm eff} = R -
\frac{1}{K^2 a} - \frac{1}{3} \frac{R^3}{a^2}$, which
deviates from the potential range $R$ only for $|a| \lesssim R$ or
very shallow wells. For van der Waals potentials, $r_{\rm eff}$ is of order $r_0$~\cite{flam99scatt}.
With the help of
$a$ and $r_{\rm eff}$, $f$ is written as~\cite{land77qm}
\begin{equation}
f(k) = \frac{1}{-\frac{1}{a} + r_{\rm eff} \frac{k^2}{2} - ik}
\label{e:scattamplitude}
\end{equation}
In the limit $k|a| \ll 1$ and $|r_{\rm eff}| \lesssim 1/k$, $f$ becomes
independent on momentum and equals $-a$. For $k|a| \gg 1$ and
$r_{\rm eff} \ll 1/k$, the scattering amplitude is $f =
\frac{i}{k}$ and the cross section for atom-atom collisions is $\sigma = \frac{4\pi}{k^2}$.
This is the so-called unitarity limit. Such a divergence of $a$ occurs whenever a new bound state is supported by the potential (see section~\ref{s:squarewell}).

\subsection{Pseudo-potentials}
\label{s:renormalization}
If the de Broglie wavelength $\frac{2\pi}{k}$ of the colliding
particles is much larger than the fine details of the interatomic
potential, $1/k \gg r_0$, we can create a simpler description by modifying the
potential in such a way that it is much easier to manipulate in
the calculations, but still reproduces the correct $s$-wave
scattering. An obvious candidate for such a ``pseudo-potential'' is
a delta-potential $\delta(\vect{r})$.

However, there is a subtlety
involved which we will address in the following. The goal is to
find an expression for the scattering amplitude
$f$ in terms of the potential $V(r) =
\frac{\hbar^2 v(r)}{m}$, so that we can try out different
pseudo-potentials, always ensuring that $f \rightarrow -a$ in the
$s$-wave limit. For this, let us go back to the Schr\"odinger
equation Eq.~\ref{e:schrodinger}. If we knew the solution to the
following equation:
\begin{equation}
    (\nabla^2 + k^2)G_k(\vect{r}) = \delta(\vect{r})
    \label{e:defGreen}
\end{equation}
we could write an integral equation for the wave function
$\Psi_{\vect{k}}(\vect{r})$ as follows:
\begin{equation}
    \Psi_{\vect{k}}(\vect{r}) = e^{i \vect{k}\cdot\vect{r}} + \int d^3 r' G_k(\vect{r}-\vect{r}')v(\vect{r}')\Psi_{\vect{k}}(\vect{r'})
\label{e:integralequation}
\end{equation}
This can be simply checked by inserting this implicit solution for
$\Psi_{\vect{k}}$ into Eq.~\ref{e:schrodinger}. $G_k(\vect{r})$ can be easily
obtained from the Fourier transform of Eq.~\ref{e:defGreen},
defining $G_k(\vect{p}) = \int d^3 r e^{-i \vect{p} \cdot \vect{r}}
G_k(\vect{r})$:
\begin{equation}
(-p^2 + k^2)G_k(\vect{p}) = 1
\end{equation}
The solution for $G_k(\vect{r})$ is
\begin{equation}
\label{e:Green}
G_{k,+}(\vect{r}) = \int \frac{d^3 p}{(2\pi)^3} \frac{e^{i \vect{p}
\cdot \vect{r}}}{k^2 - p^2 + i \eta} =
-\frac{1}{4\pi}\frac{e^{ikr}}{r}
\end{equation}
where we have chosen (by adding the infinitesimal constant $i
\eta$, with $\eta>0$ in the denominator) the solution that
corresponds to an outgoing spherical wave. $G_{k,+}(\vect{r})$ is the
{\it Green's function} of the scattering problem.
Far away from the origin, $|\vect{r}-\vect{r'}| \sim r -
\vect{r'}\cdot \vect{u}$, with the unit vector $\vect{u} =
\vect{r}/r$, and
\begin{equation}
\Psi_{\vect{k}}(\vect{r}) \sim e^{i \vect{k} \cdot \vect{r}}
- \frac{e^{i k r}}{4\pi r} \int d^3 r' e^{-i \vect{k}'\cdot
\vect{r}'} v(\vect{r}')\Psi_{\vect{k}}(\vect{r'})
\end{equation}
where $\vect{k}' = k \vect{u}$. With Eq.~\ref{e:psiasymptotic}, this invites the definition of the scattering amplitude via
\begin{equation}
f(\vect{k}',\vect{k}) = -\frac{1}{4\pi} \int d^3 r\, e^{-i \vect{k}'\cdot
\vect{r}} v(\vect{r})\Psi_{\vect{k}}(\vect{r})
\end{equation}
Inserting the exact formula for $\Psi_{\vect{k}}(\vect{r})$, Eq.~\ref{e:integralequation}, combined with Eq.~\ref{e:Green}, leads to an integral equation for the scattering amplitude
\begin{eqnarray}
f(\vect{k}',\vect{k}) = -\frac{v(\vect{k}'-\vect{k})}{4\pi}  +\int \frac{d^3 q}{(2\pi)^3} \, \frac{v(\vect{k}'-\vect{q})f(\vect{q},\vect{k})}{k^2-q^2+i\eta}
\label{e:lippmannschwinger}
\end{eqnarray}
where $v(\vect{k})$ is the Fourier transform of the potential
$v(\vect{r})$ (which we suppose to exist). This is the
Lippmann-Schwinger equation, an exact integral equation for $f$ in
terms of the potential $v$, useful to perform a perturbation
expansion. Note that it requires knowledge of
$f(\vect{q},\vect{k})$ for $q^2 \ne k^2$ (``off the energy
shell''). However, the dominant contributions to the integral do
come from wave vectors $\vect{q}$ such that $q^2 = k^2$. For
low-energy $s$-wave scattering, $f(\vect{q},\vect{k}) \rightarrow
f(k)$ then only depends on the magnitude of the wave vector
$\vect{k}$. With this
approximation, we can take $f(k)$ outside the integral. Taking the
limit $k \ll 1/r_0$, dividing by $f(k)$ and by $v_0 \equiv
v(\vect{0})$, we arrive at
\begin{eqnarray}
\frac{1}{f(k)} \approx -\frac{4\pi}{v_0} + \frac{4\pi}{v_0} \int \frac{d^3 q}{(2\pi)^3}\, \frac{v(-\vect{q})}{k^2-q^2+i\eta}
\label{e:scattampintegral}
\end{eqnarray}
If we only keep the first order in $v$, we obtain the scattering
length in {\it Born approximation}, $a = \frac{v_0}{4\pi}$. For a
delta-potential $V(\vect{r}) = V_0\, \delta(\vect{r})$, we obtain
to first order in $V_0$
\begin{equation}
V_0 = \frac{4\pi \hbar^2 a}{m}
\end{equation}
However, already the second order term in the expansion
of Eq.~\ref{e:scattampintegral} would not converge, as it involves the
divergent integral $\int \frac{d^3 q}{(2\pi)^3} \frac{1}{q^2}$. The reason
is that the Fourier transform of the $\delta$-potential does not fall off at
large momenta. Any physical potential {\it does} fall off at some large momentum, so this is not a
``real'' problem. For example, the van-der-Waals potential varies on a characteristic length scale $r_0$ and will thus have a natural momentum cut-off $\hbar/r_0$. A proper regularization of contact interactions employs the pseudo-potential~\cite{huan87} $V(\vect{r})\psi(\vect{r}) = V_0 \delta(\vect{r})\frac{\partial}{\partial r} (r \psi(\vect{r}))$. It leads exactly to a scattering amplitude $f(k) = -a/(1+ i k a)$ if $V_0 = \frac{4\pi\hbar^2 a}{m}$.

Here we will work with a Fourier transform that is equal to a
constant $V_0$ at all relevant momenta in the problem, but that
falls off at very large momenta, to make the second order term
converge. The exact form is not important. If we are to calculate
physical quantities, we will replace $V_0$ in favor of the
observable quantity $a$ using the formal prescription
\begin{equation}
\frac{1}{V_0} = \frac{m}{4\pi\hbar^2 a} - \frac{m}{\hbar^2}\int
\frac{d^3 q}{(2\pi)^3} \frac{1}{q^2} \label{e:renormalize}
\end{equation}
We will always find that the diverging term is exactly balanced by another diverging integral in the final expressions, so this is a well-defined procedure~\cite{melo93,haus99}.

Alternatively, one can introduce a ``brute force'' energy cut-off
$E_R = \hbar^2/m R^2$ (momentum cut-off $\hbar/R$), taken to be
much larger than typical scattering energies.
Eq.~\ref{e:scattampintegral} then gives
\begin{eqnarray}
\frac{1}{f(k)} \approx -\frac{4\pi}{v_0} - \frac{2}{\pi} \frac{1}{R} + \frac{2 R}{\pi} k^2 - i k
\label{e:scattampcutoff}
\end{eqnarray}
This is now exactly of the form Eq.~\ref{e:scattamplitude} with
the scattering length
\begin{eqnarray}
a = \frac{\pi}{2}\frac{R}{1+\frac{2\pi^2 R}{v_0}}
\label{e:acutoff}
\end{eqnarray}
For any physical, given scattering length $a$ we can thus find the
correct strength $v_0$ that reproduces the same $a$ (provided that
we choose $R \ll a$ for positive $a$). This approach implies an
effective range $r_{\rm eff} = \frac{4}{\pi}R$ that should be
chosen much smaller than all relevant distances. Note that as a
function of $v_0$, only one pole of $a$ and therefore only one
bound state is obtained, at $v_0 = -2\pi^2 R$.

This prompts us to discuss the relation between Eq.~\ref{e:renormalize} and
Eq.~\ref{e:lippmannschwinger}: The Lippmann-Schwinger
equation is an exact reformulation of Schr\"odinger's equation for
the scattering problem. One can, for example, exactly solve for
the scattering amplitude in the case of a spherical well
potential~\cite{bray71}. In particular, all bound states supported
by the potential are recovered. However, to arrive at
Eq.~\ref{e:renormalize}, one ignores the oscillatory behavior of
both $v(\vect{q})$ and $f(\vect{q},\vect{k})$ and replaces them by
$\vect{q}$-independent constants. As a result, Eq.~\ref{e:renormalize}, with a cut-off for the diverging integral at
a wave vector $1/R$, only allows for {\it one} bound
state to appear as the potential strength is increased (see Eq.~\ref{e:acutoff}).

We will analyze this approximation for a spherical well of depth $V$ and radius $R$.
The true scattering length for a spherical well is given by~\cite{land77qm}
\begin{equation}
    \frac{a}{R} = 1 - \frac{\tan(K R)}{K R}
\end{equation}
with $K^2 = m V/\hbar^2$.
which one can write as
\begin{eqnarray}
    \frac{a}{R} &=& 1 - \frac{\prod_{n=1}^\infty (1 - \frac{K^2 R^2}{n^2 \pi^2})}{\prod_{n=1}^\infty(1 - \frac{4K^2 R^2}{(2n-1)^2\pi^2})} \quad \left.%
\begin{array}{ll}
    \leftarrow \mbox{Zeros of }$a-R$ &\\
    \leftarrow \mbox{Resonances of }a &\\
\end{array}%
\right.
\end{eqnarray}
In contrast, Eq.~\ref{e:renormalize} with $V_0 = - \frac{4\pi}{3} V R^3$ and the ``brute force'' cut-off at $1/R$ gives
\begin{equation}
    \frac{a}{R} = \frac{K^2 R^2}{\frac{2}{\pi}K^2 R^2 - 3}
\end{equation}
The sudden cut-off strips the scattering length of all but one
zero (at $V = 0$) and of all but one resonance.
For a shallow well that does not support a bound state, the scattering length
still behaves correctly as $a = -\frac{1}{3} \frac{V}{E_R}
R$. However, the sudden cut-off
$v(\vect{q}) \approx {\rm const.}$ for $q \le \frac{1}{R}$ and 0 beyond
results in a shifted critical well depth to accommodate the first
bound state, $V = \frac{3\pi}{2} E_R$, differing from the exact
result $V = \frac{\pi^2}{4} E_R$. This could be cured by adjusting
the cut-off. But for increasing well depth, no new bound state is
found and $a$ saturates at $\sim R$, contrary to the exact result.

At first, such an approximation might be unsettling, as the
van-der-Waals potentials of the atoms we deal with contain many
bound states. However, the gas is in the ultracold regime, where
the de Broglie-wavelength is much larger than the range $r_0$ of
the potential. The short-range physics, and whether the
wave function has one or many nodes within $r_0$ (i.e. whether the
potential supports one or many bound states), is not important.
All that matters is the phase shift $\delta_s$ {\it modulo $2\pi$}
that the atomic wave packets receive during a collision. We have
seen that with a Fourier transform of the potential that is
constant up to a momentum cut-off $\hbar/R$, we can reproduce any
low-energy scattering behavior, which is described by the
scattering length $a$.  We can even realize a wide range of
combinations of $a$ and the effective range $r_{\rm eff}$ to
capture scattering at finite values of $k$. An exception is the
situation where $0 < a \lesssim r_{\rm eff}$ or potentials that
have a negative effective range. This can be cured by more
sophisticated models (see the model for Feshbach resonances in
chapter~\ref{c:feshbach}).

\subsection{Cooper instability in a Fermi gas with attractive interactions}

In contrast to bosons, the non-interacting Fermi gas does not show
any phase transition down to zero temperature. One might assume
that this qualitative fact should not change as interactions are
introduced, at least as long as they are weak. This is essentially
true in the case of repulsive interactions~\footnote{Repulsive
interactions still allow for the possibility of induced $p$-wave
superfluidity (Kohn and Luttinger~\cite{kohn65}, also
see~\cite{bara98}) however at very low temperatures $T_C \approx
E_F \exp[-13(\pi/2k_F|a|)^2]$.}. For attractive interactions, the
situation is, however, dramatically different. Even for very weak
attraction, the fermions form pairs and become superfluid, due to
a generalized from of pair condensation.

The idea of pairing might be natural, as tightly bound pairs of
fermions can be regarded as point-like bosons, which should form a
Bose-Einstein condensate. However, for weak attractive interaction
-- as is the case for the residual, phonon-induced electron-electron interaction in metals
-- it is not evident that a paired state exists. Indeed, we will
see in the following that in three dimensions there is no bound
state for two isolated particles and arbitrarily weak interaction.
However, by discussing exact solutions in 1D and 2D, where bound
states exist for weak interactions, we gain insight into how a modified
density of states will lead to bound states even in 3D --  this is
the famous Cooper instability.

\subsubsection{Two-body bound states in 1D, 2D and 3D}
\label{s:boundstates}

\begin{figure}
  \includegraphics[width=5.5in]{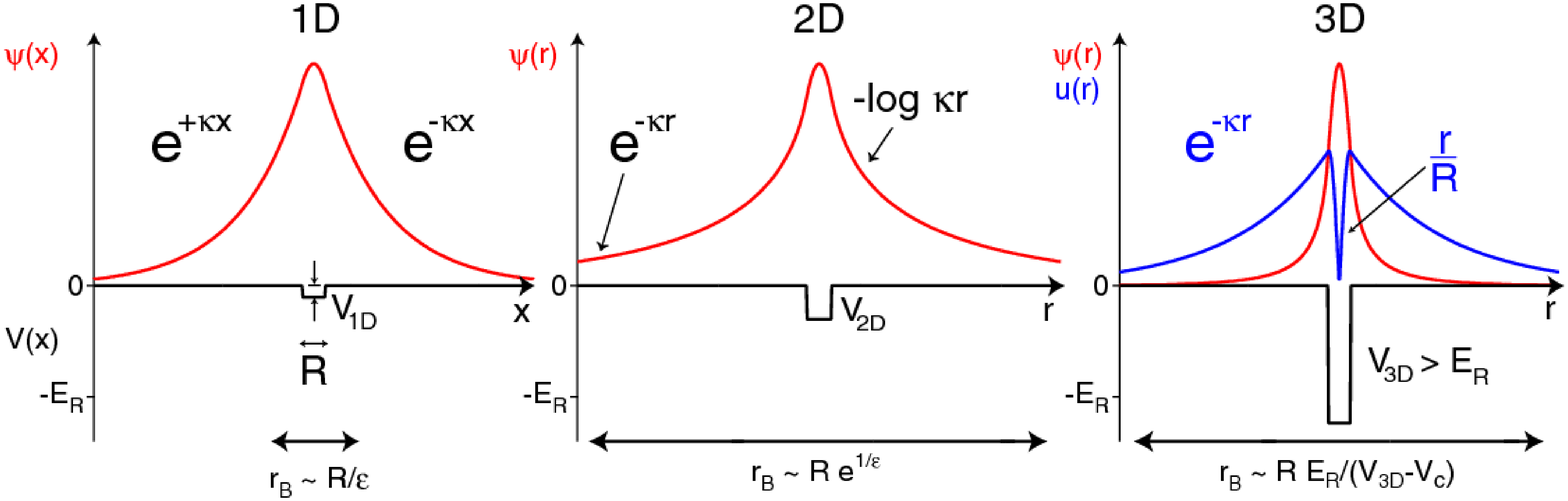}\\
  \caption[Bound state wave functions in 1D, 2D and 3D]{Bound
state wave functions in 1D, 2D and 3D for a potential well of size
$R$ and depth $V$. In 1D and 2D, bound states exist for
arbitrarily shallow wells. In terms of the small parameter
$\epsilon = V/E_R$ with $E_R = \hbar^2/m R^2$, the size of the
bound state in 1D is $R/\epsilon$. In 2D, the bound state is
exponentially large, of size $R e^{-1/\epsilon}$. In 3D, due to
the steep slope in $u(r) = r \psi(r)$, bound states can only exist
for well depths $V_{\rm 3D}$ larger than a certain threshold $V_c
\approx E_R$. The size of the bound state diverges as $R E_R /
(V_{\rm 3D}-V_c)$ for $V_{\rm 3D}>V_c$.
}\label{f:squarewell}
\end{figure}

Localizing a quantum-mechanical particle of mass $\mu = m/2$ to a certain range $R$
leads to an increased momentum uncertainty of $p \sim \hbar/R$ at
a kinetic energy cost of about $E_{R} = p^2/m = \hbar^2 / m
R^2$. Clearly, a shallow potential well of size $R$ and depth $V$
with $V/E_R \equiv \epsilon \ll 1$ cannot confine the particle
within its borders. But we can search for a bound state at energy
$|E_B| \ll E_R$ of much larger size $r_B = 1/\kappa \equiv
\sqrt{\hbar^2/m |E_B|}  \gg R$.

\begin{itemize}
\item {\bf 1D}: The bound state wave function far away from the well necessarily behaves like $e^{\pm \kappa x}$ for    negative (positive) $x$ (see Fig.~\ref{f:squarewell}a). As we traverse the well, the wave function has to change its slope by $2\kappa$ over a range $R$. This costs kinetic energy $\approx \hbar^2 \kappa/m R$
that has to be provided by the potential energy $-V$. We deduce that $\kappa \approx m R V / \hbar^2 = \epsilon/R$, where $\epsilon=V/E_R$ is a small number for a weak potential.  The size of the bound state $r_B \approx R / \epsilon$ is indeed much larger than the size of the well, and the bound state energy $E_B \approx - E_R\, \epsilon^2/2$ depends quadratically on the weak attraction $-V$. Importantly, we can {\it always} find a bound state even for arbitrarily weak (purely) attractive potentials.

    \item {\bf 2D}: For a spherically symmetric well, the Schr\"odinger equation for the radial wave function $\psi(r)$ {\it outside} the well reads $\frac{1}{r}\partial_r(r\partial_r \psi) = \kappa^2 \psi$. The solution is the modified Bessel function which vanishes like $e^{-\kappa r}$ as $r \gg 1/\kappa$ (see Fig.~\ref{f:squarewell}b). For $R\ll r \ll 1/\kappa$, we can neglect the small bound state energy $E_B \propto -\kappa^2$ compared to the
       kinetic energy and have $\partial_r(r\psi') = 0$ or $\psi(r) \approx \log (\kappa r)/\log(\kappa R)$, where $1/\kappa$ is the natural scale of evolution for
        $\psi(r)$ and we have normalized $\psi$ to be of order 1 at $R$. Note that in 2D, it is not the change in the slope $\psi'$ of the wave function
        which costs kinetic energy, but the change in $r \psi'$. {\it Inside} the well, we can assume $\psi(r)$ to be practically constant as $V \ll E_R$.
        Thus, $r \psi'$ changes from $\approx 1/\log \kappa R$ (outside) to $\approx 0$ (inside) over a distance $R$. The corresponding kinetic energy cost
         is $\frac{\hbar^2}{m r}\partial_r(r\psi')/\psi \approx \hbar^2/m R^2 \log (\kappa R) = E_R /\log (\kappa R)$, which has to be provided by the potential
          energy $-V$. We deduce $\kappa \approx \frac{1}{R}\, e^{-c E_R/V}$ and $E_B \approx -E_R\, e^{-2c E_R / V}$ with $c$ on the
          order of 1. The particle is extremely weakly bound, with its bound state energy depending exponentially on the shallow potential $-V$.
          Accordingly, the size of the bound state is exponentially large, $r_B \approx R\, e^{c E_R/V}$. Nevertheless, we can {\it always} find
          this weakly bound state, for arbitrarily small attraction.

    \item {\bf 3D}: For a spherically symmetric well, the Schr\"odinger equation for the wave function $\psi$ transforms into an effective one-dimensional problem for the
     wave function $u = r \psi$ (see Fig.~\ref{f:squarewell}c). We might now be tempted to think that there must always be a bound state in 3D, as we already found this to be the
     case in 1D. However, the boundary condition on $u(r)$ is now to vanish linearly at $r=0$, in order for $\psi(0)$ to be finite. Outside the potential well, we still have
     $u \propto e^{-\kappa r}$ for a bound state. Inside the well the wave function must fall off to zero at $r=0$ and necessarily has to change its slope from $-\kappa$ outside to
     $\sim 1/R$ inside the well over a distance $R$. This costs the large kinetic energy $\sim\hbar^2 u''/2m u \approx \hbar^2 /m R^2 = E_R$. If the well depth $V$ is smaller than a {\it critical depth} $V_c$ on the order of $E_R$, the particle cannot be bound. At $V=V_c$, the first
      bound state enters at $E=0$. As $\kappa=0$, $u$ is then constant outside the well. If the potential depth is further increased by a small amount $\Delta V \ll V_c$, $u$ again
      falls off like $e^{-\kappa r}$ for $r > R$. This requires an additional change in slope by $\kappa$ over the distance $R$, provided by $\Delta V$. So we find analogously to the
      1D case $\kappa \sim m R \Delta V / \hbar^2$. Hence, the bound state energy $E_B \approx - \Delta V^2 / E_R$ is quadratic in the ``detuning'' $\Delta V = (V-V_C)$, and the size
      of the bound state diverges as $r_B \approx R E_R / (V - V_C)$. We will find exactly the same behavior for a weakly bound state when discussing Feshbach resonances
      in chapter~\ref{c:feshbach}.
\end{itemize}

\begin{table}
\centering
\begin{tabular}{c|c|c|c}
    & 1D & 2D & 3D \\ \hline
  $V$ & $\ll E_R$ & $\ll E_R$ & $> V_c \approx E_R$ \\[2pt] \hline
  $\psi(r>R)$ & $e^{-r/r_B}$ & $K_0(\frac{r}{r_B}) = \left\{%
\begin{array}{ll}
    -\log r/r_B, & \hbox{$R \ll r \ll r_B$} \\
    e^{-r/r_B}, & \hbox{$r \gg r_B$} \\
\end{array}%
\right.$ & $\frac{e^{- r / {r_B}}}{r}$ \\[12pt]    \hline
  $r_B$ & $R \frac{E_R}{V}$ & $R\, e^{c E_R / V}$ & $R \frac{E_R}{V-V_c}$ \\[4pt]\hline
  $E_B = -\frac{\hbar^2}{m r_B^2}$ & $-V^2 / E_R$  & $-E_R e^{-2 c E_R / V}$  & $-(V-V_c)^2 / E_R$ \\
\end{tabular}
\caption{Bound states in 1D, 2D and 3D for a potential well of
size $R$ and depth $V$. $\psi(r>R)$ is the wave function outside
the well, $r_B$ is the size of the bound state, and $E_B$ its
energy ($E_R = \hbar^2/m R^2$).}\label{t:boundstate}
\label{t:boundstates}
\end{table}

The analysis holds for quite general shapes $V(r)$ of the
(purely attractive) potential well (in the equations, we only need to replace $V$ by
its average over the well - if it exists -,
$\frac{1}{R}\int_{-\infty}^\infty V(x) dx$ in 1D,
$\frac{1}{R^2}\int_0^\infty r V(r) dr$ in 2D etc.).
Table~\ref{t:boundstates} summarizes the different cases.

Applying these results to the equivalent problem of two
interacting particles colliding in their center-of-mass frame, we
see that in 1D and 2D, two isolated particles can bind for an
arbitrarily weak purely attractive interaction. Hence in 1D and
2D, pairing of fermions can be understood already at the two-particle level. Indeed,
one can show that the existence of a two-body bound state for isolated particles
in 2D is a necessary and sufficient condition for the instability of the many-body Fermi sea (Cooper instability, see below)~\cite{rand89bound}. In 3D, however, there is a
threshold interaction below which two isolated particles are
unbound. We conclude that if pairing and condensation occur for
arbitrarily weak interactions in 3D, then this must entirely be
due to many-body effects.

\subsubsection{Density of states}

\begin{table}
\centering
\begin{tabular}{c|c|c|c}
    & 1D & 2D & 3D \\ \hline

$\frac{\hbar^2}{m\Omega}\rho(\epsilon)$ & $\frac{1}{\pi}\sqrt{\frac{\hbar^2}{2m\epsilon}}$ & $\frac{1}{2\pi}$ & $\frac{1}{2\pi^2}\sqrt{\frac{2m\epsilon}{\hbar^2}}$ \\[2pt] \hline

$\frac{1}{\left|V_0\right|} = \frac{1}{\Omega}\int_{\epsilon<E_R}
d\epsilon \frac{\rho_n(\epsilon)}{2\epsilon+\left|E\right|}$ &
$\sqrt{\frac{m}{4\hbar^2 \left|E\right|}}$ &
$\frac{m}{4\pi\hbar^2}\log\frac{2E_R + |E|}{|E|}$ &
$\frac{1}{2\pi^2}\frac{m^{3/2}}{\hbar^3}(\sqrt{2E_R}
- \frac{\pi}{2} \sqrt{|E|})$
\\[4pt]\hline

$E = -\frac{\hbar^2 \kappa^2}{m}$ & $-\frac{m}{4\hbar^2} V_0^2$  & $-2 E_R\, e^{-\frac{4\pi\hbar^2}{m\left|V_0\right|}}$  & $-\frac{8}{\pi^2}E_R\,\frac{(\left|V_0\right| - V_{0c})^2}{\left|V_0\right|^2} = -\hbar^2/m a^2$ \\
\end{tabular}
\caption{Link between the density of states and the existence of a
bound state for arbitrarily weak interaction. The table shows the
density of states, $\rho(\epsilon)$, the equation relating the
bound state energy $E$ to $V_0$, and the result for $E$. It is
assumed that $E_R \gg |E|$. To compare with
table~\ref{t:boundstates} note that $|V_0| \sim V R^n$. $V_{0c} =
\sqrt{2}\,\pi^2 E_R R^3$ is the threshold interaction strength for
the 3D case. The formula for the 3D bound state energy follows from the renormalization procedure outlined in section~\ref{s:renormalization}, when expressing $V_0$ in terms of the scattering length $a$ using  Eq.~\ref{e:renormalize}.} \label{t:momentumboundstates}
\end{table}

What physical quantity decides whether there are bound states or
not? To answer this question, we formulate the problem of two
interacting particles of mass $m$ in momentum space. This allows a
particularly transparent treatment for all three cases (1D, 2D,
3D) at once, and identifies the {\it density of states} in the
different dimensions as the decisive factor for the existence of
bound states.

Searching for a shallow bound state of energy $E = -\frac{\hbar^2
\kappa^2}{m}$ ($m/2$ is the reduced mass), we start by writing the
Schr\"odinger equation for the relative wave function
$\frac{\hbar^2}{m}(\nabla^2-\kappa^2)\psi = V \psi$ in
($n$-dimensional) momentum space:
\begin{equation}
    \psi_\kappa(\vect{q}) = - \frac{m}{\hbar^2}\frac{1}{q^2 + \kappa^2} \int \frac{d^n q'}{(2\pi)^n} V(\vect{q}-\vect{q}') \psi_\kappa(\vect{q}')
\end{equation}
For a short-range potential of range $R \ll 1/\kappa$, $V(\vect{q})$ is
practically constant, $V(\vect{q}) \approx V_0$, for all relevant
$q$, and falls off to zero on a large $q$-scale of $\approx 1/R$.
For example, for a potential well of depth $V$ and size $R$, we
have $V_0 \sim - V R^n$. Thus,
\begin{equation}
    \psi_\kappa(\vect{q}) \approx  - \frac{mV_0}{\hbar^2}\frac{1}{q^2 + \kappa^2} \int_{q'\lesssim\frac{1}{R}} \frac{d^n q'}{(2\pi)^n} \psi_\kappa(\vect{q}')
\end{equation}
We integrate once more over $\vect{q}$, applying the same cut-off
$1/R$, and then divide by the common factor $\int_{q\lesssim\frac{1}{R}}
\frac{d^n q}{(2\pi)^n} \psi_\kappa(\vect{q})$. We obtain the
following equation for the bound state energy $E$:
\begin{equation}
   - \frac{1}{V_0} \;=\; \frac{m}{\hbar^2}\int_{q\lesssim\frac{1}{R}} \frac{d^n q}{(2\pi)^n} \frac{1}{q^2 + \kappa^2}\; =\; \frac{1}{\Omega}\int_{\epsilon<E_R} d\epsilon \frac{\rho_n(\epsilon)}{2\epsilon+\left|E\right|}
\label{e:densityboundstates}
\end{equation}
with the density of states in $n$ dimensions $\rho_n(\epsilon)$,
the energy cut-off $E_R = \hbar^2/m R^2$ and the volume $\Omega$
of the system (note that $V_0$ has units of energy times volume).
The question on the existence of bound states for arbitrarily weak
interaction has now been reformulated: As $|V_0| \rightarrow 0$,
the left hand side of Eq.~\ref{e:densityboundstates} diverges.
This equation has a solution for small $|V_0|$ only if the right
hand side also diverges for vanishing bound state energy $|E|
\rightarrow 0$, and this involves an integral over the density of
states. Table~\ref{t:momentumboundstates} presents the different
cases in 1D, 2D, 3D. In 1D, the integral diverges as
$1/\sqrt{|E|}$, so one can always find a bound state solution. The
binding energy depends quadratically on the interaction, as we had
found before. In 2D, where the density of states $\rho_{\rm 2D}$
is {\it constant}, the integral still diverges logarithmically as
$|E|\rightarrow 0$, so that again there is a solution $|E|$ for
any small $|V_0|$. The binding energy now depends exponentially on
the interaction and $\rho_{\rm 2D}$:
\begin{equation}
    E_{\rm 2D} = - 2 E_R \, e^{-\frac{2\Omega}{\rho_{\rm 2D} \left|V_0\right|}}
    \label{e:boundstate2D}
\end{equation}
However, in 3D the integral is finite for vanishing $|E|$, and
there is a threshold for the interaction potential to bind the two
particles.

These results give us an idea why there might be a paired state
for two fermions immersed in a medium, even for arbitrarily weak
interactions: It could be that the density of available states to the two fermions is altered due to the presence of the
other atoms. This is exactly what happens, as will be discussed in
the next section.

\subsubsection{Pairing of fermions -- The Cooper problem}
\label{s:cooperproblem}

\begin{figure}
  \centering
  \includegraphics[width=4in]{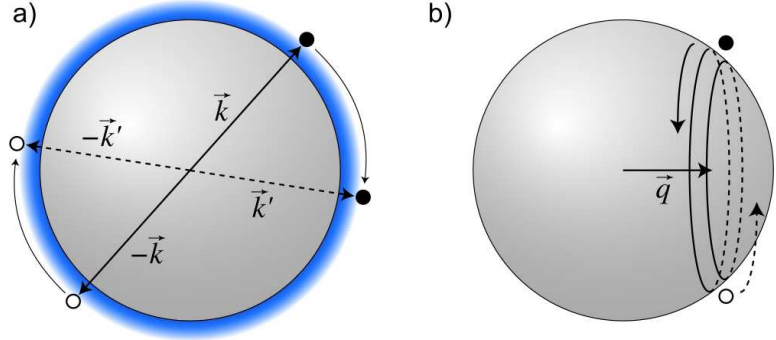}\\
  \caption[Cooper problem: Two particles scattering on top of a Fermi sea]{Cooper problem: Two particles scattering on top of a Fermi sea. a) Weakly interacting particles with equal and opposite momenta can scatter into final states in a narrow shell (blue-shaded) on top of the Fermi sea (gray shaded), which blocks possible final momentum states. b) For non-zero total momentum $2\vect{q}$, particles can scatter only in a narrow band around a circle with radius $\sqrt{k_F^2 - q^2}$.}\label{f:cooperproblem}
\end{figure}

Consider now two weakly interacting spin 1/2 fermions not in
vacuum, but on top of a (non-interacting) filled Fermi sea, the
Cooper problem~\cite{coop56}. Momentum states below the Fermi
surface are not available to the two scattering particles due to
Pauli blocking (Fig.~\ref{f:cooperproblem}a). For weak
interactions, the particles' momenta are essentially confined to a
narrow shell above the Fermi surface. The density of states at the Fermi surface is $\rho_{\rm 3D}(E_F)$, which is a constant just like in two dimensions. We should thus
find a {\it bound state} for the two-particle system {\it for
arbitrarily weak attractive interaction}.

In principle, the two fermions could form a pair at any finite
momentum. However, considering the discussion in the previous
section, the largest binding energy can be expected for the pairs
with the largest density of scattering states. For
zero-momentum pairs, the entire Fermi surface is available for
scattering, as we can see from Fig.~\ref{f:cooperproblem}a. If the
pairs have finite center-of-mass momentum $\vect{q}$, the number
of contributing states is strongly reduced, as they are confined
to a circle (see Fig.~\ref{f:cooperproblem}b). Consequently, pairs
at rest experience the strongest binding. In the following we will
calculate this energy.

We can write the Schr\"odinger equation for the two interacting
particles as before, but now we need to search for a small binding
energy $E_B = E-2E_F<0$ on top of the large Fermi energy $2E_F$ of
the two particles. The equation for $E_B$ is
\begin{equation}
    -\frac{1}{V_0} = \frac{1}{\Omega}\int_{E_F<\epsilon<E_F+E_R} d\epsilon \frac{\rho_{\rm 3D}(\epsilon)}{2(\epsilon-E_F)+\left|E_B\right|}
\label{e:Cooper}
\end{equation}
The effect of Pauli blocking of momentum states below the Fermi
surface is explicitly included by only integrating over energies
$\epsilon > E_F$.

In conventional superconductors, the natural cut-off energy $E_R$ is given by
the Debye frequency $\omega_D$, $E_R = \hbar \omega_D$,
corresponding to the highest frequency at which ions in the
crystal lattice can respond to a bypassing electron. Since we have
$\hbar \omega_D \ll E_F$, we can approximate $\rho_{\rm
3D}(\epsilon) \approx \rho_{\rm 3D}(E_F)$ and find:
\begin{equation}
    E_B = - 2 \hbar \omega_D e^{-2 \Omega/\rho_{\rm 3D}(E_F) \left|V_0\right|}\\
    \label{e:coopersuperconductor}
\end{equation}

In the case of an atomic Fermi gas, we should replace $1/V_0$
by the physically relevant scattering length $a < 0$ using the
prescription in Eq.~\ref{e:renormalize}. The equation for the
bound state becomes
\begin{equation}
    -\frac{m}{4\pi\hbar^2 a} = \frac{1}{\Omega}\int_{E_F}^{E_F+E_R} d\epsilon \frac{\rho_{\rm 3D}(\epsilon)}{2(\epsilon-E_F)+\left|E_B\right|}
- \frac{1}{\Omega}\int_0^{E_F+E_R} d\epsilon \frac{\rho_{\rm
3D}(\epsilon)}{2\epsilon} \label{e:cooperrenorm}
\end{equation}
The right hand expression is now finite as we let the cut-off $E_R
\rightarrow \infty$, the result being (one assumes
$\left|E_B\right|\ll E_F$)
\begin{equation}
    -\frac{m}{4\pi\hbar^2 a} = \frac{\rho_{\rm 3D}(E_F)}{2\Omega}
\left(-\log\left(\frac{\left|E_B\right|}{8E_F}\right) - 2\right)
\end{equation}
Inserting $\rho_{\rm 3D}(E_F) = \frac{\Omega m k_F}{2\pi^2
\hbar^2}$ with the Fermi wave vector $k_F = \sqrt{2mE_F/\hbar^2}$,
one arrives at
\begin{equation}
    E_B = - \frac{8}{e^2} E_F\, e^{-\pi/k_F \left|a\right|}
    \label{e:cooperproblem}
\end{equation}
The binding energies Eqs.~\ref{e:coopersuperconductor} and
\ref{e:cooperproblem} can be compared with the result for the
bound state of two particles in 2D, Eq.~\ref{e:boundstate2D}. The
role of the constant density of states $\rho_{\rm 2D}$ is here
played by the 3D density of states at the Fermi surface,
$\rho_{\rm 3D}(E_F)$.

The result is remarkable: Two weakly interacting fermions on top
of a Fermi sea form a bound state due to Pauli blocking. However,
in this artificial problem we neglected the interactions between
particles {\it in} the Fermi sea. As we ``switch on'' the
interactions for all particles from top to the bottom of the Fermi
sea, the preceding discussion indicates that the gas will reorder
itself into a completely new, paired state. The Fermi sea is thus
unstable towards pairing (Cooper instability). The full many-body description of such a
paired state, including the necessary anti-symmetrization of the
full wave function, was achieved by Bardeen, Cooper and Schrieffer
(BCS) in 1957~\cite{bard57}. As we will see in the next section,
the self-consistent inclusion of all fermion pairs leads to more
available momentum space for pairing. The effective density of
states is then twice as large, giving a superfluid gap $\Delta$
that differs from  $|E_B|$ (Eq.~\ref{e:cooperproblem}) by a factor
of 2 in the exponent:
\begin{equation}
    \Delta = \frac{8}{e^2} E_F\, e^{-\pi/2 k_F \left|a\right|}
\end{equation}

It should be noted that the crucial difference to the situation of two particles in vacuum in
3D is the constant density of states at the Fermi energy (and not
the 2D character of the Fermi surface). Therefore, if we were to
consider the Cooper problem in higher dimensions $n$ and have two
fermions scatter on the $(n-1)$ dimensional Fermi surface, the
result would be similar to the 2D case (due to the constant
density of states), and not to the case of $(n-1)$ dimensions.

The conclusion of this section is that Cooper pairing is a
many-body phenomenon, but the binding of two fermions can
still be understood by two-body quantum mechanics, as it is similar
to two isolated particles in two dimensions.  To first order, the many-body
physics is not the modification of interactions, but rather the
modification of the density of states due to Pauli blocking.

\subsection{Crossover wave function}
\label{s:crossoverwavefunction}
From section~\ref{s:boundstates} we know that in 3D, two fermions in isolation can form a molecule for strong enough attractive interaction. The ground state of the system should be a
Bose-Einstein condensate of these tightly bound pairs.  However,
if we increase the density of particles in the system, we will
ultimately reach the point where the Pauli pressure of the
fermionic constituents becomes important and modifies the
properties of the system.  When the Fermi energy of the
constituents exceeds the binding energy of the molecules, we
expect that the equation of state will be fermionic, i.e. the
chemical potential will be proportional to the density to the
power of 2/3. Only when the size of the molecules is much smaller than the interparticle spacing, i.e. when the binding energy largely exceeds the Fermi energy, is the fermionic nature of the
constituents irrelevant -- tightly bound fermions are spread out widely in momentum space and do not run into the Pauli limitation of unity occupation per momentum state.

For too weak an attraction there is no bound state for two
isolated fermions, but Cooper pairs can form in the medium as discussed
above. The ground state of the system turns out to be a
condensate of Cooper pairs as described by BCS theory. In contrast to the physics of molecular condensates, however,
the binding energy of these pairs is much less than the Fermi energy
and therefore Pauli pressure plays a major role.

It was realized by Leggett~\cite{legg80}, building upon work by
Popov~\cite{popo66}, Keldysh~\cite{keld68} and Eagles~\cite{eagl69}, that the crossover from the BCS- to the
BEC-regime is smooth. This is somewhat surprising since the
two-body physics shows a threshold behavior at a critical
interaction strength, below which there is no bound state for two
particles. In the presence of the Fermi sea, however, we simply
cross over from a regime of tightly bound molecules to a regime
where the pairs are of much larger size than the interparticle
spacing. Closely following Leggett's work~\cite{legg80}, and its extension to finite temperatures by Nozi\`eres and Schmitt-Rink~\cite{nozi85}, we will describe the BEC-BCS
crossover in a simple ``one-channel'' model of a potential well.
Rather than the interaction strength $V_0$ as in section~\ref{s:boundstates}, we will take the scattering length $a$ as the parameter that ``tunes'' the interaction. The relation between $V_0$ and $a$ is given by Eq.~\ref{e:renormalize} and its explicit form Eq.~\ref{e:acutoff}.
For positive $a>0$, there is a two-body bound state available at $E_B
= -\hbar^2/m a^2$ (see table~\ref{t:momentumboundstates}), while
small and negative $a<0$ corresponds to weak attraction where Cooper pairs can form in the medium. In either case, for $s$-wave interactions, the orbital part of the pair wave function
$\varphi(\vect{r}_1,\vect{r}_2)$ will be symmetric under exchange
of the paired particles' coordinates and, in a uniform system, will only depend on their
distance $\left|\vect{r}_1-\vect{r}_2\right|$. We will explore the many-body wave function
\begin{equation}
    \Psi\left(\vect{r}_1,\dots,\vect{r}_N\right) = \mathcal{A}\left\{\varphi(\left|\vect{r}_1-\vect{r}_2\right|)\chi_{12}\dots \varphi(\left|\vect{r}_{N-1}-\vect{r}_N\right|)\chi_{N-1,N}\right\}
\label{e:fermicondensatepsi}
\end{equation}
that describes a condensate of such fermion pairs, with the
operator $\mathcal{A}$ denoting the correct antisymmetrization of
all fermion coordinates, and the spin singlet $\chi_{ij} =
\frac{1}{\sqrt{2}}(\left|\uparrow\right>_i \left|\downarrow\right>_j - \left|\downarrow\right>_i \left|\uparrow\right>_j)$. In the experiment, ``spin up'' and ``spin down'' will correspond to two atomic hyperfine states.

In second quantization notation we write
\begin{equation}
    \left|\Psi\right>_N = \int \prod_i d^3 r_i \,\varphi(\vect{r}_1 - \vect{r}_2) \Psi_\uparrow^\dagger(\vect{r}_1) \Psi_\downarrow^\dagger(\vect{r}_2) \dots \varphi(\vect{r}_{N-1} - \vect{r}_N) \Psi_\uparrow^\dagger(\vect{r}_{N-1}) \Psi_\downarrow^\dagger(\vect{r}_N) \left|0\right>
\end{equation}
where the fields $\Psi_\sigma^\dagger(\vect{r}) = \sum_k
c_{k\sigma}^\dagger \frac{e^{-i \vect{k} \cdot \vect{r}}}{\sqrt{\Omega}}$. With the
Fourier transform $\varphi(\vect{r}_1-\vect{r}_2) = \sum_k
\varphi_k \frac{e^{i \vect{k} \cdot (\vect{r}_1- \vect{r}_2)}}{\sqrt{\Omega}}$ we can introduce the pair creation operator
\begin{equation}
    b^\dagger = \sum_k \varphi_k c_{k\uparrow}^\dagger c_{-k\downarrow}^\dagger
\end{equation} and find
\begin{equation}
    \left|\Psi\right>_N = {b^\dagger}^{N/2} \left|0\right>
\end{equation}
This expression for  $\left|\Psi\right>_N$ is formally identical to
the Gross-Pitaevskii ground state of a condensate of bosonic particles. However, the
operators $b^\dagger$  obey bosonic commutation relations only in
the limit of tightly bound pairs.  For the commutators, we obtain
\begin{eqnarray}
\label{e:commutators}
    \left[b^\dagger,b^\dagger\right]_- &= \sum_{k k'} \varphi_k \varphi_{k'} \left[ c_{k\uparrow}^\dagger c_{-k\downarrow}^\dagger,c_{k'\uparrow}^\dagger c_{-k'\downarrow}^\dagger \right]_- &= 0  \\
    \left[b,b\right]_- &= \sum_{k k'} \varphi^*_k \varphi^*_{k'} \left[c_{-k\downarrow} c_{k\uparrow},c_{-k'\downarrow} c_{k'\uparrow}\right]_- &= 0 \nonumber\\
\left[b,b^\dagger\right]_- &= \sum_{k k'} \varphi^*_k \varphi_{k'}
\left[c_{-k\downarrow} c_{k\uparrow},c_{k'\uparrow}^\dagger
c_{-k'\downarrow}^\dagger \right]_- &= \sum_k |\varphi_k|^2 (1 -
n_{k\uparrow} - n_{k\downarrow}) \nonumber
\end{eqnarray}

The third commutator is equal to one only in the limit where the
pairs are tightly bound and occupy a wide region in momentum
space.  In this case, the occupation numbers $n_k$ of any momentum
state $k$ are very small (see section~\ref{s:evolution} below),
and $\left[b,b^\dagger\right]_- \approx \sum_k |\varphi_k|^2 = \int {\rm d}^3 r_1\int {\rm d}^3 r_2 \,|\varphi(\vect{r}_1,\vect{r}_2)|^2 = 1$.

Working with the $N$-particle state $\left|\Psi\right>_N$ is
inconvenient, as one would face a complicated combinatoric problem
in manipulating the sum over all the $c_k^\dagger$'s (as one
chooses a certain $k$ for the first fermion, the choices for the
second depend on this $k$, etc.). It is preferable to use the
grand canonical formalism, not fixing the number of atoms but the
chemical potential $\mu$. A separate, crucial step is to define a many-body state which is
a superposition of states with different atom numbers.  In the BEC
limit, this is analogous to the use of coherent states (vs. Fock
states) in quantum optics. Let $N_p = N/2$ be the number of pairs. Then,
\begin{eqnarray}
    \label{e:coherentstate}
    \mathcal{N}\left|\Psi\right> &= \sum_{J_{\rm even}} \frac{N_p^{J/4}}{(J/2)!} \left|\Psi\right>_J &= \sum_M \frac{1}{M!} {N_p^{M/2}\; b^\dagger}^M \left|0\right> = e^{\sqrt{N_p} \;b^\dagger} \left|0\right>\nonumber \\
&= \prod_k e^{\sqrt{N_p}\;  \varphi_k c_{k\uparrow}^\dagger
c_{-k\downarrow}^\dagger} \left|0\right> &= \prod_k (1 + \sqrt{N_p}\;
\varphi_k\, c_{k\uparrow}^\dagger c_{-k\downarrow}^\dagger)
\left|0\right>
\end{eqnarray}
The second to last equation follows since the operators $c_{k\uparrow}^\dagger c_{-k\downarrow}^\dagger$ commute for different $k$, and the last equation follows from $c_k^{\dagger 2} = 0$. If we choose the constant $\mathcal{N}
= \prod_k \frac{1}{u_k} = \prod_k \sqrt{1 + N_p |\varphi_k|^2}$,
then $\left|\Psi\right>$ becomes a properly normalized state
\begin{equation}
    \left|\Psi_{\rm BCS}\right> = \prod_k (u_k + v_k c_{k\uparrow}^\dagger c_{-k\downarrow}^\dagger) \left|0\right>
    \label{e:BCSstate}
\end{equation}
with $v_k = \sqrt{N_p}\,\varphi_k u_k$ and $|u_k|^2 + |v_k|^2 = 1$. This is
the famous BCS wave function, first introduced as a variational Ansatz, later shown to be the exact solution of the simplified Hamiltonian Eq.~\ref{e:Hsimplified} (below). It is a product of wave functions referring to the occupation of pairs of single-particle momentum states, $(\vect{k},\uparrow,-\vect{k},\downarrow)$.
As a special case, it describes a non-interacting Fermi sea, with all momentum pairs occupied up to the Fermi momentum ($u_k=0,
v_k=1$ for $k<k_F$ and $u_k=1, v_k=0$ for $k>k_F$). In general, for
a suitable choice of the $v_k$'s and $u_k$'s, it describes a ``molten'' Fermi sea, modified
by the coherent scattering of pairs with zero total momentum. Pairs of momentum states are seen to be in a superposition of being fully empty and fully occupied.
The above derivation makes it clear that this wave function
encompasses the entire regime of pairing, from point bosons
(small molecules) to weakly and non-interacting fermions.

\subsection{Gap and number equation}

The variational parameters $u_k$ and $v_k$ are derived in the
standard way by minimizing the free energy $E - \mu N =
\left<\hat{H} - \mu \hat{N}\right>$. The many-body Hamiltonian for
the system is
\begin{equation}
    \hat{H} = \sum_{k,\sigma} \epsilon_k c_{k\sigma}^\dagger c_{k\sigma} + \frac{V_0}{\Omega}\sum_{k,k',q} c_{k+\frac{q}{2} \uparrow}^\dagger c_{-k+\frac{q}{2}\downarrow}^\dagger c_{k'+\frac{q}{2}\downarrow} c_{-k'+\frac{q}{2}\uparrow}
\label{e:Hamiltonian}
\end{equation}

\begin{figure}
\centering
  \includegraphics[width=3in]{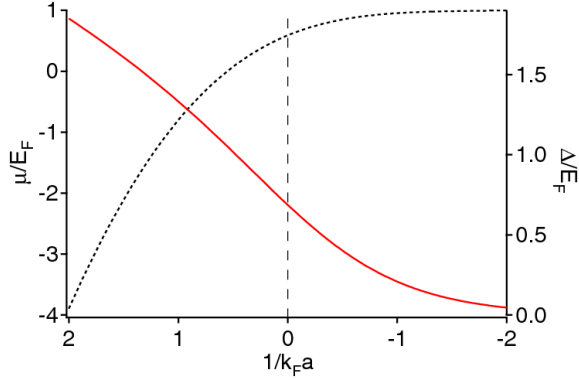}\\
  \caption[Chemical potential and gap in the BEC-BCS crossover]{Chemical potential (dotted line) and gap (straight line, red) in the BEC-BCS crossover as a function of the interaction parameter $1/k_Fa$. The BCS-limit of negative $1/k_F a$ is to the right on the graph. The resonance where $1/k_F a = 0$ is indicated by the dashed line.}\label{f:Deltamu}
\end{figure}

The dominant role in superfluidity is played by fermion pairs with
zero total momentum.   Indeed, as we have seen in
section~\ref{s:cooperproblem},  Cooper pairs with zero momentum
have the largest binding energy. Therefore, we simplify the
mathematical description by neglecting interactions between pairs
at finite momentum, i.e. we only keep the terms for $\vect{q} =
0$.
This is a very drastic simplification, as hereby density fluctuations are eliminated.  It is less critical for charged superfluids, where density fluctuations are suppressed by Coulomb interactions.  However, for neutral superfluids, sound waves  (the Bogoliubov-Anderson mode, see section~\ref{s:collexcitations}) are eliminated by this approximation.
The approximate Hamiltonian (``BCS Hamiltonian'') reads
\begin{equation}
    \hat{H} = \sum_{k,\sigma} \epsilon_k c_{k\sigma}^\dagger c_{k\sigma} + \frac{V_0}{\Omega}\sum_{k,k'} c_{k \uparrow}^\dagger c_{-k\downarrow}^\dagger c_{k'\downarrow} c_{-k'\uparrow}
\label{e:Hsimplified}
\end{equation}
The free energy becomes
\begin{eqnarray}
    \label{e:free-energy}
   \mathcal{F} = \left<\hat{H} - \mu \hat{N}\right> &=& \sum_k 2 \xi_k v_k^2 + \frac{V_0}{\Omega}\sum_{k,k'} u_k v_k u_{k'} v_{k'}\\
   \mbox{with }\;\xi_k &=& \epsilon_k - \mu \nonumber
\end{eqnarray}
Minimizing $E-\mu N$ leads to
\begin{eqnarray}
    v_k^2 &=& \frac{1}{2}\left(1 - \frac{\xi_k}{E_k}\right) \nonumber \\
    u_k^2 &=& \frac{1}{2}\left(1 + \frac{\xi_k}{E_k}\right) \nonumber \\
    \mbox{with }\;E_k &=& \sqrt{\xi_k^2 + \Delta^2}
    \label{e:ukvk}
\end{eqnarray}
where $\Delta$ is given by the {\it gap equation} $\Delta \equiv
\frac{V_0}{\Omega} \sum_k \left<c_{k\uparrow}
c_{-k\downarrow}\right> = - \frac{V_0}{\Omega} \sum_k u_k v_k = -
\frac{V_0}{\Omega} \sum_k \frac{\Delta}{2 E_k}$ or
\begin{equation}
    -\frac{1}{V_0} = \int \frac{d^3 k}{\left(2\pi\right)^3} \;\frac{1}{2E_{k}}
\end{equation}
Note the similarity to the bound state equation in free space,
Eq.~\ref{e:densityboundstates}, and in the simplified Cooper
problem, Eq.~\ref{e:Cooper}. An additional constraint is given by the {\it number equation} for the
total particle density $n = N / \Omega$
\begin{equation}
n = 2 \int \frac{d^3 k}{\left(2\pi\right)^3} \; v_k^2
\end{equation}
Gap and number equations have to be solved simultaneously to yield the two unknowns
$\mu$ and $\Delta$. We will once more replace $V_0$ by the
scattering length $a$ using prescription Eq.~\ref{e:renormalize},
so that the gap equation becomes (compare
Eq.~\ref{e:cooperrenorm})
\begin{equation}
-\frac{m}{4\pi\hbar^2 a} = \int \frac{d^3 k}{\left(2\pi\right)^3}
\left(\frac{1}{2 E_k} - \frac{1}{2 \epsilon_k}\right)
\end{equation}
where the integral is now well-defined. The equations can be
rewritten in dimensionless form with the Fermi energy $E_F =
\hbar^2 k_F^2 / 2m$ and wave vector $k_F = (3 \pi^2 n)^{1/3}$~\cite{orti05bcs}
\begin{eqnarray}
-\frac{1}{k_F a} &= &\frac{2}{\pi} \sqrt{\frac{\Delta}{E_F}}\; I_1\left(\frac{\mu}{\Delta}\right) \\
1 &=& \frac{3}{2}\left(\frac{\Delta}{E_F}\right)^{3/2} I_2\left(\frac{\mu}{\Delta}\right)\\
\mbox{with }\; I_1(z) &=& \int_0^\infty dx \;x^2 \left(\frac{1}{\sqrt{\left(x^2 - z\right)^2 + 1}} - \frac{1}{x^2}\right)\\
\mbox{and }\;  I_2(z) &=& \int_0^\infty dx \;x^2 \left(1 -
\frac{x^2 - z}{\sqrt{\left(x^2-z\right)^2 + 1}}\right)
\end{eqnarray}
This gives
\begin{eqnarray}
    -\frac{1}{k_F a} &=& \frac{2}{\pi} \left(\frac{2}{3 I_2\left(\frac{\mu}{\Delta}\right)}\right)^{1/3} I_1\left(\frac{\mu}{\Delta}\right)\\
    \frac{\Delta}{E_F} &=& \left(\frac{2}{3 I_2\left(\frac{\mu}{\Delta}\right)}\right)^{2/3}
\end{eqnarray}
The first equation can be inverted to obtain $\mu / \Delta$ as a
function of the {\it interaction parameter} $1/k_F a$, which can
then be inserted into the second equation to yield the gap $\Delta$.
The result for $\mu$ and $\Delta$ as a function of $1/k_F a$ is
shown in Fig.~\ref{f:Deltamu}. It is possible to obtain analytic
expressions for the solutions in terms of complete elliptic
integrals~\cite{mari98becbcs}.

In this derivation, we have combined the simplified Hamiltonian,
Eq.~\ref{e:Hsimplified} with the BCS variational Ansatz.  Alternatively one can
apply a decoupling (mean field) approximation to the Hamiltonian~\cite{peth02bec}. Expecting that there will be
some form of pair condensate, we assume that the pair creation and
annihilation operator only weakly fluctuates around its non-zero expectation value
\begin{eqnarray}
    C_k=  \left<c_{k\uparrow} c_{-k\downarrow}\right>= -\left< c_{k\uparrow}^\dagger
    c_{-k\downarrow}^\dagger\right>
\end{eqnarray}
chosen to be real (since the relative phase of states which differ
in particle number by two can be arbitrarily chosen).
That is, we write
\begin{equation}
c_{k\uparrow} c_{-k\downarrow} = C_k + (c_{k\uparrow} c_{-k\downarrow} - C_k)
\end{equation}
with the operator in parentheses giving rise to fluctuations that are small on the scale of $C_k$.
The gap parameter $\Delta$ is now defined as
\begin{equation}
\Delta=\frac{V_0}{\Omega} \sum_{k} C_k
\end{equation}
We only include terms in the interaction part of the Hamiltonian which
involve the $C_k$'s at least once. That is, we neglect the
correlation of fluctuations of the pair creation and annihilation
operators. One obtains
\begin{equation}
    \hat{H} = \sum_{k} \epsilon_k (c_{k\uparrow}^\dagger c_{k\uparrow}+ c_{k\downarrow}^\dagger c_{k\downarrow}) -\Delta \sum_k\left( c_{k \uparrow}^\dagger c_{-k\downarrow}^\dagger + c_{k \downarrow}
    c_{-k\uparrow} + \sum_{k'} C_{k'}\right)
\end{equation}

This Hamiltonian  is bilinear in the creation and annihilation
operators and can easily be solved by a Bogoliubov transformation~\cite{bogo58,vala58,peth02bec} from the
particle operators $c_{k\downarrow}$ and $c_{k\uparrow}$ to new
quasi-particle operators $\gamma_{k\uparrow}$ and
$\gamma_{k\downarrow}$:
\begin{eqnarray}
    \gamma_{k\uparrow} &=& u_k c_{k\uparrow} - v_k c_{-k\downarrow}^\dagger \\
    \gamma_{-k\downarrow}^\dagger &=& u_k c_{-k\downarrow}^\dagger + v_k c_{k\uparrow} \nonumber
    \label{e:quasiparticles}
\end{eqnarray}
The $u_k$ and $v_k$ are determined from the requirements that the
new operators fulfill fermionic commutation relations and that the
transformed Hamiltonian is diagonal with respect to the
quasiparticle operators.  This solution is identical to the one
obtained before for the $u_k$ and $v_k$, and the transformed
Hamiltonian becomes
\begin{equation}
\hat{H} - \mu \hat{N} = - \frac{\Delta^2}{V_0/\Omega} + \sum_k (\xi_k - E_k) + \sum_k E_k
(\gamma_{k\uparrow}^\dagger \gamma_{k\uparrow} +
\gamma_{k\downarrow}^\dagger \gamma_{k\downarrow})
\label{e:BogoliubovH}
\end{equation}
The first two terms give the free
energy $E - \mu N$ of the pair condensate, identical to Eq.~\ref{e:free-energy} when the correct $u_k$ and $v_k$ are inserted. The third term represents the
energy of excited quasi-particles, and we identify $E_k$ as
excitation energy of a quasi-particle.  The superfluid ground
state is the quasi-particle vacuum: $\gamma_{k\uparrow}
\left|\Psi\right> = 0 = \gamma_{k\downarrow} \left|\Psi\right>$.

This approach via the pairing field is analogous to the Bogoliubov
treatment of an interacting Bose-Einstein condensate: There, the
creation and annihilation operators for atoms with zero momentum
are replaced by $\sqrt{N_0}$, the square root of the number of
condensed atoms (i.e.~a coherent field). In the interaction term
of the Hamiltonian all terms are dropped that contain less than
two factors of $\sqrt{N_0}$.  In other words, the Hamiltonian
(Eq.~\ref{e:Hsimplified}) is solved by keeping only certain pair
interactions, either by using a variational pairing wave function,
or by introducing a mean pairing field. It should be noted that
these approximations are not even necessary, as the BCS
wave function can be shown to be the {\it exact} solution to the
reduced Hamiltonian $Eq.~\ref{e:Hsimplified}$~\cite{duke04review}.

\subsection{Discussion of the three regimes -- BCS, BEC and crossover}
\subsubsection{BCS limit}
\label{s:BCSlimit}
In the BCS-limit of weak attractive interaction, $k_F a
\rightarrow 0_-$, we have\footnote{This follows by substituting
$\xi = x^2 - z$ in the integrals and taking the limit
$z\rightarrow \infty$. One has $I_1(z) \approx
\sqrt{z}\left(\log(8z) - 2\right)$ and $I_2(z) = \frac{2}{3}
z^{3/2}$.}
\begin{eqnarray}
    \mu &\approx& E_F \\
    \Delta &\approx& \frac{8}{e^2} e^{-\pi/2k_F\left|a\right|}
    \label{e:BCSLimit}
\end{eqnarray}

The first equation tells us that adding a spin up and spin down
particle to the system costs a Fermi energy per particle (with the
implicit assumption that both a spin up and a spin down particle
are added, raising the total energy by $2 \mu$).  In the weakly
interacting BCS limit Pauli blocking still dominates over
interactions, and hence the particles can only be added at the
Fermi surface. The second equation is the classic result of BCS
theory for the superfluid gap\footnote{The present mean field
treatment does not include density fluctuations, which modify the
prefactor in the expression for the gap $\Delta$ ~\cite{gork61,peth02bec}.}.
Compared to the bound state energy for a single Cooper pair on top
of a non-interacting Fermi sea, Eq.~\ref{e:cooperproblem}, the gap
is larger (the negative exponent is smaller by a factor of two),
as the entire collection of particles now takes part in the
pairing\footnote{In the self-consistent BCS solution, not only the
momentum states above the Fermi surface contribute to pairing, but
also those {\it below} it, in a symmetric shell around the Fermi
momentum. In the Cooper problem the states below the Fermi surface
were excluded, reducing the effective density of states by a
factor of two.}. However, the gap is still exponentially small
compared to the Fermi energy: Cooper pairing is fragile.

The ground state energy of the BCS state can be calculated from Eq.~\ref{e:free-energy} and is
\begin{equation}
    E_{\rm G,\, BCS} = \frac{3}{5} N E_F - \frac{1}{2}\,\rho(E_F)\, \Delta^2
\end{equation}

The first term is the energy of the non-interacting normal state,
where $\frac{3}{5}E_F$ is the average kinetic energy per fermion
in the Fermi sea. The second term is the energy due to
condensation, negative as it should be, indicating that the BCS
state is energetically favorable compared to the normal state.

Although the total kinetic energy of the Fermi gas has been
increased (by populating momentum states above $E_F$), the total
energy is lower due to the gain in potential energy.  This is
valid for any kind of pairing (i.e. proton and electron forming a
hydrogen atom), since the localization of the pair wave function
costs kinetic energy.

The energy of the BCS state, $- \frac{1}{2}\,\rho(E_F)\, \Delta^2$
can be interpreted in two ways. One way refers to the wave function Eq.~\ref{e:fermicondensatepsi}, which consists of $N/2$ identical fermion pairs.  The energy per
pair is then $- \frac{3}{4} \Delta^2/E_F$.
The other interpretation refers to the BCS wave function Eq.~\ref{e:BCSstate}. It is essentially a product of a ``frozen'' Fermi sea (as $v_k \approx 1$, $u_k \approx 0$ for low values of $k$)
with a paired component consisting of $\sim \rho(E_F) \, \Delta
\sim N \Delta/E_F$  pairs, located in an energy shell of width
$\Delta$ around the Fermi energy.  They each contribute a pairing
energy on the order of $\Delta$.  The second interpretation
justifies the picture of a Cooper pair condensate. In the solution
of the Cooper problem (section~\ref{s:cooperproblem}), the pair wave function
has a peak occupation per momentum state of $\sim 1/ \rho(E_F)
\Delta$. Therefore, one can stack up  $\sim  \rho(E_F) \Delta$
pairs with zero total momentum without getting into serious
trouble with the Pauli exclusion principle and construct a
Bose-Einstein condensate consisting of $\sim \rho(E_F) \Delta$
Cooper pairs~\footnote{Similarly to the fermion pairs described by the operator $b^\dagger$, the Cooper pairs from section~\ref{s:cooperproblem} are not bosons, as shown by the equivalent of Eq.~\ref{e:commutators}. However, if there were only a few Cooper pairs, much less than $\rho(E_F) \Delta$, the occupation of momentum states $n_k$ would still be very small compared to 1 and these pairs would be to a good approximation bosons.}.

It depends on the experiment whether it reveals a pairing energy
of $\frac{1}{2} \Delta^2/E_F$ or of $\Delta$.  In RF
spectroscopy, all momentum states can be excited (see section~\ref{e:chap2RFspectroscopy}),
and the spectrum shows a gap of $\frac{1}{2} \Delta^2/E_F$ (see section~\ref{s:RFspectrum}).
Tunnelling experiments in superconductors probe the region
close to the Fermi surface, and show a pairing gap of $\Delta$.

The two interpretations for the BCS energy carry along two possible choices of the pairing wave function (see section~\ref{s:evolution}). The first one is $\varphi_k = u_k/v_k\sqrt{N_p}$, which can be shown to extend throughout the whole Fermi sea from zero to slightly above $k_F$, whereas the second one, $\psi(k) = u_k v_k$, is concentrated around the Fermi surface (see Fig.~\ref{f:excitation}).

To give a sense of scale, Fermi energies in dilute atomic gases
are on the order of a $\mu\rm K$, corresponding to $1/k_F \sim
4\,000\, a_0$. In the absence of scattering resonances, a typical 
scattering length would be about $50-100\; a_0$ (on
the order of the van der Waals-range). Even if $a < 0$, this would result
in a  vanishingly small gap $\Delta/k_B \approx 10^{-30}\dots
10^{-60}\, \rm K$. Therefore, the realization of superfluidity in
Fermi gases requires scattering or Feshbach resonances to increase
the scattering length, bringing the gas into the strongly
interacting regime where $k_F \left|a\right| > 1$ (see
chapter~\ref{c:feshbach}). In this case, the above mean field
theory predicts $\Delta > 0.22 \;E_F$ or $\Delta/k_B > 200\, \rm
nK$ for $k_F |a| > 1$, and this is the regime where current
experiments are operating.

\subsubsection{BEC limit}
\label{s:BEClimit}

In the BEC limit of tightly bound pairs, for $k_F a \rightarrow
0_+$, one finds\footnote{This result follows from the expansion of
the integrals for $z<0$ and $|z|\rightarrow \infty$. One finds
$I_1(z) = -\frac{\pi}{2}\sqrt{|z|}
-\frac{\pi}{32}\frac{1}{|z|^{3/2}}$ and $I_2(z) =
\frac{\pi}{8}\frac{1}{\sqrt{|z|}}$.}
\begin{eqnarray}
    \label{e:BEClimit}
    \mu = -\frac{\hbar^2}{2 m a^2} + \frac{\pi \hbar^2 a n}{m}\\
    \Delta \approx \sqrt{\frac{16}{3\pi}} \frac{E_F}{\sqrt{k_F a}}
\label{e:BEClimitgap}
\end{eqnarray}

The first term in the expression for the chemical potential is the
binding energy per fermion in a tightly bound molecule (see
table~\ref{t:momentumboundstates}). This reflects again the
implicit assumption (made by using the wave function in
Eq.~\ref{e:fermicondensatepsi}) that we always add {\it two}
fermions of opposite spin at the same time to the system.

The second term is a mean field contribution describing the
repulsive interaction between molecules in the gas. Indeed, a
condensate of molecules of mass $m_M = 2m$, density $n_M = n/2$
and a molecule-molecule scattering length $a_M$ will have a
chemical potential $\mu_M = \frac{4\pi \hbar^2 a_M n_M}{m_M}$.
Since $\mu_M$ is twice the chemical potential for each fermion, we
obtain from the above expression the molecule-molecule scattering
length $a_M = 2 a$. However, this result is not exact. Petrov,
Shlyapnikov and Salomon~\cite{petr04dimers} have performed an
exact calculation for the interaction between four fermions and
shown that $a_M = 0.6\, a$. The present mean field approach neglects
correlations between different pairs, or between one fermion and a
pair. If those are included, the correct few-body physics is
recovered~\cite{pier00becbcs,holl04bosefermi,hu06becbcs}.

The expression for the quantity $\Delta$ signifies neither the binding energy of molecules nor does it correspond to a gap in the excitation spectrum. Indeed, in the BEC-regime, as soon as $\mu <0$, there is no longer a gap at non-zero $k$ in the single-fermion excitation spectrum (see Fig.~\ref{f:pairwavefunction} below). Instead, we have for the quasi-particle energies $E_k = \sqrt{(\epsilon_k - \mu)^2 + \Delta^2} \approx |\mu| + \epsilon_k + \frac{\Delta^2}{2|\mu|}$. So $\Delta$ itself does not play a role in the BEC-regime, but only the combination $\Delta^2/|\mu|$ is important. As we see from Eq.~\ref{e:BEClimit},
\begin{equation}
    \frac{\Delta^2}{2|\mu|} = \frac{8}{3\pi} \frac{E_F^2}{k_F a} \frac{2 m a^2}{\hbar^2} = \frac{4}{3\pi}\frac{\hbar^2}{m} k_F^3 a = \frac{4\pi\hbar^2}{m} n\; a
\label{e:Deltasquared}
\end{equation}
which is two times the molecular mean field. In fact, we will show in section~\ref{s:excitations} that it can be interpreted here as the mean field energy experienced by a single fermion in a gas of molecules.

It might surprise that the simplified Hamiltonian Eq.~\ref{e:Hsimplified} contains interactions between two molecules or between a molecule and a single fermion at all. In fact, a crucial part of the simplification has been to explicitly {\it neglect} such three- and four-body interactions. The solution to this puzzle lies in the Pauli principle, which acts as an effective repulsive interaction: In a molecule, each constituent fermion is confined to a region of size $\sim a$ around the molecule's center of mass (see next section). The probability to find another like fermion in that region is strongly reduced due to Pauli blocking. Thus, effectively, the motion of molecules is constrained to a reduced volume $\Omega' = \Omega - c N_M a_M^3$, with the number of molecules $N_M$ and $c$ on the order of 1. This is the same effect one has for a gas of hard-sphere bosons of size $a_M$, and generally for a Bose gas with scattering length $a_M$. An analogous argument leads to the effective interaction between a single fermion and a molecule.
We see that the only way interactions between pairs, or between a pair and a single fermion, enter in the simplified description of the BEC-BCS crossover is via the anti-symmetry of the many-body wave function.

\subsubsection{Evolution from BCS to BEC}
\label{s:evolution}
\begin{figure}
\centering
  \includegraphics[width=3in]{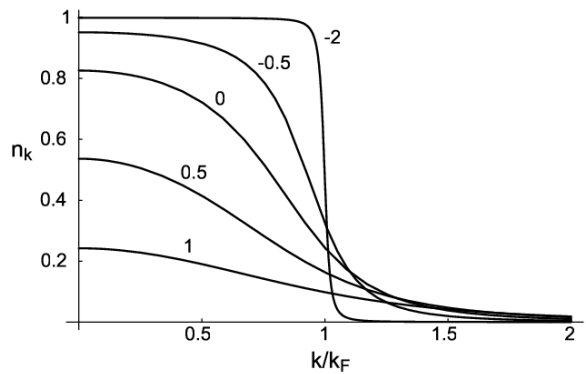}\\
  \caption[Occupation $n_k$ of momentum states $k$ in the BEC-BCS crossover]{Occupation $n_k$ of momentum states $k$ in the BEC-BCS crossover. The numbers give the interaction parameter $1/k_F a$. After~\cite{nozi85}.}\label{f:momoccupation}
\end{figure}

Our variational approach smoothly interpolates between the two
known regimes of a BCS-type superfluid and a BEC of molecules. It is a crossover, which occurs approximately between $1/k_F a = -1$ and +1 and is fully continuous. The occupation of momentum states
$n_k = v_k^2$ evolves smoothly from the step-function $\Theta(k_F
- k)$ of a degenerate Fermi gas, broadened over a width $\Delta
\ll E_F$ due to pairing, to that of $N_p$ molecules, namely the number of molecules $N_p$ times the probability $|\varphi_k|^2$ to find a molecule with momentum $k$ (we have $\varphi_k = \frac{(2 \pi a)^{3/2}}{\sqrt{\Omega}}\frac{1}{\pi}\frac{1}{1+k^2 a^2}$) (see
Fig.~\ref{f:momoccupation}). It is also interesting to follow the
evolution of the ``Cooper pair'' wave function\footnote{Note that
this definition is not equal to the Fourier transform of the pair
wave function $\varphi(\vect{r})$ introduced in
Eq.~\ref{e:fermicondensatepsi}, which would be $v_k/u_k\sqrt{N_p}$. The
definition given here is the two-point correlation function. Both
definitions for the Cooper pair wave function show a sharp feature,
either a peak or an edge at the Fermi surface, of width $\sim
\delta k$, thus giving similar behavior for the real space
wave function.} both in $k$-space, where it is given by
$\left<\Psi_{BCS}\right|c^\dagger_{k\uparrow}
c^\dagger_{-k\downarrow}\left|\Psi_{BCS}\right> = u_k v_k$, and in
real space, where it is
\begin{eqnarray}
\psi(\vect{r}_1,\vect{r}_2) &=&
\left<\Psi_{BCS}\right|\Psi_{\uparrow}^\dagger(\vect{r}_1)\Psi_{\downarrow}^\dagger(\vect{r}_2)\left|\Psi_{BCS}\right>
= \frac{1}{\Omega}\sum_k u_k v_k e^{-i \vect{k} \cdot (\vect{r}_1-\vect{r}_2)} \nonumber \\
&=&\frac{1}{\Omega}\sum_k \frac{\Delta}{2E_k}\, e^{-i \vect{k} \cdot
(\vect{r}_1-\vect{r}_2)} \label{e:cooperpairwavefunction}
\end{eqnarray}

\begin{figure}
\centering
 \includegraphics[width=5.5in]{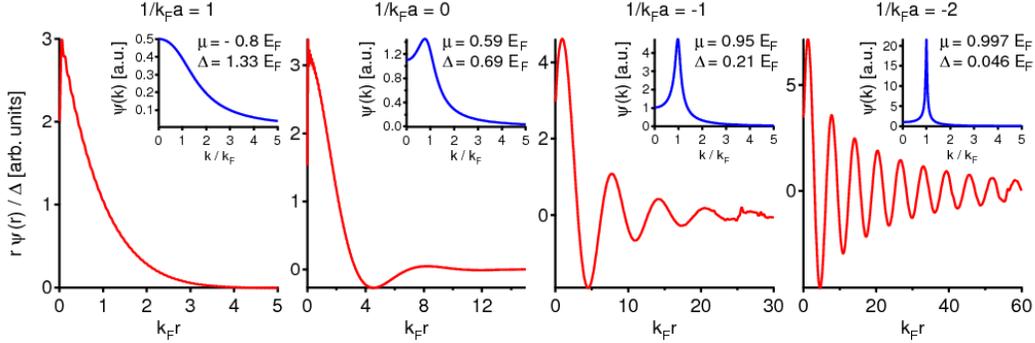}\\
  \caption[Evolution of the spatial pair wave function $\psi(r)$ in the BEC-BCS crossover]{Evolution of the spatial pair wave function $\psi(r)$ in the BEC-BCS crossover. The inset shows the Fourier transform $\psi(k)$, showing clearly that in the BCS-limit, momentum states around the Fermi surface make the dominant contribution to the wave function. In the crossover, the entire Fermi sphere takes part in the pairing. In the BEC-limit, $\psi(k)$ broadens as the pairs become more and more tightly bound. $\psi(r)$ was obtained via numerical integration of $\int_{-\mu}^\infty d\xi \frac{\sin(r \sqrt{\xi + \mu})}{\sqrt{\xi^2 + \Delta^2}}$ (here, $\hbar = 1 = m$), an expression that follows from Eq.~\ref{e:cooperpairwavefunction}.}\label{f:pairwavefunction}
\end{figure}

\begin{figure}
\centering
 \includegraphics[width=3in]{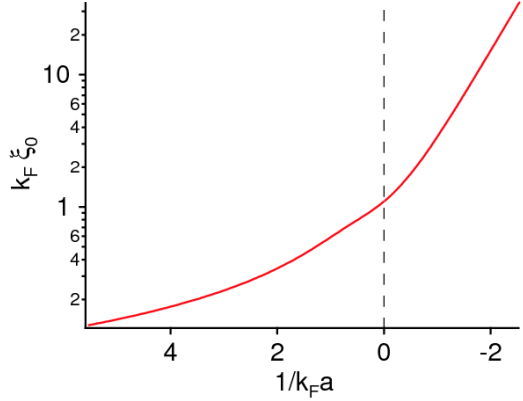}\\
  \caption[From tightly bound molecules to long-range Cooper pairs]{From tightly bound molecules to long-range Cooper pairs. Evolution of the pair size $\xi_0  =\sqrt{\frac{\left<\psi(\vect{r})\right|r^2\left|\psi(\vect{r})\right>}{\left<\psi(\vect{r})|\psi(\vect{r})\right>}}$ as a function of the interaction parameter $1/k_F a$. On resonance (dashed line), the pair size is on the order of the inverse wave vector, $\xi_0(0) \sim \frac{1}{k_F}$, about a third of the interparticle spacing.}\label{f:pairsize}
\end{figure}

In the BCS limit, the pairing occurs near the Fermi surface $k=
k_F$, in a region of width $\delta k \sim \frac{\partial
k}{\partial \epsilon} \delta \epsilon \approx \frac{\Delta}{\hbar
v_F}$, where $v_F$ is the velocity of fermions at the Fermi
surface. Therefore, the spatial wave function of Cooper pairs has
a strong modulation at the inverse wave vector $1/k_F$, and an
overall extent of the inverse width of the pairing region, $\sim
1/\delta k \sim \frac{\hbar v_F}{\Delta} \gg 1/k_F$. More
quantitatively, Eq.~\ref{e:cooperpairwavefunction} gives (setting
$r = \left|\vect{r}_1-\vect{r}_2\right|$)~\cite{bard57}
\begin{equation}
    \psi(r) = \frac{k_F}{\pi^2 r}\frac{\Delta}{\hbar v_F} \sin(k_F r)\, K_0\left(\frac{r}{\pi \xi_{BCS}}\right)\; \stackrel{r\rightarrow \infty}{\sim}\;\sin\left(k_F r\right)\, e^{-r/(\pi \xi_{BCS})}
\end{equation}
where $K_0(k r)$ is the modified Bessel function that falls off as
$e^{-k r}$ at infinity. We have encountered a similar exponential
envelope function for a two-body bound state (see
table~\ref{t:boundstate}). The characteristic size of the Cooper
pair, or the {\it two-particle correlation length} $\xi_0$, can be
defined as $\xi_0^2
=\frac{\left<\psi(\vect{r})\right|r^2\left|\psi(\vect{r})\right>}{\left<\psi(\vect{r})|\psi(\vect{r})\right>}$,
and this gives indeed $\xi_0 \sim 1/\delta k$,
\begin{equation}
    \xi_0 \approx \xi_{BCS} \equiv \frac{\hbar v_F}{\pi \Delta} \gg 1/k_F \quad \mbox{in the BCS-limit}
\end{equation}

In the BEC limit, $u_k v_k \propto \frac{1}{1+(k a)^2}$, and
so
\begin{equation}
    \psi(\vect{r}_1,\vect{r}_2) \sim \frac{e^{- \left|\vect{r}_1 - \vect{r}_2\right|/a}}{\left|\vect{r}_1 - \vect{r}_2\right|}
\end{equation}
which is simply the wave function of a molecule of size $\sim a$
(see table~\ref{t:boundstate}). The two-particle correlation
length\footnote{This length scale should be distinguished from the
{\it coherence length} $\xi_{phase}$ that is associated with
spatial fluctuations of the order parameter. The two length scales
coincide in the BCS-limit, but differ in the BEC-limit, where
$\xi_{phase}$ is given by the healing length $\propto
\frac{1}{\sqrt{n a}}$. See~\cite{pist94xi} for a detailed
discussion.} is thus $\xi_0 \sim a$.
Figs.~\ref{f:pairwavefunction} and~\ref{f:pairsize} summarize the
evolution of the pair wave function and pair size throughout the
crossover.

\subsection{Single-particle and collective excitations}
\label{s:excitations}

Fermionic superfluids can be excited in two ways: Fermi-type excitations of single atoms or Bose-like excitations of fermion pairs. The first is related to pair breaking, the second to density fluctuations -- sound waves.

\subsubsection{Single-particle excitations}
\label{s:singleparticleexcitations}

The BCS-state $\left|\Psi_{\rm BCS}\right>$ describes a collection
of pairs, each momentum state pair
$(\vect{k}\uparrow,-\vect{k}\downarrow)$ having probability
amplitude $u_k$ of being empty and $v_k$ of being populated.  We
now calculate the energy cost for adding a {\it single} fermion in state
$\vect{k}\uparrow$, which does not have a pairing partner, i.e.
the state $-\vect{k}\downarrow$ is empty.  This requires a kinetic
energy $\xi_k$ (relative to the chemical potential). For the other
particles, the states $(\vect{k}\uparrow,-\vect{k}\downarrow)$
are no longer available,  and according to Eq.~\ref{e:free-energy}
the (negative) pairing energy is increased by $-2 \xi_k v_k^2 - 2
\frac{V_0}{\Omega} u_k v_k \sum_{k'} u_{k'} v_{k'}$, which equals
$-\xi_k (1-\frac{\xi_k}{E_k}) + \frac{\Delta^2}{E_k} = E_k -
\xi_k$ (see Eq.~\ref{e:ukvk}). The total cost for adding one
fermion is thus simply $\xi_k + (E_k - \xi_k) = E_k$ (again relative to $\mu$, i.e. this is the cost in free energy). In the same
way, one calculates the cost for removing a fermion from the
BCS-state (e.g. deep in the Fermi sea), and leaving behind an
unpaired fermion in state $-\vect{k}\downarrow$.  The result is
again $E_k$. This shows that adding or removing a particle creates
a quasi-particle with energy $E_k$, as we had found already via
the Bogoliubov transformation Eq.~\ref{e:BogoliubovH}. For
example, the quasi-particle excitation
\begin{equation}
    \gamma_{k\uparrow}^\dagger\left|\Psi_{\rm BCS}\right> = c_{k\uparrow}^\dagger\prod_{l\ne k}\left(u_l + v_l c_{l\uparrow}^\dagger c_{-l\downarrow}^\dagger\right)\left|0\right>
\end{equation}
correctly describes the removal of the momentum pair at $(\vect{k}\uparrow,-\vect{k}\downarrow)$, and the addition of a single fermion in $\vect{k}\uparrow$.

\begin{figure}
\centering
 \includegraphics[width=5.5in]{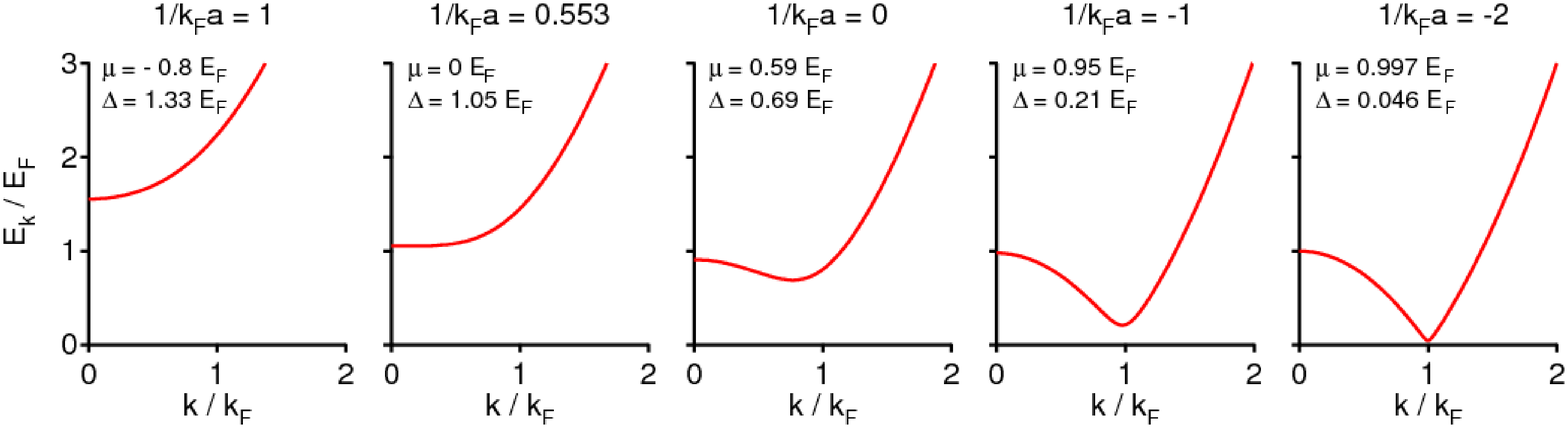}\\
  \caption[Evolution of the single-particle excitation spectrum in the BEC-BCS crossover]{Evolution of the single-particle excitation spectrum in the BEC-BCS crossover. On the BEC-side, for $\mu < 0$, the minimum required energy to add a particle is $\sqrt{\mu^2 + \Delta^2}$ and occurs at $k=0$. This qualitatively changes at $1/k_F a = 0.553$ where $\mu = 0$. For $\mu>0$, the minimum energy is $\Delta$ and occurs at $k = \sqrt{2 m \mu/\hbar^2}$.}\label{f:excitation}
\end{figure}

Fig.~\ref{f:excitation} shows the single-particle excitation
energy $E_k$ for different interaction strengths in the BEC-BCS
crossover. For $\mu > 0$, the minimum energy required to remove a
particle from the condensate occurs for $\xi_k = \mu$ and is
$\Delta$, which gives $\Delta$ the name of the superfluid gap. One
dramatic consequence of this gap is that it prevents single
fermions to enter the superfluid, resulting in phase separation in
imbalanced Fermi mixtures~\cite{shin06phase}.
For $\mu < 0$ the minimum
energy to remove a particle becomes $\sqrt{\mu^2 + \Delta^2}$ and occurs
for $k = 0$.

To excite the system without adding or removing particles can be
done in two ways:
One can remove a particle,  requiring an energy $E_k$, and add it
back at energy cost $E_{k'}$, thus creating two unpaired particles
with momenta $k$ and $k'$.  The second possibility is to excite a
fermion pair in $(\vect{k}\uparrow,-\vect{k}\downarrow)$ into the
state orthogonal to the ground state, which can be written
\begin{equation}
      \gamma_{k\uparrow}^\dagger\gamma_{-k\downarrow}^\dagger\left|\Psi_{\rm BCS}\right> = (v_k - u_k c_{k\uparrow}^\dagger c_{-k\downarrow}^\dagger)\prod_{l\ne k}\left(u_l + v_l c_{l\uparrow}^\dagger c_{-l\downarrow}^\dagger\right)\left|0\right>
\end{equation}

Instead of the pairing energy $\xi_k-E_k$ for that state, the energy for such an excitation is $\xi_k + E_k$, that is, this excited pair state lies an energy $2E_k$ above the BCS ground state.
The minimum energy required to excite the system, without changing the particle number, is thus $2\Delta$ in the BCS limit, and $2\sqrt{\mu^2 + \Delta^2}$ in the BEC-limit.
In the latter case, from Eq.~\ref{e:BEClimit}, one has
\begin{equation}
    2\sqrt{\mu^2 + \Delta^2} \approx |E_B| - \mu_M + \frac{2\Delta^2}{|E_B|}
\end{equation}
The first two terms $|E_B| -\mu_M = \hbar^2/ma^2 - \mu_M$ give the energy required to remove a molecule (the positive mean field $\mu_M$ pushes this energy closer to threshold). The last term will then correspond to the energy required to add two unpaired fermions into the system. From our discussion in section~\ref{s:BEClimit}, we expect that this should cost twice the mean field energy $\mu_{\rm BF} = \frac{4\pi\hbar^2 a_{\rm BF} n_M}{m_{\rm BF}}$ of a fermion interacting with a cloud of bosons, the molecules. Here, $a_{\rm BF}$ is the Boson-Fermion scattering length and $m_{\rm BF} = 2 m_B m_F/(m_B+m_F) = 4/3\,m$ is twice the reduced mass of the boson-fermion system. With the help of Eq.~\ref{e:Deltasquared} we equate
\begin{equation}
    \frac{\Delta^2}{|E_B|} = \frac{4\pi\hbar^2 a n}{m} \equiv \frac{4\pi\hbar^2 a_{\rm BF} n_M}{m_{\rm BF}} = \frac{3\pi\hbar^2 a_{\rm BF} \frac{n}{2}}{m}
\end{equation}
and obtain the Boson-Fermion scattering length at the mean field level,
\begin{equation}
    a_{\rm BF} = \frac{8}{3}\, a
\end{equation}
The exact value $a_{\rm BF} = 1.18\, a$ has been obtained already 50 years ago~\cite{skor57abf}.

\subsubsection{RF excitation into a third state}
\label{s:RFspectrum}
The hyperfine structure of ultracold atoms offers more than just two states ``spin up'' and ``spin down''. This allows for a new type of single-particle excitation, not available for electrons in superconductors, namely the transfer of, say, a spin up fermion into a third, empty state, $\left|3\right>$, via a radiofrequency (RF) transition (see sections~\ref{e:chap2RFspectroscopy} and~\ref{s:rfspectroscopy}). We have all the tools ready to calculate the excitation spectrum for RF spectroscopy in the case where the third state does not interact with atoms in the spin up or spin down states. Due to its long wavelength, Doppler shifts are negligible and the RF excitation flips the spin from $\left|\uparrow\right>$ to $\left|3\right>$ and vice versa regardless of the momentum state of the atom, and without momentum transfer. The RF operator is thus
\begin{equation}
    \hat{V} = V_0 \sum_k c^\dagger_{k3} c_{k\uparrow} + c^\dagger_{k\uparrow}c_{k3}
\end{equation}
where $V_0$ is the strength of the RF drive (the Rabi frequency $\omega_R = 2V_0/\hbar$) taken to be real.
As the third state is initially empty, only the first part contributes when acting on the initial state. To calculate the action of the spin flip $c^\dagger_{k3} c_{k\uparrow}$ on the BCS state, we express $c_{k\uparrow} = u_k \gamma_{k\uparrow} + v_k \gamma^\dagger_{-k\downarrow}$ in terms of the Bogoliubov quasi-particle operators (Eq.~\ref{e:quasiparticles}). As the BCS-state is the quasi-particle vacuum, $\gamma_{k\uparrow} \left|\Psi_{\rm BCS}\right> = 0$, and one has
\begin{eqnarray}
\label{e:RFaction}
    c^\dagger_{k3} c_{k\uparrow}\left|\Psi_{\rm BCS}\right> = v_k c^\dagger_{k3} \gamma^\dagger_{-k\downarrow} \left|\Psi_{\rm BCS}\right> \qquad \mbox{and thus}\\
    \hat{V}\left|\Psi_{\rm BCS}\right> = V_0 \sum_k v_k c^\dagger_{k3} \gamma^\dagger_{-k\downarrow} \left|\Psi_{\rm BCS}\right>
\label{e:fullRFaction}
\end{eqnarray}
When the RF excitation removes the particle from the BCS state, it creates a quasi-particle with a cost in total energy of $E_k - \mu$ (see section~\ref{s:singleparticleexcitations}). The energy cost for creating the particle in the third state is, apart from the bare hyperfine splitting $\hbar \omega_{\uparrow 3}$, the kinetic energy $\epsilon_k$. In total, the RF photon has to provide the energy
\begin{equation}
    \label{e:dispersion}
    \hbar \Omega(k) = \hbar \omega_{\uparrow 3} + E_k + \epsilon_k - \mu
\end{equation}
Fermi's Golden Rule gives now the transition rate $\Gamma(\omega)$ at which particles leave the BCS state and arrive in state $\left|3\right>$ ($\omega$ is the RF frequency).
\begin{equation}
   \Gamma(\omega) \equiv \frac{2\pi}{\hbar} \sum_f \left|\left<f\left|\hat{V}\right|\Psi_{\rm BCS}\right>\right|^2 \delta\left(\hbar\omega - E_f\right)
\end{equation}
where the sum is over all eigenstates $\left|f\right>$ with energy $E_f$ (relative to the energy of the BCS state). The relevant eigenstates are just the states calculated in Eq.~\ref{e:RFaction}: $\left|k\right> \equiv c^\dagger_{k3} \gamma_{-k\uparrow}\left|\Psi_{\rm BCS}\right>$ of energy $\hbar \Omega(k)$. The sum over final states becomes a sum over momentum states, and, according to Eq.~\ref{e:fullRFaction}, the matrix element is $V_0 v_k$.
The condition for energy conservation, $\hbar\omega = \hbar\Omega(k)$, can be inverted via Eq.~\ref{e:dispersion} to give $\epsilon_k$ in terms of $\omega$. The delta function then becomes $\delta(\hbar\omega-\hbar\Omega(k)) = \frac{1}{\hbar}\frac{d \epsilon_k}{d\Omega} \,\delta(\epsilon_k - \epsilon(\omega))$. With $\frac{d \Omega}{d \epsilon_k} = \frac{\xi_k}{E_k}+1 = 2 u_k^2$, we obtain the simple expression
\begin{equation}
    \Gamma(\omega) = \frac{\pi}{\hbar} V_0^2 \,\rho(\epsilon_k)\, \left.\frac{v_k^2}{u_k^2}\right|_{\epsilon_k = \epsilon(\omega)} = \pi N_p\, V_0^2 \,\rho(\epsilon_k)\,\left. |\varphi_k|^2\;\right|_{\epsilon_k = \epsilon(\omega)}
\end{equation}
This result shows that RF spectroscopy of the BCS state directly measures the fermion pair wave function $\varphi_k$ (see Eq.~\ref{e:fermicondensatepsi} and Eq.~\ref{e:coherentstate}). Note that it is $\varphi_k = v_k/u_k\sqrt{N_p}$, rather than the Cooper pair wave function $\psi_k = u_k v_k$, that appears here. While the two coincide in the BEC-limit of tightly bound molecules (apart from the normalization with $\sqrt{N_p}$), they are quite different in the BCS regime, where $\varphi_k$ extends throughout the entire Fermi sea, while $\psi_k$ is peaked in a narrow range around the Fermi surface (see Fig.~\ref{f:pairwavefunction}). This goes back to the two possible interpretations of the BCS state discussed in section~\ref{s:BCSlimit}, either as a condensate of $N/2$ fermion pairs (Eq.~\ref{e:fermicondensatepsi}) or as the product of a Fermi sea and a condensate of Cooper pairs (Eq.~\ref{e:BCSstate}). In RF spectroscopy, the first point of view is the natural choice, as the RF interaction couples to all momentum states in the entire Fermi sea.

We now discuss the spectrum itself. From here on, frequencies $\omega$ are given relative to the hyperfine frequency $\omega_{\uparrow3}$. From Eq.~\ref{e:dispersion} we see that the minimum or threshold frequency required to excite a particle into state $\left|3\right>$ is
\begin{eqnarray}
    \hbar\omega_{\rm th} = \sqrt{\mu^2 + \Delta^2} - \mu \rightarrow \left\{%
\begin{array}{lll}
    \frac{\Delta^2}{2E_F} & \hbox{in the BCS-limit} \\
    0.31 E_F & \hbox{on resonance} \\
    |E_B| = \frac{\hbar^2}{m a^2} & \hbox{in the BEC-limit} \\
\end{array}%
\right.
\end{eqnarray}
In either limit, the threshold for RF spectroscopy thus measures the binding energy of fermion pairs (apart from a prefactor in the BCS-limit, see section~\ref{s:BCSlimit}), and {\it not} the superfluid gap $\Delta$, which would be the binding energy of Cooper pairs described by $\psi_k$.

To obtain the spectrum explicitly, we calculate $\epsilon(\omega) = \frac{1}{2\hbar\omega}(\hbar \omega - \hbar \omega_{\rm th})(\hbar \omega + \hbar \omega_{\rm th}+ 2\mu)$ and $\left.v_k^2/u_k^2\right|_{\epsilon_k = \epsilon(\omega)} = \Delta^2/(\hbar \omega)^2$. With $\rho(E_F) = 3N/4E_F$ the spectrum finally becomes
\begin{equation}
    \Gamma(\omega) = \frac{3\pi}{4\sqrt{2}\hbar}\, \frac{N \,V_0^2\Delta^2}{E_F^{3/2}} \,\frac{\sqrt{\hbar\omega - \hbar\omega_{\rm th}}}{\hbar^2\omega^2}\sqrt{1 + \frac{\omega_{\rm th}}{\omega} + \frac{2\mu}{\hbar\omega}}
\end{equation}
In the BEC-limit, this reduces to (see Eq.~\ref{e:BEClimitgap})
\begin{equation}
\label{e:gammaBEC}
    \Gamma_{\rm BEC}(\omega) = \frac{4}{\hbar} N_M \,V_0^2 \,\sqrt{|E_B|}\frac{\sqrt{\hbar\omega - |E_B|}}{\hbar^2\omega^2}
\end{equation}
This is exactly the dissociation spectrum of $N_M = N/2$ non-interacting molecules (compare to the Feshbach association spectrum Eq.~\ref{e:Chap5FermiGolden} in section~\ref{c:feshbach} for $a_{\rm bg}\rightarrow 0$).
\begin{figure}[ht]
\centering
 \includegraphics[width=3in]{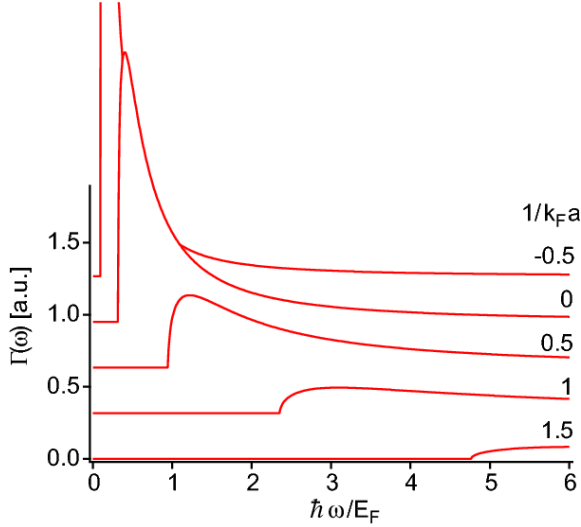}\\
  \caption[RF spectra in the BEC-BCS crossover]{RF spectra in the BEC-BCS crossover for transitions into a third, empty and non-interacting state. The threshold changes smoothly from the binding energy of molecules in the BEC-regime to the binding energy of fermion pairs $\sim \Delta^2/E_F$ in the BCS-regime.}\label{f:Gamma}
\end{figure}
Fig.~\ref{f:Gamma} shows the RF spectra for various values of the interaction strength in the BEC-BCS crossover. Qualitatively, the shape of $\Gamma(\omega)$ does not change much, always staying close to the characteristic asymmetric shape of a molecular dissociation spectrum like Eq.~\ref{e:gammaBEC}, with the pair binding energy as the only relevant energy scale.
For example, the spectrum has a maximum at
\begin{equation}
    \hbar \omega_{\rm max} = \frac{1}{3}\left(-4\mu+ \sqrt{16 \mu^2 + 15 \Delta^2}\right) \rightarrow \left\{%
\begin{array}{ll}
    \frac{5}{8}\frac{\Delta^2}{E_F} = \frac{5}{4}\hbar\omega_{\rm th}, & \hbox{in the BCS-limit} \\
0.40 E_F = 1.26\, \hbar\omega_{\rm th} , & \hbox{on resonance} \\
\frac{8}{3}|E_B|, & \hbox{in the BEC-limit} \\\end{array}%
\right.
\end{equation}
which is always on the order of the fermion pair binding energy. The spectrum falls off like $1/\omega^{3/2}$ at large frequencies. This is due to the asymptotic momentum distribution $v_k^2 \sim 1/k^4$. The long tails lead to a divergence of the mean transition frequency, so that the sum rule approach in~\ref{s:clockshifts} does not give a sensible result here. The divergence is removed if the third state interacts with atoms in state $\left|\downarrow\right>$. Note that the BCS formalism neglects interactions between spin up and spin down that are already present in the normal state, which may contribute additional shifts and broadening of the spectra. For the superfluid $^6$Li system in the $\left|1\right>$ and $\left|2\right>$ states, the accessible final state $\left|3\right>$ has strong interactions with state $\left|1\right>$. Therefore, the experimental spectra in the resonance region~\cite{chin05rf,shin07rf,schu07pair} are qualitatively different from the idealized spectra presented here (see~\cite{schu07tbp}).
For recent theoretical studies on RF spectroscopy, incorporating final state interactions, see~\cite{baym07,punk07rf,pera08rf}.

\subsubsection{Collective excitations}
\label{s:collexcitations}

In addition to single-particle excitations, we have to consider
collective excitations related to density fluctuations or sound
waves~\footnote{The reduced BCS Hamiltonian Eq.~\ref{e:Hsimplified} does not contain density fluctuations. One needs to work with the Hamiltonian Eq.~\ref{e:Hamiltonian}.}.  Sound modes have a linear dispersion relation $E_k=\hbar
c_s k$.  In the weakly-interacting BEC-limit, the  speed of sound
is given by the Bogoliubov solution $c_s = \sqrt{\mu_M/m_M} =
\sqrt{4\pi \hbar^2 a_M n_M / m_M}$. For stronger interactions, the
Lee-Huang-Yang expansion becomes important (Lee, Huang and Yang,
1957) which increases the speed of sound by a factor $1+16
\sqrt{\frac{n_M a_M^3}{\pi}}$~\cite{huan87}. The Bogoliubov sound mode
finds its analogue in the BCS-regime, where it is called the
Bogoliubov-Anderson mode, propagating at the speed of sound $v_F /
\sqrt{3}$, with $v_F = \hbar k_F / m$ the Fermi
velocity\footnote{This speed of sound can be calculated using the
hydrodynamic equation $c = \sqrt{\frac{\partial P}{\partial
\rho}}$, $\rho = m \, n$ and the pressure of a normal Fermi gas $P
= \frac{2}{3} \frac{E}{V} = \frac{2}{5} E_F \,n = \frac{2}{5}
\frac{\hbar^2}{2m} (3\pi^2)^{2/3} n^{5/3}$. Thus, the sound mode
is already present in the normal Fermi gas, the main effect of
superfluid pairing being to push low-lying single-particle
excitations up in energy, which would otherwise provide damping.
The low temperature limit for the normal gas is peculiar.  On the
one hand, the damping vanishes at zero temperature, on the other
hand, the sound mode cannot propagate, as collisions are absent and the gas
can no longer maintain local equilibrium.}. The connection between
the BEC and BCS results is smooth, as expected and found
by~\cite{melo93,ohas03coll,comb06coll}.

Sound waves are described by hydrodynamic equations, which are
identical for superfluid hydrodynamics and inviscid, classical collisional
hydrodynamics.  For trapped clouds of finite size, collective
modes are the solutions of the hydrodynamic equations for the
geometry of the trapped cloud.  In a harmonic trap, the size of
the cloud depends on the square root of the chemical potential (not counting binding energies), just like
the speed of sound.  As a result, the lowest lying collective
excitations are proportional to the trap frequency and, to leading
order, independent of the density and size of the cloud~\cite{gior07fermi}.

Modes for which the velocity field has zero divergence are called
surface modes.  Their frequencies are independent of the density
of states and do not change across the BCS-BEC crossover, as long
as the system is hydrodynamic. However, the frequencies are
different from the collisionless regime (where all frequencies are
integer multiples of the trap frequencies) and can be used to
distinguish hydrodynamic from collisionless behavior.
In contrast, compression modes depend on the compressibility of
the gas and therefore on the equation of state.  On both the BEC
and BCS side and at unitarity, the chemical potential is
proportional to a power of the density $\mu \propto n^\gamma$. The
frequency of breathing modes depends on $\gamma$, which has been
used to verify that $\gamma=1$ on the BEC side and
$\gamma=2/3$ at unitarity~\cite{kina04sfluid,bart04coll,altm07precision}.

For an extensive discussion of collective modes we refer to the contributions of S. Stringari and R. Grimm to these proceedings.

\subsubsection{Landau criterion for superfluidity}
\label{s:landau}

The Landau criterion for superfluidity gives a critical velocity
$v_c$, beyond which it becomes energetically favorable to transfer
momentum from the moving superfluid (or the moving object) to
excitations~\cite{lifs80statphys2}. As a result, superfluid flow is damped.
Creating an excitation carrying plane-wave momentum $\hbar\vect{k}$ costs an energy $E_k + \hbar\vect{k}\cdot\vect{v}$ in the rest frame (Doppler shift). The minimum cost occurs naturally for creating an excitation with $\vect{k}$ antiparallel to the velocity $\vect{v}$ of the superfluid. This is only energetically favored if $E_k - \hbar k v < 0$, leading to Landau's criterion for the critical velocity:
\begin{equation}
    v_c = \min_k \frac{E_k}{\hbar k}
 \label{e:landaucriterion}
\end{equation}
The minimum has to be taken over all possible excitations,
including single-particle excitations, collective excitations
(and, for certain geometries, such as narrow channels and small
moving objects, excitation of vortex pairs).

\begin{figure}
\centering
 \includegraphics[width=2.7in]{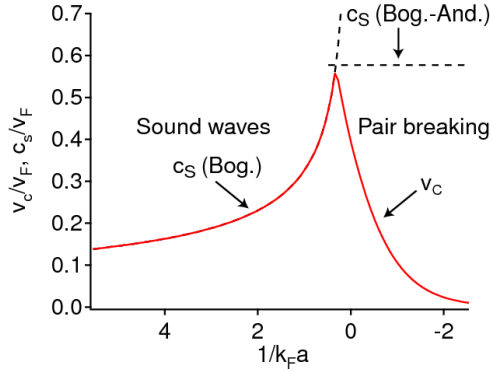}\\
  \caption[Critical velocity $v_c$ in the BEC-BCS crossover]{Critical velocity $v_c$ in the BEC-BCS crossover. The relevant excitations in the BEC-regime correspond to Bogoliubov (Bog.) sound waves with speed of sound $c_s = \sqrt{\frac{\mu}{m}} = \frac{v_F}{\sqrt{3\pi}} \sqrt{k_F a}$. This sound mode eventually becomes the Bogoliubov-Anderson (Bog.-And.) mode in the BCS-regime, with $c_s = \frac{v_F}{\sqrt{3}}$. The evolution is smooth~\cite{melo93,ohas03coll,comb06coll}, but only the limiting cases are shown here. In the BCS-regime the excitations with the lowest critical velocity are single-particle excitations that break a Cooper pair. Here, $v_c \approx \frac{\Delta}{\hbar k_F}$. After~\cite{comb06coll}.}\label{f:velocity}
\end{figure}

On the BCS side, the single-particle excitation spectrum derived
above gives a  critical velocity of
\begin{equation}
    v_{c,\rm BCS} = \min_k \frac{E_k}{\hbar k} = \sqrt{\left(\sqrt{\mu^2 + \Delta^2} - \mu\right)/m} \quad \stackrel{\frac{\Delta}{\mu} \rightarrow 0_+}{\rightarrow} \quad\frac{\Delta}{\hbar k_F}
\end{equation}
An object that is dragged through the superfluid faster than this
velocity will break fermion pairs.  For sound excitations, the
Landau criterion gives the speed of sound as critical velocity. In
a simple approximation (neglecting possible coupling between
single particle and collective excitations), the critical velocity
for the superfluid is given by the smaller of the two velocities.
On the BEC side, where the pairs are tightly bound, the speed of
sound determines the critical velocity, whereas on the BCS side,
the critical velocity comes from pair breaking.

For the BEC-side, it has been shown in~\cite{comb06coll} that for
small momenta $k \ll 1/a$ that do not resolve the composite nature
of the molecules, the expression for the Bogoliubov-dispersion
$E_{k,\rm BEC} = \sqrt{(\frac{\hbar^2 k^2}{2 m_M}+\mu_M)^2 -
\mu_M^2}$ remains valid even well into the crossover region. This
allows us to determine the speed of sound in an approximate way,
which is shown in Fig.~\ref{f:velocity}. Notable is the sharp peak
in the critical velocity around resonance which reflects the
rather narrow transition from a region where excitation of sound
limits superfluid flow, to a region where pair breaking dominates.
At the same time, the onset of dissipation switches from low $k$'s
(sound) to high $k$'s (pair breaking).  It is near the Feshbach
resonance that the superfluid is most
stable~\cite{sens06vort,comb06coll}. This makes the critical
velocity one of the few quantities which show a pronounced peak
across the BEC-BCS crossover, in contrast to the chemical
potential, the transition temperature (except for a small
hump), the speed of sound and the frequencies of shape
oscillations, which all vary monotonically.

\subsection{Finite temperatures}
At finite temperature, the superfluid state has thermal
excitations in the form of the quasiparticles introduced in
Eq.~\ref{e:quasiparticles}.  These quasiparticles modify the gap
and number equations for the BCS state from which we derive an
expression for the superfluid transition temperature.

\subsubsection{Gap equation at finite temperature}

At finite temperature, the expectation value for the pairing field
$C_k=  \left<c_{k\uparrow} c_{-k\downarrow}\right>$ becomes
\begin{equation}
    \left<c_{k\uparrow} c_{-k\downarrow}\right> = -u_k v_k \left(1 - \left<\gamma_{k\uparrow}^\dagger \gamma_{k\uparrow}\right> -\left<\gamma_{k\downarrow}^\dagger \gamma_{k\downarrow}\right>\right)
\end{equation}
As the quasi-particles are fermions, they obey the Fermi-Dirac
distribution $\left<\gamma_{k\uparrow}^\dagger
\gamma_{k\uparrow}\right> = \frac{1}{1+e^{\beta E_k}}$. The
equation for the gap $\Delta = \frac{V_0}{\Omega} \sum_k
\left<c_{k\uparrow} c_{-k\downarrow}\right>$ thus becomes
(replacing $V_0$ as above by the scattering length $a$)
\begin{equation}
\label{e:finiteTgap}
-\frac{m}{4\pi\hbar^2 a} = \int \frac{d^3 k}{\left(2\pi\right)^3}
\left(\frac{1}{2 E_k} \tanh\left(\frac{\beta E_k}{2}\right) -
\frac{1}{2 \epsilon_k}\right)
\end{equation}

\subsubsection{Temperature of pair formation}

\begin{figure}
\centering
\includegraphics[width=3.2in]{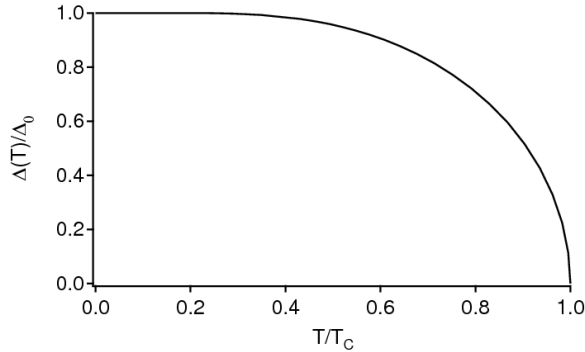}\\
  \caption[Temperature dependence of the gap $\Delta$]{Temperature dependence of the gap in the BCS regime. $\Delta(T)$ is normalized by its value $\Delta_0$ at zero temperature, and temperature is given in units of $T^* \approx T_C$.}\label{f:gap}
\end{figure}

We are interested in determining the temperature $T^* = 1/\beta^*$
at which the gap vanishes. In the BCS-limit, this procedure gives
the critical temperature for the normal-to-superfluid transition.
Setting $\Delta = 0$ in the gap equation, one needs to
solve~\cite{nozi85,drec92,melo93}
\begin{equation}
-\frac{m}{4\pi\hbar^2 a} = \int \frac{d^3 k}{\left(2\pi\right)^3}
\left(\frac{1}{2 \xi_k} \tanh\left(\frac{\beta^* \xi_k}{2}\right)
- \frac{1}{2 \epsilon_k}\right)
\end{equation}
simultaneously with the constraint on the total number of atoms.
Above the temperature $T^*$, we have a normal Fermi gas with a
Fermi-Dirac distribution for the occupation of momentum states, so
the number equation becomes
\begin{equation}
\label{e:numbertstar}
    n = 2 \int \frac{d^3 k}{\left(2\pi\right)^3} \frac{1}{1+e^{\beta^* \xi_k}}
\end{equation}
In the BCS-limit, we expect $\mu \gg k_B T^*$ and thus find $\mu
\approx E_F$. Inserted into the gap equation, this gives the
critical temperature for BCS superfluidity
\begin{equation}
    T_{\rm BCS}^* = T_{C,\rm BCS} = \frac{e^\gamma}{\pi} \frac{8}{e^2} e^{-\pi/2k_F\left|a\right|} = \frac{e^\gamma}{\pi} \Delta_0
    \label{e:TCBCS}
\end{equation}
with Euler's constant $\gamma$, and $e^\gamma \approx 1.78$. Here,
we distinguish $\Delta_0$, the value of the superfluid gap at zero
temperature, from the temperature dependent gap $\Delta(T)$. From Eq.~\ref{e:finiteTgap} one
can show that
\begin{equation}
    \Delta(T) \approx \left\{%
\begin{array}{ll}
    \Delta_0 - \sqrt{2\pi \Delta_0\, k_B T}\, e^{-\Delta_0/k_B T}, & \qquad \mbox{for } T \ll T_C \\
    \sqrt{\frac{8 \pi^2}{7 \zeta(3)}}\, k_B T_C \, \sqrt{1 - \frac{T}{T_C}}, & \qquad \mbox{for } T_C - T \ll T_C \\
\end{array}%
\right. \label{e:gapfiniteT}
\end{equation}
The full temperature dependence is shown in Fig.~\ref{f:gap}.

In the BEC-limit, the chemical potential $\mu = E_B /2 = -
\hbar^2/2m a^2$ is again given by half the molecular binding
energy as before, and the temperature $T^*$ is found to be
\begin{equation}
    T_{\rm BEC}^* \approx \frac{1}{3}\frac{\left|E_b\right|}{ W\left[\left(\frac{\pi}{6}\right)^{\frac{1}{3}}\frac{\left|E_b\right|}{2E_F}\right]}
\end{equation}
where $W(x)$ is the Lambert $W$-function, solution to $x = W e^W$ with expansion $W(x) \approx \ln(x) - \ln(\ln(x))$ (useful for $x\gtrsim 3$).

$T_{\rm BEC}^*$ is {\it not} the critical temperature for the superfluid
transition but simply the temperature around which pairs start to
form. The factor involving $W(x)$ has its origin in the entropy of the
mixture of molecules and free fermions, which favors unbound fermions and
lowers $k_B T^*$ below the binding energy $E_b$. There is no phase
transition at $T^*$, but a smooth crossover.

\subsubsection{Critical temperature}
\label{s:crittemp}

\begin{figure}
\centering
 \includegraphics[width=2.5in]{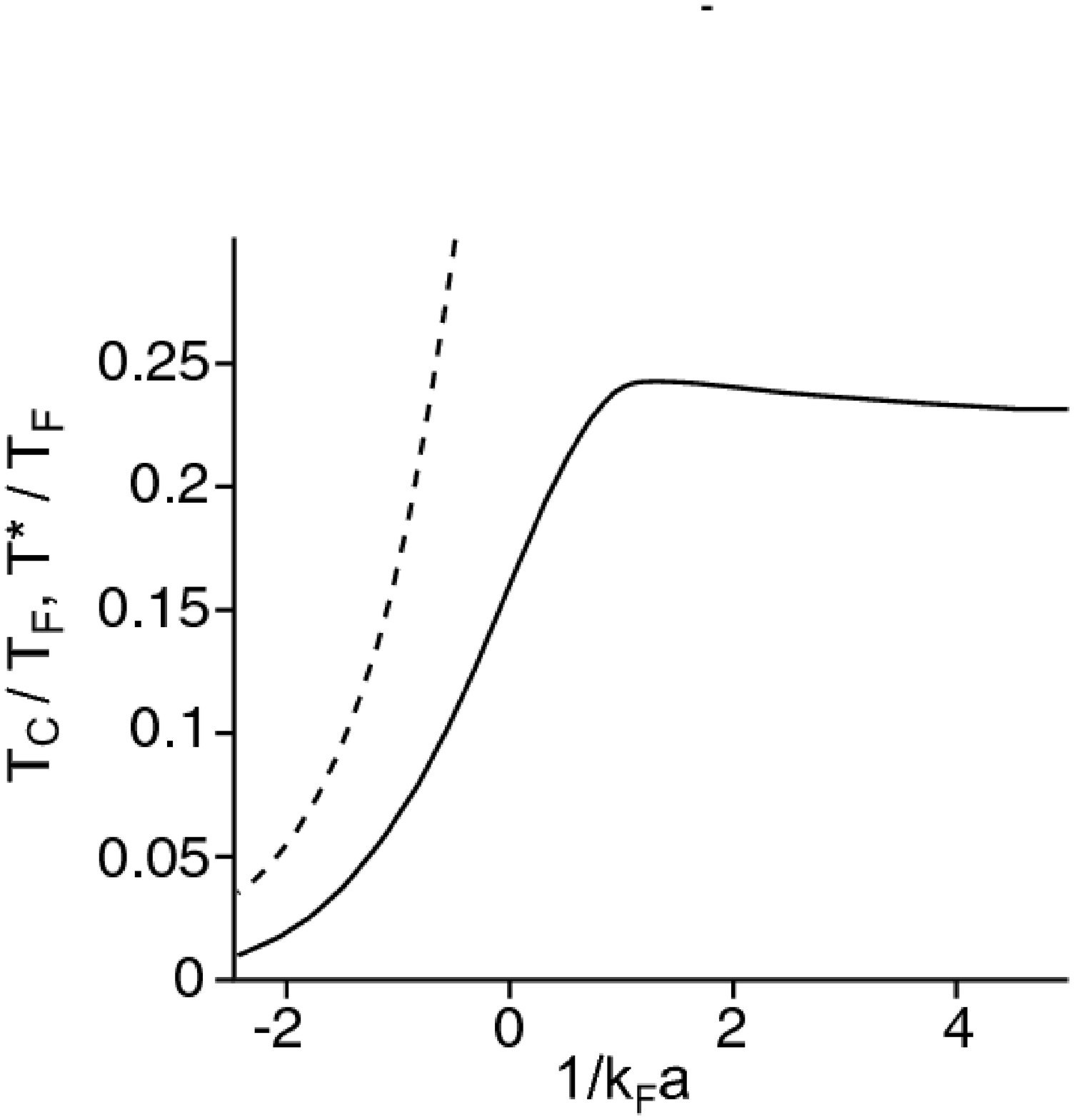}\\
  \caption[Superfluid transition temperature $T_C$ and pair creation temperature $T^*$]{Superfluid transition temperature $T_C$ and pair creation temperature $T^*$ (dashed line) in the BEC-BCS crossover. In the BEC regime, $T_C$ corresponds to the BEC transition temperature for a gas of molecules. In the BCS regime, the critical temperature depends exponentially on the interaction strength, drastically reducing $T_C$. $T_C$ extracted from~\cite{haus07bcsbec}, $T^*$ calculated from Eq.~\ref{e:numbertstar}.}\label{f:criticalT}
\end{figure}

Determining $T_C$, the temperature at which long-range order is
established, requires an additional term in the number equation, namely the inclusion of non-condensed pairs~\cite{nozi85,drec92,melo93,hu06becbcs,haus07bcsbec}. The result is that in the deep BEC-regime, the
critical temperature is simply given by the non-interacting value
for the BEC transition of a gas of molecules at density $n_M =
n/2$ and mass $m_M = 2m$,
\begin{equation}
    T_{C,{\rm BEC}} = \frac{2 \pi \hbar^2}{m_M} \left(\frac{n_M}{\zeta(\frac{3}{2})}\right)^{2/3} = \frac{\pi\hbar^2}{m} \left(\frac{n}{2\zeta(\frac{3}{2})}\right)^{2/3} = 0.22 E_F
\end{equation}
This result holds for weakly interacting gases. For stronger
interactions, there is a small {\it positive} correction
$T_C/T_{C,{\rm BEC}} = 1 + 1.31 n_M^{1/3} a_M$, with $a_M = 0.60
a$~\cite{baym99tc,baym00tc,baym01tc,arno01tc,kash01tc}. On the
BCS-side, the critical temperature should smoothly connect to the
BCS result given above. This implies that there must thus be a
local maximum of the critical temperature in the
crossover~\cite{haus07bcsbec}. The value of $T_C$ at unitarity has
been calculated
analytically~\cite{nozi85,melo93,hu06becbcs,haus07bcsbec}, via
renormalization-group methods~\cite{nish06epsilon} and via
Monte-Carlo simulations~\cite{bulg06TC,buro06TC}. The result is
$T_C = 0.15-0.16 T_F$~\cite{haus07bcsbec,buro06TC}. Note that
these values hold for the homogeneous case, with $k_B T_F =
\hbar^2 (6\pi^2 n)^{2/3}/2m$. In the trapped case, they apply
locally, but require knowledge of the local $T_F$ and therefore
the Fermi energy in the center of the trap.  This requires
knowledge of the central density $n_U$ as a function of
temperature and the global Fermi energy $E_F = \hbar \bar{\omega}
(3 N)^{1/3}$. Using, as a first approximation, the
zero-temperature relation $n_U = \xi^{-3/4} n_F$ from
Eq.~\ref{e:Unitaritydensity}, with $n_F$ the density of a
non-interacting Fermi gas of the same number of atoms, gives $k_B
T_{C,{\rm Unitarity}} = 0.15 \epsilon_F(\vect{0}) = 0.15
\frac{1}{\sqrt{\xi}} E_F \approx 0.23 E_F$. Fig.~\ref{f:criticalT}
shows the behavior of $T_C$ as a function of the interaction
strength.

\subsubsection{``Preformed'' pairs}

In the region between $T_C$ and $T^*$, we will already find bound
pairs in the gas, which are not yet condensed. In the BCS-limit,
where $T^* \rightarrow T_C$, condensation occurs at the same time
as pairing, which, as we see now, is no longer true for stronger
interactions. Deep on the molecular side, it is of course not
surprising to find thermal molecules above $T_C$. However, the
qualitative picture of thermal (i.e. non-condensed) pairs still
holds in the entire crossover region from $-1 < 1/k_F a < 1$.
These uncondensed pairs are sometimes called ``preformed''
(pairing occurs before condensation) and also occur in a part of
the phase diagram of High-T$_C$ superconductors, the Nernst regime
of the pseudogap~\cite{lee06hightc}.

\subsection{Long-range order and condensate fraction}
\label{s:condensatewavefunction}

In this and the following section, we discuss in detail the condensate and superfluid fractions. In dilute gas BECs, the difference between the two quantities is negligible, but their distinction is crucial in the BEC-BCS crossover and in the BCS limit.

Fritz London proposed in 1938 that superfluidity is a quantum
mechanical phenomenon. It should thus be possible to encode the
properties of the superfluid in a macroscopic wave function that
depends only on one or a few coordinates. In the case of Bose
gases, it is the 1-particle density matrix that describes
superfluidity~\cite{penr56,lifs80statphys2},
\begin{equation}
    \rho_1(\vect{r},\vect{r}') = \left<\Psi_B^\dagger(\vect{r})\Psi_B(\vect{r}')\right>
\end{equation}
where $\Psi_B^\dagger(\vect{r})$ is the creation operator for a
boson at point $\vect{r}$.
The sum of all eigenvalues of this matrix is equal to the number
of particles $N$.  The criterion for Bose-Einstein condensation,
as first introduced by Onsager and Penrose~\cite{penr56}, is the
existence of exactly one macroscopic eigenvalue, i.e. with a value
of order $N$.  Such a macroscopically occupied state implies
long-range order, signalled by an off-diagonal ($\vect{r} \ne
\vect{r}'$) matrix element that does not vanish for large
distances $|\vect{r}-\vect{r}'|$.
\begin{equation}
\lim_{|\vect{r}-\vect{r}'|\rightarrow \infty}\rho_1(\vect{r},\vect{r}') = \psi_B(\vect{r})\psi_B^*(\vect{r}')
\end{equation}
where $\psi_B(\vect{r})$ is the macroscopic wave function or order
parameter describing the Bose superfluid~\footnote{For a
discussion of the relation between condensation and superfluidity,
see chapter~\ref{c:expobservation}.}. ``Macroscopic'' means that
the number of condensed bosons $N_0 = \int {\rm d}^3r
\,|\psi_B(\vect{r})|^2$ is extensive, i.e. large compared to 1, or
more precisely that the condensate fraction $N_0/N$ is finite in
the thermodynamic limit. $n_0(\vect{r}) \equiv
|\psi_B(\vect{r})|^2$ is the density of the condensed gas. Thus,
an absorption image of a weakly interacting Bose-Einstein
condensate directly reveals the magnitude of the wave function.
This has led to the direct visualization of coherence between two
Bose condensates~\cite{andr97int}, spatial coherence within a
condensate~\cite{bloc00coh} and of vortex
lattices~\cite{madi00,abos01latt,hodb02vort,enge02}.
    
For {\it fermionic} gases, the 1-particle density matrix can {\it
never} have a macroscopic matrix element, as the occupation number
of a particular quantum state cannot exceed unity due to the Pauli
principle. After our discussion of fermionic pairing, it does not
come as a surprise that for fermionic superfluids, long-range
order shows up in the 2-particle density
matrix~\cite{lifs80statphys2,camp97ODLRO}
\begin{equation}
    \rho_2(\vect{r}_1,\vect{r}_2,\vect{r}_1',\vect{r}_2') = \left<\Psi_\uparrow^\dagger(\vect{r}_1)\Psi_\downarrow^\dagger(\vect{r}_2)\Psi_\downarrow(\vect{r}_2')\Psi_\uparrow(\vect{r}_1')\right>
\end{equation}
where we added spin labels corresponding to the case of $s$-wave pairing.
Analogous to the Bose case, we can check for the presence of a {\it pair condensate} by increasing the distance between the primed and the unprimed coordinates, that is between the two centers of mass $\vect{R} = (\vect{r}_1 + \vect{r}_2)/2$ and $\vect{R}'=(\vect{r}'_1+\vect{r}'_2)/2$. If there is long-range order, one will find a macroscopic ``off-diagonal'' matrix element
\begin{equation}
\lim_{|\vect{R} - \vect{R}'|\rightarrow \infty}\rho_2(\vect{r}_1,\vect{r}_2,\vect{r}_1',\vect{r}_2') = \psi(\vect{r}_1,\vect{r}_2)\psi^*(\vect{r}_1',\vect{r}_2')
\end{equation}
The function $\psi(\vect{r}_1,\vect{r}_2) = \left<\Psi_\uparrow^\dagger(\vect{r}_1)\Psi_\downarrow^\dagger(\vect{r}_2)\right>$ is a macroscopic quantity in BCS theory. It is equal to the Cooper pair wave function discussed above and given by the Fourier transform of the pairing field $-C^*_k = \left<c_{k\uparrow}^\dagger c_{-k\downarrow}^\dagger\right>$.
The density of condensed fermion pairs $n_0(\vect{R})$ is obtained from $|\psi(\vect{r}_1,\vect{r}_2)|^2$ by separating center of mass $\vect{R}$ and relative coordinates $\vect{r}= \vect{r}_2 - \vect{r}_1$ and integrating over $\vect{r}$:
\begin{equation}
    n_0(\vect{R}) = \int {\rm d}^3 r\,|\psi(\vect{R}-\vect{r}/2,\vect{R}+\vect{r}/2)|^2
\end{equation}
More accurately, $n_0$ is the total density of pairs $n/2$, times the average occupancy of the paired state.
With $-C_k^* = u_k v_k \tanh(\frac{\beta E_k}{2})$, we can calculate the condensate density in a uniform system~\cite{sala05,orti05bcs}:
\begin{equation}
    n_0 = \frac{1}{\Omega} \sum_k u_k^2 v_k^2 \tanh^2\left(\frac{\beta E_k}{2}\right)\; \stackrel{T\rightarrow 0}{=} \;
\frac{1}{\Omega} \sum_k u_k^2 v_k^2 =
\frac{3\pi\sqrt{2}}{32}\,n\,\frac{\Delta}{E_F} \sqrt{\frac{\mu + \sqrt{\mu^2 + \Delta^2}}{E_F}}
    \label{e:condfrac}
\end{equation}
The condensate fraction $2n_0/n$ is non-vanishing in the
thermodynamic limit $\Omega \rightarrow \infty, n\rightarrow {\rm
const}$, and therefore macroscopic. It is shown in
Fig.~\ref{f:n0vsnSF} as a function of temperature in the
BCS-regime, and in Fig.~\ref{f:condfractheo} of
section~\ref{s:condensatefraction}, where it is compared to
experimental results in the BEC-BCS crossover. In the BEC-limit,
with the help of Eq.~\ref{e:BEClimit}, $n_0$ becomes the density
of molecules or half the total atomic density $n$, as expected,
\begin{equation}
    n_0 = n/2
\end{equation}
corresponding to a condensate fraction of 100\%. As interactions
increase, the Bogoliubov theory of the interacting Bose gas
predicts that the zero-momentum state occupation decreases and
higher momentum states are populated. This quantum depletion is
$\frac{8}{3}\sqrt{\frac{n_M a_M^3}{\pi}}$ for a molecular gas of
density $n_M$ and scattering length $a_M$. At $k_F a$ = 1, this
would reduce the condensate fraction to 91\% (using $a_M = 0.6
\,a$). The mean field ansatz for the BEC-BCS crossover cannot
recover this beyond-mean field correction proportional to $\sqrt{n
a^3}$. Indeed, the only way the repulsion between two molecules is
built into the mean field theory is via the Pauli exclusion
principle for the constituent fermions. Rather,
Eq.~\ref{e:condfrac} predicts a depletion proportional to $n a^3$,
which underestimates the true quantum depletion. Monte-Carlo
studies are consistent with the Bogoliubov
correction~\cite{astr05cond}. On resonance, Eq.~\ref{e:condfrac}
predicts a condensate fraction $2 n_0/n = 70\%$, whereas the
Monte-Carlo value is 57(2)\%~\cite{astr05cond}. In the BCS-regime,
where $\mu \approx E_F$ and the gap is exponentially small, one
finds from Eq.~\ref{e:condfrac}:
\begin{equation}
\label{e:condfracBCS}
    n_0 = \frac{m k_F}{8\pi\hbar^2}\, \Delta = \frac{3\pi}{16} \,n\,\frac{\Delta}{E_F}
\end{equation}
The condensate fraction thus decreases exponentially with the
interaction strength, like the gap $\Delta$. This strong depletion
is entirely due to the Pauli principle, which can be seen from
Eq.~\ref{e:condfrac}. Fully occupied ($u_k = 0$) and unoccupied
($v_k = 0$) momentum states do not contribute to the condensate
fraction. The bulk contribution comes from states in only a narrow
energy range of width $\sim\Delta$ around the Fermi surface, as
they are in a superposition of being occupied (with amplitude
$v_k$) or unoccupied (amplitude $u_k$). Their total number is
$\sim N\Delta/E_F$ (see Eq.~\ref{e:condfracBCS}).
The phase of this superposition state (the relative phase between the complex numbers $u_k$ and $v_k$, the same for all $\vect{k}$) defines the macroscopic phase of the superfluid state. Indeed, introducing a global phase factor $e^{i \alpha}$ into the BCS state,  $\left|\psi_{\rm BCS}\right> = \prod_k (u_k  +  e^{i \alpha} v_k c_{k\uparrow}^\dagger c_{-k\downarrow}^\dagger) \left|0\right>$ is equivalent to a rotation of the ``coherent state'' in Eq.~\ref{e:coherentstate} by an angle $\alpha$, from $\exp\left(\sqrt{N_p}\; b^\dagger\right)\left|0\right>$ to $\exp\left(\sqrt{N_p}\; e^{i \alpha} b^\dagger\right) \left|0\right>$.   This is in direct analogy with BEC and the optical laser.

\subsection{Superfluid density}

\begin{figure}
\centering
 \includegraphics[width=3in]{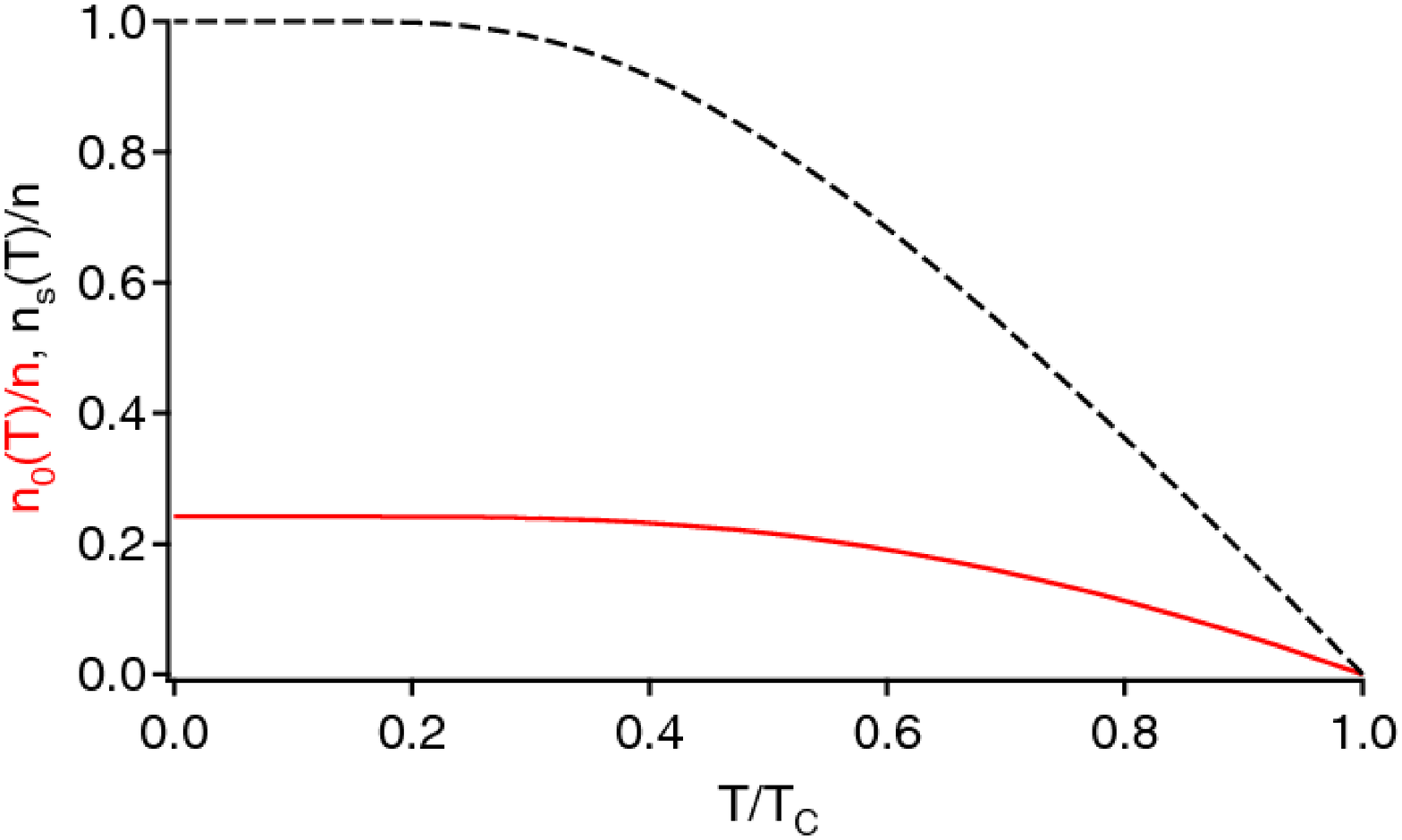}\\
  \caption[Condensate density versus superfluid density]{Condensate density $n_0$ (straight line) and superfluid density $n_s$ (dashed line) versus temperature in the BCS-regime ($1/k_F a = -1$). The superfluid fraction is 100\% at $T=0$, while the condensate fraction saturates at 24\%. Note that both densities vanish linearly with temperature (within mean-field theory) as they approach $T_C$.}\label{f:n0vsnSF}
\end{figure}

It is important to distinguish the density of condensed fermion pairs $n_0$, which is smaller than the total density even at zero temperature, from the superfluid density $n_s$. The superfluid density is the part of the system that does not respond to external rotation or shear motion. At zero temperature the entire system is superfluid and thus $n_s = n$. As discussed above, one encounters this difference between $n_0$ and $n_s$ already in BECs~\cite{gior97}. Fig.~\ref{f:n0vsnSF} compares the two quantities for the BCS-regime.

The distinction between superfluid and normal density $n_n$ provides the basis of the two-fluid hydrodynamic model of superfluids and superconductors~\cite{tisz38,land41}.
To obtain $n_s$, one can place the system in a long tube that is slowly set in motion with velocity $\vect{v}$~\cite{legg75}. According to Landau's criterion (\ref{s:landau}), as long as $v < v_c$, no new excitations from the superfluid are created, so the superfluid stays at rest. However, due to friction with the walls of the tube, the collection of already existing excitations will be dragged along by the tube. The total momentum density $\vect{P}$ of the system is thus only due to this normal gas of excitations, $\vect{P} = n_n m \vect{v}$, which defines the normal density $n_n$. The superfluid density is then $n_s = n - n_n$.

We have seen in section~\ref{s:excitations} that there are two
types of excitations in a fermionic superfluid: Excitations of
fermionic (quasi-)particles, related to pair breaking, and bosonic
excitations of pairs of fermions. Both types will contribute to
the normal
density~\cite{lifs80statphys2,enge97becbcs,andr03becbcs,tayl06sf}.
Single-particle excitations are frozen out for temperatures well
below the characteristic temperature $T^*$ for pair formation. In
a molecular BEC, $T_C \ll T^*$, and fermions are strongly bound.
The only relevant thermal excitations are thus due to
non-condensed pairs. For $k_B T \ll \mu_M$, the excitations are
dominantly phonons. In the BCS-regime, $T_C = T^*$, and the normal
density contains both single-fermion excitations from broken pairs
as well as bosonic excitations of pairs of fermions, the
Bogoliubov-Anderson sound mode (see~\ref{s:collexcitations}). Near
$T_C$, which is close to $T^*$ in the BCS regime, single fermion
excitations dominate. At low temperatures $k_B T \ll \Delta$, they
are frozen out and the contribution from sound waves dominates. At
intermediate temperatures, the two types of excitations are
coupled, leading to damping of the sound waves~\cite{tayl06sf}.

The normal density is obtained from the total momentum of the gas of excitations, that moves with velocity $\vect{v}$ with respect to the stationary superfluid part. In the reference frame moving with the normal gas, the excitation energies $\epsilon$ of the superfluid are Doppler shifted to $\epsilon - \hbar\vect{k}\cdot\vect{v}$~\cite{lifs80statphys2}. The momentum is thus
\begin{equation}
    \vect{P}_{B,F} = \sum_k \hbar \vect{k}\, f_{B,F}(\epsilon - \hbar\vect{k}\cdot\vect{v})
\end{equation}
where the subscripts $B$ or $F$ correspond to the bosonic and fermionic contribution, respectively, and $f_{B,F}(\epsilon)$ is the Bose-Einstein and Fermi-Dirac distribution, resp. For small velocities, this gives
\begin{equation}
    \vect{P}_{B,F} = \sum_k \hbar \vect{k} (\hbar\vect{k}\cdot\vect{v}) \left(-\frac{\partial f_{B,F}}{\partial \epsilon}\right) = \frac{1}{3} \vect{v} \sum_k \hbar^2 k^2 \left(-\frac{\partial f_{B,F}}{\partial \epsilon}\right)
\end{equation}
The last equation follows from spherical symmetry, obeyed by the energy levels and the gap $\Delta$ in an $s$-wave fermionic superfluid. It implies that $\vect{P}$ is in the direction of $\vect{v}$ and allows to replace $(\vect{k}\cdot\frac{\vect{v}}{v})^2$ by its angular average, $\frac{1}{3} k^2$. The final formula for the normal density is, with $\epsilon_k = \hbar^2 k^2 / 2m$~\cite{lifs80statphys2},
\begin{equation}
    n_n^{B,F} = \frac{2}{3} \frac{1}{\Omega}\sum_k \epsilon_k \left(-\frac{\partial f_{B,F}}{\partial \epsilon}\right)
\end{equation}
\paragraph{Contribution from sound waves}
Sound waves have $\epsilon = \hbar c_s k$ and thus
\begin{equation}
   n_n^B = -\frac{\hbar}{3 m c_s} \int \frac{{\rm d}^3 k}{(2\pi)^3} k^2 \frac{\partial f_{B}}{\partial k} =  \frac{2\pi^2}{45} \frac{k_B^4 T^4}{m\hbar^3 c_s^5} = \frac{\pi^4}{120}\, n \left(\frac{k_B T}{E_F}\right)^4 \left(\frac{v_F}{c_s}\right)^5
\end{equation}
In the crossover and on the BCS-side, $c_s \approx v_F$, and so $n_n^B/n \approx (k_B T/E_F)^4$, a small contribution that dominates only for $k_B T \ll \Delta$ (see below).

\paragraph{Contribution from fermionic quasi-particle excitations}
Fermionic quasi-particle excitations have $\epsilon = E_k$. Spin up and spin down excitations both contribute, giving a normal density
\begin{equation}
    n_n^F = \frac{4}{3} \frac{1}{\Omega}\sum_k \epsilon_k \left(-\frac{\partial f_F}{\partial E_k}\right) = \frac{1}{3\Omega}\sum_k \beta \epsilon_k \frac{1}{\cosh^2\left(\frac{\beta E_k}{2}\right)}
\end{equation}
Via partial integration, it is not hard to see that for $\Delta = 0$, $n_n^F = n$, that is, the entire system is normal and consists exclusively of thermally excited quasi-particles. This is because in mean-field BCS theory, $\Delta = 0$ implies that $T>T^*$, the temperature for pair formation. Below $T^*$, both quasi-particles and thermal pairs contribute to the normal gas. Below $T_C$, the superfluid density $n_s$ becomes non-zero.
In the BCS-regime, we find~\cite{abri75,lifs80statphys2}
\begin{eqnarray}
    n_n^F =& n\sqrt{\frac{2\pi \Delta_0}{k_B T}}\; e^{-\Delta_0/k_B T}  &\qquad \hbox{for } T\ll T_C \\
    n_s =& n\,\frac{7\zeta(3)}{4\pi^2 T_C^2} \Delta^2 = 2\,n\left(1-\frac{T}{T_C}\right) & \qquad  \hbox{for } T \approx T_C
\label{e:ns}
\end{eqnarray}
Close to $T_C$, the superfluid density is proportional to the square of the gap. This provides a natural normalization of the superfluid order parameter in the next section.
The exponential suppression of the quasi-particle contribution at low temperatures is characteristic for a gapped excitation spectrum. At temperatures $k_B T \ll \Delta$, bosonic sound waves dominate the normal component.
In the BEC-regime, the role of the excitation gap is played by $|\mu|$, which is (half) the binding energy of molecules. Hence, already far above $T_C$, fermionic excitations are frozen out and exponentially suppressed like $e^{-|\mu|/k_B T}$. Bosonic excitations dominate at all temperatures $T<T_C$.

\subsection{Order parameter and Ginzburg-Landau equation}
\label{s:orderparameter}
A Bose superfluid is described by $\psi_B(\vect{r})$, the macroscopic wave function or order parameter. For fermionic superfluids, $\psi(\vect{r}_1,\vect{r}_2)$ is the wave function for fermions bound in Cooper pairs in the condensate. Then the function
\begin{equation}
    \psi_C(\vect{R}) \equiv \psi(\vect{R},\vect{R})
\end{equation}
describes the motion of the center of mass of these pairs and lends itself as the order parameter for a fermionic superfluid.
In a uniform system, $\psi_C(\vect{R})$ is a constant proportional to the gap $\Delta$:
\begin{equation}
\label{e:orderparam}
    \psi_C(\vect{R}) = \frac{1}{\Omega}\sum_k \left<c_{k\uparrow}^\dagger c_{-k\downarrow}^\dagger\right> = - \frac{1}{V_0} \Delta
\end{equation}
where we have used the gap equation~\footnote{In the BEC-regime, one needs to include thermal molecules in the number equation if $\Delta$ is to vanish at $T = T_C$~\cite{nozi85,melo93}.}. This can be extended to a non-uniform system in which the density and $\Delta(\vect{R})$ does not vary rapidly (local density approximation).
One should point out that it is the presence of a non-zero order parameter, defined via the two-particle density matrix, that signals superfluidity, not the presence of a gap in the excitation spectrum. Gapless superfluidity might occur when the quasi-particle excitations are different for spin up and spin down fermions, one branch touching zero (for example $E_{k\downarrow} = 0$) close to a second-order phase transition to the normal state. Such breaking of time-reversal symmetry leading to gapless superconductivity can occur for example in thin superconducting films in a magnetic field, or in the presence of magnetic impurities~\cite{abri60,tink04sc}.

Close to $T_C$, the order parameter will be small, and after Ginzburg and Landau one can expand the free energy of the superfluid in terms of the small parameter $\psi_C(\vect{r})$. From here, one derives the famous Ginzburg-Landau equation for the order parameter~\cite{abri75,lifs80statphys2}
\begin{equation}
    -\frac{\hbar^2 \nabla^2}{2 m^*}\psi_C + a\,\psi_C + b \left|\psi_C\right|^2\psi_C = 0
\label{e:GinzLan}
\end{equation}
The Ginzburg-Landau theory was developed for superconductors on purely phenomenological grounds in 1950, before the advent of BCS theory. $m^*$ was introduced as the mass of the ``superelectrons'' carrying the supercurrent. It is conventional to choose $m^* = 2m$, the mass of a fermion pair. However, this choice modifies the normalization of $\psi_C$ from Eq.~\ref{e:orderparam} to $\left|\psi_C\right|^2 = n_s/2$, one-half the superfluid density~\footnote{The origin of this normalization is the free energy density $F$ of a superflow with velocity $v_s = \hbar \nabla\phi/2 m$, where $\phi$ is the phase of the wave function $\psi_C = \left|\psi_C\right|\,e^{i \phi}$. By definition of the superfluid density, $F = \frac{1}{2} m v_s^2\, n_s = \frac{1}{8} \hbar^2 \left|\nabla \phi\right|^2/m$, but in terms of $\psi_C$ we have $F = \frac{1}{2} \hbar^2 \left|\psi_C\right|^2 \left|\nabla \phi\right|^2/m^*$. From $m^* = 2 m$ follows $\left|\psi_C\right|^2 = n_s/2$.}. This is consistent with Eq.~\ref{e:ns}, which shows that $n_s \propto \Delta^2$ close to $T_C$. Note that one could have equally well normalized $|\psi_C|^2 = n_0$ via the density of condensed fermion pairs, as this also vanishes like $n_0 \propto T-T_C \propto \Delta^2$ at $T_C$. This would, however, change the mass $m^*$ into $2\, m\, n_0/n_s$.
The parameter $b$ has to be positive, otherwise one could gain energy by making $\left|\psi_C\right|$ arbitrarily large.
In a uniform system, the squared magnitude of the order parameter, in the superfluid state, is $\left|\psi_C\right|^2 = -a/b$, which should start from zero at $T=T_C$ and then grow. Taylor expansion gives $a(T) = a'(T-T_C)/T_C$.
The Ginzburg-Landau equation was later {\it derived} from BCS theory by Gorkov. With the choice $m^* = 2 m$, his derivation gives~\cite{abri75} $a' = \frac{6(\pi T_C)^2}{7\zeta(3)E_F}$ and $b = a'/n$.

The Ginzburg-Landau equation has exactly the form of a non-linear Schr\"odinger equation for the center-of-mass wave function of a fermion pair. In the BEC-regime at $T=0$, a rigorous microscopic theory, which does not require a small order parameter is the Gross-Pitaevskii equation describing the condensate of molecules. It is formally identical to Eq.~\ref{e:GinzLan} if we set $-a = \mu_M$, the chemical potential of molecules, and $b = 4\pi\hbar^2 a_M / m_M$, describing the interactions between molecules. In a uniform system and at $T = 0$, $-a = b\,n_M$, as $|\psi_C|^2 = n_M = n/2$ in the BEC-regime.

For a non-uniform system, Eq.~\ref{e:GinzLan} defines a natural length scale over which the order parameter varies, the Ginzburg-Landau coherence length
\begin{equation}
    \xi_{\rm GL}(T) = \sqrt{\frac{\hbar^2}{4 m \,\left|a\right|}} = \left\{%
\begin{array}{ll}
    0.74\;\xi_{\rm BCS}\;\left(\frac{T_C}{T_C - T}\right)^{1/2} & \hbox{in BCS-regime} \\
\xi_{\rm BEC}\;\left(\frac{T_C}{T_C - T}\right)^{1/2} & \hbox{in the BEC-regime} \\\end{array}%
\right.
\end{equation}
$\xi_{\rm BEC}=\sqrt{\frac{1}{8\pi a_M n_M}}$ is the healing length of the molecular condensate.
$\xi_{\rm GL}$ becomes very large close to the critical
temperature, and in particular it can be large compared to the
BCS-coherence length $\xi_{\rm BCS} = \hbar v_F / \pi \Delta_0$,
defined above via the zero-temperature gap $\Delta_0$. Spatial
variations of the wave function $\psi_C$ then occur at a length scale
much larger than the size of a Cooper pair, and in this regime, the wave function can be described by a {\it local}
equation, although the pairs are extended~\cite{stin97ginzburg}. While the G.-L. equation was originally derived close to $T_C$, assuming a small order parameter, its validity can be extended to all temperatures under the only condition that $\Delta(\vect{r})$ varies slowly compared to $\xi_{\rm GL}(0)$~\cite{wert63}. This condition is less and less stringent as we cross-over into the BEC-regime of tightly bound molecules.

Note that very close to $T_C$, fluctuations of the order parameter
are large and the G.-L. equations are no longer valid. The size of
this critical region is given by $|T-T_C|/T_C \ll (T_C/E_F)^4$ in
the BCS-regime, and $|T-T_C|/T_C \ll k_F a$ in the
BEC-regime~\cite{melo93}. The correlation length then diverges as
$(T_C - T)^{-\nu}$ and the superfluid density vanishes
as~\cite{lifs80statphys2} $(T_C-T)^{(2-\alpha)/3}$ with universal
critical exponents $\alpha$ and $\nu$, instead of the linear
behavior $\propto T_C-T$ implied in the Ginzburg-Landau theory
($\alpha \approx 0$ and $\nu \approx 0.67$ for a complex scalar
order parameter in 3D~\cite{jose66,klei99crit}).

\paragraph{Detecting the order parameter}

One appealing feature of dilute gas experiments is the ability to
directly visualize the order parameter. In the BEC-limit, the
entire gas is condensed. As with atomic BECs, density profiles of
the molecular gas then directly measure the condensate density
$n_0$. In particular, the contrast of interference fringes and of
vortex cores approaches 100\%. However, in the BCS regime, the
condensate fraction decreases. Furthermore, pairs dissociate in
ballistic expansion. This can be avoided by ramping towards the
BEC-regime during expansion. As described in
section~\ref{c:expobservation}, it has been possible to observe
condensates and vortices across the entire BEC-BCS crossover.

\subsection{Crossing over from BEC to BCS}

Throughout the BEC-BCS crossover, all quantities vary smoothly, many
of them even monotonously with $1/k_F a$.  Still, the question has
often been raised in what region(s) of the crossover qualitative
changes occur. When the initial observations of condensation of
fermion pairs were
announced~\cite{grei03molbec,joch03bec,zwie03molBEC,bart04,rega04,zwie04rescond},
the value $1/k_F a=0$ was regarded as special, since this value
separates the regimes where two atoms in isolation will or will
not form a weakly bound pair.  Observations at $1/k_F a>0$ were
classified as molecular condensates, and those at $1/k_F a<0$ as
fermionic condensates.  However, it is clear that the absence or
presence of an extremely weakly bound two-body state (with $a\gg
1/k_F$) does not affect the many-body system, since many-body
pairing is dominant in this regime. In the following section, we
summarize all qualitative criteria we are aware of, which define
specific values of $1/k_F a$ where qualitative changes in physical
properties occur. Of course, different criteria lead to different
values.  It appears that for the case of a broad Feshbach
resonance, all important qualitative changes occur in the window
$0.2 < 1/k_F a < 0.9$. We therefore suggest that one should refer
to molecular BEC only in the regime $1/k_F a >1$.  Although BCS
theory seems to be qualitatively correct already for $1/k_F a <
0$, we refer to the whole region $1/ k_F |a| <1$ as the crossover
region,
in accordance with most other authors.   It seems most natural to
use the word fermionic condensates for the regime with $1/ k_F a
<-1$ and apply it to superconductors, superfluid $^3$He and the
atomic Fermi gases. The big and unique accomplishment of the field
of ultracold atoms has been the creation of the first crossover
condensates, which connect two regimes that could be studied only
separately before. In this crossover regime, bosonic and fermionic descriptions are merged or co-exist.

\begin{itemize}
    \item {\it Excitation spectrum.} At $\mu = 0$, the character of single-particle excitations changes (see Fig.~\ref{f:excitation}): For $\mu <0$, the minimum excitation energy lies at $k = 0$, while for $\mu >0$, the minimum occurs at non-zero momenta, around $k = k_F$ in the BCS-regime. In the BCS mean-field solution, this point lies at $1/k_F a = 0.55$, a more refined theory gives $1/k_F a = 0.41$~\cite{haus07bcsbec}.

    \item {\it Critical velocity.} In the BEC-limit, the critical velocity is due to excitations of sound waves, while  in the BCS-regime, it is determined by pair breaking. Both types of excitations become more costly closer to resonance, and consequently there is a maximum in the critical velocity that occurs at $1/k_F a \approx 0.3$ (see Fig.~\ref{f:velocity}).

    \item {\it Normal density.} Close to $T_C$ and in the BEC-limit, $n_n$ is dominated by bosonic excitations, thermal fermion pairs, while in the BCS-limit it is broken pairs that contribute mostly to the normal density. The point where the two contributions are equal lies at about $1/k_F a \approx 0.2$, and the crossover between the two occurs rapidly, between $1/k_F a \approx 0$ and $0.4$~\cite{andr03becbcs}.

    \item {\it Balanced superfluidity.} The pairing gap $\Delta$
    in the BCS-regime presents a natural barrier for excess atoms
to enter the superfluid. A dramatic consequence of this is the
observable phase separation in a trap between an equal superfluid
mixture of spin up and spin down atoms and a normal imbalanced gas
surrounding it~\cite{shin06phase}. In fact, on the BCS-side this
is the consequence of a first order phase transition between the
balanced superfluid and the normal gas~\cite{shee06phase}. Only if
the chemical potential difference $\mu_\uparrow - \mu_\downarrow$
between the spin up and spin down species becomes larger than
$2\Delta$ (or $2\sqrt{\mu^2 + \Delta^2}$ if this occurs at $\mu <
0$), can unpaired atoms (quasi-particles) enter the superfluid.
From that point on the situation can be described as an
interacting Bose-Fermi mixture of bosonic molecules and single
unpaired fermions. This is likely to occur around $\mu =
0$~\cite{Son05}, so again on the BEC-side of the resonance, around
$1/k_F a = 0.41$.

   \item {\it Absence of unpaired minority fermions.}
    As we already introduced imbalanced Fermi systems, we can go
to the extreme case of a single spin down atom emersed in a sea of
spin up atoms. A natural question to ask is: Will the single spin
down fermion still form a ``monogamous'' molecular pair with a
spin up fermion (BEC limit), or will it rather interact with an
entire collection of majority atoms (``polaron'' or polygamous
pairing)? This intriguing question has recently been studied via a
diagrammatic Monte-Carlo calculation, and a critical interaction
strength $1/k_F a = 0.90(2)$ separating the two regimes of
``pairing'' has been found~\cite{prok07polaron}.

\item {\it Pair condensate in the presence of a Fermi sea.} On the BCS side and on resonance, a large population imbalance destroys superfluidity (Clogston-Chandrasekhar-limit, see section~\ref{s:imbalance}). It is a feature of the BEC limit that a small number of molecules can condense even in the presence of a large Fermi sea of one of the two spin components. This ``BEC'' property is probably lost around the point where the chemical potential $\mu_\uparrow+\mu_\downarrow$ becomes positive, as a very small molecular BEC will have $\mu \approx 0$. It is likely that a necessary and sufficient criterion for having a BEC is the existence of a ``monogamous'' molecular paired state, so again $1/k_F a \approx 0.9$.

\item {\it Critical temperature.} On the BEC-side, the critical temperature is given by the value for a non-interacting gas of bosonic molecules, $T_{C,{\rm BEC}} = 0.22\, T_F$ (see section~\ref{s:crittemp}). For increasing interactions,
the critical temperature first {\it increases}, before it drops to
$T_{C,{\rm Unitarity}} \approx 0.15\, T_F$ at unitarity and then to
exponentially small values on the BCS-side of the resonance. There
is thus a (low-contrast) maximum of the critical temperature in
the crossover, which lies around $1/k_F a \approx 1.3
$~\cite{haus07bcsbec}.

\item {\it Pair size.} Another crossover occurs in the pair size, which can be smaller or larger than the interparticle spacing. With the definition for the pair size given above (section~\ref{s:evolution}), we found $\xi_0 \approx 1/k_F$ on resonance, $1/3$ of the interparticle spacing. At $k_F a \approx -0.9$, $\xi_0 \approx n^{-1/3}$. Of course, different definitions of an average size can easily differ by factors of $2$ or $3$, and it is not clear whether the pair size should be compared to $n^{-1/3}$ or to $1/k_F$. However, it is clear that long-range Cooper pairs as found in superconductors - with many other particles fitting in between - are only encountered for $1/k_F a \ll -1$.

\item {\it Narrow Feshbach resonance.} For a narrow Feshbach resonance (see section~\ref{s:narrowfeshbach}), the crossover from closed channel dominated molecular BEC to open channel BCS-type superfluidity occurs at $1/k_F a \ll -1$.

\item {\it Equation of state exponent (Fig.~\ref{f:gamma}).}  The
exponent $\gamma$ in the approximate equation of state $\mu(n)\sim n^\gamma$, as calculated in the BEC-BCS model, has a (low-contrast) minimum at $1/k_F a \approx -0.45$.
$\gamma$ changes from the bosonic value ($\gamma = 1$) on the BEC-side to the fermionic value ($\gamma = 2/3$) on the BCS-side. However, since universality demands $\gamma = 2/3$ already on resonance, the crossover region is located mainly between $1/k_F a = 1$ and $1/k_F a = 0$. Near the Feshbach resonance, but still on the BEC side, the
equation of state is then already fermionic ($\gamma \approx 2/3$), which is a strong reason not to call this system a molecular BEC.

\end{itemize}

\section{Feshbach resonances}
\label{c:feshbach}

Feshbach resonances are crucial for the study of strongly
interacting fermions.  Typical scattering lengths in alkali atoms
are on the order of the van der Waals range $r_0 \approx 50-100\,
a_0$. Common interparticle spacings in ultradilute gases are
$n^{-1/3} \sim 10\, 000\, a_0$, corresponding to $k_F = (2\,500
\,a_0)^{-1}$. For such small interaction strengths $k_F
\left|a\right| \sim 0.03$, the critical temperature for achieving
fermionic superfluidity is exponentially small, $T_C \approx
10^{-23}\, T_F$. Clearly, one requires a way to enhance the
interatomic interactions, for example via scattering resonances.

Early on, $^6$Li was considered as an exception and as a promising
candidate to achieve fermionic
superfluidity~\cite{stoo96superfluid}, as its triplet scattering
length was found to be unusually large and negative, about $-2\,
000\, a_0$~\cite{abra97}. The reason is that the (triplet or electron-spin aligned) interatomic potential
of $^6$Li could, if it were just a bit deeper, support an additional bound state, so low-energy collisions are almost
resonant.  What first seemed to be special for $^6$Li, namely a
large negative scattering length, can now be created in
many two-atom systems by tuning the scattering length near a {\it
Feshbach resonance}. These resonances occur as a bound state in the interatomic potential is tuned into resonance with the energy of two colliding atoms. This tuning is possible via an applied
magnetic field if the magnetic moment of the bound state differs
from that of the two unbound atoms.

In this chapter, we provide a quantitative description of Feshbach
resonances. Our goal is to provide a thorough discussion of the
conditions on which the closed channel molecular state can be
eliminated, so that the physics is reduced to potential scattering
(so-called single channel scattering).  This is the case of the
so-called broad Feshbach resonance.  We start first by summarizing
the features of scattering resonances in a single channel, by
using the attractive spherical well as an exactly solvable
example, and then present a model for Feshbach resonances.

\subsection{History and experimental summary}
\label{s:history}

Herman Feshbach introduced a formalism to treat nuclear scattering
in a unified way~\cite{fesh58,fesh62}. In elastic collisions, for
example, a free nucleon colliding with a target nucleus can
undergo resonant scattering. This occurs  whenever the initial
scattering energy is equal to that of a ``closed channel'' bound
state between the nucleon and nucleus in the absence of the
incoming ``scattering channel''. A ``closed channel'' has a higher
asymptotic energy than the ``incoming'' or initial scattering energy
and inelastic decay into such a channel is energetically
forbidden. The Feshbach formalism allowed to treat scattering
entirely in the ``open channel'' by introducing an effective
potential that described coupling into and out of the closed
channel. In atomic physics, a related type of resonance is encountered for example in highly excited atoms and ions, where a discrete autoionized state is coupled to a continuum of scattering states. Various aspects of such resonances were studied by Fano~\cite{fano61phaseshifts}.

Feshbach resonances at zero energy are realized by tuning an
external magnetic field.  This was predicted for hydrogen in
1976~\cite{stwa76} and for cold alkali atoms in
1993~\cite{ties93}. In cold atom experiments, the initial emphasis
was on modification of elastic and inelastic atomic
collisions~\cite{inou98,cour98fesh,robe98}, but it soon turned out
that Feshbach resonances  opened a new avenue towards ultracold
molecules: Instead of cooling the molecules themselves, it became
possible to create them cold by associating ultracold atoms. The
first observation of a Feshbach resonance in ultracold
atoms~\cite{inou98,sten98stro} showed strong losses in the atomic
signal that were attributed~\cite{abel99,mies00,yuro99} to the
formation of ultracold, highly vibrationally excited molecules.
However, it was predicted that these molecules, formed out of two
bosonic atoms, would undergo fast vibrational relaxation into more
tightly bound molecular states. Still, in experiments on
$^{85}$Rb, the presence of the molecules, as short-lived as they
were (lifetime $\sim 100\,\rm\mu$s), could be detected via
coherent beats between the free atomic and the bound molecular
state~\cite{donl02mol}. Studies of the decay of {\it fermionic}
gases close to a Feshbach
resonance~\cite{diec02fesh,bour03,rega03lifetime} held a peculiar
surprise: The maximum atom loss was not centered on resonance, but
was shifted towards regions where the Feshbach molecular state was
already quite deeply bound. The gas close to resonance was {\it
stable}~\cite{diec02fesh,ohar02science, bour03,rega03lifetime}, in
stark contrast to the bosonic case. This molecular state could be
{\it reversibly} populated via a magnetic field sweep across
resonance~\cite{rega03mol}, at a conversion efficiency exceeding
90\%~\cite{cubi03,hodb05mol}. Most importantly, it was found to be
long-lived~\cite{cubi03,joch03lith,stre03,rega03lifetime}, with
lifetimes between about 100 ms (for $^{40}$K) and several 10 s
(for $^6$Li). This is to be compared to the molecular lifetimes on
the order of only 5 ms observed in bosonic
gases~\cite{herb03cs_mol,xu03na_mol,durr04mol}. The remarkable
stability of fermion dimers near Feshbach resonances is directly
linked to the Pauli principle~\cite{petr04dimers}: The
characteristic size of dimers is $a$, the scattering length for
atom-atom collisions. A relaxation into more deeply bound
molecular states of size $r_0$ (roughly the van der Waals-range)
requires at least three fermions to be within a distance $r_0$
from each other. As two of them necessarily have the same spin,
the relative wave function has to be antisymmetric, i.e. it has a
node when the relative distance $r=0$ and varies proportional to
$kr$ for small values of $r$, where $k\sim 1/a$ is the
characteristic momentum spread of the dimer. This suppresses
relaxation processes by a certain power of $(k r_0) \sim (r_0/a)$.
For dimer-dimer scattering, the power is 2.55
~\cite{petr04dimers}. What is crucial for this suppression is the
Pauli principle and the large ratio between initial and final
size.  For bosons, the reverse is true, i.e. the relaxation rate
diverges with
$a^{3.5...4}$~\cite{fedi96reco,niel99boserecomb,esry99recomb,beda00boserecomb,petr04bose},
although the overlap integral between initial and final state
decreases. The ratio between good to bad collisions can be very
high for fermion dimers near Feshbach resonances, since in
contrast to inelastic collisions, elastic scattering is not
suppressed.

For an extensive discussion of dimer stability, we refer the reader to the lecture notes of G. Shlyapnikov in these proceedings.

\subsection{Scattering resonances}
\label{s:squarewell}

We summarize first some results for the attractive spherical well
potential which are derived in many text books.  Our model for
Feshbach resonances will have a region in detuning around the
resonance, where the interaction and the scattering look very
similar to the case of the spherical well.

A three-dimensional spherical well potential of  radius $R$ and depth
$V$ has scattering states with energy $E>0$ and also bound states
with energy $E<0$ when the depth is larger than a critical value
$V_c$. We define $E = \hbar^2 k^2/m$ for $E>0$ and wave
vector $k$, $|E| = \hbar^2 \kappa^2/m$ for $E<0$, $V \equiv \hbar^2 K^2/m$ and $E_R = \hbar^2/m
R^2$. The critical well depth is $V_c = \frac{\pi^2}{4} E_R$.  New bound states
appear when $K_n R = (2n+1)\pi/2$  at $V_n = (2n+1)^2 V_c$.

In the ultracold regime $E \ll E_R$, or eqv. $k R \ll 1$, and for $E \ll
V$, scattering states have the same radial wave function inside
the well as bound states with $|E| \ll V$: $u(r<R) = A\sin(K
r)$. Outside the well, $u(r)$ for scattering states is of the form $u(r>R) = \sin(k r +
\delta_{\rm s})$. Matching value and slope of $u(r>R)$ and $u(r<R)$ at $r=R$ fixes the phase shift $\delta_{\rm s}$ via the condition: $k \cot(kR + \delta_{\rm s}) = K \cot(K R)$. The scattering length is
\begin{equation}
    \label{e:scattlen}
    a = -\lim_{k \ll 1/R} \frac{\tan \delta_{\rm s}}{k} = R \left(1 - \frac{\tan{K R}}{K R}\right)
\end{equation}

By expanding $k \cot \delta_{\rm s}= -1/a + \frac{1}{2} r_{\rm eff} k^2 +\dots$, we obtain
the effective range $r_{\rm eff} = R - 1/(K^2 a) - \frac{1}{3} R^3/a^2$, which is small and on the
order of $R$, unless the well is very shallow $K^2 R a \ll 1$, or $a$ is smaller then $R$.  We will see below
that Feshbach resonances lead to {\it negative} values of $r_{\rm eff}$ that can be large.

The scattering length in Eq.~\ref{e:scattlen} diverges whenever a
new bound state enters the potential. This relationship applies to
any potential scattering of finite range $R$: A diverging
scattering length signifies that the phase shift $\delta_{\rm s}$
due to the potential well is approaching $\pi/2$.
At $R$, we then have a normalized
slope $u'(r)/u(r) \approx k / \tan\delta_{\rm s} = -1/a$ for the
scattering wave function. For {\it positive} $a$, this can just as
well be continued by a bound state wave function $e^{-\kappa r}$
with matching slope, which gives $\kappa = 1/a$. So apart from the
scattering solution, we find a new bound state solution of
Schr\"odinger's equation at negative energy
\begin{equation}
    E_B = - \hbar^2/m a^2\qquad \mbox{for } a > 0.
\end{equation}

Away from resonances, the scattering length is close to the
``background'' scattering length $R$. Close to a resonance (at $V_n
= \hbar^2 K_n^2 / m$), the scattering length diverges as
\begin{equation}
    a \approx \frac{1}{K R (K-K_n)} \approx \frac{2 \hbar^2}{m R (V - V_n)}
\end{equation}
and for $a>0$ the bound state energy depends on $V$ like
\begin{equation}
    E_B = - \frac{\hbar^2}{m a^2} = - \frac{1}{4} \frac{(V - V_n)^2}{E_R}
\end{equation}
This general behavior for weakly bound states was found already in
chapter~\ref{s:boundstates}: The binding energy depends
quadratically on the ``detuning'', and the scattering length is inversely
proportional to the ``detuning''. The beauty and power of
Feshbach resonances is that this detuning is now controlled by an
externally applied field.

\subsection{Feshbach resonances}
\label{s:feshbachmodel}

\begin{figure}
\centering
  \includegraphics[width=4in]{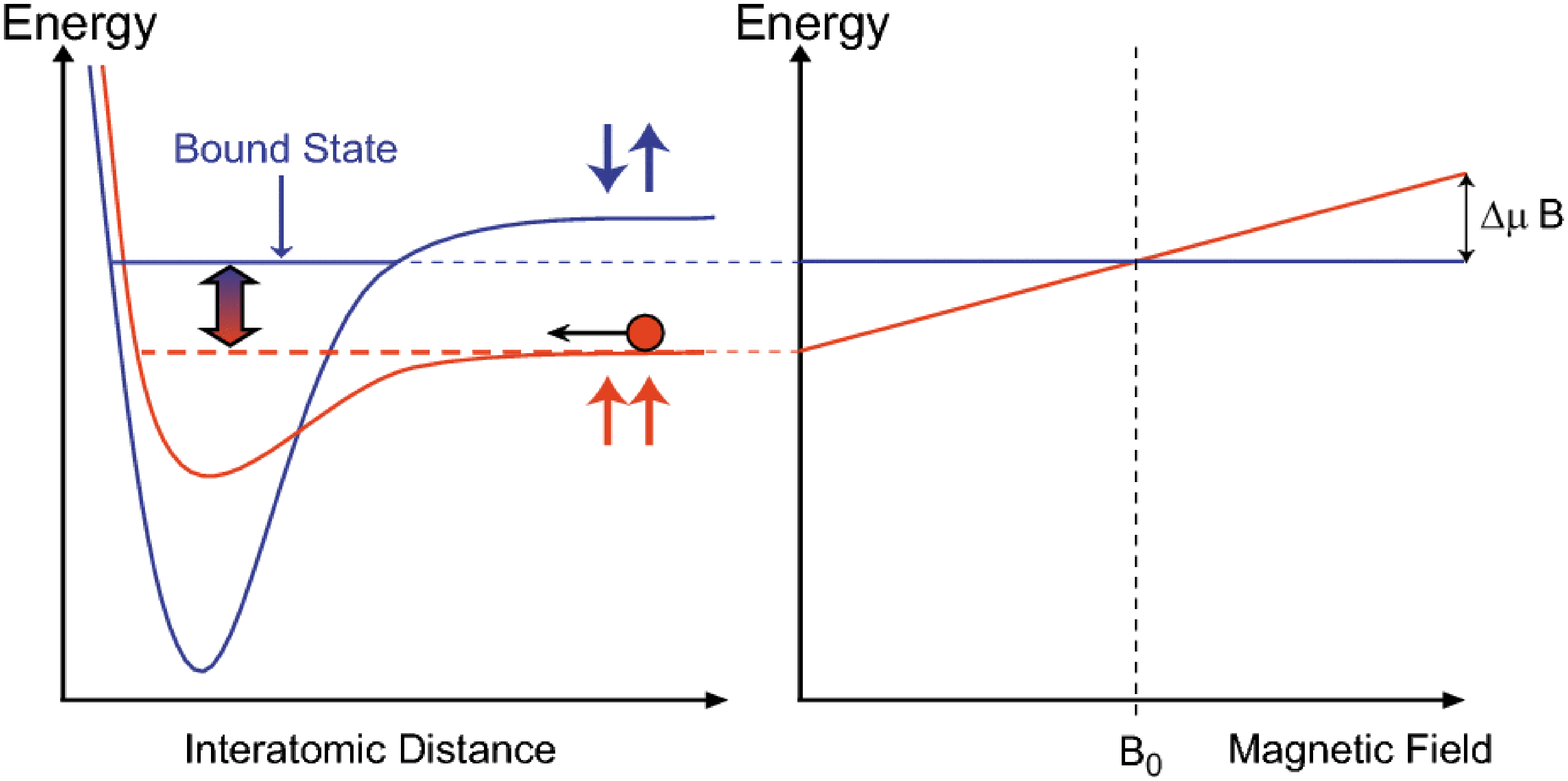}\\
  \caption[Origin of Feshbach resonances]{Origin of Feshbach resonances. Atoms entering for example in the triplet potential are coupled to a singlet bound molecular state. By tuning the external magnetic field, this bound state can be brought into resonance with the incoming state (at $B_0$ in the graph on the right).}\label{f:Feshbachpotentials}
\end{figure}

We now turn to realistic interaction potentials between alkali
atoms. Here, the interaction actually depends on the internal
structure of the two colliding atoms, namely on the relative spin
orientation of their valence electrons, singlet or triplet. In
Fig.~\ref{f:Feshbachpotentials} for example, the atoms enter in a
triplet configuration. If there was no coupling between the
singlet $V_S$ and the triplet potential $V_T$, the atoms would
simply scatter off each other in $V_T(r)$, acquiring some certain,
fixed phase shift. However, the {\it hyperfine interaction}
$V_{\rm hf}$ is not diagonal in the total electronic spin
$\vect{S} = \vect{s}_1 + \vect{s}_2$ of the two atoms and thus
provides a coupling between singlet and triplet potentials~\cite{moer95res}:
\begin{eqnarray}
    V_{\rm hf} &=& a_{\rm hf} \left(\vect{s}_1 \cdot \vect{i}_1 + \vect{s}_2 \cdot \vect{i}_2\right) \nonumber \\
           &=& \frac{a_{\rm hf}}{2} \vect{S} \left(\vect{i}_1 + \vect{i}_2\right) + \frac{a_{\rm hf}}{2} \left(\vect{s}_1 - \vect{s}_2\right) \left(\vect{i}_1 - \vect{i}_2\right) \nonumber \\
           &=& V_{\rm hf}^+ + V_{\rm hf}^-
\end{eqnarray}
with the hyperfine constant $a_{\rm hf}$ and the nuclear spins
$\vect{i}_{1,2}$ of the two atoms.

The coupling $V_{\rm hf}^-$ connects singlet and triplet states
since the operator $\vect{s}_1 - \vect{s}_2$ is antisymmetric in 1 and 2, and
therefore couples symmetric (triplet) electronic spin states to
antisymmetric (singlet) states. It is thus fully off-diagonal in
the singlet/triplet basis, implying that  coupling matrix elements
are on the order of unity. $V_{\rm hf}^-$ should thus have
matrix elements on the order of $a_{\rm hf}$.

The singlet potential is a ``closed channel'', meaning that singlet
continuum states are not available as final scattering states by
energy conservation. A Feshbach resonance occurs when the state that the atoms collide in (the ``incoming'' state) is resonant with a bound state in this singlet potential. The energy difference
between the incoming and the Feshbach bound state can be tuned via
an applied magnetic field, due to their different magnetic moments
(see Fig.~\ref{f:Feshbachpotentials}).

\begin{figure}
\centering
  \includegraphics[width=3.5in]{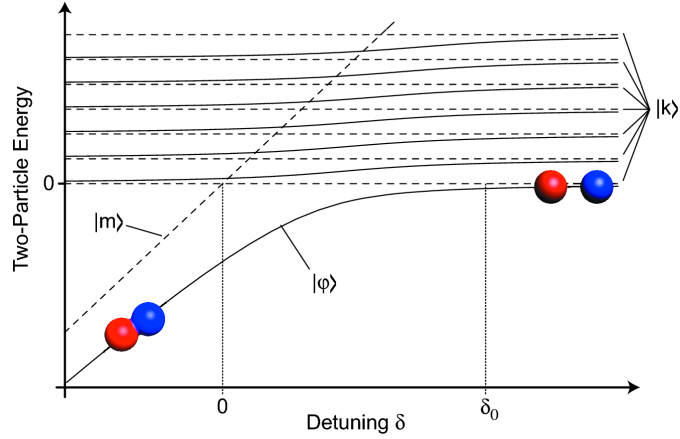}\\
  \caption[Simple model for a Feshbach resonance]{Simple model for a Feshbach resonance. The dashed lines show the uncoupled states: The closed channel molecular state $\left|m\right>$ and the scattering states $\left|k\right>$ of the continuum. The uncoupled resonance position lies at zero detuning, $\delta = 0$. The solid lines show the coupled states: The state $\left|\varphi\right>$ connects the molecular state $\left|m\right>$ at $\delta \ll 0$ to the lowest state of the continuum above resonance. At positive detuning, the molecular state is ``dissolved'' in the continuum, merely causing an upshift of all continuum states as $\varphi$ becomes the new lowest continuum state. In this illustration, the continuum is discretized in equidistant energy levels. In the continuum limit (Fig.~\ref{f:genericfeshbach}), the dressed molecular energy reaches zero at a finite, shifted resonance position $\delta_0$.}\label{f:feshbachsketchA}
\end{figure}

\subsubsection{A model for Feshbach resonances}

\begin{figure}
\centering
  \includegraphics[width=3.5in]{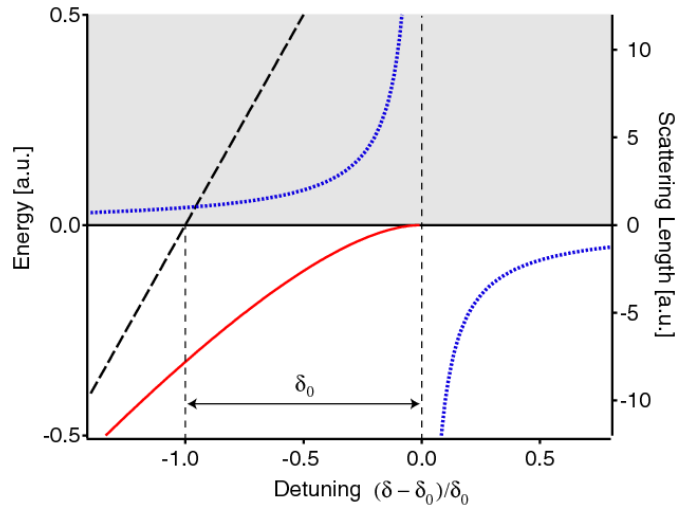}\\
  \caption[Bound state energy and scattering length close to a Feshbach resonance]{
  Bare, uncoupled molecular state (dashed line), coupled, bound molecular state (solid line) and scattering length (dotted line) close to a Feshbach resonance. The shaded area represents the continuum of scattering states, starting at the collision threshold at $E=0$. Interaction between the molecular state and the continuum shifts the position of the resonance by $\delta_0$ from the crossing of the uncoupled molecular state with threshold. Note the quadratic behavior of the bound state energy with detuning $(\delta - \delta_0)$ close to resonance.}\label{f:genericfeshbach}
\end{figure}

Good insight into the Feshbach resonance mechanism can be gained
by considering two coupled spherical well potentials, one each for
the open and closed channel~\cite{duin04feshbachreview}. Other
models can be found in~\cite{kohl06feshbachreview,shee06phase}.
Here, we will use an even simpler model of a Feshbach resonance,
in which there is only one bound state of importance
$\left|m\right>$ in the closed channel, the others being too far
detuned in energy (see Fig.~\ref{f:feshbachsketchA}). The
continuum of plane waves of relative momentum $\vect{k}$ between
the two particles in the incoming channel will be denoted as
$\left|k\right>$. In the absence of coupling, these are
eigenstates of the free hamiltonian
\begin{eqnarray}
    H_0 \left|k\right> &=& 2\epsilon_k \left|k\right> \qquad \mbox{with } \epsilon_k = \frac{\hbar^2 k^2}{2m} >0 \nonumber \\
    H_0 \left|m\right> &=& \delta \left|m\right> \qquad \;\,\,\mbox{with } \delta \lessgtr 0
\end{eqnarray}
where $\delta$, the bound state energy of the ``bare'' molecular
state, is the parameter under experimental control. We consider interactions explicitly only between
$\left|m\right>$ and the $\left|k\right>$'s. If necessary,
scattering that occurs exclusively in the incoming channel can be
accounted for by including a phase shift into the scattering
wave functions $\left|k\right>$ (see Eq.~\ref{e:psiopenchannel}), i.e. using
$\psi_k(r) \sim \sin(k r + \delta_{\rm bg})/r$, where $\delta_{\rm bg}$ and $a_{\rm bg} = -\lim_{k\rightarrow 0} \tan\delta_{\rm bg}/k$ are the (so-called background) phase shift and scattering length, resp., in the open channel. 
First, let us
see how the molecular state is modified due to the coupling to the
continuum $\left|k\right>$. For this, solve
\begin{eqnarray}
    H \left|\varphi\right> &=& E \left|\varphi\right> \nonumber \\
    \mbox{ with } \left|\varphi\right> &=& \alpha \left|m\right> + \sum_k c_k \left|k\right>
\end{eqnarray}
for $E<0$, where $H = H_0 + V$ and the only non-zero matrix
elements of $V$ are $\left<m\right|V\left|k\right> =
g_k/\sqrt{\Omega}$ and their complex conjugates (we will take
$g_k$ to be real). $\Omega$ is the volume of the system and
introduced in this definition for later convenience. We quickly
find
\begin{eqnarray}
    (E - 2\epsilon_k)\, c_k &=& \frac{g_k}{\sqrt{\Omega}}\, \alpha \nonumber\\
    (E - \delta)\, \alpha &=& \frac{1}{\sqrt{\Omega}}\sum_k g_k\, c_k = \frac{1}{\Omega}\sum_k \frac{g_k^2\,\alpha}{E - 2\epsilon_k} \nonumber \\
    \mbox{and thus }\qquad E - \delta &=& \frac{1}{\Omega} \sum_k \frac{g_k^2}{E - 2\epsilon_k}
    \label{e:boundstateequation}
\end{eqnarray}
We only consider low-energy $s$-wave scattering, where the range of
the potential $r_0$ is much smaller than the de Broglie wave
lengths, $r_0 \ll 1/k$. The closed channel molecular state $\left|m\right>$ will have a size $R$ on the order of $r_0$, that the de Broglie waves of colliding atoms cannot resolve. The couplings $g_k$ will not vary much for such low-energy collisions with $k \ll 1/R$. One can thus take $g_k \approx g_0$ constant, up to a natural cut-off $E_R = \hbar^2 / m R^2$, and $g_k = 0$ beyond. We then find:
\begin{eqnarray}
    \left|E\right| + \delta &=& \frac{g_0^2}{\Omega} \int_0^{E_R} d\epsilon \frac{\rho(\epsilon)}{2\epsilon+|E|}
\end{eqnarray}
The integral on the RHS is identical to that in the bound state
equation in one-channel scattering,
Eq.~\ref{e:densityboundstates}. There, the LHS was simply the
inverse scattering strength $-1/V_0$. The two-channel problem
introduces an energy-dependence in the strength of the
potential, $V_0 \rightarrow g_0^2/(E-\delta)$. The integral gives
\begin{eqnarray}
 \left|E\right| + \delta    &=& \frac{g_0^2\rho(E_R)}{\Omega}\left\{1 - \sqrt{\frac{|E|}{2E_R}} \arctan\left(\sqrt{\frac{2E_R}{|E|}}\right)\right\} \\
       &\approx& \left\{%
\begin{array}{ll}
    \delta_0 - \sqrt{2 E_0 \left|E\right|} & \hbox{for } |E| \ll E_R \\
    \delta_0\frac{2}{3}\frac{E_R}{|E|} & \hbox{for } |E| \gg E_R  \\
\end{array}%
\right.   \nonumber \\
    \mbox{ with } \qquad \delta_0 &\equiv& \frac{g_0^2}{\Omega} \sum_k \frac{1}{2\epsilon_k} = \frac{4}{\pi} \sqrt{E_0 E_R} \nonumber\\
    \mbox{ and } \qquad E_0 &\equiv& \left(\frac{g_0^2}{2\pi}\left(\frac{m}{2\hbar^2}\right)^{3/2}\right)^2 \nonumber
    \label{e:E0definition}
\end{eqnarray}
 $E_0$ is an energy scale associated with the coupling constant $g_0$.
As illustrated in Fig.~\ref{f:feshbachsketchA}, for positive
detuning $\delta > \delta_0$ the original molecular state is
``dissolved'' in the continuum. Due to the coupling of the molecular
state with the continuum, the resonance position is shifted by
$\delta_0$. For $\delta - \delta_0 < 0$, we find a true bound
state at
\begin{equation}
    E = \left\{%
\begin{array}{ll}
       - E_0 + \delta - \delta_0 + \sqrt{(E_0 - \delta + \delta_0)^2 - (\delta - \delta_0)^2} & \hbox{for } |E| \ll E_R\\
\frac{\delta}{2} - \sqrt{\frac{\delta^2}{4} + \frac{2}{3}\delta_0 E_R}, & \hbox{for } |E| \gg E_R \\\end{array}%
\right. \label{e:Feshbachboundstate}
\end{equation}
The ``dressed'' bound state energy $E$ is shown in
Fig.~\ref{f:genericfeshbach}. Far away from the resonance region,
for $\delta \ll -(E_R^3 E_0)^{1/4}$, one finds $E \approx \delta$,
thus recovering the original bound state. On the other hand, close
to resonance, the energy $E \approx - \frac{1}{2} {\left(\delta -
\delta_0\right)^2} /{E_0} = - \frac{8}{\pi^2}E_R{\left(\delta -
\delta_0\right)^2} /{\delta_0^2}$ depends quadratically on the
detuning $\delta-\delta_0$, as expected.

\paragraph{Scattering amplitude}

To find the scattering amplitude, we solve Schr\"odinger's
equation for $E>0$. A small imaginary part $i \eta$ with $\eta>0$
is added to the energy to ensure that the solution will correspond
to an outgoing wave. The goal is to see how the coupling to the
molecular state affects scattering in the incoming channel. In an
approach formally equivalent to the solution for bound states,
Eq.~\ref{e:boundstateequation}, we find for the amplitudes $c_k$
in the open channel:
\begin{equation}
    (E - 2\epsilon_k)\, c_k = \frac{g_k}{\sqrt{\Omega}}\, \alpha = \sum_q \frac{g_k}{\sqrt{\Omega}}\frac{1}{E - \delta + i\eta}\frac{g_q}{\sqrt{\Omega}}\, c_q \equiv \sum_q V_{\rm eff}(k, q)\, c_q
\end{equation}
By eliminating the closed-channel molecular amplitude $\alpha$
from the equations, the scattering problem is now entirely
formulated in the open channel. The molecular state causes an
effective interaction $V_{\rm eff}$ that corresponds to two atoms
colliding and forming a molecule (matrix element
$\frac{g_q}{\sqrt{\Omega}}$), spending some small amount of time
(of order $\hbar/(E-\delta)$) in the molecular state
(propagator $\frac{1}{E - \delta + i\eta}$) and exiting again as
two unbound atoms (matrix element $\frac{g_k}{\sqrt{\Omega}}$).

The $s$-wave scattering amplitude can now be obtained using the
general expression we found in chapter~\ref{s:elasticcollisions},
equation~\ref{e:scattampintegral}. We only need to insert the
effective potential $V_{\rm eff}(k,q)$ in place of
$V(\vect{k}-\vect{q})/\Omega$. The problem is simplified by
setting as before all $g_k = g_0$ for $E < E_R$ and $g_k = 0$ for
$E > E_R$. The replacement is
\begin{equation}
V_0 \rightarrow V_{\rm eff}\, \Omega = \frac{g_0^2}{E - \delta}
\end{equation}
as we had found for the bound state problem, and
Eq.~\ref{e:scattampintegral} becomes
\begin{equation}
    \frac{1}{f_0(k)} \approx -\frac{4\pi\hbar^2}{m g_0^2} (E - \delta) + 4\pi \int \frac{d^3 q}{(2\pi)^3}\, \frac{1}{k^2-q^2+i\eta}
\end{equation}
The integral on the RHS is identical to what one encounters in
one-channel scattering: it generates the necessary $- i k $ (see
Eqs.~\ref{e:scattamplitude} and~\ref{e:scattampcutoff}),
it determines the resonance position by introducing a shift $-4\pi
\int \frac{d^3 q}{(2\pi)^3}\, \frac{1}{q^2} =
-\frac{2}{\pi}\frac{\sqrt{2}}{R}$ and it contributes to the effective
range $r_{\rm eff}$ with a term $\propto R$ (which we neglect in
the following, as $E_R$ is taken to be the largest energy scale in
the problem). All the physics of the two-channel model is
contained in the first term, which includes the molecular state
energy $\delta$ and a term proportional to $E \propto k^2$ that
will give another contribution to the effective range. Using
$-4\pi \int \frac{d^3 q}{(2\pi)^3}\, \frac{1}{q^2} = -
\frac{4\pi\hbar^2 \delta_0}{m g_0^2}$ and replacing $g_0$ in favor
of $E_0$ via Eq.~\ref{e:E0definition}, we have
\begin{equation}
    \frac{1}{f_0(k)} \approx -\sqrt{\frac{m}{2\hbar^2 E_0}} (\delta_0 - \delta) -\frac{1}{2}\sqrt{\frac{2\hbar^2}{m E_0}} k^2 - i k
\end{equation}
The scattering amplitude is now in the general form of
Eq.~\ref{e:scattamplitude}, and we can read off the scattering
length and the effective range of the model:
\begin{eqnarray}
\label{e:effrange}
    a &=& \sqrt{\frac{2 \hbar^2 E_0}{m}}\frac{1}{\delta_0 - \delta} \\
    r_{\rm eff} &=& -\sqrt{\frac{2\hbar^2}{m E_0}}
\end{eqnarray}
The scattering length, shown in Fig.~\ref{f:genericfeshbach},
diverges at the {\it shifted} resonance position $\delta =
\delta_0$. Not surprisingly, we recover $E = -\hbar^2 / m a^2$ for
the bound state energy close to resonance for positive $a>0$, as
it should be (see Eq.~\ref{e:Feshbachboundstate}).

In the experiment, the Feshbach resonance occurs for a certain
magnetic field $B_0$. With the magnetic moment difference
$\Delta\mu$ between the incoming state and the closed (uncoupled)
molecular state, we have $\delta - \delta_0 = \Delta \mu (B -
B_0)$ (taking $\Delta\mu$ to be constant). Including the
background scattering length $a_{\rm bg}$ for collisions that
occur entirely in the open channel, the scattering length can be
written in its usual form
\begin{equation}
    a = a_{\rm bg} \left(1 - \frac{\Delta B}{B - B_0}\right) \qquad \mbox{with } \Delta B = \sqrt{\frac{2 \hbar^2 E_0}{m}}\frac{1}{\Delta\mu\; a_{\rm bg}}
\label{e:Feshbachscattlen}
\end{equation}

\subsection{Broad versus narrow Feshbach resonances}
Initially it was assumed that all Feshbach resonances represent a
novel mechanism for fermionic pairing and superfluidity.
Treatments of ``resonance superfluidity''~\cite{holl01} and
``composite Fermi-Bose superfluid''~\cite{timm01} explicitly
introduced coupled atomic and molecular fields as an extension of
the standard BEC-BCS crossover theory presented in
chapter~\ref{c:BECBCStheory}. On the other hand, as discussed in
the previous chapter, the closed channel molecular state can be
formally eliminated from the description by introducing an
effective potential acting on the atoms in the open channel.  We
will see in this section that sufficiently close to the Feshbach
resonance, and for sufficiently small Fermi energies, the physics
is indistinguishable from a single channel model such as the
attractive spherical well (with $R \ll 1/k_F$) discussed in
section~\ref{s:squarewell} or a (suitably regularized) contact
interaction.  It turns out that this simple description applies to
the experimental studies in $^6$Li and $^{40}$K. In these cases,
that involve so-called broad Feshbach resonances, the resonance
simply provides a knob to turn $1/k_F a$ continuously from large
positive values to negative values. The physics is independent of
the nature of the molecular state. Therefore, these system are
universal, i.e. they are ideal realizations of the
``standard'' BEC-BCS crossover physics described in
chapter~\ref{c:BECBCStheory}, and not a new form of ``Feshbach'' or
``resonance'' superfluidity.

\subsubsection{Energy scales}

To address the question on the range of parameters where the
molecular state does play a role, we consider the energy scales in the
problem. They are $E_0$, the energy scale associated with the
coupling strength, the detuning from resonance $\delta-\delta_0$, as a function
of which we want to study the system, and $E_F$, the Fermi energy.
As we will show, the ratio $E_F/E_0$ of the Fermi energy to the
coupling energy scale is the parameter that decides whether the
physics around the resonance is universal ($E_0 \gg E_F$) or
whether the closed molecular channel still plays a role ($E_0 \ll
E_F$)~\cite{comb03fesh,bruu04eff,bruu04univ,shee06phase}. With
Eq.~\ref{e:effrange} this can be equivalently expressed as $k_F
r_{\rm eff} \ll 1$, that is, universality requires the effective
range of the potential to be much smaller than the interparticle
distance. In principle, we have two more energy scales, the cutoff
energy scale $E_R$ and the shift $\delta_0$. $E_R$ is much larger
than the Fermi energy, as we deal with dilute gases where the
interatomic distance is large compared to the range of the
potential. Then, the shift $\delta_0 \sim \sqrt{E_R E_0}$ is much
larger than $E_F$ if $E_0 \gg E_F$, and does not lead to an
additional criterion.

\subsubsection{Criterion for a broad resonance}

The criterion $E_0 \gg E_F$ for a broad resonance is found in
several different ways, each of which is insightful.

\paragraph{BEC-side}

For a spherical well potential, the bound state energy is given by
the universal relation $E_B = - \hbar^2 / m a^2$ (as long as $|E_B|
\ll E_R$). This signifies that the character of the molecular
state is entirely described by the scattering length, a property
of the scattering states in the open channel. However, for the
two-channel Feshbach model discussed above, this relation holds
only for $\delta_0-\delta \ll E_0$ (see
Eq.~\ref{e:Feshbachboundstate}) or equivalently for $\hbar^2 / m a^2
\ll E_0/2$. To observe the universal version of the BEC-BCS
crossover presented in the last chapter, the bound state should
obey the universal behavior already when $k_F a \approx 1$. This
yields the condition $E_F \ll E_0$ for the ``BEC''-side of the
resonance.

A more quantitative way to see this is by calculating the
contribution of the closed channel molecule to the ``dressed''
molecular state
\begin{equation}
    \left|\varphi\right> = \alpha \left|m\right> + \sum_k c_k \left|k\right>
\end{equation}
This can be calculated from the magnetic moment of
$\left|\varphi\right>$, relative to two free atoms: Bare molecules
have a relative magnetic moment $\Delta \mu$, so
$\mu_{\left|\varphi\right>} = \alpha^2 \Delta\mu$. One finds with
Eq.~\ref{e:Feshbachscattlen}: $\mu_{\left|\varphi\right>} =
\frac{\partial E_B}{\partial B} = \sqrt{\frac{2 |E_B|}{E_0}}\,
\Delta \mu$. When the binding energy $|E_B|$ becomes comparable to
the Fermi energy $E_F$ near resonance, the closed channel
contribution $\alpha^2 = \sqrt{\frac{2 |E_B|}{E_0}} $ should
already be negligible in order for the physics to be dominated by
the open channel. Again, this gives the criterion $E_F \ll E_0$.

\begin{figure}
    \centering
    \includegraphics[width=5.4in]{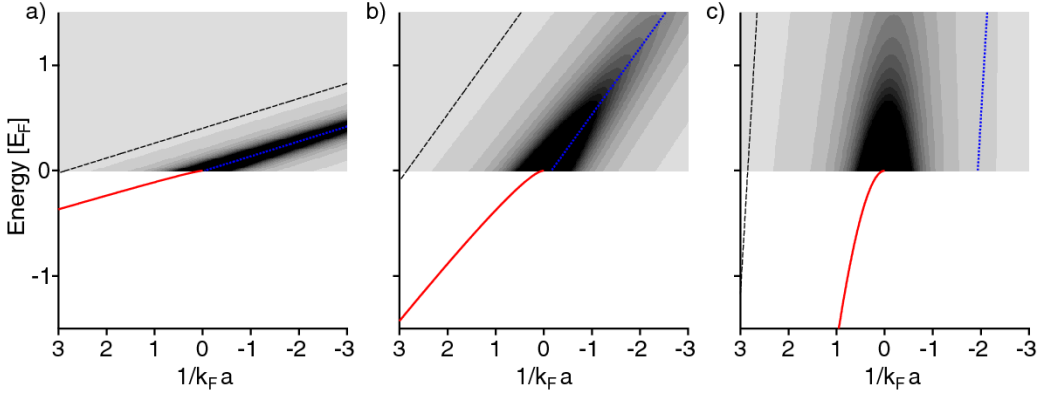}
    \caption[Broad versus narrow Feshbach resonances]{Broad versus narrow Feshbach resonances. Shown are the two-body bound state energy (straight line, in units of $E_F$), and scattering cross section contours (in gray shades) for various Feshbach coupling strengths $E_0$. The dashed line is the bare molecular state (here, $E_R = 20 E_F$, thus $\delta = 0$ at $1/k_F a = \frac{2}{\pi}\sqrt{E_R/E_F} = 2.85$). The dotted line marks the position of finite-energy maxima of the cross section for $\hbar^2/m a^2 > E_0/2$ (see text). a) $E_0 = 0.005\, E_F$, b) $E_0 = 0.1\, E_F$, c) $E_0 = 15\, E_F$. In c), bound state and scattering cross section contours closely approach those for a contact potential. A similar figure can be found in~\cite{shee06phase}.}
    \label{f:crosssection}
\end{figure}

\paragraph{BCS-side}
\label{s:bcsside}
For $\delta>\delta_0$ the molecular bound state has disappeared,
but the closed channel molecule still leaves its mark in the
scattering cross section~\cite{shee06phase} ($E = \hbar^2 k^2/m$):
\begin{eqnarray}
    \sigma(E) &=& 4\pi|f(E)|^2 = 4\pi r_{\rm eff}^2 \frac{E_0^2}{\left(E - \delta + \delta_0\right)^2 + 2 E_0 E} \nonumber \\
    &=& 4\pi r_{\rm eff}^2 \frac{E_0^2}{(E - E_{\rm res})^2 + \Gamma^2/4} \label{e:crosssection}\\
\mbox{with } E_{\rm res} &=& \delta - \delta_0 - E_0 \nonumber \\
\mbox{and } \Gamma^2 &=& 4 E_0^2 \left(\frac{2\left(\delta -
\delta_0\right)}{E_0} - 1\right) \label{e:gammaresonance}
\end{eqnarray}
For $\delta - \delta_0 > E_0$ (equivalently $\hbar^2/m a^2 >
E_0/2$), $E_{\rm res}>0$ and a resonance appears in the scattering
cross section at finite energies. This resonance is just the
(shifted) bare molecular state that has acquired a finite lifetime
$\hbar\Gamma^{-1}$ due to the coupling to the continuum. While the
width $\Gamma$ increases with detuning like $\sim
\sqrt{\delta-\delta_0}$ for $E_{\rm res} \gg E_0$, the relative
width $\Gamma/E_{\rm res} \sim 1/\sqrt{\delta-\delta_0}$ {\it
decreases}, so that the relative position of the molecular state
resonance can in principle be measured more and more accurately in
scattering experiments.  For large values of $\delta$, the
expression Eq.~\ref{e:gammaresonance} for $\Gamma$ approaches the result of Fermi's
Golden Rule, $\Gamma= 2 \pi\, g_0^2\, \rho(\epsilon)/\Omega$,
where $\epsilon=(\delta-\delta_0)/2$ is the energy of the fragments
after molecular dissociation, and $\rho(\epsilon) \propto
(\delta-\delta_0)^{1/2}$ is the density of final states.  This
relation was verified by observing the decay of molecules after a
rapid ramp across the resonance~\cite{muka04}. Clearly, in this
region, the molecular state has a life of its own.

The closed channel molecular state causes a finite-energy scattering resonance on the ``BCS''-side starting at
$\delta - \delta_0 = E_0$ that is non-universal. In single-channel scattering
off a delta-potential or a spherical well potential, no such
finite-energy resonances exist.  We require for universal behavior
that such resonances do not occur within the BEC-BCS crossover,
i.e. within the strongly interacting regime, where $k_F |a| \gtrsim 1$,
or equivalently $\delta - \delta_0 \lesssim E_F$. This leads again to
the condition $E_0 \gg E_F$.

Fig.~\ref{f:crosssection} summarizes these findings. The BEC-BCS
crossover occurs for $-1 \lesssim 1/k_F a \lesssim 1$. If $E_0 \gg E_F$, the
``dressed'' molecular state is almost completely dissolved in the
open channel continuum throughout the crossover and the details of
the original molecular state $\left|m\right>$ do not play a role
(case of a ``broad'' Feshbach resonance). The binding energy is
given by $E_B = -\hbar^2/m a^2$ on the BEC-side, and the scattering
cross section has the universal form $\sigma(k) = 4\pi\frac{a^2}{1  + k^2 a^2}$. On the
other hand, if $E_0$ is comparable to $E_F$, then the molecular
state affects the many-body physics and it needs to be included in
the description of the gas (case of a ``narrow'' Feshbach
resonance)~\cite{palo04res,simo05fesh,shee06phase}.

\paragraph{On resonance}

A stringent criterion for universal behavior requires that all
scattering properties for detunings $\delta - \delta_0 \le E_F$
and for energies $E<E_F$ are identical to the case of a
(regularized) delta-potential or a localized spherical well
potential, where the scattering amplitude $f$ is given by
$1/f=-1/a - ik$, i.e. the contribution of the effective range,
$\frac{1}{2}r_{\rm eff} k^2$ to $1/f$ is negligible.  For the
total cross section, which is proportional to $|f|^2$
(Eq.~\ref{e:crosssection}), the effective range correction is
negligible for $k\ll 1/r_{\rm eff}$, or equivalently, $E_F \ll
E_0$. However, the real part of $f$, which determines the mean
field energy for a dilute gas, depends more strongly on the
effective range:  it is equal to $-1/a k^2$ for large $a$, but
approaches a constant $r_{\rm eff}/2$ if $r_{\rm eff} \ne 0$. This
does not spoil universality, though, as the mean field energy
associated with such a small effective scattering length, $\propto
\hbar^2 n r_{\rm eff}/m$, has to be compared to the many-body
interaction energy, $\beta E_F \propto \hbar^2 k_F^2/m \propto
\hbar^2 n / m k_F$ (see chapter~\ref{c:BECBCStheory}), which
dominates as long as $k_F \ll 1/r_{\rm eff}$ or $E_0>E_F$.

\paragraph{Magnetic moment}

Finally, we want to come back to the schematic description of the
Feshbach resonance in Fig.~\ref{f:feshbachsketchA} that uses
discrete states in a finite volume $L^3$.  Each energy curve has
two avoided crossings. For small coupling, the slope in between
these crossings is still given by the magnetic moment of the
molecule, i.e. in this region, population can be purely in the
closed channel.  This picture is lost when many states couple,
i.e. the resonant coupling $g_k/L^{3/2}$ is larger than the level
spacing, which is about $E_L=\hbar^2/m L^2$.  In order to still
maintain some character of the closed channel molecule, one must
have $g_k<\hbar^2/m L^{1/2}$ or $E_0 < E_L$.  However, any
finite volume approximation has to choose $L$ at least comparable
to the interatomic spacing, or equivalently, the zero-point energy
$E_L$ has to be less than the Fermi energy. Therefore, for the
case of a broad resonance with $E_0>E_F$, the simple picture of
two-level avoided crossings no longer applies, the molecular state
gets ``smeared'' out and ``distributed'' over many open channel
states.

\subsubsection{Coupling energy scale}

We can relate the coupling energy scale $E_0$ to experimentally
observable parameters. Using Eq.~\ref{e:Feshbachscattlen} and the
definition for $E_0$, Eq.~\ref{e:E0definition}, one has
\begin{equation}
    E_0 = \frac{1}{2} \frac{(\Delta\mu \Delta B)^2}{\hbar^2 / m a_{\rm bg}^2}
\end{equation}
The fraction of the dressed molecular wave function that is in the
deeply bound state $\left|m\right>$ is
\begin{equation}
\alpha^2 = \sqrt{\frac{2 E_B}{E_0}} = 2 \sqrt{\frac{E_F}{E_0}}
\frac{1}{k_F a}
\end{equation}

For the resonance used in the experiments by D. Jin on $^{40}$K,
$E_0/k_B \approx 1 \, \rm mK$, which should be compared to a
typical Fermi energy of $E_F/k_B = 1\, \mu\rm K$. This resonance is
thus broad~\cite{szym05}. Still, at $k_F a = 1$ the fraction of the wave
function in the closed channel molecule is $\alpha^2 \approx 6\%$.
This might possibly explain the shorter lifetime of the gas of
molecules $^{40}\rm K_2$ close to resonance~\cite{rega03lifetime}
as compared to the case in $^6\rm Li_2$~\cite{bour04phd}. For the
wide Feshbach resonance in $^6$Li, one has $E_0/k_B \approx 50
\,\rm K$, an unusually broad resonance. The strongly interacting regime where $1/k_F |a| < 1$ is thus completely in the universal regime.
The simple relation $E_B = -\hbar^2 / m
a^2$ holds to better than 3\% already at a magnetic field of 600 G,
230 G away from resonance, while the strongly interacting regime
is entered only above $\approx 750$ G. Indeed, the closed channel
contribution to the dressed molecular state has been measured in
the group of R. Hulet~\cite{part05} to be less than 1\% at
magnetic fields beyond 600 G and less than $10^{-3}$ throughout
the entire strongly interacting regime beyond $k_F |a| \approx 1$.

\subsubsection{Narrow Feshbach resonance}
\label{s:narrowfeshbach}

The Feshbach resonance in $^6$Li at 543 G, in turn, has $E_0/k_B
\approx 1 \,\mu \rm K$ and is thus narrow. In the case of a narrow
resonance, the many-body physics is qualitatively different from
the BEC-BCS crossover picture since molecular states will be
populated even above the resonance.  However, we have just shown
how the molecular states have ``disappeared'' or have become scattering
resonances.  So how does many-body physics modify these results of
two-body physics?

For a narrow resonance and detunings $\delta - \delta_0 < 0$, all
fermion pairs are still tightly bound in the closed channel
molecular state, where they form a condensate. For $0 <\delta -
\delta_0 < 2 E_F$, the molecular condensate coexists with a
BCS-type fermionic superfluid. Here, the molecular state (unstable
in vacuum above threshold, represented by the resonances in the
scattering cross section in Fig.~\ref{f:crosssection}) is stabilized by Pauli
blocking, as the outgoing momentum states are occupied by fermions
in the BCS-state. Equilibrium between fermions and molecules
requires that the chemical potential of the fermions is $\mu =
(\delta-\delta_0)/2$. This means that the molecular state ``shaves
off'' all fermions above $\mu$ (they form molecules), and the Fermi
sea is only filled up to this energy~\cite{falc04}. Only for $\delta -
\delta_0 > 2 E_F$ is the molecular state no longer occupied and we are left
with a BCS-type superfluid. However, since the resonance is
narrow, the interactions for $\delta - \delta_0 > 2 E_F \gg E_0$
will be very small, $k_F \left|a\right| < \sqrt{\frac{E_0}{E_F}}
\ll 1$, rendering the observation of such a state very difficult.

The transition from the narrow to the broad resonance requires a
more complete two-channel description (see~\cite{shee06phase} and references therein), where even the
two-body scattering physics is modified by the Fermi sea.
One example is the transition from the narrow to the broad case right
on resonance. In the two-body picture, no scattering resonances
(and therefore identifiable molecular states) exist. This remains true for a small Fermi sea, with $E_F \ll E_0$, that cannot appreciably affect the open channel states. However, as the Fermi energy becomes comparable to the coupling $E_0$, more and more $k$-states are occupied and Pauli blocked, and the closed channel molecular state can no longer completely dissolve in the continuum states. For $E_F \gg E_0$, the closed channel molecular state is present in its ``undressed'' form, and one expects a condensate of these ``protected'' closed channel molecules to coexist with a Fermi sea.
For an extensive discussion of one and two-channel descriptions, we refer the reader to the contribution of M. Holland to these lecture notes.

\subsection{Open channel resonance and the case of \li}

\begin{figure}
\centering
  \includegraphics[width=3.5in]{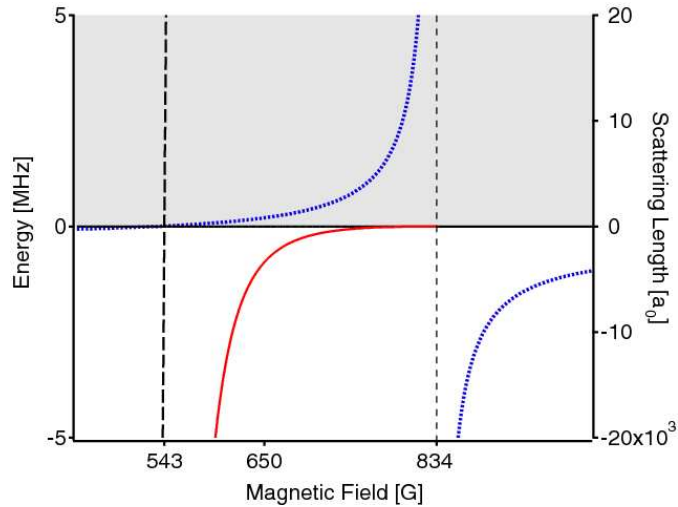}\\
  \caption[Feshbach resonances in $^6$Li]{Feshbach resonances in
 $^6$Li between the two lowest hyperfine states $\left|F,m\right>
 = \left|1/2,1/2\right>$ and $\left|1/2,-1/2\right>$. A wide
 Feshbach resonance occurs at 834.15 G. The resonance position is
 shifted by an unusually large amount of $\sim 300$ G from the
 crossing of the uncoupled molecular state at 543 G (thick dashed line). A second,
 narrow Feshbach resonance occurs right at 543 G, shifted by less
 than 200 mG.  The solid line shows the energy of the bound molecular state, and the dotted line the scattering length.}\label{f:Lifeshbach}
\end{figure}

\li\ stands out compared to all other fermionic atoms studied thus
far by its enormously broad Feshbach resonance. It is this fact
that has allowed direct evaporation of the gas at a fixed magnetic
field directly into a molecular condensate, an experiment almost
as straightforward in principle as Bose-Einstein condensation of
bosonic atoms in a magnetic trap. Lithium is the fermion of choice
at Duke, Rice, Innsbruck, ENS and MIT, and also in a growing
number of new experimental groups.

Fig.~\ref{f:Lifeshbach} shows the $s$-wave scattering length for
collisions between the two lowest hyperfine states of \li,
$\left|F,m\right> = \left|1/2,1/2\right>$ and
$\left|1/2,-1/2\right>$. The prominent feature is the broad
Feshbach resonance centered around $B_0 = 834.15$ G. The resonance
is approximately described by Eq.~\ref{e:Feshbachscattlen} with
$a_{\rm bg} = -1\,405\, a_0$, $\Delta B = 300$
G~\cite{bart04fesh}. These values are very untypical when compared
with scattering lengths and Feshbach resonance widths in other
alkali atoms. Background scattering lengths are typically on the
order of $\pm 100\, a_0$ or less, roughly the range of the van der
Waals-potential. Widths of other observed Feshbach resonances are
two, rather three orders of magnitude smaller than $\Delta B$.
Clearly, the broad Feshbach resonance in \li\ is a special case.

\begin{figure}
\centering
  \includegraphics[width=3.3in]{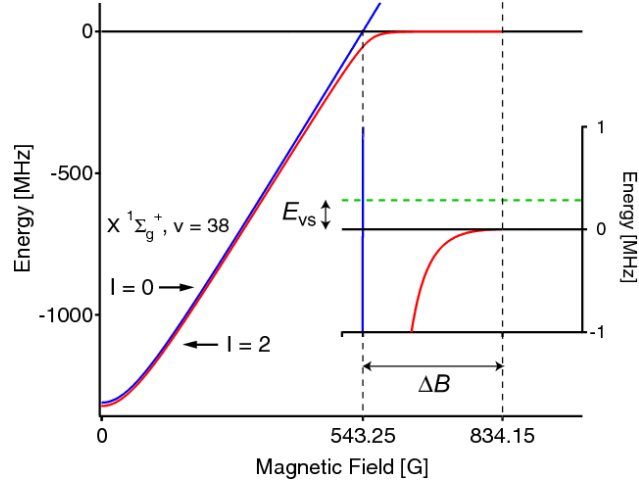}\\
  \caption[Bound state energies for $^6$Li$_2$ in a magnetic field]{Bound state energies for $^6$Li$_2$ in a magnetic field. The most weakly bound state of the singlet potential, $X ^1 \Sigma_g^+, v = 38$, splits into two hyperfine components with total nuclear spin $I=0$ and $I=2$. The state with $I=0$ is almost not coupled to the triplet scattering continuum, causing the narrow resonance at 543.2 G. In turn, the state $I=2$ is very strongly coupled and leads to the broad resonance at 834 G, a shift of $\Delta B \approx 300$ G.
The strong coupling is caused by the large background scattering length $a_{\rm bg}$ in the triplet
potential. The dashed line in the inset shows the associated energy of the ``virtual'', almost bound state $E_{\rm vs}
\approx h \cdot 300$ kHz, very close to threshold ($E=0$).}\label{f:Lifeshbachenergies}
\end{figure}

The unusually large background scattering length of \li\ that
approaches $-2\,100 \, a_0$ at high fields, signals a resonance
phenomenon even away from the wide Feshbach resonance. Indeed, if
the triplet potential of $^6$Li were just about $\hbar^2/m a_{\rm
bg}^2 \approx h\cdot 300\,\rm kHz$ deeper, it would support a new
bound state. This ``missing'' potential depth should be compared
to {\it typical} spacings between the highest lying bound states
of the van der Waals potential, several tens of GHz. The resulting
very large background scattering length modifies the free
continuum states $\left|k\right>$ in a simple but important way:
It increases the probability for the two colliding atoms to be
close to each other. This leads to a much better wave function
overlap between the free continuum states and the closed channel
bound state --- in the language of molecular spectroscopy, one has
a much larger Franck-Condon factor.

In the following, we want to show this quantitatively by directly
calculating the coupling strength $g_k = \left<m|V|k\right>$ as a
function of the background scattering length. The states
$\left|k\right>$ are eigenstates of the Hamiltonian $H_0$, which
includes the scattering potential in the open channel. Outside
that potential, the wave function $\psi_k(\vect{r}) =
\left<\vect{r}|k\right>$ becomes
\begin{equation}
\label{e:psiopenchannel}
    \psi_k(\vect{r}) = \frac{1}{\sqrt{\Omega}} \frac{\sin(kr + \delta_{\rm bg})}{kr}
\end{equation}
For a background scattering length much larger than the range of the potential $a_{\rm bg} \gg r_0$, we can neglect the short-range behavior of $\psi$ at $r\lesssim r_0$. The chosen normalization ensures the closure relation $\sum_k c_k \left<k'|k\right> = c_k'$ to hold.
The closed channel molecular state will be taken to be of the form
\begin{equation}
    \psi_m(\vect{r}) = \frac{1}{\sqrt{2\pi R}}\frac{e^{-r/R}}{r}
\end{equation}
This also neglects the short-range behavior of $\psi_m(\vect{r})$ for $r \lesssim r_0$, permissible if the size of the molecule is much larger than the interatomic potential, $R \gg r_0$.

\begin{figure}
\centering
  \includegraphics[width=4.8in]{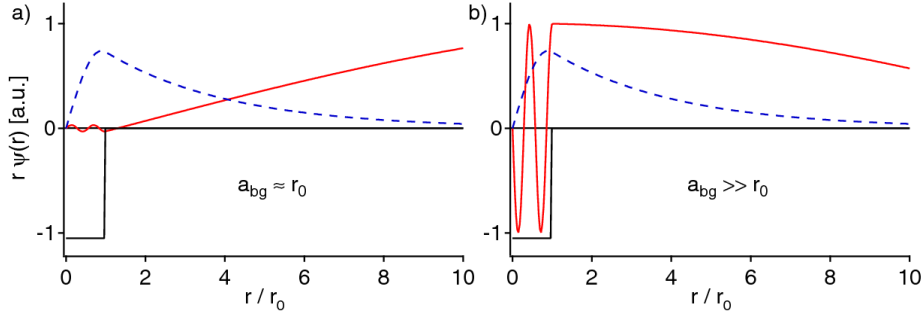}\\
  \caption[Influence of the background scattering length on the Feshbach coupling]{Influence of the background scattering length on the Feshbach coupling. Shown is a spherical well example, with a well of radius $r_0$. Solid line: Open channel radial wave function $r \psi(r)$. Dashed line: Molecular state in the closed channel, with an assumed smaller well depth. a) ``typical'' background scattering length, $a_{\rm bg} \approx r_0$, b) large scattering length,  $a_{\rm bg} \gg r_0$. The probability to overlap with the bound state is resonantly enhanced.}\label{f:background}
\end{figure}

Fig.~\ref{f:background} shows the situation for two different
background scattering lengths. For $k \ll 1/R$ we have $r \psi_k(\vect{r})
\approx \frac{1}{\sqrt{\Omega}} (r-a_{\rm bg})$.  For  $a_{\rm bg}
\lesssim R$, the probability for two colliding particles to be
within a range $R$ of each other is simply $\sim R^3/\Omega$, as
it would be for non-interacting particles.  However, for $a_{\rm
bg} \gtrsim R$, the probability increases to $\sim a_{\rm bg}^2
R/\Omega$. The coupling to the closed channel molecular state
should be enhanced by the same factor, so we expect
$\left|\left<m|V|k\right>\right|^2 \propto a_{\rm bg}^2$. A simple
calculation gives
\begin{eqnarray}
\left<m|V|k\right> &=& V_{\rm hf} \int {\rm d}^3r\, \psi_m^*(\vect{r})\psi_k(\vect{r}) \\
      &=& \frac{V_{\rm hf}}{k}\;\sqrt{\frac{8\pi R}{\Omega}}\;\frac{\sin\delta_{\rm bg} + k R \cos\delta_{\rm bg}}{1 + k^2 R^2}
\end{eqnarray}
where $V_{\rm hf}$ is the amplitude of the hyperfine interaction $V_{\rm hf}^-$ between the open and closed channel.
In the $s$-wave limit of large de Broglie wavelengths, we can approximate $\tan\delta_{\rm bg} \approx -k\, a_{\rm bg}$. We then get
\begin{eqnarray}
    g_k^2 = \Omega \left|\left<m|V|k\right>\right|^2 = 8\pi \left|V_{\rm hf}\right|^2 R^3\;\frac{\left(1- a_{\rm bg}/R\right)^2}{\left(1+k^2 a_{\rm bg}^2\right)\left(1+k^2 R^2\right)^2}
\end{eqnarray}
$g_k^2$ is just proportional to the background scattering cross section $\sigma_{\rm bg} = 4\pi a_{\rm bg}^2/(1 + k^2 a_{\rm bg}^2)$, including its $k$-dependence and the unitarity limit:
\begin{equation}
    g_k^2 = 2 \left|V_{\rm hf}\right|^2 R\, \sigma_{\rm bg} \frac{\left(1 - R/a_{\rm bg}\right)^2}{\left(1 + k^2 R^2\right)^2}
\end{equation}

In section~\ref{s:bcsside} we have seen how $g_k^2$ determines the lifetime of a molecular state placed in the continuum at energy $E$. In our model, this lifetime becomes, by Fermi's Golden Rule,
\begin{eqnarray}
\label{e:Chap5FermiGolden}
    \Gamma(E) &=& \frac{2\pi}{\hbar}\sum_k \left|\left<m|V|k\right>\right|^2 \delta(2\epsilon_k - E)
           = \frac{\pi}{\hbar} \frac{1}{\Omega}\,\rho\left(E/2\right)\; g_{k(E/2)}^2 \\
           &=& \frac{4}{\hbar} \left|V_{\rm hf}\right|^2 \;\left(1- \frac{a_{\rm bg}}{R}\right)^2\frac{\sqrt{E} \sqrt{E_R} \, E_{\rm bg}}{\left(E_{\rm bg}+ E\right)\left(E_R+E\right)^2}
\end{eqnarray}
where $E_{\rm bg} = \hbar^2/m a_{\rm bg}^2$.
The latter is the {\it exact} same expression that one obtains for a bound-free radio-frequency transition that dissociates (or associates) a molecule into (from) two free atoms~\cite{chin05rf} (valid in the threshold regime $a_{\rm bg}, R\gg r_0$). One merely has to replace the hyperfine coupling by the Rabi coupling $\frac{1}{2}\hbar \Omega_R$. The dependence on $\sqrt{E}$ is the usual Wigner threshold law. This reiterates the analogy between Feshbach resonances and photo or RF association.

The $k$-dependent coupling $g_k^2$ rolls off over a characteristic
range $1/R$, as expected. So the natural cut-off energy is $E_R =
\hbar^2/m R^2$, as before. For $k \ll 1/R$, we obtain
\begin{equation}
    g_0^2 = 2\left|V_{\rm hf}\right|^2 R\, \sigma_{\rm bg} \left(1 - R/a_{\rm bg}\right)^2
\end{equation}
with the limiting cases
\begin{eqnarray}
    g_0^2 =& 8\pi\left|V_{\rm hf}\right|^2 R^3\qquad &\mbox{for $a_{\rm bg} \ll R$} \\
    g_0^2 =& 8\pi\left|V_{\rm hf}\right|^2 R\, a_{\rm bg}^2\qquad &\mbox{for $a_{\rm bg} \gg R$}
\end{eqnarray}
The resonant limit $a_{\rm bg} \gg R$ is valid if $k a_{\rm bg} \ll 1$. In the strongly interacting regime where $k a_{\rm bg} \gg 1$, the Feshbach coupling is unitarity limited to $g_0^2 = 8 \pi\left|V_{\rm hf}\right|^2 R /k^2$.

We thus arrive at the conclusion that a large background scattering length resonantly enhances the coupling to the closed channel molecular state. This is simply because of the increased probability for two particles colliding in the open channel to be near each other.

Since $g_0^2$ determines the effective range of the Feshbach scattering amplitude, Eq.~\ref{e:effrange}, we can say that the background scattering length tunes the effective range. In fact, for $a_{\rm bg} \gg R$, we find
\begin{equation}
    r_{\rm eff} = - \frac{\hbar^2}{m a_{\rm bg}^2} \frac{E_R}{\left|V_{\rm hf}\right|^2} R
\end{equation}
The larger the background scattering length, the smaller the effective range. The criterion for a broad Feshbach resonance, $E_0 \gg E_F$, now reads
\begin{equation}
    E_F \ll \frac{2\left|V_{\rm hf}\right|^4}{E_R}\frac{m^2 a_{\rm bg}^4}{\hbar^4}
\end{equation}
Knowing that $^{40}$K does not have an unusually long background scattering length, this relation implies that for reasonably strong couplings $V_{\rm hf}$ (a sizeable fraction of the hyperfine splitting), but only ``standard'' background scattering lengths, molecular sizes $R$ and for typical Fermi energies, Feshbach resonances are broad.

We can now easily calculate the energy shift $\delta_0$ due to this enhanced Feshbach coupling. This directly gives the magnetic field shift $B_0 - B^* = \delta_0/\Delta\mu$ between the Feshbach resonance position at $B_0$, where the dressed bound state energy vanishes, and the magnetic field $B^*$ where the uncoupled molecular state would cross threshold. In $^6$Li, $B_0 = 834\,\rm G$, whereas $B^* = 543\,\rm G$ (in fact, an almost uncoupled, second closed channel molecular state causes a narrow resonance at $B^*$, see Fig.~\ref{f:Lifeshbachenergies}). With the definition in Eq.~\ref{e:E0definition}, and again in the limit $1/k \gg a_{\rm bg} \gg R$,
\begin{equation}
\label{e:shift}
    \delta_0 =  \frac{4}{\pi}\sqrt{E_0 E_R} = \frac{1}{\sqrt{2}\pi^2}\frac{m\, g_0^2 }{\hbar^2 R} = \frac{4\sqrt{2}}{\pi}\left|V_{\rm hf}\right|^2 \frac{m a_{\rm bg}^2}{\hbar^2}
\end{equation}

With the known magnetic field shift in $^6$Li, using $a_{\rm bg} \approx 2100\, a_0$ and $\Delta\mu = 2\mu_B = 2.8 \,\rm MHz/G$, we can now obtain an estimate of the hyperfine coupling strength, $V_{\rm hf} \approx h \cdot 10\,\rm MHz$. This is indeed a ``typical'' coupling strength: The hyperfine
constant for \li\ is $a_{\rm hf} \approx h \cdot 150\,\rm MHz$, setting an upper bound on the matrix element $V_{\rm hf}$ which is less than $a_{\rm hf}$ due to Clebsch-Gordan coefficients. Of course, our model neglects short range physics that may affect $g_0^2$.

Some authors~\cite{marc04res,kemp04} arrive at the same conclusion of Eq.~\ref{e:shift} by introducing a ``virtual state'' at the energy $E_{\rm vs} \sim \hbar^2/m a_{\rm bg}^2$ above threshold and replacing the interaction of the molecular state with the scattering continuum by an effective interaction between molecular and virtual state only. In this language, second-order perturbation theory predicts an energy shift of $\delta_0 \approx \left|V_{\rm hf}\right|^{2}/E_{\rm vs} = \left|V_{\rm hf}\right|^{2} m a_{\rm bg}^2/\hbar^2$, exactly as we have obtained above. We point out that there is no finite energy scattering resonance
associated with this ``imaginary'' state. Rather, it signifies that
if the potential were deeper by $\sim E_{\rm vs}$, it would
support a new bound state just below threshold.

To summarize: the history of interactions in Fermi gases has gone full circle. At first, $^6$Li was thought to be a great candidate for fermionic superfluidity because of its large and negative background scattering length. Then it was realized that scattering lengths can be tuned at will close to a Feshbach resonance -- so essentially any fermionic atom that could be laser cooled became a good candidate (as Feshbach resonances have so far been found for any atom, whenever experimentalists started to search for them). But in the end, it is still $^6$Li that is the most robust choice, and this indeed {\it because} of its large scattering length -- since this is what enhances the Feshbach coupling and makes the resonance abnormally -- fantastically -- large.

\section{Condensation and superfluidity across the BEC-BCS crossover}
\label{c:expobservation}

In this section, we present experimental results on condensation
and superfluid flow across the BEC-BCS crossover.  We will start
with some general remarks on different signatures for
superfluidity, give some background on vortices and describe the
experimental methods to observe condensation and vortex lattices
in gases of fermionic atoms together with the results achieved.

\subsection{Bose-Einstein condensation and superfluidity}
Two phenomena occurring at low temperature have received special
attention:  Bose-Einstein condensation and superfluidity.  An
interesting question is how the two are related. K. Huang has
pointed out~\cite{huan95} that Bose-Einstein condensation is not
necessary for superfluidity, but also not sufficient.  This is
illustrated by the examples in table~\ref{t:condensation}.

\begin{table}
  \centering
  \begin{tabular}{p{.25\linewidth}p{.2\linewidth}p{.1\linewidth}p{.1\linewidth}p{.3\linewidth}}
 &   System & BEC & Superfluid & \\[5pt]
 &   3D & $\surd$ & $\surd$  &\\
 &   Ideal gas & $\surd$ & $\O$ & \\
 &   2D, $T\ne 0$ & $\O$ & $\surd$  &\\
 &   2D, $T= 0$ & $\surd$ & $\surd$ & \\
 &   1D, $T= 0$ & $\O$ & $\surd$&
  \end{tabular}
  \caption[Condensation versus superfluidity]{Condensation versus
superfluidity.  Condensation and superfluidity are two different,
but related phenomena.  The ideal gas is Bose condensed, but not
superfluid. In lower dimensions, fluctuations can destroy the
condensate, but still allow for
superfluidity.}\label{t:condensation}
\end{table}

The ideal Bose gas can undergo Bose-Einstein condensation, but it does
not show superfluid behavior since its critical velocity is zero.
Superfluidity requires interactions.  The opposite case
(superfluidity without BEC) occurs in lower dimensions.  In 1D at
$T=0$~\cite{lieb631dbose,lieb631dboseb} and in 2D at finite temperature, superfluidity occurs~\cite{bish782dSF}, but
the condensate is destroyed by phase fluctuations~\cite{hohe67longrange,bere72BKT,kost73BKT}.  In 2D at zero
temperature, there is both a condensate and superfluidity~\cite{schi712dbec}.

In 3D, condensation and superfluidity occur together. An
interesting case that has been widely discussed are bosons in a
random potential. For weak disorder and weak interactions, there
is an unusual regime where the superfluid fraction is smaller than
the condensate fraction~\cite{huan92,gior94disorder}. It appears
that some part of the condensate is pinned by the disorder and
does not contribute to the superfluid flow.  However, the
extrapolation to strong disorder and the conclusion that the
system can be Bose condensed without being
superfluid~\cite{huan92,huan95} is not
correct~\cite{astr02disorder,yuka07disorder}.  The condensate and
superfluid fraction disappear together when the disorder is
sufficiently strong~\cite{astr02disorder,yuka07disorder}.

A very comprehensive discussion on the relation between
superfluidity and BEC is presented in the Appendix of Ref.~\cite{bloc07review}.  When condensation is generalized to
quasi-condensation in lower dimensions the two phenomena
become equivalent.  It is shown that superfluidity plus finite
compressibility are sufficient conditions for either condensation
or quasi-condensation.  The reverse is also true, i.e.
condensation or quasi-condensation are necessary for
superfluidity.  Here, superfluidity is defined by the rigidity of
the system against changes in the phase of the boundary condition
and condensation by the presence of a macroscopic eigenvalue of
the density matrix, which, for translationally invariant systems,
implies off-diagonal long range order.  Quasi-condensates are
local condensates without long range order.

This discussion on bosons applies directly to a gas of bosonic molecules
created at a Feshbach resonance. For a Fermi gas, the examples for
lower dimensions apply as well~\cite{hohe67longrange} (for a
discussion of superconductivity in 2D films and arrays,
see~\cite{tink04sc}). The example of the non-interacting Bose
gas, however, does not carry over: A non-interacting Fermi gas does not form
a pair condensate. The effect of disorder on a BCS superfluid is
complex. The pair condensate survives in the presence of local
impurities (weak disorder), with the order parameter and $T_C$
unchanged~\cite{ande59dirty}, while condensate fraction and
superfluid density are reduced~\cite{orso07disorder}.

These examples lead to the conclusion that experimentalists need
to study both condensation and superfluidity!

\subsection{Signatures for superfluidity in quantum gases}
\label{s:superfluidsignatures}
What constitutes an observation of superfluidity? Even
theoretically, superfluidity is defined in several different ways.  The most frequent definition employs the concept of rigidity against
phase-twisting~\cite{legg73helium,fish73super}.  In some definitions, even a non-interacting
BEC qualifies as a superfluid~\cite{lieb02bose}.

From the experimentalists' point of view, superfluidity consists of
a host of phenomena, including phase coherence, transport without
dissipation, an excitation spectrum which results in a non-zero
value of Landau's critical velocity (usually a phonon spectrum),
the Meissner effect, the existence of quantized vortices, and a reduction of
the moment of inertia.  After the discovery of Bose-Einstein condensation in 1995, it still took until 1999 before researchers agreed that superfluidity was established, through the observation of
vortices~\cite{matt99vort,madi00} and a critical velocity in a stirred
condensate~\cite{rama99,onof00sup}. The general consensus was that
the experimental verification of superfluidity required the
observation of some aspect of superfluid flow that would
not be possible in a classical system.  Therefore, neither the
hydrodynamic expansion of a condensate was regarded as evidence
(since collisionally dense classical clouds would behave in the
same way), nor the observation of phonon-like excitations, nor the
interference of condensates, which established phase coherence only
for a stationary cloud. The observation of a critical velocity
~\cite{rama99,onof00sup} provided  evidence for superfluid flow, although the
contrast between the behavior in the superfluid and normal regime
did not even come close to the drop in resistivity or viscosity
that was observed when superconductors or superfluids were
discovered. Long-lived flow in the form of vortices has been
regarded as a smoking gun for superfluids.  However, vortices can
be long lived even in classical liquids~\cite{sche99}. What sets the
superfluid apart is the quantization of vortices and the fact
that the ground state with angular momentum is necessarily a state
with vortices.  The emergence of vortex arrays and vortex
lattices~\cite{madi00,abos01latt,hodb02vort,enge02} after driving
surface excitations~\cite{onof00} dramatically demonstrated both
properties. Although there is no rigorous derivation showing that
ordered lattices of uniformly charged vortices prove superfluidity, we are not aware of any system
or observation that could provide a counter example.

The reduction of the moment of inertia is another distinguishing
feature of superfluid flow. It can be observed through the
so-called scissors mode, a collective excitation created by a
sudden rotation~\cite{guer99scissors,
mara00scis,ming01}, or by observing the expansion of a rotating
superfluid~\cite{edwa02,hech02,modu03}. Both methods have been
regarded as a way to {\it directly} observe superfluidity.
However, studies with normal Fermi gases have impressively
demonstrated that both features, originally regarded as a unique
signature of superfluids, occur already for normal gases deep in
the hydrodynamic regime where dissipation is extremely
small~\cite{clan07irrot,wrig07finite}.

It appears that superfluid and low-viscosity collisional
hydrodynamics can only be distinguished if there is sufficient
time for the small but finite viscosity in the normal phase to create
vorticity, a velocity field with $\nabla \times \vect{v} \ne 0$
whereas the superfluid will always continue to be irrotational
(unless quantized vortices are nucleated). For instance, if the
flow field has equilibrated with a slowly rotating container, then
collective excitations will reveal the difference between a
superfluid and a normal fluid~\cite{cozz03}.

The last example shows how the physics of  strong interactions can ``obscure'' the seemingly dramatic
transition to a superfluid state. The normal state of a Fermi gas around the Feshbach
resonance is already almost ``super'' due to its very low viscosity.
Many experiments uncovered the unique properties of this strongly interacting gas,
and eventually its transition into the superfluid state. The
observation of anisotropic expansion~\cite{ohar02science} was
initially believed to provide evidence for superfluidity. However,
such an expansion was also observed in a normal strongly
interacting Fermi gas~\cite{ohar02science,bour03,rega03fesh} and
was predicted to occur even at $T=0$ since Pauli blocking is no
longer effective during expansion~\cite{gupt04coll}.

Can the damping of modes distinguish between superfluid and normal
flow?  In a simple picture, damping in collisional hydrodynamics
increases with lower temperature, because Pauli blocking lowers
the collision rate and increases the mean free path. In contrast,
damping of superfluid hydrodynamics decreases with lower
temperature, because the normal density, which provides friction,
decreases. The observation of such a decrease of the damping rate
of collective excitations was regarded as evidence for
superfluidity~\cite{kina04sfluid}.  Later, however, it was found
that a similar positive slope of damping vs. temperature occurs
for a normal strongly interacting Fermi gas~\cite{kina05damping}.

The observation of a ``paring gap'' in RF spectroscopy was
regarded as strong evidence for
superfluidity~\cite{chin04gap,kinn04}, mainly based on the
theoretical interpretation of the experimental data. However,
experiments with population imbalanced Fermi gases showed that RF
spectra of normal and superfluid clouds are identical, and that RF
spectroscopy cannot distinguish between the two phases, at least
not at the current level of resolution~\cite{schu07pair}. The
reason is the presence of strong pair correlations in the normal
phase and possibly also strong interactions in the final state
used for spectroscopy, which were not included in the models used
to interpret the data.

When collective excitations were studied as a function of the
scattering length, intriguing sudden peaks in the damping rate
were observed~\cite{bart04coll,kina04hydr}.  The conjecture is
that this may reflect a resonance of the collective mode with the
pairing energy $\Delta$, and damping would occur due to pair
breaking. This phenomenon remains to be systematically studied.
The observation of pair correlations across the BEC-BCS crossover
was consistent with predictions of a theory for the superfluid
state~\cite{part05}, but it seems that similar pair correlations
also exist in the normal state~\cite{grei04corr}.  Finally, a kink
was observed when the specific heat~\cite{kina05heat} or the
entropy~\cite{luo07entropy} were determined as a function of the
temperature or the energy of the cloud. Due to the signal-to-noise
ratio, these kinks could be distilled only by separately fitting
the low- and high-temperature regions.

The discussion above has summarized many aspects of superfluid
systems, some of which are shared with strongly interacting normal
gases. In the following sections we will focus on the two
phenomena that do not occur in a normal gas: condensation and the
formation of quantized vortices in rotating superfluids.

\subsection{Pair condensation below the Feshbach resonance}
\begin{figure}[ht]
\begin{center}
\includegraphics[width=3in]{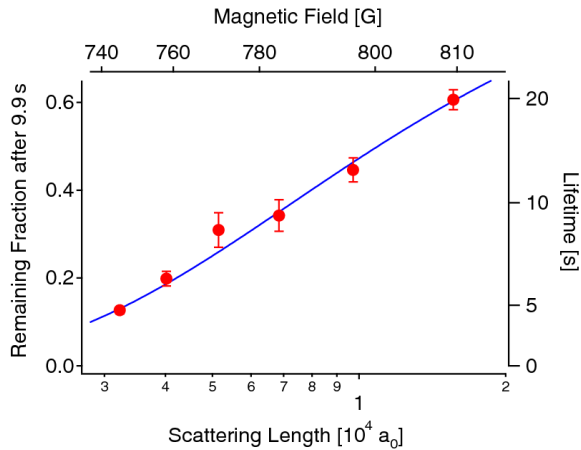}
\caption[Lifetime of molecules in partially condensed
clouds]{Lifetime of molecules in partially condensed clouds. The
cloud with initially about $1 \times 10^6$ molecules was held for
9.9 s (initial density about $5 \times 10^{12}\,\rm cm^{-3}$,
slightly varying with the interaction strength). Shown is the
remaining fraction as a function of scattering length. The
lifetime $1/\Gamma$ is calculated under the simplifying assumption
of a pure exponential decay $e^{-\Gamma t}$. The line is a fit
with a power law for $\Gamma = c a^{-p}$, giving $p = -0.9$. The
clouds were partially condensed (up to 80\% condensate fraction at
the largest scattering length), and all measurements were done in
the strongly interacting regime where $a>1/k_F$, so the expression
for the relaxation rate differs from the prediction for weakly
interacting, thermal molecules ($\Gamma = c\,
a^{-2.55}$)~\cite{petr04dimers}.}\label{f:lifetime}
\end{center}
\end{figure}
The successful creation of ultracold molecules out of ultracold
atoms via Feshbach resonances in gases of
fermions~\cite{rega03mol,cubi03,joch03lith,stre03} and
bosons~\cite{donl02mol,herb03cs_mol,xu03na_mol,durr04mol} brought
the goal of Bose-Einstein condensation of molecules into close
reach.  Indeed, molecular samples in cesium close
to~\cite{herb03cs_mol} and in sodium clearly
within~\cite{xu03na_mol} the regime of quantum degeneracy were
generated.  However, their lifetime was too short to observe an
equilibrium Bose-Einstein condensate. Molecules formed of fermions
turned out to have a much longer lifetime due to greater stability
against inelastic decay (see chapter~\ref{c:exptechniques} and
section~\ref{s:history}). Within a few months, this favorable property
allowed the successful Bose-Einstein condensation of molecules, or
more precisely of strongly interacting fermion pairs~\cite{grei03molbec,joch03bec,zwie03molBEC,bart04,bour04coll,part05}.
\begin{figure}[ht]
\begin{center}
\includegraphics[width=2.45in]{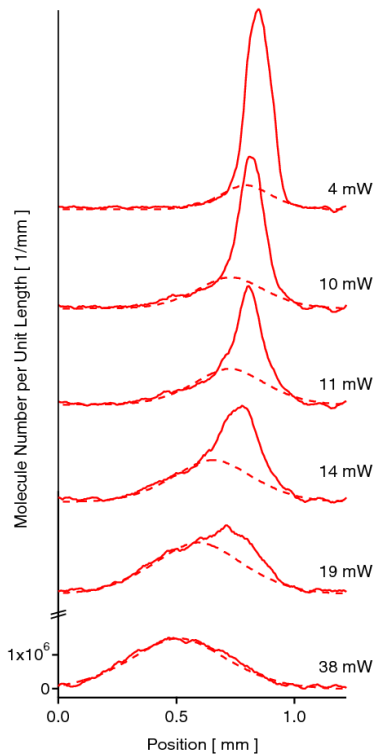}
\caption[Bimodal density distribution emerging in a cloud of
molecules]{Bimodal density distribution emerging in a cloud of
molecules. Shown are radially integrated profiles of absorption
images such as those in Fig.~\ref{f:moleculeBEC}, as a function of
final laser power. The dashed lines are fits to the thermal
clouds.}\label{f:molbecwaterfall}
\end{center}
\end{figure}
\begin{figure}[t]
\begin{center}
\includegraphics[width=4.5in]{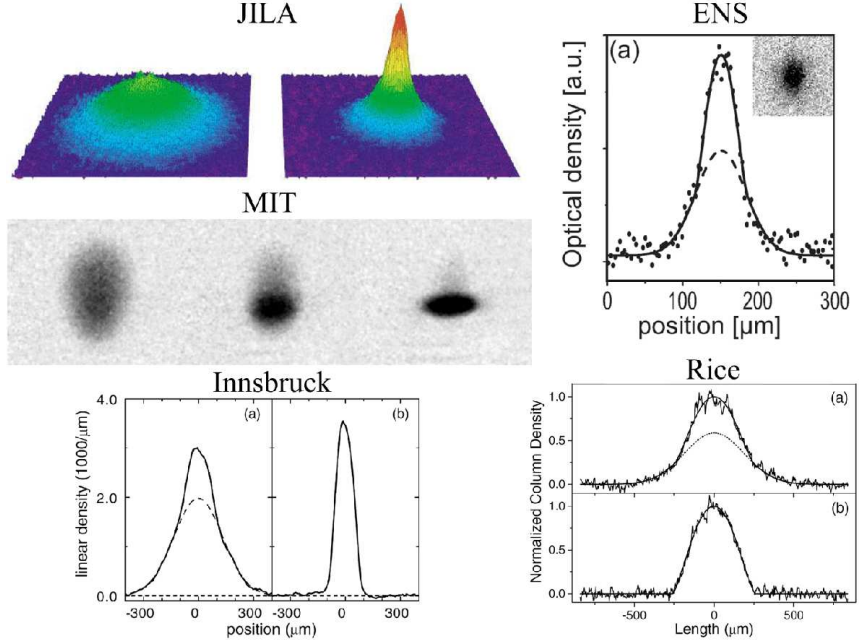}
\caption[Observation of Bose-Einstein condensation of
molecules]{Observation of Bose-Einstein condensation of molecules. The gallery shows bimodal density distributions observed after expansion and molecule dissociation at JILA~\cite{grei03molbec}, after expansion, dissociation and zero-field imaging at MIT~\cite{zwie03molBEC} and at the ENS~\cite{bour04coll}, and in-situ profiles from Innsbruck~\cite{bart04} and Rice~\cite{part05}.}\label{f:moleculeBEC}
\end{center}
\end{figure}
\begin{figure}[th]
\begin{center}
\includegraphics[width=3.5in]{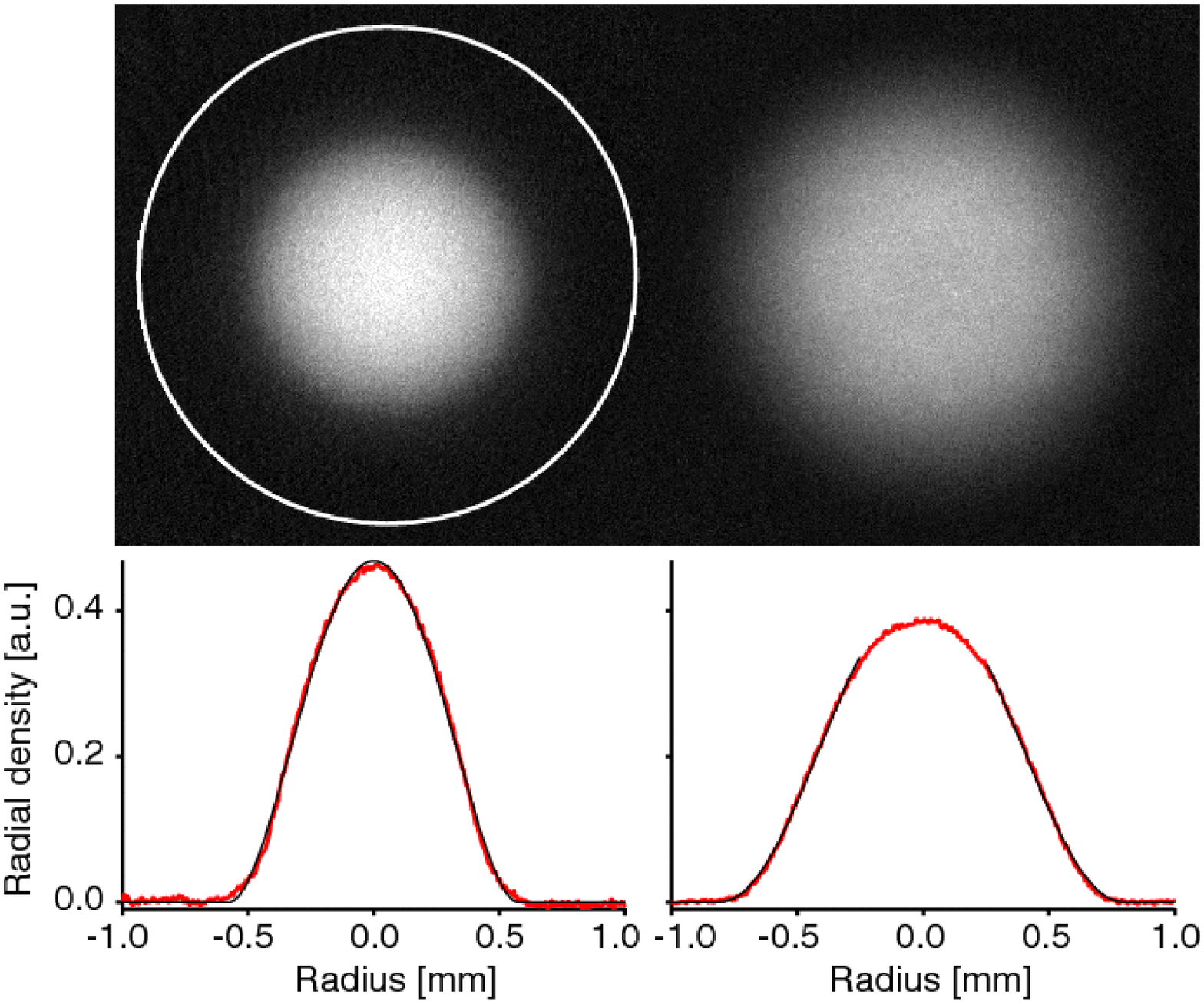}
\caption[Comparison between a molecular BEC and a degenerate Fermi
sea]{Comparison between a molecular BEC (a) and a degenerate Fermi
sea (b). The condensate containing $N_m = 6\times 10^6$ molecules
is in the strongly interacting regime at a magnetic field of 780 G
($1/k_F a = 0.6$). Its expanded size is almost as large (factor $\sim 0.7$)
as an expanded non-interacting Fermi gas containing $N_m$ atoms, indicated
by the white circle. Image and profile b) show an essentially
spin-polarized Fermi sea (minority component of $<2\%$ not shown)
containing $N = 8 \times 10^6$ atoms at the same field. The images
were taken after 12 ms expansion with the probe light aligned with
the long axis of the cigar-shaped clouds.
}\label{f:BECFermiprofiles}
\end{center}
\end{figure}

In the case of \li, the long lifetime of molecules~\cite{cubi03}
(see Fig.~\ref{f:lifetime}) enables us to evaporate the Fermi
mixture at a fixed magnetic field, just like cooling a cloud of
bosonic atoms towards BEC. As the mixture is cooled by ramping
down the trapping laser power, molecules form as the temperature
becomes comparable to the binding energy. Accordingly, the atomic
signal observed in zero-field imaging vanishes: We can see this in
Fig.~\ref{f:feshbachlocation} for fields below resonance, where
essentially no atomic signal is measured. Below a certain
temperature, one observes the striking onset of a bimodal density
distribution, the hallmark of Bose-Einstein condensation (see
Figs.~\ref{f:moleculeBEC} and~\ref{f:molbecwaterfall}). The
emergence of the bimodality was actually accentuated by an
anharmonic trapping potential where a shallow minimum of the
potential was offset with respect to the deeper potential which
held the thermal cloud. Fig.~\ref{f:moleculeBEC} shows the gallery
of molecular Bose-Einstein condensates observed at
JILA~\cite{grei03molbec}, MIT~\cite{zwie03molBEC}, Innsbruck~\cite{bart04},
at the ENS~\cite{bour04coll} and at Rice~\cite{part05}.

In contrast to weakly interacting Bose gases, the condensate peak
is not much narrower than the thermal cloud, indicating a large
mean-field energy of the BEC, comparable to $k_B$ times the
condensation temperature. As we move closer to the Feshbach
resonance, the size of the condensate grows to be almost that of a
degenerate Fermi gas (see Figure~\ref{f:BECFermiprofiles}).
The average distance $n_M^{-1/3}$ between molecules becomes comparable to the molecular size in free space, given
approximately by the scattering length. Thus we have entered
the strongly interacting regime of the BEC-BCS crossover where two-body pairing is modified by Pauli pressure.

\subsection{Pair condensation above the Feshbach resonance}
\label{s:paircondensation}
When the Feshbach resonance is approached, the bimodality of the
cloud becomes almost undetectable.  There is no strong spatial signature of the phase
transition, even at a much better signal-to-noise ratio
than in the initial observations (see Fig.~\ref{f:dilemma} in section~\ref{s:resonantgases}).  We have observed
weak signatures in these spatial profiles (see section~\ref{s:directunitarity} below). A
second difficulty with fermion pair condensates on the BCS-side is
the instability of the pairs during expansion. When the gas
becomes more dilute the pair binding energy can decrease below
($k_B$ times) the local temperature, causing pairs to break
during time of flight. See section~\ref{s:vortexexpansion} below for a study of
this effect using vortices.

To extend the study of pair condensation from below to above the
Feshbach resonance, a new detection method was needed.  Such a
method was introduced by the JILA group~\cite{rega04} and later
adapted to \li\ by our group~\cite{zwie04rescond}.  The rapid ramp
technique is discussed in detail in section~\ref{s:rapidramp} and also in the
contribution of D. Jin to these proceedings.  The concept of this
technique is to prevent the fragile fermion pairs from
dissociating by sweeping the magnetic field towards the BEC-side
of the resonance, thereby transforming them into stable molecules
(see Fig.~\ref{f:rampingfields}). This is done in the moment the
trap is switched off for expansion. If each fermion pair is
transferred into a tightly bound molecule, the momentum
information of the original pair is preserved. Time-of-flight
analysis of the resulting molecules should thus allow one to infer the
momentum distribution of pairs in the gas above resonance.  The momentum distribution might be broadened by the residual mean-field interaction of molecules after the ramp.  However, these interactions are greatly
reduced by sweeping sufficiently far away from the Feshbach resonance into the weakly interacting regime where $k_F a \ll 1$.

\begin{figure}[ht]
\begin{center}
\includegraphics[width=5in]{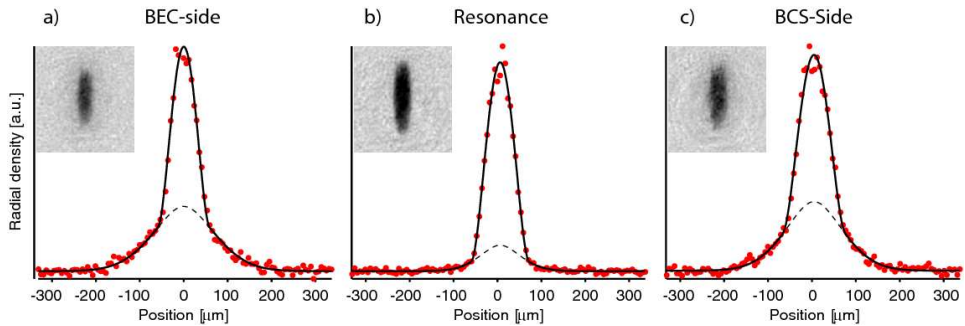}
\caption[Fermion pair condensates]{Fermion pair condensates. Axial
density of the atomic cloud after switching off the optical trap,
a rapid ramp to zero field (in $< 100 \,\mu s$), further expansion
(for $10$ ms), and dissociation of the resulting molecules by
ramping back across resonance. The initial field $B_0$, the number
of fermion pairs $N$, the condensate fraction and the interaction
parameter $1/k_F a$ where a) 745 G, 700 000, 47\%, 1.2; b) 835 G,
$1.4 \times 10^6$, 81\%, 0.0 (resonance); c) 912 G, $1 \times
10^6$, 49\%, -0.5.}\label{f:fermipairBEC}
\end{center}
\end{figure}

This technique enabled us to demonstrate fermion pair
condensation in the entire BEC-BCS crossover. Sample images and
profiles of the resulting molecular clouds are shown in
Fig.~\ref{f:fermipairBEC}. The drastically reduced interaction
results in a clear separation of the condensate from the ``thermal''
or uncondensed part of the cloud~\footnote{At zero field, the
scattering length between molecules should be on the order of the
singlet scattering length of lithium atoms, which is about 40
$a_0$. The exact value is not known. In fact, the residual
mean-field interaction at zero field is so low that the condensate
practically does not expand if the rapid ramp is performed
immediately after switching off the trap. For this reason, it is
sometimes beneficial to let the cloud expand by some amount {\it
before} the rapid ramp is performed. This converts some of the
interaction energy in the cloud into kinetic energy, which allows
one to ``choose'' the final expanded size of the molecular
condensate.}.

\begin{figure}[ht]
\begin{center}
\includegraphics[width=3.0in]{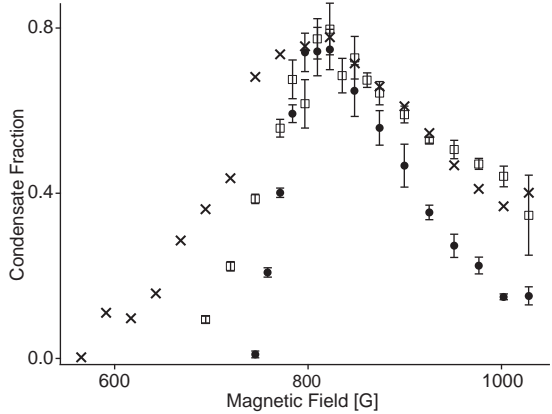}
\caption[Condensate fraction as a function of magnetic
field]{Condensate fraction in the BEC-BCS crossover as a function
of the magnetic field before the rapid ramp. The symbols
correspond to different hold times, 2 ms (crosses), 100 ms
(squares) and 10 s (circles). From~\cite{zwie04rescond}.}\label{f:condfracvsB}
\end{center}
\end{figure}

The condensate fraction was determined by fitting a bimodal
distribution to the profiles like those in
Fig.~\ref{f:fermipairBEC}, a parabola for the central dense part
and a gaussian for the thermal background (see
chapter~\ref{c:analysis}). Remarkably large condensate fractions
were found throughout the entire BEC-BCS crossover, with a peak of
80\% at $B \approx 820\,\rm G$, close to the resonance, but still
on its BEC-side (see Fig.~\ref{f:condfracvsB}).

The high condensate fraction is a hint that the pairs in the
strongly interacting regime on the BCS-side of the resonance are
still smaller than the interparticle spacing, not larger, as one would expect for
conventional Cooper pairs.  An intuitive assumption is that
during the magnetic field sweep an atom preferably forms a
molecule with its nearest neighbor \footnote{This should happen as
long as the relative momentum is not larger than the inverse
distance, i.e. the neighbor populates the same phase space cell.
This is the case in the regime of quantum degeneracy, and it is
experimentally confirmed by the almost 100\% conversion from
atoms into molecules (see~\ref{s:makingmolecules}).}.  In the case of localized pairs, molecules
are then formed from the original ``Cooper partners''.  In the case of
delocalized Cooper pairs, molecules might rather form out of
uncorrelated atoms, resulting in a thermal cloud after the ramp.

In accord with this argument, BEC-BCS crossover theory predicts
that the pair size $\xi$ will be smaller than the interparticle
spacing $n^{-1/3}$ up to $k_F a \approx -1$ (see
section~\ref{s:evolution}).  So far no experiment on Fermi
gases has shown condensation or superfluidity in a regime where
$k_F |a|$ ($a<0$) is significantly less than 1 and hence where
pairing is truly long-range.  Observing superfluidity for $k_F a
< -1$ would require exponentially lower temperatures of $T/T_F < 0.28\, e^{-\pi/2k_F|a|} \ll 0.06$ and,
furthermore, the sweep technique may no longer allow the observation of pair condensation.

\begin{figure}[ht]
\begin{center}
\includegraphics[width=3in]{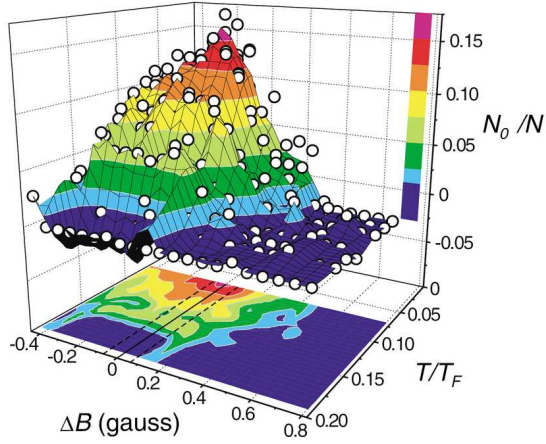}
\caption[Condensate fraction in the BEC-BCS crossover in $^{40}$K]{Condensate fraction in the BEC-BCS crossover obtained by the JILA group using $^{40}$K, as a function of degeneracy $T/T_F$ and magnetic field (interaction strength) around the Feshbach resonance. From~\cite{rega04}.}\label{f:JILAphasediagram}
\end{center}
\end{figure}

A simple theoretical model (see below) agrees with the high
condensate fraction, but the latter is in stark contrast to the
maximum fraction of about 14\% found in experiments with
$^{40}$K~\cite{rega04} (see Fig.~\ref{f:JILAphasediagram}). The
reason for this discrepancy might be related to the shorter
lifetime of the Fermi mixture in $^{40}$K close to resonance, on
the order of 100 ms~\cite{rega03lifetime}. In addition, technical issues particular to $^{40}$K may play a role including strong losses during the probing procedure.

In our experiments, the condensates were found to be very
long-lived. For a hold time of 10 s, the condensate fraction on
resonance was observed to be still close to its initial value.  In
fact, these lifetimes can compare very favorably to those found
for atomic Bose-Einstein condensates. On the BEC-side, the
condensate decayed more rapidly due to the increasing rate of
vibrational relaxation of the molecules away from resonance. The
decay of the condensate fraction on the BCS-side
can be caused by heating and atom loss due to inelastic collisions or a larger sensitivity to
fluctuations of the trapping fields.

\begin{figure}[ht]
\begin{center}
\includegraphics[width=3.5in]{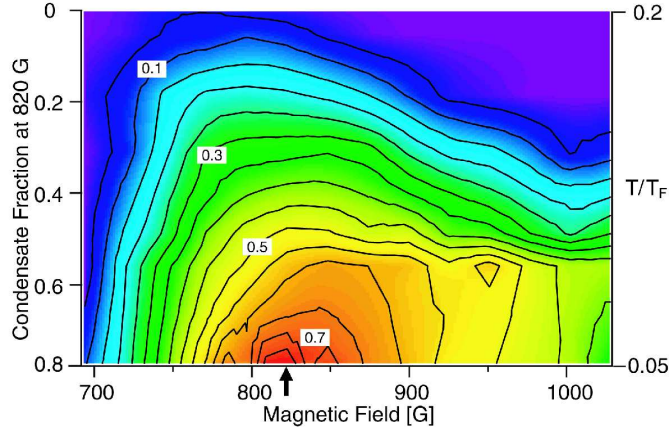}
\caption[Condensate fraction as a function
of magnetic field and temperature in $^6$Li.]{Condensate fraction as a function of magnetic field and
temperature in the MIT experiments on $^6$Li. Condensates are obtained in the entire BEC-BCS
crossover. The highest condensate fraction and highest onset
temperature are obtained on the BEC-side close to resonance. As a
model-independent measure of temperature, the condensate fraction
at 822 G (see arrow) is used as the vertical axis. The Feshbach
resonance lies close to this point, at 834 G. From~\cite{zwie04rescond}.}\label{f:phasediagram}
\end{center}
\end{figure}

\subsubsection{Comparison with theory}
\label{s:condensatefraction}

Figs.~\ref{f:JILAphasediagram} and ~\ref{f:phasediagram} show ``phase diagrams'' for the
condensate fraction as a function of temperature and interaction
strength. Several theoretical
studies~\cite{ho04proj,avde05,chen06phase} have confirmed the
general behavior of the ``critical temperature'' of the observed
condensation phenomenon in $^{40}$K and in $^6$Li.

In section~\ref{c:BECBCStheory}, we discussed the BEC-BCS
crossover theory which predicts a condensate fraction (see Eq.~\ref{e:condfrac}) of
\begin{equation}
    n_0 \,=\, \frac{N_0}{\Omega} \,=\, \frac{1}{\Omega} \sum_k u_k^2 v_k^2 \,= \frac{m^{3/2}}{8 \pi \hbar^3} \Delta^{3/2} \sqrt{\frac{\mu}{\Delta} + \sqrt{1+\frac{\mu^2}{\Delta^2}}}
\end{equation}
\begin{figure}[ht]
\begin{center}
\includegraphics[width=3in]{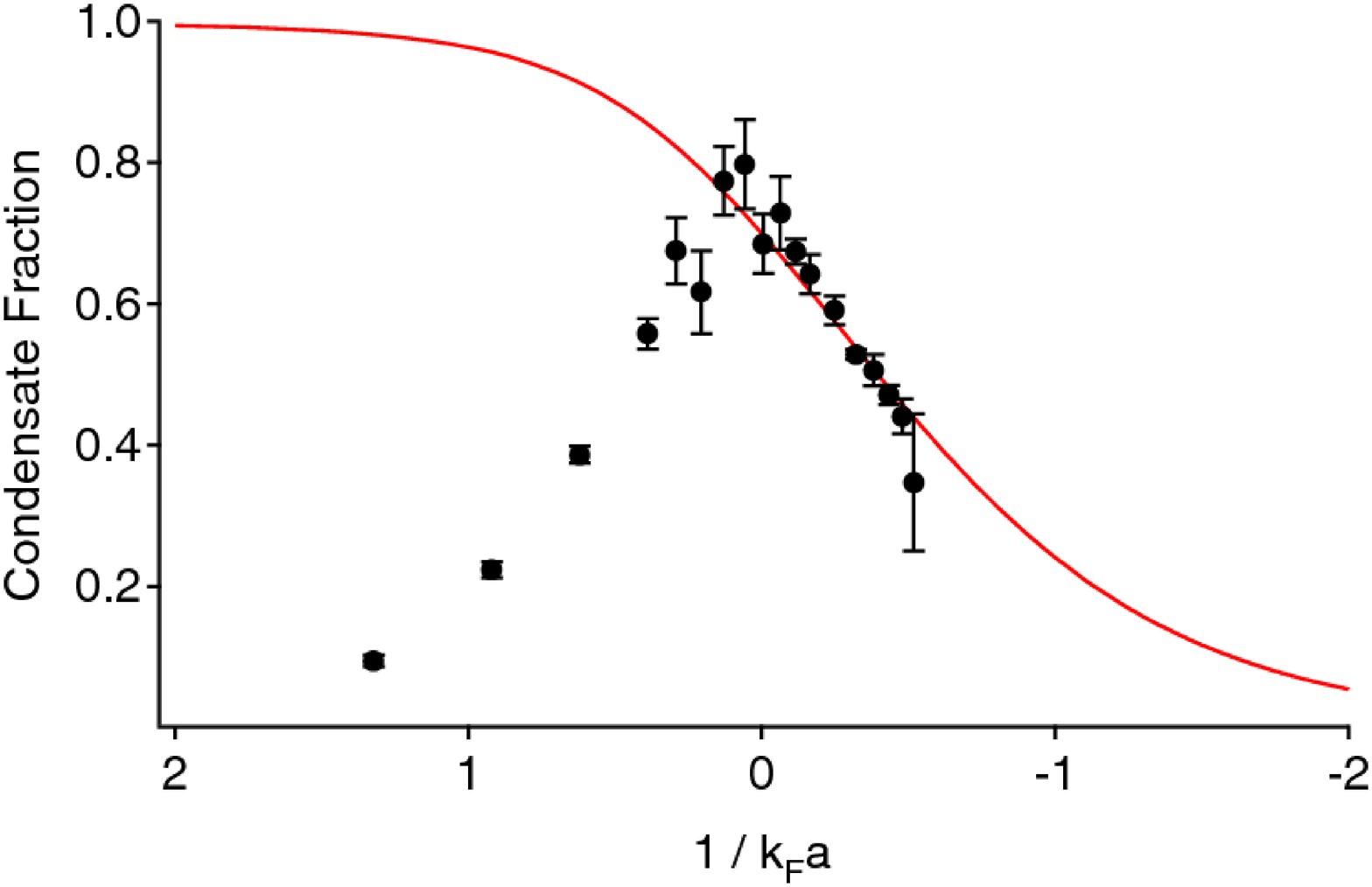}
\caption[Condensate fraction as a function of the interaction
strength]{Condensate fraction as a function of the interaction
strength in the BEC-BCS crossover. The circles show the 100 ms
data of Fig.~\ref{f:condfracvsB}. The interaction strength is
calculated using the known scattering length as a function of
magnetic field and the experimental value $1/k_F = 2\,000\, a_0$.
The curve shows the variational BCS prediction for the condensate
fraction. On the BEC-side, heating due to vibrational relaxation
leads to fast decay on the condensate. Figure adapted
from~\cite{sala05} using
Eq.~\ref{e:condfrac}.}\label{f:condfractheo}
\end{center}
\end{figure}
Fig.~\ref{f:condfractheo} compares the variational BCS prediction
to our results.  The very close agreement must be considered
fortuitous since the simple crossover theory is only
qualitatively correct near resonance.  Furthermore,  it is not
clear how accurately the observed molecular condensate fraction
after the ramp reflects  the pair condensate fraction before the
ramp.

The strongest confirmation that the bimodal density distributions
observed after the ramp are an indicator of a phase transition
comes from the direct detection of condensation in population
imbalanced clouds (see section~\ref{s:directimbalance}). In the
next section we summarize experimental evidence that the
condensate fraction cannot change strongly during the sweep time.
On the other hand, some evidence has been
reported~\cite{zwie04form}, that the conversion efficiency into
molecules is higher for condensates. This effect increases the
condensate fraction during the sweep, but the effect is
small~\cite{zwie04form}.
As discussed in section~\ref{s:rapidramp}, the system's dynamics
during the sweep poses a difficult challenge to theory, due to the
presence of several timescales for coherent and incoherent
evolution.

\subsubsection{Formation Dynamics}
\label{s:formation}
\begin{figure}[htb]
\begin{center}
\includegraphics[width=3.5in]{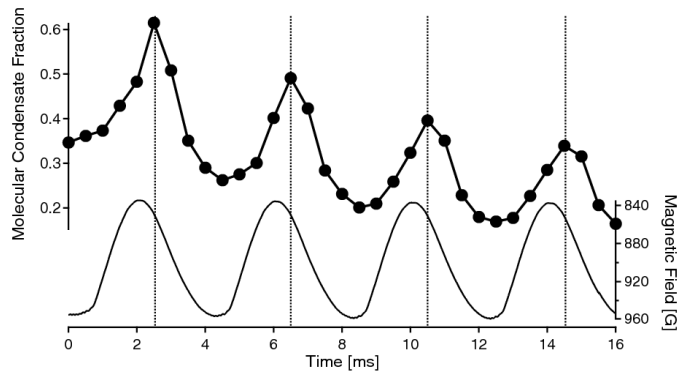}
\caption[Relaxation time of fermionic pair condensates]{Relaxation time of fermionic pair condensates. Shown is the delayed response of the condensate fraction to a 250 Hz magnetic field modulation on the BCS side of the Feshbach resonance. From~\cite{zwie04form}.}\label{f:delay}
\end{center}
\end{figure}
The underlying assumption for the rapid sweep technique is that
the momentum distribution of pairs is not changing during the
sweep time.  Evidence for this was obtained both in the JILA and
MIT experiments, where it was shown that the condensate fraction
did not change when the sweep rate was varied or when the density
was changed.  However, since in both experiments the sweep time
through the strongly interacting regime was comparable to the
inverse Fermi energy, dynamics during the sweep could not be fully
ruled out.  We addressed this issue by directly measuring the
relaxation time of the strongly interacting Fermi system.  This
was done by modulating the magnetic field (and therefore the
scattering length) and observing the growth or decay of the
condensate.  In the experiment, we used a periodic modulation of
the magnetic field and measured the phase shift of the induced
condensate modulation (see Fig.~\ref{f:delay}).

The observed relaxation time of 500 $\mu$s was much longer than
the time to sweep through the strongly interacting regime (10
$\mu$s).  Therefore, it should not be possible for the condensate
fraction to change noticeably during the sweep. Still, a possible loophole is
some exotic mechanism for such a change that is not captured by the simple relaxation models assumed in the
experimental tests.

In conclusion, the rapid ramp to the BEC-side has proven to be a
very valuable tool for the detection of condensation in the
BEC-BCS crossover. Moreover, the ramp provides us with a way to
preserve the topology of the pair wave function on the BCS-side.
This allows the observation of vortex lattices in the entire
BEC-BCS crossover, as will be discussed in
section~\ref{s:vortexobservation}. In the following, we will show
that condensates can be detected without any ramp, by direct
absorption imaging.

\subsection{Direct observation of condensation in the density profiles}
The hallmark of Bose-Einstein condensation in atomic Bose gases
was the sudden appearance of a dense central core in the midst of
a large thermal cloud~\cite{ande95,davi95bec}. This direct
signature in the density distribution derives from a clear
separation of energy scales in weakly interacting gases. The
condensate's repulsive mean-field $\mu \propto n a$ is much
smaller than the critical temperature (times $k_B$) at which
condensation occurs, $T_C \propto n^{2/3}$: The gas parameter $n
a^3$ is much less than 1 (about $4 \times 10^{-6}$ for $^{23}$Na
condensates). In a harmonic trap, the different energy scales
directly translate into the different sizes of a thermal cloud,
$R_{\rm th} \propto \sqrt{T}$, and of a condensate $R_C \propto
\sqrt{\mu}$. This is the situation we encounter with weakly
interacting molecular clouds in Fermi mixtures on the ``BEC''-side
of the Feshbach resonance. However, as the interactions between
molecules are increased by moving closer to the Feshbach
resonance, the size of the molecular condensate grows and the
bimodal feature close to $T_C$ becomes almost invisible. In
strongly interacting Fermi gases, the separation of energy scales
is no longer given. On resonance, the size of the condensate is
governed by $\mu \approx 0.5 E_F$, while $k_B T_C \approx 0.15 E_F
\ll E_F$, so that the normal cloud's size is dominated not by
temperature, but by the Fermi energy. The question arises whether
the condensate still leaves a trace in the cloud as the gas
undergoes the phase transition.

\subsubsection{Anomalous density profiles at unitarity}\footnote{The results of this section have not been published elsewhere.}
\label{s:directunitarity}
\begin{figure}[t]
\begin{center}
\includegraphics[width=4in]{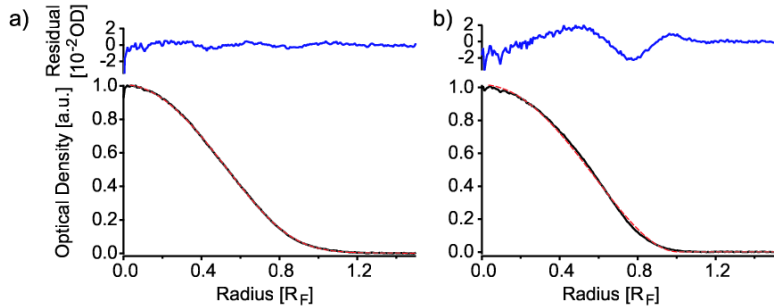}
\caption[Density profiles of an equal Fermi mixture on
resonance]{Density profiles of an equal Fermi mixture on
resonance. The temperature in a) was $T/T_F \approx 0.15$, whereas
in b) it was $T/T_F \approx 0.09$. Temperatures were determined
from the thermal molecular cloud after the rapid ramp, and might
not be quantitatively accurate. Both gas clouds contained a
condensate after the rapid ramp to the BEC-side. The condensate
fraction was: a) 7\%, b) 60\%.}\label{f:directsamples}
\end{center}
\end{figure}
We have indeed found a faint signature of condensation in density
profiles of the unitary gas on resonance after expansion. To a
very good approximation, the trapping potential was cylindrically
symmetric (see section~\ref{s:opticaltrap}). This allowed us to obtain
low-noise profiles via azimuthal averaging.

Sample profiles are shown in Fig.~\ref{f:directsamples}. To
observe a {\it deviation} from the shape of a non-interacting
Fermi cloud, an unconstrained finite-temperature fit is performed
on the profiles. The relevant information is now contained in the
residuals of such a fitting procedure.

The fit residuals deviate at most by 2\% from the non-interacting
Fermi shape. This explains why this effect has not been observed
in earlier experiments. Despite of the rather small deviation,
the non-interacting fit is affected by the ``kinks''. This draws
into question whether the ``effective temperature'' typically
obtained from such fits to the whole profile, is a well-defined
quantity. For a well controlled determination of an effective
temperature, only the profile's wings should be fit, where the gas
is normal.
\begin{figure}[thb]
\begin{center}
\includegraphics[width=2.5in]{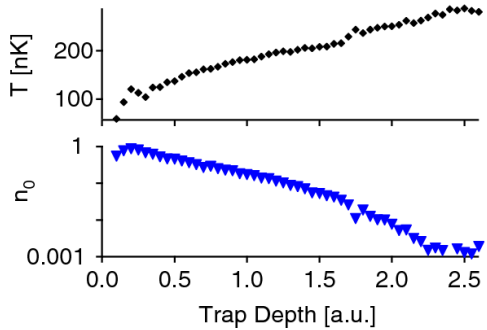}
\caption[Condensate fraction and temperature as a function of
evaporation depth.] {Condensate fraction $n_0$ and temperature as
a function of evaporation depth on resonance. The condensate
starts to form around a trap depth of $U\sim 2.2$ corresponding to
4.2 $\mu$K. The temperature was determined from the thermal wings
of expanding molecular clouds after the rapid ramp. The Fermi
temperature decreased slowly from 1.5 $\mu$K for $U = 2.6$ to 1.4
$\mu$K at $U = 0.4$, and dropped quickly due to atom spilling
below $U = 0.2$. All measurements were done after recompression
into a deeper trap with $U = 2.0$.}\label{f:chisquare}
\end{center}
\end{figure}

\begin{figure}[bht]
\begin{center}
\includegraphics[width=5.3in]{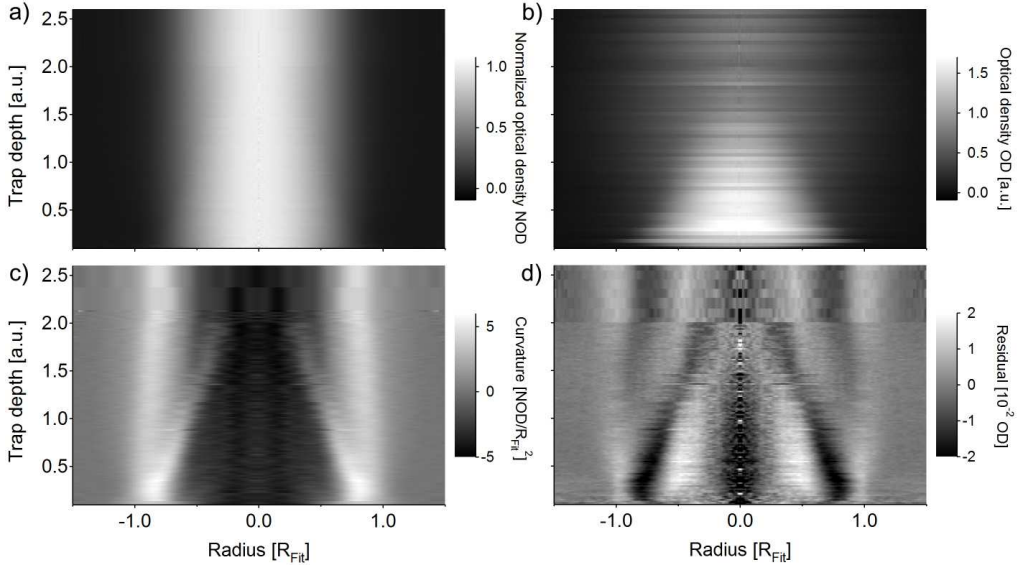}
\caption[Density profiles, their curvature and their fit-residuals
on resonance]{Density profiles, their curvature and their
fit-residuals on resonance. a) Density profiles on resonance as a
function of trap depth. There is no sign of a phase transition at
this resolution. b) After the rapid ramp to the BEC-side and
expansion, a condensate is clearly visible below a certain trap
depth. c) The curvature of the density profiles on resonance in a)
carries a signature of the condensate. No field ramp is required.
d) The fit residuals for a finite temperature Thomas-Fermi fit.
Also here, the condensate's imprint in the density profile is
clearly visible. To obtain the curvature, the noisy central region
of $\pm 0.1 R_F$ in each profile was replaced by a fit.}\label{f:densityplots}
\end{center}
\end{figure}

A convenient way to graph fit residuals as a function of
temperature is by means of a ``density'' plot of gray shades, with
white and black corresponding to positive or negative deviations
of the measured profile from the fit. This is shown in
Fig.~\ref{f:densityplots}. Also included in this figure is the
information on the density profiles and their curvature. While the
profiles themselves do not appear to change with temperature (trap
depth) on the scale of the plot, we observe an intriguing
structure appearing in the residuals at an evaporation depth of
about $U = 2\,\mu$K. The curvature of the density profiles shows a
similar qualitative behavior.

To indicate that the observed feature indeed stems from the
superfluid, we also include a density plot of the profiles
obtained with the rapid ramp method from section~\ref{s:paircondensation} above. This
allows to clearly separate the condensate and thermal cloud in
expansion. The condensate fraction is included in
Fig.~\ref{f:chisquare}, and shows that the condensate appears
around a trap depth of 4.2 $\mu$K. We observe that a small
condensate does not leave a strong signature in the gas cloud,
unlike the case of weakly interacting Bose gases. Only when the
condensate has grown to an appreciable size (about 20\% in our
data) does it significantly deform the density profiles.

At the lowest temperatures and high condensate fractions, the
quality of Fermi fits improves again, indicating that now a large
fraction of the gas is in the superfluid state. The size of the
cloud is then $R_{\rm TF} = 0.83\,R_F$, which gives $\xi(0) \approx
0.47$, in accord with other experiments and theory (see section
~\ref{s:unitarity} above).

To conclude, the density profiles in resonantly interacting Fermi
gases are modified in the presence of a superfluid core. Such
features have been predicted by several
authors~\cite{holl01,pera04temp,ho04uni,staj05dens}, but had
previously been too small to be observable. In the next section, we will demonstrate how an imbalance in the spin up versus spin down population in the gas greatly enhances the visibility of the condensate and leads to a striking signature of condensation.

\subsubsection{Direct observation of the onset of condensation in Fermi mixtures with unequal spin populations}
\label{s:directimbalance}

\begin{figure}[th]
\begin{center}
\includegraphics[width=4.5in]{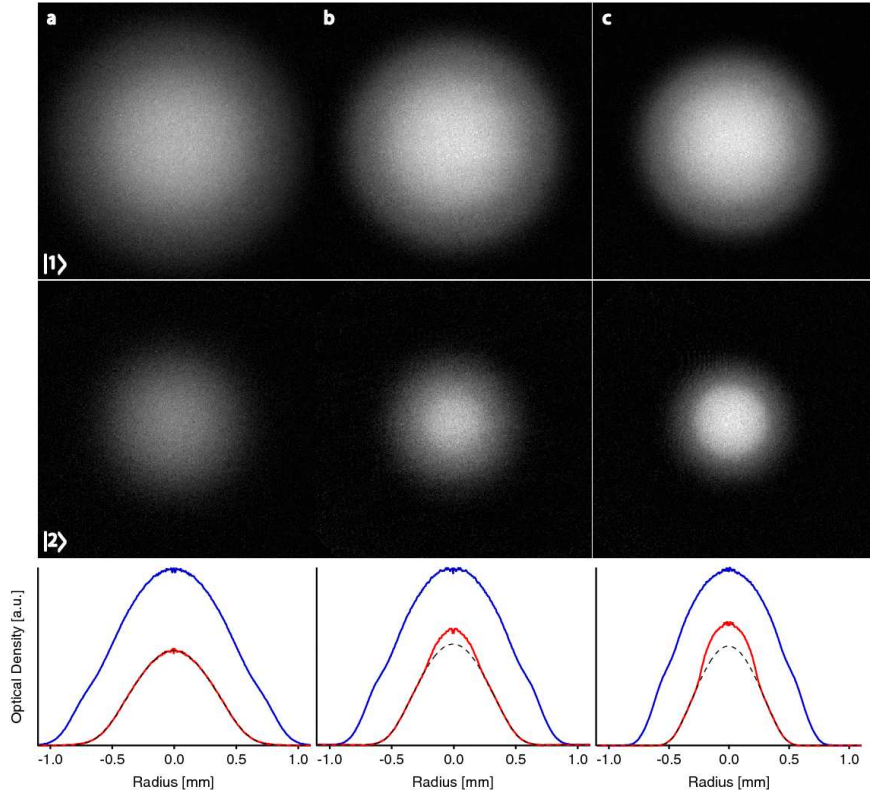}
\caption[Direct observation of condensation in imbalanced clouds on resonance]{Direct observation of condensation in imbalanced clouds on resonance. The upper row shows majority clouds, the lower row minority clouds, for an imbalance of $\delta = 60\%$. The dashed line is a fit to the wings of the minority cloud to a Thomas-Fermi profile, clearly missing the central feature. Temperature was varied by lowering the trapping power. To within 20\%, temperatures can be obtained from the ballistically expanding wings of the majority cloud. We have $T/T_F = 0.14$ (a), 0.09 (b) and 0.06 (c). Here, $k_B T_F$ is the Fermi temperature of an equal mixture containing the same total atom number. The figure shows data from ~\cite{zwie06direct}.}\label{f:directresonance}
\end{center}
\end{figure}

\begin{figure}[th]
\begin{center}
\includegraphics[width=4.5in]{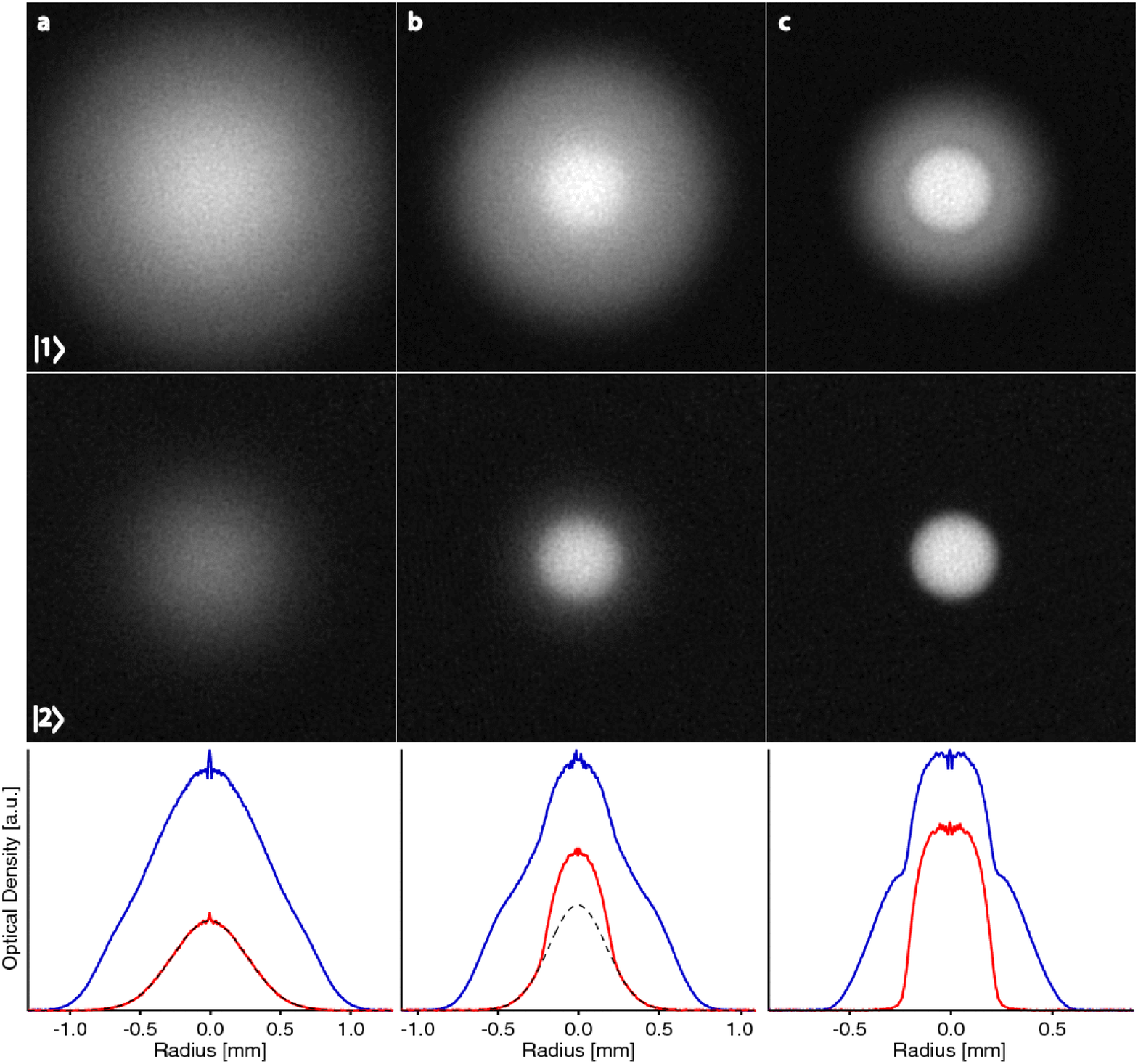}
\caption[Condensation in a strongly interacting, imbalanced Fermi mixture on the BEC-side]{Condensation in a strongly interacting, imbalanced Fermi mixture on the BEC-side, at $B = 780$ G or $1/k_F a \approx 0.5$. In this regime one may start to describe the imbalanced gas as a strongly interacting atom-molecule mixture. Unlike on resonance, essentially all minority atoms are part of condensed fermion pairs. Temperatures were $T/T_F = 0.2$ (a), $T/T_F = 0.12$ (b) and $T/T_F \le 0.05$ (c). The figure shows data from ~\cite{zwie06direct}.}\label{f:directBEC}
\end{center}
\end{figure}
We have seen that a balanced mixture of spin up and spin down fermions at unitarity does not show a strong signature of condensation. The reason is that on resonance, due to the symmetry in atom numbers, only one energy scale is available, the Fermi energy.
In stark contrast, breaking the symmetry in atom numbers and working with
Fermi mixtures with unequal spin populations produces a direct and
striking signature of the superfluid phase transition in the
spatial density profiles both in expansion~\cite{zwie06direct} and
in trap~\cite{shin06phase}. A similar situation has been
encountered in Bose-Einstein condensation, where breaking the
symmetry of a spherical trap resulted in dramatic anisotropic
expansion of the condensate, now a hallmark of the BEC phase
transition.

Part of the reason for the direct signature is the new hierarchy
of energy scales. The normal majority and minority cloud sizes are
governed by the respective Fermi energies (rather, chemical
potentials) $\mu_{\uparrow}$ and $\mu_{\downarrow}$, while the
cloud size of fermion {\it pairs} is governed by the average
chemical potential, $(\mu_\uparrow+\mu_\downarrow)/2$. The deeper
reason for the sudden change of the spatial profile at the phase
transition is that fermionic superfluids (around resonance and in
the BCS-regime) do not tolerate unpaired fermions, at least at
zero temperature. The superfluid gap presents an energy barrier
for these ``singles'' to enter the superfluid. This leads to a
superfluid central region of equal spin populations surrounded by
the polarized normal phase. The two regimes are separated by a
first order phase transition, at which the density imbalance
jumps. The presence of such a sudden change in the density
distribution allowed the first direct observation of the phase
transition, without the need for sweeps to the BEC side of the
Feshbach resonance~\cite{zwie06direct}.

We present here, side by side, the density profiles of an
imbalanced Fermi mixture at unitarity
(Fig.~\ref{f:directresonance}) and on the BEC-side of the Feshbach
resonance (Fig.~\ref{f:directBEC}). In the BEC-regime,
the sharp phase boundary between a balanced
superfluid and the normal region no longer occurs. It is replaced
by an expulsion of the normal cloud of unpaired atoms from the
molecular condensate, which can be understood from a mean-field
picture. The imbalanced gas has turned into a Bose-Fermi mixture
of molecules and unpaired fermions. We know from
section~\ref{s:excitations} that molecules repel unpaired fermions
with a ``Bose-Fermi'' scattering length $a_{\rm BF} =
1.18\,a$~\cite{skor57abf}. As a result, unpaired fermions
experience a ``Mexican-hat'' potential $V(\vect{r}) +
\frac{4\pi\hbar^2}{a_{\rm BF}}{m_{\rm BF}} n_M(\vect{r})$ in the
presence of molecules at density $n_M$.

The physics of imbalanced fermionic superfluids is discussed in
section~\ref{s:imbalance}.  The detailed analysis of the spatial
density profiles for population imbalanced Fermi clouds is still
an area of current research, and will not be covered in these
lecture notes.

The sudden change in the density profile of imbalanced mixtures -- as a function of temperature at fixed imbalance, or as a function of imbalance at fixed temperature --
occurred simultaneously with the appearance of a condensate peak
after a rapid ramp to the BEC side~\cite{zwie06direct,shin06phase}.   This
provides strong confirmation that condensates observed via the
rapid ramp technique truly mark a phase transition. In the
following section, we will present the demonstration that these
condensates are indeed superfluid --- the observation of vortex
lattices.

\subsection{Observation of vortex lattices}
\label{s:vortexobservation}
The most dramatic demonstration of superfluidity in Bose-Einstein
condensates is the observation of vortex lattices in rotating
systems (see section~\ref{s:vortexaspects}).  It was a natural
goal to repeat such experiments for ultracold Fermi gases and to
demonstrate superfluidity due to fermionic pairing.  In this
section, based on the Ph.D. thesis of one of the authors, we
include details on the experimental techniques and results that
were not included in the original publication~\cite{zwie05vort}.

Before we discuss the experimental realization, we will summarize some
basic properties of vortices.  In particular, we will show how a
macroscopic wave function can accommodate vortices, and emphasize
that it is not the existence of vortices, but rather the
quantization of circulation, that is unique to superfluids and
superconductors.

\subsubsection{Some basic aspects of vortices}
\label{s:vortexaspects} Superfluids are described by a macroscopic
wave function $\psi(\vect{r})$ which is zero in the normal state
and non-zero in the superfluid state, so it qualifies as the {\rm
order parameter} of the superfluid phase transition (see
section~\ref{s:orderparameter}). As a wave function, it is a complex quantity
with a magnitude and phase $\phi$
\begin{equation}
    \psi(\vect{r}) = \left|\psi(\vect{r})\right| e^{i \phi(\vect{r})}
\end{equation}

The velocity of the superfluid is the gradient of its phase,
\begin{equation}
    \vect{v} = \frac{\hbar}{m^*} \nabla \phi
    \label{e:velocity}
\end{equation}
where $m^*$ is the mass of the bosonic entities forming the
superfluid. In the case of fermionic superfluids, we have $m^* =
2m$, where $m$ is the fermion mass. Integrating
Eq.~\ref{e:velocity} around a closed loop inside the superfluid,
we immediately arrive at the Onsager-Feynman quantization
condition~\cite{onsa49quant,feyn53,feyn54},
\begin{equation}
    \oint \vect{v} \cdot d\vect{l} = n \frac{h}{m^*}
    \label{e:quantization}
\end{equation}
with integer $n$. If the superfluid wave function has no nodal
lines and the loop fully lies in a simply connected region of
space, we must have $n=0$. However, Eq.~\ref{e:quantization} can
be fulfilled with $n\ne 0$ if the wave function contains a {\it
vortex}, that is, a flow field that depends on the vortex core
distance $r$ like $v \sim 1/r$. At the location of the vortex, the
wave function  vanishes, it has a nodal line. This is the way a
superfluid can carry angular momentum. In case of cylindrical
symmetry (with the vortex core at the center), the angular
momentum per boson or fermion pair is quantized in units of
$\hbar$. Note that vortices are a property of the superfluid in
the ground state at given angular momentum. This is in marked
contrast to {\it classical} vortices, which exist only in
metastable or non-equilibrium situations. Vortex patterns will
ultimately decay into rigid body rotation whenever the
viscosity is non-zero.

Vortices of equal charge repel each other.  This immediately
follows from kinetic energy considerations.  Two vortices on top
of each other double the velocities and quadruple the energy.  Two
vortices far separated have only twice the energy of a single
vortex.  As a result, vortices with charge $|n|> 1$ will quickly
decay into singly charged vortices~\cite{shin04doublevortex}.  If
many vortices are created, they minimize the total kinetic energy
of the cloud by arranging themselves into a regular hexagonal
lattice, called Abrikosov lattice~\cite{abri57}.

How can quantized vortices nucleate?   Vortices cannot suddenly
appear within the condensate, as the angular momentum contained
within a closed loop inside the condensate cannot abruptly jump.
Rather, the nodal lines have to enter the condensate from a {\it
surface}, where the condensate's wave function is zero. This
surface can also be the surface of a stirrer, if it fully expels
the condensate.

One pathway to generate vortices is to excite {\it surface} modes.
They are generated by moving a boundary condition (stirrer or
container walls) faster than the local critical velocity $v_c$ for
such excitations~\cite{angl01vort}. Which surface excitations are
efficiently created depends on the shape of the
stirrer~\cite{dalf01crit,rama01nuc}, or, in the case of a rotating
container, the roughness of the container walls. Accordingly, the
necessary critical angular velocity $\Omega_c$ to nucleate
vortices will  depend on the stirrer's shape. Note that $\Omega_c$
can be much higher than the {\it thermodynamic} critical angular
velocity $\Omega_{th}$. The latter is the angular velocity at
which, in the rotating frame, the ground state of the condensate
contains a single vortex. But simply rotating the condensate at
$\Omega_{th}$ will not lead to this ground state, because a vortex
has to form on the surface where its energy is higher than in the
center presenting an energy barrier.  Driving a surface excitation
provides the necessary coupling mechanism to ``pump'' angular
momentum into the condensate, which can subsequently relax into a
state containing vortices.

\subsubsection{Realization of vortices in superconductors and superfluids}
The Lorentz force on charged particles due to a magnetic field is
equivalent to the Coriolis force on neutral particles due to
rotation.  Therefore, a magnetic field does to a superconductor
what rotation does to a neutral superfluid.  Weak magnetic fields
are completely expelled by a superconductor (the Meissner effect),
analogous to a slow rotation with angular velocity less than
$\Omega_{th}$ for which the neutral superfluid does not acquire
angular momentum.  For higher magnetic fields, quantized magnetic
flux lines, vortices,  penetrate the superconductor.

Quantized circulation in superfluid $^4$He was observed by Vinen
in 1958~\cite{vine58} by measuring the frequency of a thin wire's
circular motion placed at the center of the rotating superfluid.
Quantized magnetic flux was measured by Deaver and
Fairbanks~\cite{deav61} and Doll and N\"abauer in
1961~\cite{doll61} by moving a thin superconducting cylinder of
tin toward and away from a conducting coil and measuring the
electromotive force induced in the coil as a function of applied
field. Entire Abrikosov lattices of magnetic flux lines were
observed by using ferromagnetic particles that were trapped at the
lines' end-points (Tr\"auble and Essmann~\cite{trau67},
Sarma~\cite{sarm67}, independently in 1967). The direct
observation of vortex lattices in superfluid $^4$He was achieved
in 1979 by Yarmchuk, Gordon and Packard~\cite{yarm79} by imaging
ions trapped in the core of the vortex lines. In gaseous
Bose-Einstein condensates, single vortices were created by a phase
imprinting technique~\cite{matt99vort}, and vortex lattices were
created by exposing the condensate to a rotating
potential~\cite{madi00,abos01latt,hodb02vort,enge02}. Using the
method of the vibrating wire, the presence of quantized
circulation was confirmed for the fermionic superfluid $^3$He in
1990 by Davis, Close, Zieve and Packard~\cite{davi91he3}. The MIT
work described here represents the first direct imaging of
vortices in a fermionic superfluid. It is worth adding that
glitches in the frequency of pulsars, fast rotating neutron stars,
have been attributed to the spontaneous decay of vortex lines
leaving the neutron pair superfluid~\cite{alpa95neutron,donn91}.

\subsubsection{Experimental concept}
\begin{figure}[t]
\begin{center}
\includegraphics[width=5.3in]{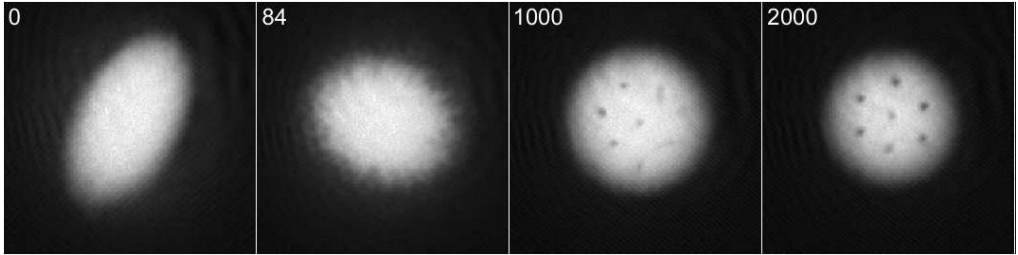}
\caption[Fate of a quadrupole oscillation in a rotating atomic
Bose-Einstein condensate]{Fate of a quadrupole oscillation in a
rotating atomic Bose-Einstein condensate. The images show a sodium
condensate in the magnetic trap after stirring slightly above the
quadrupole resonance (at 52 Hz, trapping frequencies $\nu_\perp =
73$ Hz and $\nu_z = 18$ Hz) and equilibrating for a certain time
$t$ (time given in ms). First, the condensate rotates in the form
of a perfect quadrupolar collective excitation. After about 100
ms, density depletions looking like vortex cores start to appear
at the edges of the condensate. Between 500 ms and 1 s, some of
these penetrate into the condensate as vortex lines, which arrange
themselves into an ordered lattice after about 1-2 s.}
\label{f:sodiumquadrupole}
\end{center}
\end{figure}
For weakly interacting Bose gases in magnetic traps,
the techniques for setting the cloud in rotation are well
established~\cite{onof00,madi00,abos01latt,halj01vort,hodb02vort}.
In~\cite{halj01vort,hodb02vort}, the initially axially symmetric
magnetic potential is deformed into an ellipse in the radial
plane, which is then set in rotation. In~\cite{madi00,onof00}, an
asymmetric optical dipole potential is superimposed with a
cylindrically symmetric magnetic trap, again resulting in an
elliptically deformed potential. In these cases, the role of the
``rotating container walls'' needed to nucleate vortices is played
by the smooth elliptical deformation. This potential can excite
only a specific surface excitation of the condensate, a rotating
quadrupole mode. This collective excitation carries angular
momentum $m= \pm 2$ (the axial component of angular momentum) and
can only be excited around a resonant angular
frequency~\footnote{While a collective excitation carrying angular
momentum $m$ has an energy $\hbar \omega_\perp \sqrt{m}$, it is
$m$-fold symmetric and is thus excited at a frequency $\Omega =
\omega_\perp / \sqrt{m}$, see~\cite{stri96coll}.} $\Omega_Q =
\omega_\perp / \sqrt{2}$, where $\omega_\perp$ is the radial
trapping frequency~\footnote{In the presence of an elliptic
deformation, one needs to replace $\omega_\perp$ by
$\sqrt{(\omega_x^2 + \omega_y^2)/2}$, where $\omega_{x,y}$ are the
trapping frequencies in the direction of the long and short axis
of the ellipse.}. The excited quadrupole mode will eventually
decay (via a dynamical instability) into vortices~\cite{madi01}
(see Fig.~\ref{f:sodiumquadrupole}).

In the MIT experiments~\cite{onof00,abos01latt,abos02_form}, two
(or more) small, focused laser beams were symmetrically rotated
around the cloud. Vortices could be created efficiently over a
large range of stirring frequencies~\cite{abos01latt,rama01nuc}.
The small beams presented a sharp obstacle to the superfluid, most
likely creating vortices {\it locally} at their
surface~\cite{rama01nuc}, corresponding to high angular momentum
excitations with low critical angular velocities. This is the
strategy followed in our experiment on rotating Fermi gases.

A major technical challenge was to create a trapping potential
which had a high degree of cylindrical symmetry.  In Bose gases,
this was provided by a magnetic trap, a TOP trap
in~\cite{madi00,enge02} that can be accurately adjusted for a very
symmetric potential, or a Ioffe-Pritchard trap
in~\cite{abos01latt} that has a high degree of cylindrical
symmetry built-in.  For fermions, one had to engineer an optical
trap with a very round laser beam for optical trapping and
carefully align it parallel to the symmetry axis of the magnetic
saddle point potential, formed by the magnetic Feshbach fields and
gravity. In this ``sweet spot'', gravity is balanced by magnetic
field gradients, and the only remaining force acting on the atoms
is from the laser beam (see section~\ref{s:opticaltrap}).

Experimentally, the trapping potential was designed using a sodium BEC as a test
object.  This had the advantage that the
experimental parameters for the creation of vortex
lattices were well-known. After this had been accomplished, the next challenge was
to identify the window in parameter space that would allow to observe vortices in a Fermi gas. It was not evident whether such a window existed at all, where heating during the
stirring would not destroy the superfluid, and where the decay
rate of vortices would be slow enough to allow their
crystallization and observation.

\subsubsection{Experimental setup}
\label{s:vortexexpsetup}

\begin{figure}[t]
\begin{center}
\includegraphics[width=5.3in]{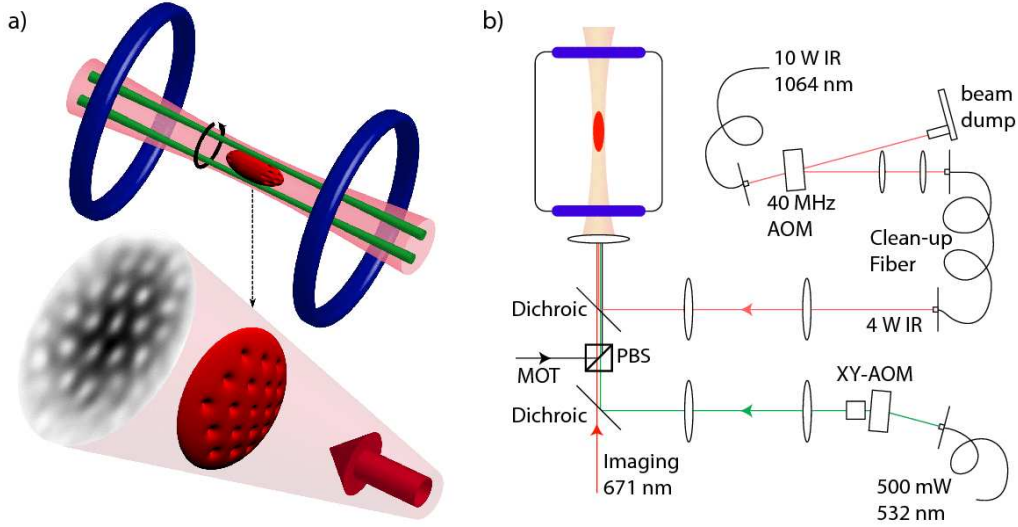}
\caption[Experimental setup for the observation of vortices in a
Fermi gas]{Experimental setup for the observation of vortices in a
Fermi gas. a) Sketch of the geometry. The atomic cloud (in red) is
trapped in a weakly focused optical dipole trap (pink). The coils
(blue) provide the high magnetic offset field to access the
Feshbach resonance as well as the axial confinement (additional
curvature coils not shown). Two blue-detuned laser beams (green)
rotate symmetrically around the cloud. An absorption image of the
expanded cloud shows the vortices. b) Optical setup for the vortex
experiment. The laser beam forming the dipole trap is spatially
filtered using a high-power optical fiber.  Care is taken not to
deteriorate the quality of the gaussian beam's roundness when
passing through several lenses after the fiber exit. The stirring
beam (green) passes through two crossed AOMs that deflect it in
the transverse (XY) plane. These beams are overlapped with the
imaging light by dichroic mirrors. The light for the
magneto-optical trap (MOT) is overlapped on a polarizing beam
splitter cube (PBS).}\label{f:experimentalsetup}
\end{center}
\end{figure}

\begin{figure}[t]
\begin{center}
\includegraphics[width=5.3in]{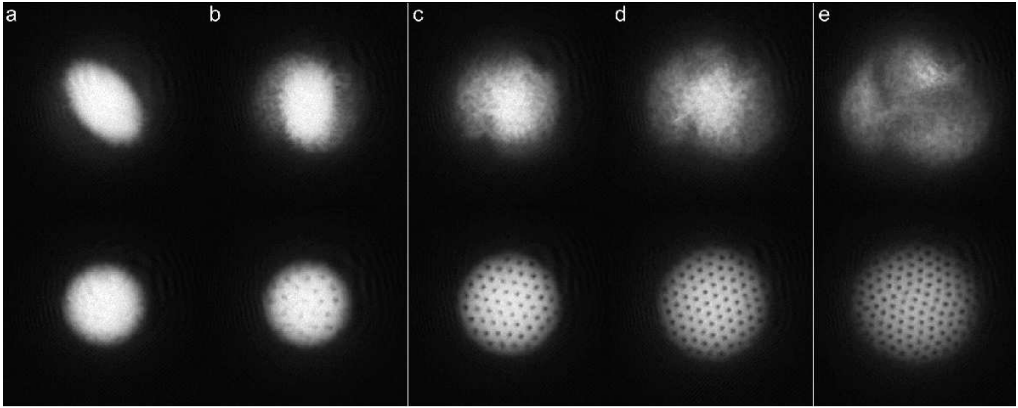}
\caption[Vortex nucleation for violent stirring in an atomic
BEC]{Vortex nucleation for violent stirring in an atomic BEC. The
upper row shows expansion images of sodium condensates after 500
ms of stirring at the quadrupole frequency, for different laser
powers of the stirring beam. The lower row shows the resulting BEC
after 300 ms of equilibration time. This suggests that the
condensate has to be severely excited to generate many vortices.
From left to right, the laser power was increased for each
subsequent image by a factor of two. The cloud was held in a
magnetic trap.} \label{f:powerdependence}
\end{center}
\end{figure}

The experimental setup is presented in
Fig.~\ref{f:experimentalsetup}.  It was tested and optimized using
sodium Bose-Einstein condensates.  Figure \ref{f:powerdependence}
shows how we determined the parameters for stirring and
equilibration using a sodium BEC in a magnetic trap. The next step
was to repeat these experiments in an optical trap, initially
without high magnetic field, optimizing the shape of the optical
trap.  Then magnetic fields were added.  This required a careful
alignment of the optical trap to the magnetic saddle point. We
estimate the residual ellipticity of the transverse potential to
be less than $2\%$ \footnote{Of course, this cannot compare with
the almost perfect roundness of a (magnetic) TOP trap, with
residual ellipticity of less than 0.1\%~\cite{halj01vort}.}.
Magnetic fields were left on during the expansion.  For lithium
this was crucial: molecules at the initial densities are
only stable against collisions at magnetic fields close to the
Feshbach resonance. During the expansion at high magnetic field,
the cloud could tilt, revealing small misalignments between the
optical trap and the saddle point.  Even more importantly, additional steps were necessary to ensure vortex visibility after expansion into the saddle point potential, as discussed below.

\begin{figure}[t]
\begin{center}
\includegraphics[width=3.5in]{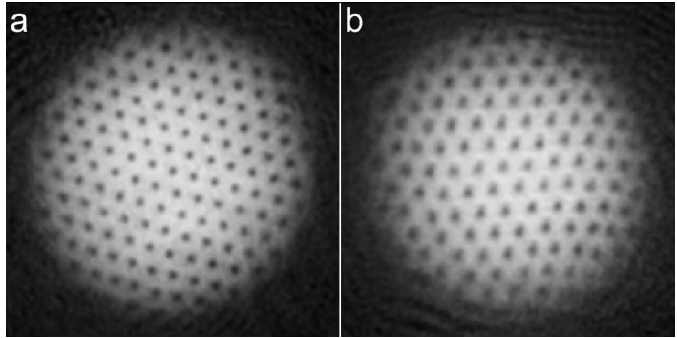}
\caption[Vortex lattice in a Bose-Einstein condensate of sodium
atoms in the magnetic and the optical trap]{Vortex lattice in a
Bose-Einstein condensate of sodium atoms in the magnetic trap
(left) and the optical trap (right image). The optical trap
(highest number obtained $\sim$ 120 vortices) can favorably
compare with the magnetic trap (highest number in our experiment
$\sim 150$ vortices).}\label{f:sodiumlattice}
\end{center}
\end{figure}

Finally, large vortex lattices containing about 120 vortices were
created in sodium Bose-Einstein condensates  both in magnetic
and optical traps with similar lifetimes (see
Fig.~\ref{f:sodiumlattice}), and we were ready to proceed with
lithium. This required two changes to the trap geometry:  (1)
lithium is lighter, and keeping the saddle point of the combined
magnetic and gravitational potential in place requires reducing the field curvature. (2) The higher
chemical potential of lithium required a larger beam waist to
obtain the correct trap depth for evaporative cooling (see equation~\ref{e:waist} in section~\ref{s:opticaltrap}).

\paragraph{Expansion of vortex lattices}
In the trap, the vortex size is on the order of the healing length
(for an atomic or a molecular BEC) or of the inverse Fermi
wavevector (for a strongly interacting Fermi gas), about 200 nm.
This small size is prohibitive for in-situ detection using optical
techniques. Fortunately, angular momentum conservation allows vortices to survive the expansion of the
condensate, which we can thus use as a ``magnifying glass''.
However, only in simple geometries is the expansion a faithful
magnification. Complications arise due to the expansion into a
saddle potential.

Hydrodynamic expansion into a saddle point potential has been
discussed in chapter~\ref{c:analysis}.   How does the vortex core size change
during expansion? There are two regimes one can simply understand,
the initial hydrodynamic expansion and the ballistic expansion at
long times of flight. In the first part of the expansion, the
mean-field $\mu \propto n a$ changes so slowly that the condensate
wave function can still react to the change in density: Adjustments
on the healing length scale $\xi$ - about the size of a vortex in
equilibrium - can occur at a rate $\hbar/m \xi^2 = \mu / \hbar$.
As long as the rate of change of $\mu$ - essentially the rate of
change of the radial Thomas-Fermi radius $R_r$ - is smaller than $\mu / \hbar$, the vortex core
can still adjust in size to the local mean-field. It thus grows as
$\xi \propto 1/\sqrt{n(t) a} \propto R_r(t)
\sqrt{\frac{R_a(t)}{a}}$ where $R_a(t)$ is the axial Thomas-Fermi radius after expansion time $t$. If $R_a$ does not vary
appreciably, $\xi / R_r(t)$ will remain {\it constant} during the
expansion, the vortex core grows just as the size of the
condensate, and the magnification is faithful.

Once the rate of change of $\mu(t)$ becomes comparable to
$\mu(t)/\hbar$, the condensate can no longer adiabatically adapt
to the lowering density. The characteristic expansion rate being
$\omega_r$, this occurs when $\mu(t) \approx \hbar \omega_r$. For
much longer expansion times, we are in the limit of ballistic
expansion. Here, each particle escapes outward with the given
velocity (in free space) or, in the case of the saddle potential,
with a radial acceleration proportional to its distance from the
origin. This simply rescales the radial dimension, and thus
stretches the vortex core and the cloud size by the same factor.
Again the magnification is faithful.

\begin{figure}[t]
\begin{center}
\includegraphics[width=5.3in]{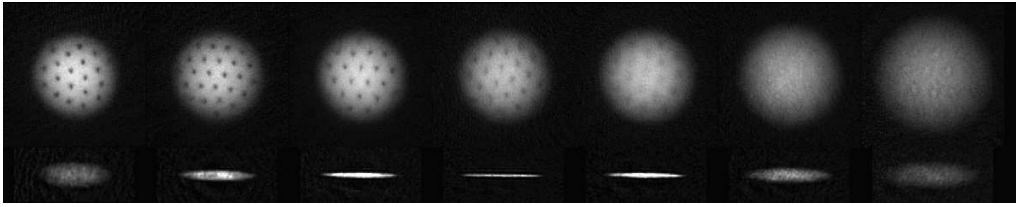}
\caption[Decrease of vortex visibility for a sodium condensate
expanding into a magnetic saddle potential]{Decrease of vortex
visibility for a sodium condensate expanding from the optical trap
into a magnetic saddle potential. Top and bottom row show axially
and radially integrated optical densities, respectively. The
saddle potential is confining in the axial and anti-confining in
the radial direction. As the condensate expands radially, it
collapses in the axial dimension, a direct consequence of
hydrodynamic flow. The vortex cores shrink and collapse onto
themselves, thereby filling in completely and forming ring-like
structures (see text for details). For the images, the magnetic
field curvature ($\nu_z = 26\,\rm Hz$) was switched off after,
from left to right, 0, 2, 3, 4, 5, 7 and 10 ms. The total time of
flight was constant at 35 ms.} \label{f:sodiumworms}
\end{center}
\end{figure}

However, in our experiment we are not in the quasi-2D regime where
$R_a \gg R_r$. The saddle potential ``squishes'' the cloud in the
axial dimension, as the decreasing mean-field no longer stabilizes
the condensate's axial size. According to the above estimate, the
vortex cores will {\it shrink} in comparison to the cloud size by
a factor $\propto \sqrt{R_a(t)}$. We can see the effect on a
sodium condensate in our optical trap in Fig.~\ref{f:sodiumworms},
where the axial curvature was left on for longer and longer times during expansion.

To work around this problem, the magnetic field curvature was
quickly reduced after releasing the cloud from the trap, by ramping down the current in the curvature
coils (in about 1 ms). As this increases the overall offset field (the curvature bias field is aligned opposite to the Feshbach bias), the current in the Feshbach coils is decreased
accordingly, so as to leave the offset field $B_0$ - and the
interaction parameter of the Fermi mixture - constant. The radial
expansion is sped up even further in comparison to the axial
evolution by actively ``squishing'' the cloud about 3 ms before
release. This is done by simply ramping up the power in the
optical trapping beam by a factor of 4. Not only does this
increase the radial trapping frequency, but it also excites a
``breathing'' mode in the condensate. The result is that the
condensate expands almost twice as fast radially as without these steps.

\subsubsection{Observation of vortex lattices}
\label{s:vortices}
\begin{figure}[t]
\begin{center}
\includegraphics[width=3.5in]{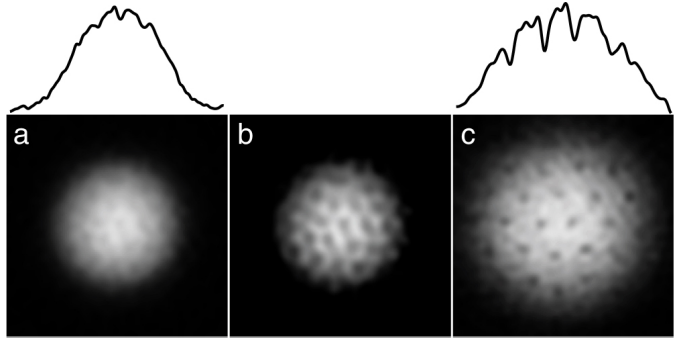}
\caption[Vortex lattice in a $^6\rm Li_2$ molecular
condensate]{Vortex lattice in a $^6\rm Li_2$ molecular condensate.
In a), stirring, equilibration and imaging of the vortex lattice
all took place at a fixed field of 766 G ($1/k_F a = 1.3$). Image a) shows the very first clear signature we observed.
The vortex core depletion is barely 10\%. b) A Fourier filter
applied to a) clearly shows the Abrikosov vortex lattice. c) The
improved scheme of ``squishing and release'' (see text), as well
as a sudden reduction of the interaction strength led to an
improved vortex contrast.
From~\cite{zwie05vort}.}\label{f:firstlithium}
\end{center}
\end{figure}

The search for vortices in Fermi gases started on the BEC side of
the Feshbach resonance.  We hoped that far on the molecular side,
in the regime where $k_F a$ is small, the situation would be fully
analogous to the case of atomic condensates. However, too far away
from resonance, the molecules (which are in the highest
vibrational state of the interatomic potential!) can undergo rapid
vibrational relaxation via three-body collisions, leading to
heating and trap loss. The lifetime of the gas needs to be longer
than the vortex nucleation and equilibration time (typically 1 s). If the entire preparation of vortex lattices is to happen at the same magnetic field, this limits the smallest values of $k_F a$ one can study ($k_F a
\gtrsim 2$ in our experiment).

On the other hand, closer to resonance, the vortex cores become
smaller. The core size in the BEC regime is given by the healing
length $\xi \propto \frac{1}{k_F} \frac{1}{\sqrt{k_F a}}$ and
decreases for increasing $k_F a$, eventually saturating in the
unitarity regime at a value on the order of $1/k_F$.  Furthermore,
closer to resonance, quantum depletion becomes important: The
condensate density $n_0$ is no longer equal to the total density,
and the vortex core loses contrast.

Fortunately, it turned out that a window existed, and at a field
of $766$ G ($1/k_F a = 1.3$), we were successful: After stirring
the cloud for 800 ms and letting the cloud equilibrate in 400 ms,
we observed a vortex lattice in the density profile
(Fig.~\ref{f:firstlithium}). This established superfluidity for
fermion pairs.

Starting from here, different methods were developed
to improve the vortex contrast. Not surprisingly it turned out that reducing the interaction strength had the largest impact. In the moment the vortex lattice is released from the
trap, the magnetic field is lowered to fields around 700 G
($1/k_F a \approx 3$ initially, further increasing during
expansion). If the condensate still has time to react to this
change in scattering length, the vortex size $\xi \propto
R_r(t)/\sqrt{a}$ will increase relative to the condensate's radius
(the expression for $\xi$ is valid in the BEC-limit, and assumes
radial expansion, see previous section). On the other hand, we found
that the ramp should not move too far into the weakly interacting
regime: The condensate would simply not expand anymore as
practically all the repulsive mean-field has been taken out of the
cloud. We also explored delaying the ramp until some expansion has
taken place.  However, if the delay was too long, the condensate
had reduced its density to the point where it was not able to
adjust quickly enough to the new interaction strength and increase
its vortex size.  For this, the ``reaction rate'' of the
condensate wave function, $\mu / \hbar$ at the final field should be
faster than the rate of change of $\mu$, that is, the rate of
change of $R_r(t)$.

\begin{figure}[t]
\begin{center}
\includegraphics[width=5in]{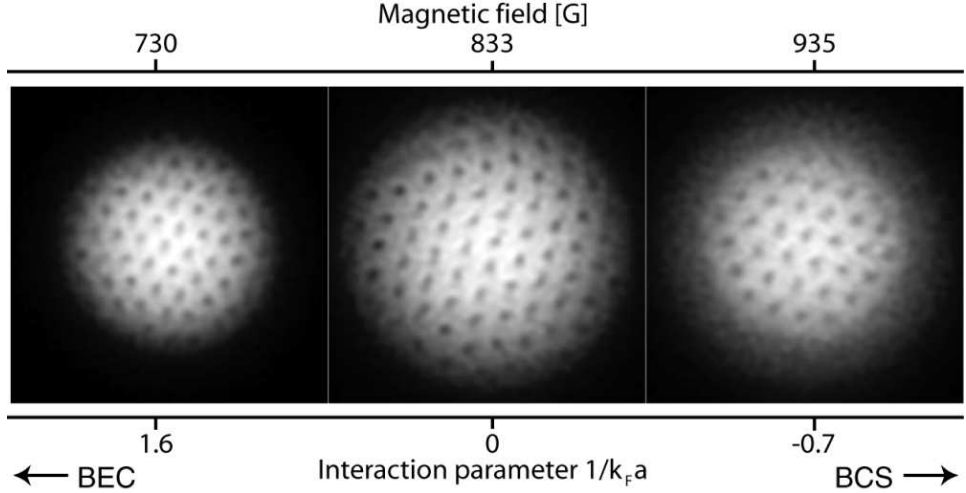}
\caption[Observation of vortices in a strongly interacting Fermi
gas]{Observation of vortices in a strongly interacting Fermi gas,
below, at and above the Feshbach resonance.   This establishes
superfluidity and phase coherence in fermionic gases.  After a
vortex lattice was created at 812 G, the field was ramped in 100
ms to 792G (BEC-side), 833G (resonance) and 853G (BCS-side), where
the cloud was held for 50 ms. After 2ms of ballistic expansion,
the magnetic field was ramped to 735G for imaging (see text for
details). The field of view of each image is 880 $\mu$m x 880
$\mu$m. More recent version of Fig. 3 in~\cite{zwie05vort}.}\label{f:BECBCSvortices}
\end{center}
\end{figure}

This technique of observing vortices worked on both sides of the
Feshbach resonance. Stirring, equilibration and initial expansion
could be performed at magnetic fields between 740 G ($1/k_F a =
2$) and 860 G ($1/k_F a = -0.35$), before switching to the
BEC-side during expansion for imaging. The observation of ordered
vortex lattices above the Feshbach resonance at 834 G, on the
BCS-side of the resonance, establishes superfluidity and phase
coherence of fermionic gases at interaction strengths where there
is no two-body bound state available for pairing.

The sweep down to 735 G solved another potential difficulty in
detecting vortices on the BCS side. The condensate fraction is reduced by quantum depletion on the BEC-side, and on the BCS side it is only $\sim \Delta / E_F$ (see~\ref{s:condensatewavefunction}). In a simplified picture, it is only this
``coherent''  part of the atomic density which vanishes at the
vortex core, reducing the contrast (for more elaborate treatments, see~\cite{bulg03,sens06vort,chie06vortex}).  By sweeping to the weakly interacting BEC regime,
the low contrast vortices on the BCS-side are transformed into BEC-type vortices with
high contrast.

Since a ramp is involved in the detection of vortex lattices, the relevant time scales need to be analyzed, as in the case of the observation of condensation via rapid ramps. The conclusion is that vortex lattices cannot form during the 10 ms of expansion at the imaging field, on the BEC-side of the resonance. We observed that the vortex lattice
needs many hundreds of milliseconds to form in the stirred cloud.
This is the same time scale found for the lattice formation in
atomic BECs~\cite{madi01,abos02_form}. This time scale was found
to be independent of temperature~\cite{abos02_form} and seems to
represent an intrinsic time scale of superfluid hydrodynamics, dependent only on the trapping frequencies. It
is also in agreement with a theoretical study of vortex formation
in strongly interacting Fermi gases~\cite{toni05}.
When a thermal cloud is slowly cooled through the transition
temperature~\cite{halj01vort}, the condensate first forms without
a vortex.  As the condensate grows, vortices are nucleated at the
surface and then enter the condensate~\cite{angl01vort}. When a
thermal cloud is suddenly cooled, a condensate with phase
fluctuations will form~\cite{angl99,shva02non_eq_bec} which can
arrange themselves into a vortex tangle. In either case, one would
expect a crystallization time of at least several hundred
milliseconds before a regular vortex lattice would emerge. Also,
it takes several axial trapping periods for the vortex tangle to
stretch out. Even if these time scales were not known, it is not
possible to establish a regular vortex lattice with long-range
order in a gas that expands at the speed of sound of the trapped
gas. Opposing edges of the expanding cloud simply cannot
``communicate'' fast enough with each other.

The regularity of the lattice proves that all vortices have the
same vorticity. From their number, the size of the cloud and the
quantum of circulation $h/2m$ for each vortex, we can estimate the
rotational frequency of the lattice. For an optimized stirring
procedure, we find that it is close to the stirring frequency.

\subsubsection{Vortex number and lifetime}
The number of vortices that could directly be created on the
BCS-side was rather low in the first experiments, as the stirring
seems to have had an adverse effect on the stability of the pairs.
This corresponds to the expectation that the gas is more robust on
the BEC-side, where the lowest excitations are sound waves, while
on the BCS-side it is pair breaking. To optimize the vortex number
on the BCS-side, first a large vortex lattice was produced close
to resonance, at 812 G, before ramping the magnetic field beyond
the Feshbach resonance. In this way, large numbers of vortices
could be obtained in the entire BEC-BCS crossover (see
Fig.~\ref{f:BECBCSvortices}).

\begin{figure}[t]
\begin{center}
\includegraphics[width=5.2in]{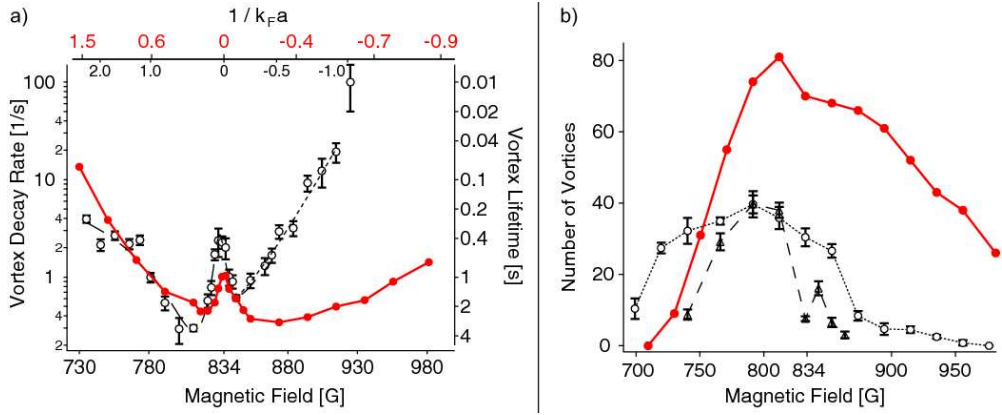}
\caption[Lifetime and number of vortices in the BEC-BCS
crossover]{1/e-Lifetime (a) and number (b) of vortices in the
BEC-BCS crossover. Vortices are long-lived across the entire
BEC-BCS crossover. A narrow dip in the lifetime on resonance is
presumably due to pair breaking (see text). Open symbols are data from April 2005~\cite{zwie05vort}.
Optimization of the system using a deeper trap resulted in
improved characteristics on the BCS-side, shown by the full
symbols (data from June 2005). In b), the triangles give the
number obtained by stirring and equilibrating both at the given
field. Stirring at 812 G and subsequently ramping to the final
field for equilibration resulted in the data shown as
circles. Figure adapted from~\cite{zwie05vort}, incorporating more recent results.}\label{f:vortexlifetimeandnumber}
\end{center}
\end{figure}

By monitoring the number of vortices after a variable delay time,
the vortex lifetime was determined
(Fig.~\ref{f:vortexlifetimeandnumber}a).  The vortex lifetime
around the Feshbach resonance is on the same order of what was
found for atomic BECs. This displays the high degree of
metastability of vortices in superfluids.  One picture for vortex
decay assumes that thermal excitations (or the normal component)
provide friction between some residual trap anisotropies and the
rotating superfluid~\cite{zhur01diss,fedi02vortex}. The difference
in lifetimes observed in two different data sets
(Fig.~\ref{f:vortexlifetimeandnumber}a) can be explained by
changes in the trap geometry, different atom number and
temperature. A deeper trap can hold large Fermi clouds on the
BCS-side better, leading to a longer lifetime in this regime.

There is a peculiar dip in the lifetime on resonance, which may be
caused by the coupling of external motion to internal degrees of
freedom.  One possibility is a resonance between the pair binding
energy and the rotation frequency.  This requires pairs with a
very small binding energy, which should exist only in the far
outside wings of the cloud.  For example, the (two-body) molecular
binding energy at 830 G, 4 G away from resonance, is only $k_B
\times 3 \, \rm nK$ or $h \times 60\,\rm Hz$. This is on the order
of $\hbar \times$ the rotation frequency $\Omega$ of the vortex
lattice. If the molecules rotate around the trap, trap
anisotropies may excite them resonantly and cause dissociation. On
the BCS-side, any possible resonance may be suppressed by density
dependent broadening. Of course, the fraction of pairs at such low
binding energies is very small, and they can contribute to damping
in a major way only if surface effects are important.

\subsubsection{A rotating bucket}\footnote{The results of this section have not been published elsewhere.}

\begin{figure}[t]
\begin{center}
\includegraphics[width=5.2in]{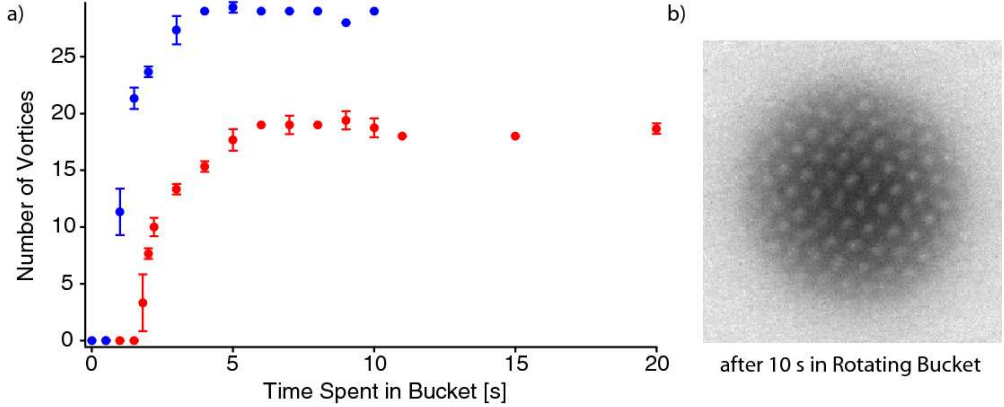}
\caption[Rotating bucket for superfluid Fermi gases]{Rotating
bucket for superfluid Fermi gases. a) Vortex number vs time spent
in the rotating trap. After am equilibration time, the number of
vortices stays constant. The final vortex number depends on the
power of the green stirring beam, indicating slippage due to
residual friction with the ``container walls''. The lower and upper
curve correspond to lower and higher green beam power. b)
Absorption image of an optimized vortex lattice, containing about
75 vortices, after 10 s hold time in the rotating bucket. The
magnetic field for all data was 812 G, corresponding to $1/k_F a
\approx 0.2$.}\label{f:rotatingbucket}
\end{center}
\end{figure}

All the experiments described so far first set the cloud in
rotation using the stirring beam and then let the gas equilibrate
in the stationary trap into a vortex lattice. In the stationary
trap, the vortex lattice is of course only metastable (lasting as
long as the angular momentum $L_z$ is conserved), whereas in a
trap rotating at a constant angular frequency $\Omega$, the vortex
lattice would be the true ground state. Mathematically speaking,
both situations are described by the Hamiltonian $H - \Omega L_z$,
where the second term is, in the latter case, the usual
transformation to a frame rotating at frequency $\Omega$, whereas
in the first case, the Lagrangian multiplier $\Omega$ enforces the
conservation of $L_z$.

In some previous experiments with BECs the rotating anisotropy
needed to be switched off before an ordered vortex lattice could
form~\cite{abos01latt}, possibly because the rotating laser beam was not
moving smoothly enough to allow equilibration into a vortex
lattice.  In our setup, it has become possible to observe vortex
lattices in the presence of a rotating bucket and even increase the
lifetime of the vortex lattice by maintaining the rotating drive.

The experiment was performed at 812 G in a trap with radial
trapping frequency $\nu_R = 90$ Hz. The two stirring laser beams
(power in each beam $\approx$ 100 $\mu$W, waist ${\rm w} = 16\, \mu$m)
created only a weak potential of about 20 nK each on the cloud
(mean-field $\mu \approx 400$ nK). They were rotated around the
cloud at a frequency of 70 Hz. For imaging, the atoms were
released from the combined trap, the confining optical potential
plus the repulsive stirring beam. We found that it was possible to
stabilize a vortex lattice containing 19 vortices for 20 s (see
Fig.~\ref{f:rotatingbucket}), limited only by the computer memory
controlling the experiment.

The final vortex number depended on the laser power of the
stirrer. Increasing the power in the stirring laser by 60\%
increased the equilibrium vortex number to 29. This suggests that
the equilibrium vortex lattice, at least at the weaker laser
power, had not reached the angular velocity of the stirrer. It
appears that the drive was necessary to compensate for friction
with some residual trap anisotropy.  At later stages of the
experiment, we were able to stabilize 75 vortices for 10 s in a
deeper trap with $\nu_r = 120$ Hz.

These experiments are analogous to the pioneering ``rotating
bucket'' experiments on $^4$He~\cite{yarm79}, where it was
possible to maintain a rotating superfluid containing four vortices
for eleven hours, only limited by the eventual exhaustion of the refrigerator helium supply.

\subsubsection{Superfluid expansion of a rotating gas}
\label{s:vortexexpansion}
\begin{figure}[t]
\begin{center}
\includegraphics[width=5.3in]{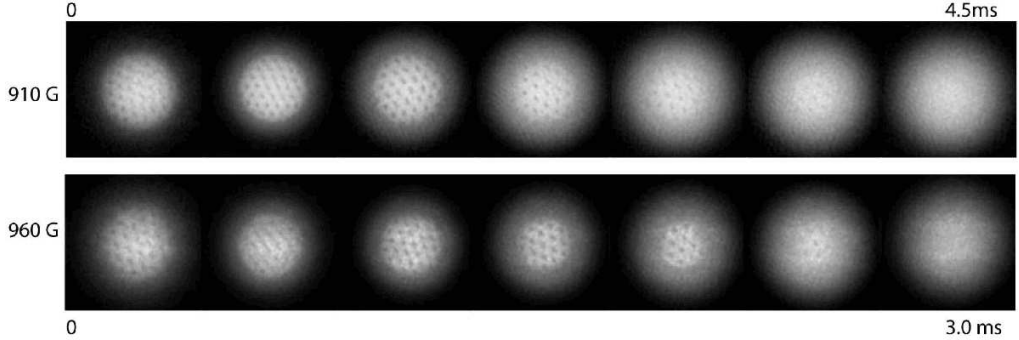}
\caption[Superfluid expansion of a strongly interacting rotating
Fermi gas.]{Superfluid expansion of a strongly interacting
rotating Fermi gas. Shown are absorption images for different
expansion times on the BCS-side of the Feshbach resonance at 910 G
(0,1,2,3,3.5,4, and 4.5 ms) and at 960 G (0, 0.5, 1, 1.5, 2, 2.5,
and 3 ms), before ramping to the BEC-side for further expansion
and imaging. The total time-of-flight was constant. The vortices
served as markers for the superfluid parts of the cloud.
Superfluidity survived during expansion for several ms, and was
gradually lost from the low-density edges of the cloud towards its
center. The field of view of each image is 1.2 mm $\times$ 1.2
mm. From~\cite{schu06pair}.}\label{f:pairbreaking}
\end{center}
\end{figure}

What will happen when a fermionic superfluid expands?  First, it
should follow the hydrodynamic equations for a superfluid flow~\cite{ante07}.  However, ultimately, pairing is a many-body effect,
and when the cloud becomes very dilute, at finite temperature, the pairs will eventually break and
superfluidity will be quenched.  This is different from the
situation for a BEC and at unitarity.  Since  phase space density
is preserved during expansion, $T/T_F$ or $T/T_C$ is, here, constant and
the gas remains superfluid~\footnote{For interacting condensates,
there is a small variation of critical phase space density as
function of particle density, and therefore the gas can cross the
phase transition during adiabatic expansion~\cite{houb97crit}.}

By using rotating Fermi gases, vortices serve now as a convenient
marker for the superfluid phase.  When the superfluid is quenched,
the  vortices will disappear~\cite{schu06pair}.  We allow the
Fermi gas to expand on the BCS-side for a certain time $t_{\rm
BCS}$, then ramp down to the BEC-side for further expansion and
imaging.  Vortices can be observed only when the gas is still a
superfluid at the moment of the magnetic field sweep. The total
expansion time is kept constant.

It is found that superfluid flow initially persists during the
expansion. Then, vortices start to disappear first at the edges of
the cloud, then, for longer BCS-expansion $t_{\rm BCS}$, further
inwards until the last vortex disappears at the cloud's center
(see Fig.~\ref{f:pairbreaking}). The time $t_{\rm BCS}$ for which
the last vortex disappears, increases the closer we are to
resonance, that is, the larger the interaction strength and the
stronger the fermion pairs are bound. Vortices and therefore
superfluid flow in free expansion were observed up to expansion times, for example, of 2.5 ms at 960 G and of 5 ms at 910 G.

\begin{figure}[t]
\begin{center}
\includegraphics[width=3in]{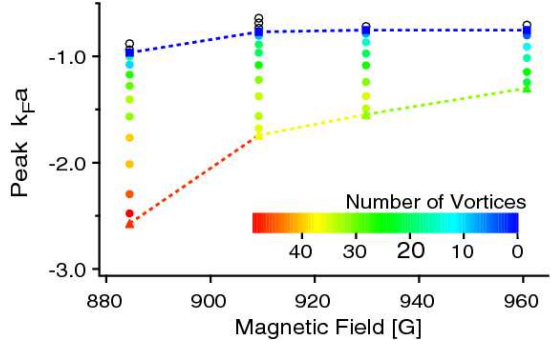}
\caption[Central interaction strength during superfluid
expansion]{(Color online) Central interaction strength $k_F a$
during superfluid expansion. Starting at a central $k_F a$ in the
optical trap, vortices survive up to an interaction strength $k_F
a \approx -0.8$, almost independent of the magnetic field
(scattering length $a$). Filled circles correspond to partially
superfluid, open circles to normal clouds. The observed number of
vortices is color coded. From~\cite{schu06pair}.}\label{f:peakKFa}
\end{center}
\end{figure}

By varying the magnetic field and thus the scattering length, we
find that the last vortex always disappears at about the same
value of the interaction parameter $k_F a \approx -0.8$ (see
Fig.~\ref{f:peakKFa}). The simplest explanation for this
observation is that we cross the phase transition line during
expansion. While $T/T_F$ is an adiabatic constant for the
expansion, $T/T_C$ is not, as $T_C/T_F$ depends exponentially on
the density. As the density decreases, the critical temperature in
the outer regions of the cloud eventually drops below $T$,
superfluidity is lost starting from the edges inwards.

We can estimate the critical interaction strength for this
breakdown to occur. At our lowest temperatures, $T/T_F = 0.05$. The formula for $T_C$ due to
Gorkov and Melik-Barkhudarov~\cite{gork61} gives $T_C =
\frac{e^\gamma}{\pi} \left(\frac{2}{e}\right)^{7/3} T_F\,
e^{-\pi/2 k_F |a|} = 0.28\, T_F\, e^{-\pi/2 k_F |a|}$. This
formula should be valid in the BCS-regime where $k_F |a| \lesssim
1$. The equation
\begin{equation}
        1 = \frac{T}{T_C} = \frac{T}{T_F} \, \frac{T_F}{T_C} \approx 0.05\, \frac{T_F}{T_C} = 0.18 \, e^{\pi/2 k_F |a|}
\end{equation}
gives a critical interaction strength $k_F \left|a\right| = 0.9$,
close to the observed value.  For a discussion of other mechanisms
which can explain the observed disappearance of vortices, we refer
to the original publication~\cite{schu06pair}. On resonance we have
$1/k_F a = 0$ during the whole expansion, and the gas should
remain superfluid.  Indeed, in the experiment we do not see
evidence for a sudden quenching of vortices, but rather a gradual
loss in contrast. Vortices could be discerned at total densities
as low as $1.2 \times 10^{11}\,\rm cm^{-3}$, thus providing
evidence for fermionic pairing and superfluidity at average interatomic distances of
2 $\mu$m.

\section{BEC-BCS crossover:  Energetics, excitations, and new systems}
\label{c:otherstudies}

In the previous chapter, we focused on two key observations in the
BEC-BCS crossover, pair condensation and vortices, that provide
direct access to the phenomena of coherence, a macroscopic
wave function, and superfluid flow.  A host of other studies have
been performed in this regime, which we summarize here.  We divide
these experiments into three different categories:
characterization of the equilibrium state by energy, entropy and
momentum distribution, dynamic measurements addressing collective
excitations, sound and the critical velocity for superfluidity,
and thirdly new systems, where the original two-component fermion
system has been modified, either by an optical lattice, or by
imbalanced populations.  Both of these modifications add another
``dimension'' to the system, which is the lattice depth and the
imbalance.

\subsection{Characterization of the equilibrium state}

\subsubsection{Energy measurements}
\label{s:energymeasurements}
The total energy of the cloud determines how large the cloud is in
the harmonic oscillator potential, or how fast it expands after
switching off the trap~\footnote{The binding energy of molecules must be subtracted, as this internal energy cannot be converted into external, mechanical energy.}. On resonance,  the virial theorem provides
a simple relationship between cloud size and total energy.  Using
the universality hypothesis that the only relevant energy scale is
the Fermi energy, it follows that the potential energy is half the
total energy, as for a
non-interacting gas~\cite{thom05virial}.  In a homogenous system (and locally for a trapped system), the
energy content of an interacting Fermi gas is parameterized as $E
= (1+\beta)E_F$ where $\beta E_F$ is the contribution of
interactions.  For unitarity limited interactions, $\beta$ is an
important universal parameter characterizing the ground state of
strongly interacting fermions.

The total energy of an interacting Fermi gas at or close to
resonance was derived from measurements of the cloud size either
in trap or after
expansion~\cite{ohar02science,gehm03stab,bart04,kina05heat,part06phase,stew06pot}.
In an interesting variant, the Paris group applied a rapid
switching of the magnetic field to zero, which was faster than the
trap period~\cite{bour03}.  In this case, the interaction energy
could be removed before it had been converted to kinetic energy.
By comparing expansion with immediate or delayed magnetic field
switching, the interaction energy could be directly measured. This
work showed  a surprising behavior of the interaction energy in a
wide region of magnetic fields below the Feshbach resonance. This
behavior was later explained by the formation of molecules and was
probably the first hint that these molecules would be stable.

Current experiments (including a measurement of the speed of
sound~\cite{jose07sound}) give $\beta \approx 0.58$, different
experiments agree to within 10\%, and most importantly agree with
theoretical predictions (via analytical
methods~\cite{bake99neutron,stee00power,heis01,pera04bcsbec,hu06becbcs,haus07bcsbec,hu07universal},
Monte-Carlo calculations~\cite{carl03,astr04,carl05} and
renormalization group methods~\cite{nish06epsilon}). A table
summarizing all experimental and theoretical determinations of
$\beta$ can be found in the contribution of C. Salomon to these
proceedings. The fact that the same value of $\beta$ was found
for $^6$Li and $^{40}$K is an impressive confirmation of
universality~\cite{hu07universal} and is a powerful demonstration
for the use of ultracold atoms as a model system for many-body
physics.

\begin{figure}[ht]
\begin{center}
\includegraphics[width=3.5in]{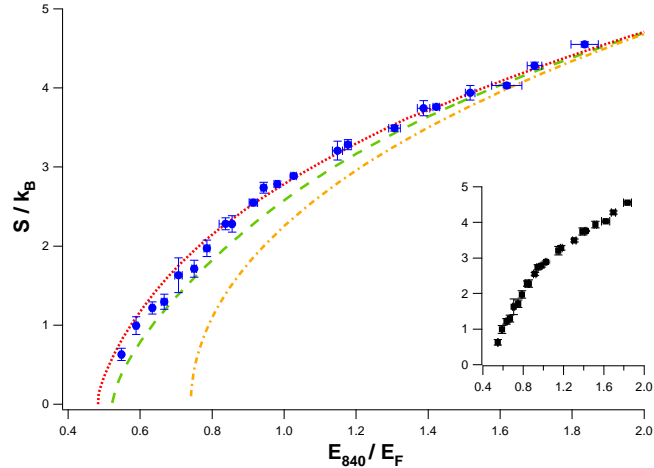}
\caption[Measured entropy per particle of a strongly interacting
Fermi gas]{Measured entropy per particle of a strongly interacting
Fermi gas at 840 G versus its total energy per particle (blue
dots). Various theoretical predictions are compared. Reprinted
from~\cite{luo07entropy}.}\label{f:entropy}
\end{center}
\end{figure}

It is possible to obtain the entropy of the cloud from size or
energy measurements.  For this, the magnetic field is
adiabatically swept to transform the system into a weakly
interacting Fermi gas. The observed size or energy in this regime
gives the entropy, since their relation is accurately known for an
ideal Fermi gas.  This allows the determination of entropy vs.\
energy for the strongly interacting gas~\cite{luo07entropy} (see
Fig.~\ref{f:entropy}).  The results of this study agreed well with
the predictions from Monte Carlo simulations that vary smoothly
across the phase transition. By using a split power law fit, a
value for a critical energy $E_c$ had been obtained. However,
since the fit did not address the behavior in the critical region,
it is not clear how accurately the split power law fit can
determine the transition point.

When energy measurements were combined with empirical thermometry
and theoretical corrections, the dependence of the heat capacity
of the Fermi gas on temperature could be deduced~\cite{kina05heat}.

Further discussion of energy measurements can be found in the contributions of C. Salomon and S. Stringari to these lecture notes.

\subsubsection{Momentum distribution}
The momentum distribution of the atoms in the cloud can be
determined by releasing them from the trap and simultaneously
switching the scattering length to zero.  Such studies have been performed in
both $^{40}$K~\cite{rega05} and $^6$Li (see C.
Salomon's contribution to these proceedings).

Far on the BCS side, one finds the momentum distribution of an
ideal Fermi gas in a harmonic trap.  On the BEC side, the momentum
distribution approaches the squared magnitude of the molecular
wave function's Fourier transform.  The crossover region smoothly
interpolates between these two limits.  The modification of the
momentum distribution due to the superfluid phase transition is
too small to be discernable in these
measurements~\cite{chen06momentum}.  Momentum distributions are
discussed in the contributions of D. Jin and C. Salomon to these
proceedings.

\subsubsection{Molecular character}
Near a Feshbach resonance, the closed channel molecular state responsible for the resonance is
mixed with the continuum of scattering states in the open channel.  In the case of $^6$Li,
those two channels have singlet and triplet character,
respectively.  Close to the Feshbach resonance, the loosely bound molecular state becomes completely dominated or ``dressed'' by the open channel.
This was confirmed by applying a molecular probe to a cold
Fermi gas, thereby exciting atom pairs to an electronically excited, singlet
molecular state at a rate that was proportional to their closed
channel character $Z$~\cite{part05}.

The wave function describing the dressed molecule or fermion pair can be written~\cite{part05}
\begin{equation}
    \left|\psi_p\right> = \sqrt{Z} \left|\psi_{v = 38}(S = 0)\right> + \sqrt{1 - Z} \left|\phi_a(S=1)\right>
\end{equation}
where $\left|\psi_{v=38}(S=0)\right>$ denotes the closed channel, singlet molecular state, and $\left|\phi_a(S=1)\right>$ the open channel, triplet contribution, with relative probability amplitude $\sqrt{Z}$ and $\sqrt{1-Z}$, respectively. In the singlet
channel, only the  $v=38$ vibrational state is relevant due to its
near-resonant energy.

\begin{figure}[ht]
\begin{center}
\includegraphics[width=3in]{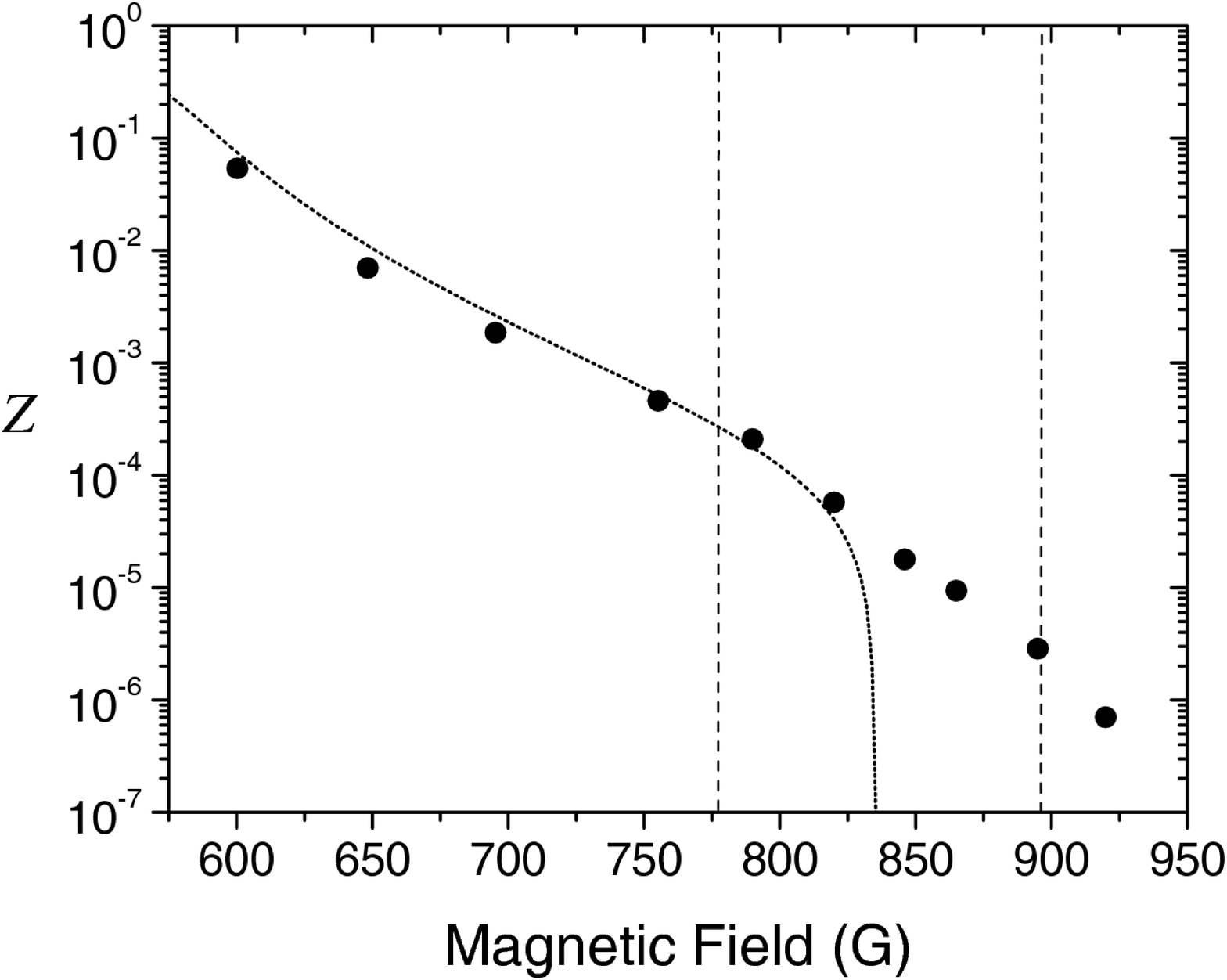}
\caption[Closed channel character
$Z$ of lithium atom pairs vs magnetic field.]{Closed channel character
$Z$ of lithium atom pairs versus magnetic field. The dotted line shows
the closed channel character of the bound molecular state below
the Feshbach resonance. Reprinted from~\cite{part05}.}\label{f:riceZ}
\end{center}
\end{figure}

By monitoring trap loss during the excitation, $Z$ could be
determined (see Fig.~\ref{f:riceZ}) and it was verified that the
Feshbach resonance in \li\ is indeed broad, that is, the closed
channel contribution to the pair wave function is negligible
throughout the crossover region.

In the two-channel BEC-BCS crossover description, $Z$ is
proportional to $\Delta^2$~\cite{part05}, and one might hope that
by measuring $Z$ one actually measures the magnitude of the
macroscopic order parameter. However, a spectroscopic
determination of the singlet character is a {\it local} probe,
sensitive only to $g^{2}(\vect{0})$, the two-particle correlation
function at zero distance. As such, it can measure local pair
correlations, but not global phase coherence or condensation of
pairs. In fact, the BEC-BCS crossover allows for pair correlations
above $T_C$, that can be seen in RF spectra (see
section~\ref{s:rfspectroscopy}). A closed channel character of
these pairs has indeed been identified even in the normal phase,
on both sides of the Feshbach resonance~\cite{grei04corr}.

\subsection{Studies of excitations}
To explore a new form of matter, one should probe its response to
perturbations. In this section we summarize experimental studies
on excitations of Fermi gases in the BEC-BCS crossover region.
These studies include sound waves (resonant standing waves and
sound propagation), observation of the critical velocity for
superfluid flow, and single particle excitations probed via RF
spectroscopy.

\subsubsection{Collective excitations} By a sudden or periodic
modulation of the trapping potential, eigenmodes of the trapped
cloud can be excited.  The eigenfrequencies can be sensitive to
the equation of state, $\mu(n)$, and thus characterize the
interactions in the system. Collective excitations were among the
first properties studied in the case of atomic BECs. They provided
stringent tests of the Gross-Pitaevskii-equation governing those
condensates~\cite{jin96coll,mewe96coll,jin97,stam98coll,dalf99rmp}, and posed challenges to finite-temperature theories.

In the case of fermions, collective excitations were studied
already for weakly interacting gases in the ``pre-Feshbach era''~\cite{gens01coll}.  After the realization of fermion pair
condensation, extensive studies were carried out at Duke and
Innsbruck. These are discussed in detail in the contribution of R.
Grimm to this volume.

Chapter~\ref{c:analysis} discusses the equations of motion for
strongly interacting gases, including collisional and superfluid
hydrodynamics. To obtain the response to a small perturbation of
the confining (harmonic) potential, one can directly use
Eq.~\ref{e:eulerclass} in the case of classical hydrodynamics, or
Eq.~\ref{e:euler} for superfluid hydrodynamics, and linearize the
equations for small oscillatory changes of the cloud radius. For
instance, a ``breathing'' mode of a cigar-shaped cloud can be
excited by suddenly squeezing the cigar in both radial directions.
By only squeezing one radial dimension, one excites a ``standing
quadrupole'' mode. Depending on the symmetry of the excited mode,
different eigenfrequencies are found that depend more or less
strongly (or do not depend at all) on the equation of state. The
latter is parameterized by the exponent $\gamma$ in $\mu(n)
\propto n^\gamma$.  Both the Duke~\cite{kina04sfluid} and
Innsbruck~\cite{bart04coll} group have confirmed the value of
$\gamma=2/3$ at the Feshbach resonance, which is the same as for a
weakly interacting Fermi gas.  A precision study of collective
oscillations on the BEC side of crossover has verified the famous
Lee-Yang-Huang correction to the equation of state of a strongly
interacting Bose gas~\cite{altm07precision}.

We refer the reader to section~\ref{s:superfluidsignatures} for a discussion of further
experimental studies of collective excitations.  They
characterized the strongly interacting and superfluid regimes as a
function of scattering length and temperature and revealed an
intriguing (and not yet fully understood) picture of hydrodynamic
behavior and smooth or sudden transitions to collisionless
dynamics.

\subsubsection{Speed of sound}
Density perturbations propagate at the speed of sound. In
section~\ref{s:collexcitations} we discussed that the Bogoliubov
sound mode on the BEC side smoothly evolves into the
Bogoliubov-Anderson mode on the BCS side.

\begin{figure}[ht]
\begin{center}
\includegraphics[width=3.5in]{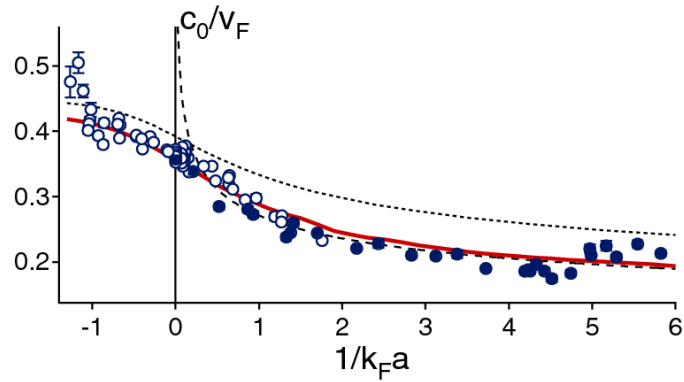}
\caption[Sound velocity normalized to $v_F$  versus the
interaction parameter, $1/k_Fa$.]{Sound velocity normalized by the Fermi velocity $v_F$ versus the
interaction parameter, $1/k_Fa$.  Black dotted curve: mean-field
theory based on the Leggett ground state (see~\ref{s:landau}).  Gray
(red) solid curve: quantum Monte Carlo calculation~\cite{astr07private}. Black dashed curve:
Thomas-Fermi theory for a molecular BEC. Reprinted from~\cite{jose07sound}.}\label{f:dukesound}
\end{center}
\end{figure}

A laser beam focused into the center of the cloud can create a
localized density perturbation, which then propagates along a
cigar-shaped atom cloud~\cite{andr97prop}.  Using this technique, the
Duke group has recently measured the speed of sound in a Fermi gas
across the BEC-BCS crossover and found very good agreement with
Monte-Carlo predictions~\cite{jose07sound} (see Fig.~\ref{f:dukesound}).

\subsubsection{Critical velocity}
Superfluid flow breaks down above a critical velocity.  This
critical velocity is a threshold velocity for creating
excitations.  For density fluctuations, it is the speed of sound,
discussed in the previous paragraph,  and this provides the
critical velocity on the BEC side. It monotonously increases
towards resonance. On the BCS side, as discussed in
section~\ref{s:landau}, pair breaking becomes the dominant
mechanism. The pairing energy is largest near resonance, resulting
in a maximum of the critical velocity around resonance.

\begin{figure}[ht]
\begin{center}
\includegraphics[width=3in]{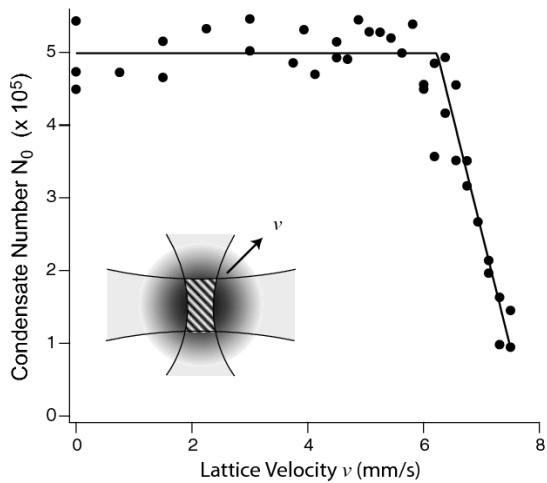}
\caption[Onset of dissipation for superfluid
fermions in a moving optical lattice.]{Onset of dissipation for superfluid
fermions in a moving optical lattice. (Inset) Schematic of the
experiment in which two intersecting laser beams produced a moving
optical lattice at the center of an optically trapped cloud
(trapping beams not shown). Number of fermion pairs which remained
in the condensate after being subjected to an optical lattice of
depth 0.2 $E_F$ for 500 ms, at a magnetic field of 822 G ($1/k_Fa
= 0.15$). An abrupt onset of dissipation occurred above a critical
velocity. Reprinted from~\cite{mill07critical}.}\label{f:criticalvelocity}
\end{center}
\end{figure}

This has been recently observed at MIT~\cite{mill07critical}.   By
recording the onset of dissipation in a Fermi cloud exposed to a
weak one-dimensional lattice moving at a variable velocity,
critical velocities were obtained.  When the magnetic field was
varied, they showed a peak near resonance.  In these experiments
the lattice was created by two focused laser beams crossing at an
angle of about 90 degrees exposing only the central part of the
cloud to the moving lattice (Fig.~\ref{f:criticalvelocity}). When the whole cloud was
excited by the moving lattice, much lower critical velocities were
found, most likely due the breakdown of superfluidity in the low
density spatial wings of the cloud.  Using larger depths of the
moving lattice, smaller values of the critical velocity were
found.  This shows that the lattice is not only a way to probe the
Fermi gas, it is also a way to create new systems with interesting
properties (see section~\ref{s:newsystems}).

\subsubsection{RF spectroscopy}
\label{s:rfspectroscopy}
Single-particle excitations can reveal the nature of pairing.  On
the BEC side, the excitation of a single atom requires breaking a
molecular bond, thus providing information about the binding
energy.  On the BCS side, single particle excitations reveal the
superfluid energy gap and give access to the microscopic physics
underlying these Fermi mixtures.

\begin{figure}[h]
\begin{center}
\includegraphics[width=3.5in]{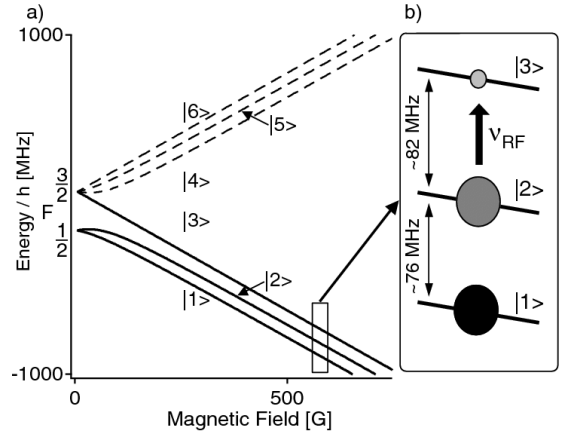}
\caption[RF spectroscopy in $^6$Li]{RF spectroscopy using a
three-level system in $^6$Li. a) Hyperfine structure of
lithium. b) Close-up on the three lowest hyperfine states involved
in RF spectroscopy. Typically, the Fermi mixture is prepared
in state $\left|1\right>$ and $\left|2\right>$, and the RF pulse
drives the transition to the initially empty state
$\left|3\right>$. Figure adapted
from~\cite{gupt03rf}.}\label{f:rfspectroscopy}
\end{center}
\end{figure}

In condensed matter samples, the presence of an excitation gap is
clearly seen in tunnelling experiments between a superconductor
and a normal metal, divided by a thin insulating barrier. The
tunnel effect allows individual electrons to pass through the
barrier. For this to occur, electrons must first be excited from
the pair condensate, which costs an energy $\Delta$. For applied
voltages smaller than $\Delta/e$, no tunnelling occurs.
Abstracting from the tunnelling example, what is required to
measure an excitation spectrum is the coupling (= tunnelling)
between the initial many-body state of interest and a
well-characterized reference state (= metal). In atomic Fermi
gases, this situation can be established to some degree using RF
spectroscopy. Starting with a Fermi mixture of atoms in, say, the
hyperfine states $\left|1\right>$ and $\left|2\right>$, an RF
pulse couples atoms from state $\left|2\right>$ into an empty
state $\left|3\right>$. If state $\left|3\right>$ is
non-interacting with states $\left|1\right>$ and $\left|2\right>$,
it serves as a reference state. This situation is illustrated in
Fig.~\ref{f:rfspectroscopy}. It is important to note that, because the final state is empty, the RF pulse
can excite the entire Fermi sea, and not just atoms at the Fermi surface as in tunnelling experiments.

In chapter~\ref{c:exptechniques} we presented RF spectroscopy as
an experimental tool,  summarized the basics of RF spectroscopy in
a two-level system (where no line shifts due to interactions can
be observed), and discussed simple (weak interactions, mean field
shift) and exact (sum rule for average shift) limits for the case
of three levels. In section~\ref{s:RFspectrum}, we calculated the spectrum for an RF excitation of spin up ($\left|2\right>$) atoms out of the BCS state and into the empty state $\left|3\right>$.

In this section, we want to summarize how this ``RF tool'' has been
applied to strongly interacting Fermi systems.  The full
interpretation of these results is still an open question.

\paragraph{RF spectroscopy of normal Fermi gases}
For an isolated atom, the resonant frequency $\omega_{23}$ for
this transition is known to an extreme accuracy. In the presence
of a surrounding cloud of interacting atoms, however, the
transition can be shifted and broadened. The shifts can originate
from the atom experiencing the ``mean-field'' of the surrounding
gas. Pairing between fermions can lead to additional frequency
shifts, as the RF field has to provide the necessary energy to
first ``break'' the pair before the excited atom can be transferred
into the final state (this picture implies that final state
interactions are negligible). Broadening can be inhomogeneous, for
example due to averaging over a range of densities in a trapped
sample, or intrinsic (homogeneous), reflecting the local
correlations (and thus pairing) between atoms.

\begin{figure}[h]
\begin{center}
\includegraphics[width=3.5in]{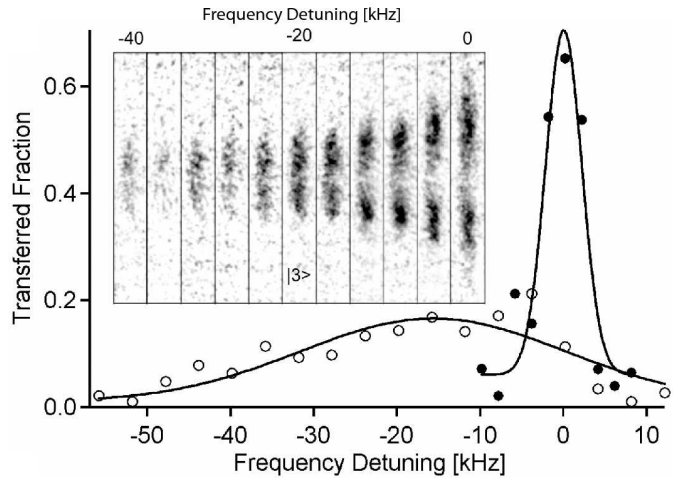}
\caption[Mean field clock shift in an interacting Fermi
mixture]{Mean field clock shift in an interacting Fermi mixture.
RF spectroscopy is performed on the transition from
$\left|2\right> \rightarrow \left|3\right>$ in the presence (open circles) and
absence (filled circles) of atoms in state $\left|1\right>$. The broadening
reflects the inhomogeneous density distribution in the trap, as
can be verified via absorption images of atoms in the final state
shortly after the RF pulse. Atoms in the low-density wings have
smaller shifts than at the center at high density. Figure adapted
from~\cite{gupt03rf}.}\label{f:meanfield}
\end{center}
\end{figure}

When a $\left|2\right>$ atom is transferred into state
$\left|3\right>$ in the presence of a cloud of $\left|1\right>$
atoms, its ``mean-field'' interaction energy changes: The final
state interacts differently with $\left|1\right>$ than the initial
state. This leads to a clock shift in the RF spectrum (see
chapter~\ref{c:exptechniques}). The first experiments on RF
spectroscopy in Fermi gases observed such ``mean-field''
interaction shifts close to a Feshbach
resonance~\cite{rega03fesh,gupt03rf} (see Fig.~\ref{f:meanfield})
and demonstrated the tunability of interactions around such
resonances. Furthermore, it was found that near the Feshbach
resonance, the mean-field shifts did not diverge~\cite{gupt03rf},
contrary to the simple picture that shifts should be proportional
to the difference in scattering lengths. In fact, in the case of
$^6$Li, mean-field shifts were found to be practically absent
close to the Feshbach resonances. In this experiment, both the
initial and the final state were strongly interacting with state
$\left|1\right>$ and it was supposed that the two, unitarity
limited energy shifts cancel out in the transition. This
interpretation was recently confirmed in~\cite{baym07}, where it
was found that average clock shifts depend on the {\it inverse} of
scattering lengths and thus become small near Feshbach resonances.

RF spectroscopy was used to detect and to study Feshbach
molecules. Potassium molecules formed via a sweep across the
Feshbach resonance were detected by RF spectroscopy
~\cite{rega03mol}.  The molecular line was shifted with respect to
the atomic resonance by the molecular binding energy. Bound-bound
and bound-free transitions in $^6$Li were used to precisely
determine the position of the Feshbach resonance and other
scattering properties~\cite{bart04fesh,chin05rf}.

\paragraph{RF spectroscopy of superfluid Fermi gases}
After the arrival of fermion pair condensates, the Innsbruck group
traced the evolution of the molecular spectrum all the way across
resonance~\cite{chin04gap} (see also the article by R. Grimm in
these proceedings). Although two-body physics no longer supports a
bound state beyond the resonance, the spectra were still shifted
and broad, providing evidence for a pairing gap in the strongly
interacting Fermi mixture. As expected for fermionic pairing, the
observed feature scaled with the Fermi energy, whereas an
``atomic'' peak, observed at higher temperature, was narrow and
showed only a small shift from the resonance position for isolated
atoms. Theoretical modelling suggested that the gas was in the
superfluid regime at the lowest temperatures, where the ``atomic''
peak had fully disappeared~\cite{kinn04pairing}. However, the
interpretation of the spectra relied on a theory that neglected
interactions in the final state, between states $\left|3\right>$
and $\left|1\right>$ (such spectra were calculated in section~\ref{s:RFspectrum}). Recent theoretical
work~\cite{baym07,punk07rf,pera08rf} and also experimental studies
by the MIT group~\cite{schu07tbp} have demonstrated the importance
of such final state interactions.

\begin{figure}[h]
\begin{center}
\includegraphics[width=3in]{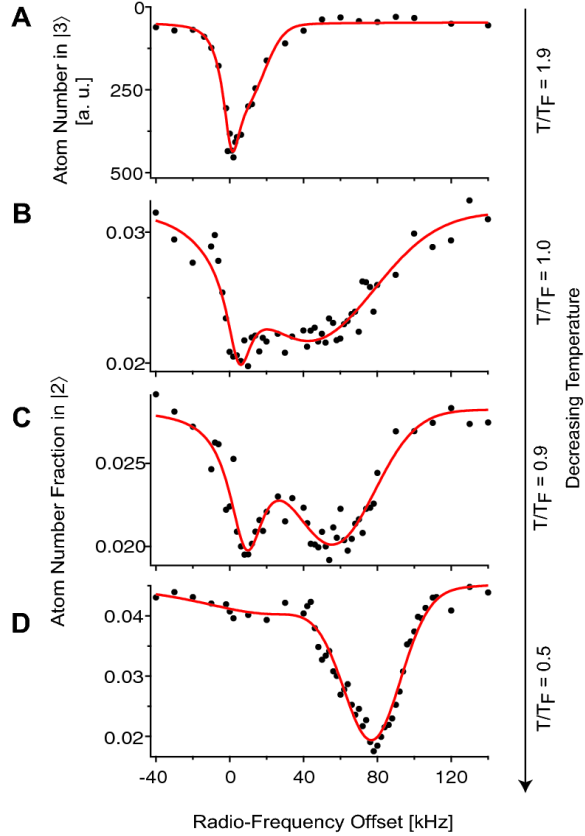}
\caption[Radio-frequency spectroscopy of the minority component in
a strongly interacting mixture of fermionic
atoms.]{Radio-frequency spectroscopy of the minority component in
a strongly interacting mixture of fermionic atoms at an imbalance
of 0.9, clearly above the Chandrasekhar-Clogston limit of
superfluidity (see section~\ref{s:imbalance}). As the temperature is lowered,
the spectrum shows the transition from an ``atomic peak'' (A, with
some asymmetry and broadening) to an almost pure ``pairing'' peak
(D). Figure adapted from
Ref.~\cite{schu07pair}.}\label{f:minorityrf}
\end{center}
\end{figure}

Furthermore, using fermion mixtures with population imbalance it
was shown that RF spectra of the $\left|2\right> \rightarrow
\left|3\right>$ transition do not change as the gas undergoes the
superfluid to normal transition~\cite{schu07pair} (see
Fig.~\ref{f:minorityrf}).  The gas can be normal without any
``atomic peak''  in the RF spectrum, in contrast to earlier
interpretations of such ``pure'' pairing spectra.
The conclusion is that
RF spectra probe correlations and pairing only locally and, at the
current level of sensitivity, cannot distinguish between a normal
and a superfluid phase.

One important technical advance is the introduction of spatially
resolved RF ``tomography''~\cite{shin07rf} that allows the
reconstruction of local spectra free of inhomogeneous broadening
(see also section~\ref{s:tomographic}). For the resonantly
interacting superfluid, this method was used to demonstrate a true
frequency gap and a sharp onset of the spectrum at a frequency
shift corresponding to about 40\% of the Fermi energy.

Final state interactions are currently a major limitation in the
interpretation of RF spectra.  In lithium, for the resonant
superfluid in states $\left|1\right>$ and $\left|2\right>$ the
final state interaction between states $\left|1\right>$ and
$\left|3\right>$ has a large scattering length of $a_{13} = -3300
a_0$.  Recent results show that Fermi mixture initially in states
$\left|1\right>$ and $\left|3\right>$ provide clearer spectra,
presumably because the final state $\left|2\right>$ is less
strongly interacting with either state $\left|1\right>$ or
$\left|3\right>$ ($a_{23} = 1100 a_0$ and $a_{12} = 1400 a_0$ at
the $\left|1\right>$-$\left|3\right>$ resonance at $B = 690$ G)~\cite{schu07tbp}.

\subsection{New systems with BEC-BCS crossover}
\label{s:newsystems}
The field of physics stays vibrant by creating new systems to find
new phenomena.  Two major extensions of the BEC-BCS crossover in
an equal mixture of two fermionic states are the addition of
optical lattices and population imbalanced Fermi mixtures.

\subsubsection{Optical lattices}
Early studies of $^{40}$K in an optical lattice were carried
out at Z\"urich~\cite{kohl05fermilattice}.  The band structure and Fermi surfaces for
non-interacting fermions and interacting fermion mixtures were
observed and a normal conductor and band insulator were
realized. For a discussion of these experiments, see the contribution of T. Esslinger to these proceedings.

\begin{figure}[h]
\begin{center}
\includegraphics[width=3in]{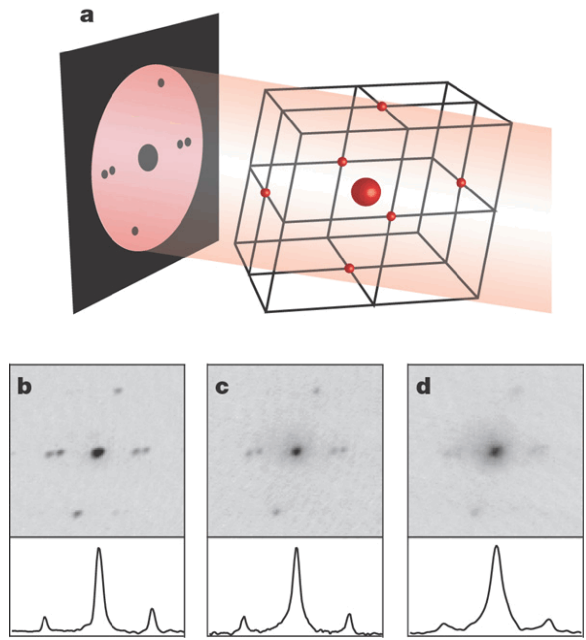}
\caption[Observation of high-contrast interference
of fermion pairs released from an optical lattice.]{Observation of high-contrast interference
of fermion pairs released from an optical lattice below and above
the Feshbach resonance. a, The orientation of the reciprocal
lattice, also with respect to the imaging light. b­d, Interference
peaks are observed for magnetic fields of 822 G (b), 867 G (c) and
917 G (d). The lattice depth for all images is 5 recoil energies.
The field of view is 1 mm.  Density profiles through the vertical
interference peaks are shown for each image. Figure reprinted from~\cite{chin06}.}\label{f:latticesuperfluid}
\end{center}
\end{figure}

Loading a superfluid fermion mixture into a weak optical lattice
should not destroy superfluidity. The only effect of the
lattice is to replace the bare mass by an effective mass. Evidence
for superfluid behavior was recently observed at MIT~\cite{chin06}.
When the fermionic cloud was released from the lattice, and a
rapid magnetic field sweep converted atom pairs into molecules,
the characteristic lattice interference pattern was observed, the
signature of long-range coherence providing indirect evidence for
superfluidity (Fig.~\ref{f:latticesuperfluid}).  Delayed rapid switching of the magnetic
field out of the strongly interacting region was necessary to
prevent collisions during ballistic expansion, which would have
destroyed the interference pattern.

For deeper lattice depths, the interference pattern disappeared.
This is analogous to the superfluid-to-Mott-insulator transition
in bosons, but now in the regime of strong interactions, which
will need a multi-band picture for its full description~\cite{duan05lattice}.

\subsubsection{Population imbalanced Fermi mixtures}
\label{s:imbalance}
The subject of superfluidity with population imbalanced Fermi
gases is almost as old as BCS theory itself.  Over the last three years, with
its realization in ultracold gases, it became a new frontier with
major theoretical and experimental activities, and would deserve a
review article on its own.  Here we can only summarize some basic
aspects of this rich system. For an extensive discussion of imbalanced gases see the contribution of F. Ch\'evy to these lectures notes.

Often, breaking a symmetry provides additional insight into a
system, even into its symmetric configuration.  Breaking the
equality of the two spin populations has given us new and very
direct ways to observe the superfluid phase transition, without
need for magnetic field sweeps~\cite{zwie06direct,shin06phase} (see Figs.~\ref{f:directresonance} and~\ref{f:directBEC} in section~\ref{s:directimbalance}, and Fig.~\ref{f:phasecontrast} in section~\ref{s:techniques}). In addition, it opened a way to measure the
absolute temperature for a strongly interacting system, using
the non-interacting wings of the majority component as a thermometer (see section~\ref{s:temperature}).  Scientifically, imbalance is another way (besides temperature, and scattering length) to probe how stable the
superfluid is.  In addition, in an imbalanced gas, there is no longer a smooth crossover between the BEC- and the BCS-regimes. Instead, a first-order transition takes place: If the attractive interactions become too weak, the
superfluid state becomes normal. This is in contrast to the population
balanced case, where superfluidity occurs for arbitrarily small
interactions. The stronger the imbalance, the smaller is the
window of superfluidity.

Or phrased differently, on the BCS side (and also on the BEC side
close to the resonance) there is a critical imbalance, beyond which superfluidity breaks down. This quantum phase transition~\footnote{The name ``quantum'' is deserved as it occurs at zero temperature. Sometimes, the term ``quantum phase transition'' is reserved for second order phase transitions.} is called the Chandrasekhar-Clogston limit, which we
will derive in the next paragraph.

Imbalance introduces a much richer phase diagram.  The quantum
phase transition at zero temperature continues as a first order
phase transition at finite temperature up to tricritical point,
where it becomes second order~\cite{pari07finitetemp}.  At high imbalance, the system
is normal even at zero temperature, and one can study a highly
correlated normal phase without the complications of thermal
excitations~\cite{schu07pair}.  Imbalance also offers new
opportunities to investigate pairing.  Using RF spectroscopy, our
group is currently studying differences in the RF spectrum of
the majority and minority component.  One question, which we
address, is whether the majority atoms show a bimodal spectrum
reflecting a paired and unpaired component.  This will occur in a
molecular picture far on the BEC side, but one may expect that on
the BCS side the distinction between unpaired and paired atoms is
blurred or vanishes.

\paragraph{Chandrasekhar-Clogston limit}
The Chandrasekhar-Clogston limit follows from a simple model.  A
two component Fermi gas can either be in a normal state with two
different Fermi energies for the two components, or in a
superfluid state that requires balanced populations. The
superfluid is stabilized by the condensation energy density
$\delta E_s$, which we will later set equal to the BCS result
$\frac{1}{2}\rho_F \Delta^2$ where $\Delta$ is the superfluid gap,
and $\rho_F \equiv \rho(E_F)/\Omega$ is the density of states (per spin state and per volume) at the
Fermi energy.
A balanced superfluid region can only be created by expelling
majority atoms which requires extra kinetic energy.  It is the
interplay between superfluid stabilization energy and kinetic
energy which determines the phase boundary.

We consider $N_{\uparrow}$ and $N_{\downarrow}$ fermions in a box
of volume $\Omega$.  Since the imbalance can only be accommodated
in the normal phase, we assign a volume fraction $\eta$ to it, and
the rest of the volume is superfluid.  The superfluid and
(average) normal densities, $n_s$ and $n_n$, may be different, and
are constrained by the constant total number of atoms
\begin{equation}
N=N_{\uparrow}+N_{\downarrow}=2 n_s (1-\eta)\,\Omega + 2 n_n\, \eta\, \Omega.
\end{equation}
The  energy density of the superfluid gas is
\begin{equation}
    E_s = 2 \frac{3}{5} E_F[n_s]\, n_s - \delta E_s
\end{equation}
where $E_F[n] \propto n^{2/3}$ is the Fermi energy of a Fermi gas at density $n$. For the normal phase we have
\begin{eqnarray}
E_n &=&\frac{3}{5} \left\{E_F\left[n_n+\frac{\Delta n}{\eta}\right]
\left(n_n+\frac{\Delta n}{\eta}\right) + E_F\left[n_n-\frac{\Delta n}{\eta}\right] \left(n_n-\frac{\Delta n}{\eta}\right)\right\} \\
&\approx& 2 \frac{3}{5} E_F[n_n]\, n_n + \frac{1}{\rho_F}\left(\frac{\Delta n}{\eta}\right)^2
\end{eqnarray}

where we have assumed that the density difference $\Delta
n=(N_{\uparrow}-N_{\downarrow})/2\Omega$ is much smaller than the
total average density per spin state
$n=(N_{\uparrow}+N_{\downarrow})/2\Omega$.

The total energy is minimized as a function of $\eta$ and
$n_s-n_n$. For $\eta=1$ the whole system becomes normal.  The
calculation is simplified if we introduce a Lagrangian multiplier $\mu$ to account for the constraint on the total number of atoms, and then minimize the total free energy $E_{tot} - \mu N$ with respect to $\eta$, $n_s$ and $n_n$.
\begin{eqnarray}
\label{e:totalenergy}
E_{tot} -\mu N&=& \left(2 \frac{3}{5}  E_F[n_s] n_s -\delta E_s\right) \Omega\, (1-\eta)\nonumber\\
&+&\left(2
\frac{3}{5} E_F[n_n] n_n + \frac{1}{\rho_F}\left(\frac{\Delta n}{\eta}\right)^2\right)\Omega\, \eta\nonumber\\
&-&\mu\; 2\Omega\, \left((1-\eta) n_s + \eta\, n_n\right)
\end{eqnarray}

Using $\frac{3}{5} E_F[n_s] n_s - \frac{3}{5} E_F[n_n] n_n = E_F[n_{\rm av}]
(n_s-n_m)$ where $n_{\rm av}$ is between $n_n$ and $n_s$, and
setting the $\eta$ derivative to zero, we obtain
\begin{equation}
\label{e:eta}
\eta^2 = \frac{(\Delta n)^2}{\rho_F \left\{\delta E_s+2 \left(\mu- E_F[n_{\rm av}]\right)\left(n_s
- n_n\right)\right\}}
\end{equation}

Setting the other two derivatives (with respect to $n_s$ and
$n_n$) of Eq.~\ref{e:totalenergy} to zero provides expressions for the density
difference and the chemical potential. They show that both
${\mu-E_F[n_{\rm av}]}$ and $n_s - n_n$ scale linearly with $\delta E_s$, i.e. that
the second term in the denominator of Eq.~\ref{e:eta} is second order in
$\delta E_s$ and negligible for weak interactions (BCS limit).
Substituting $\frac{1}{2}\rho_F \Delta^2$ for $\delta E_s$, one obtains for the
normal volume fraction
\begin{equation}
    \eta = \frac{\sqrt{2}\Delta n}{ \rho_F \Delta}
\end{equation}
This becomes one for $\Delta n = \rho_F \Delta / \sqrt{2}$, which
is the Chandrasekhar-Clogston limit for the superfluid to normal
quantum phase transition.

In case of superconductors, the number imbalance can be created by
a magnetic field $B$ (assuming that the Meissner effect is
suppressed, e.g. in heavy fermion or layered
superconductors~\cite{casa04}) $\Delta n = \rho_F \mu_B B$  which
leads to the Chandrasekhar-Clogston limit in its original form
$\mu_B B_c=\Delta/ \sqrt{2}$ where $B_c$ is the critical magnetic
field~\cite{chan62,clog62}.  Using Eq.~\ref{e:TCBCS}, one obtains
$B_{c} = 18 500\,{\rm G}\, \frac{T_C}{{\rm K}}$, much larger than
the thermodynamic critical field of a conventional superconductor,
which is $B_c = \sqrt{\mu_0 \rho_F} \Delta \approx 50\,{\rm
G}\,\frac{T_C}{{\rm K}}$. 
This shows that conventional superconductors will be quenched by
orbital effects of the magnetic field (Meissner effect and flux
quanta), and not by spin effects (Chandrasekhar-Clogston limit).

\begin{figure}[tb]
\begin{center}
\includegraphics[width=5.3in]{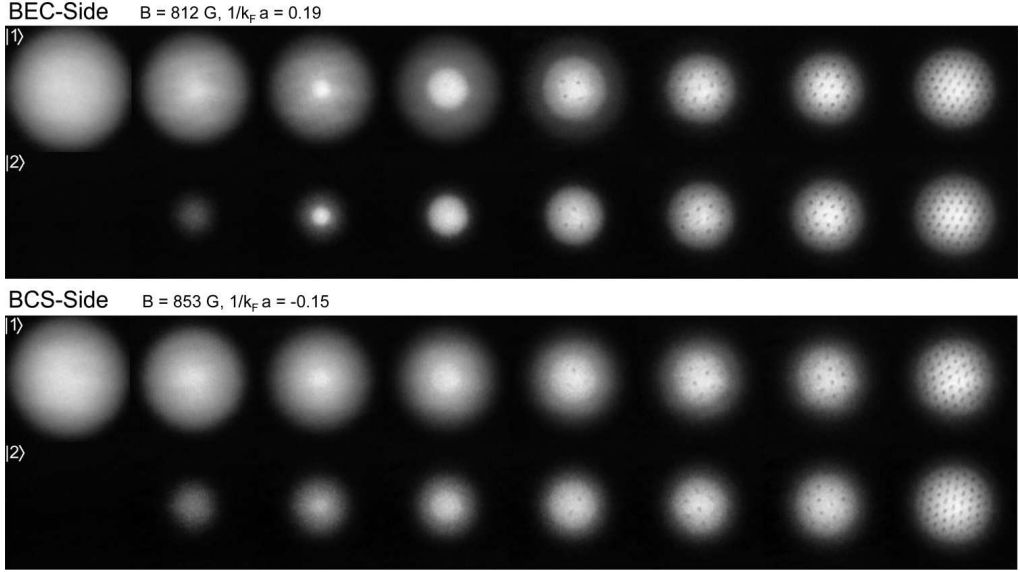}
\caption[Observation of vortices in a strongly interacting Fermi
gas with imbalanced spin populations.]{Observation of vortices in
a strongly interacting Fermi gas with imbalanced spin populations.
For the 812 G data, the population imbalance
$(N_\uparrow-N_\downarrow)/(N_\uparrow+N_\downarrow)$ was (from
left to right) 100, 90, 80, 62, 28, 18, 10 and 0\%. For the 853 G
data, the mismatch was 100, 74, 58, 48, 32, 16, 7 and 0\%.
From~\cite{zwie05imbalance}.}\label{f:imbalancedvortex}
\end{center}
\end{figure}

\paragraph{Experimental observations}

\begin{figure}[t]
\begin{center}
\includegraphics[width=3in]{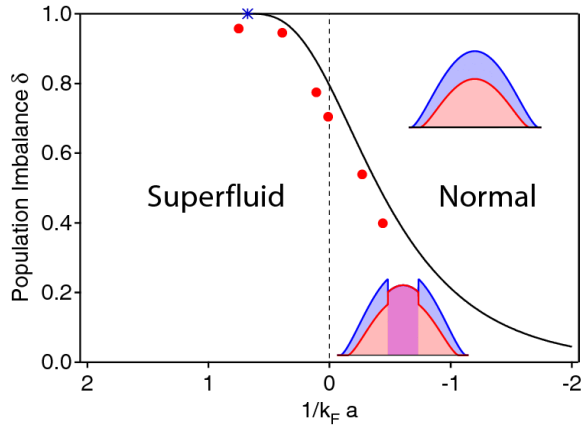}
\caption[Critical population
imbalance as a function of the interaction strength.]{Critical population
imbalance $\delta = (N_\uparrow-N_\downarrow)/(N_\uparrow+N_\downarrow)$ between the two spin states for which the
superfluid-to-normal transition is observed. The profiles indicate the distribution of the gas in the harmonic trap. Data from~\cite{zwie05imbalance}.}\label{f:imbalancephasediagram}
\end{center}
\end{figure}

The proof for the occurrence of superfluidity in imbalanced gases
was obtained, as in the balanced case, by the observation of
superfluid flow in the form of vortices
(Fig.~\ref{f:imbalancedvortex}).  Since vortices can be difficult
to create and observe near the phase boundaries, the superfluid
phase diagram has been mapped out by using pair condensation as an
indicator for superfluidity.  The phase diagram
(Fig.~\ref{f:imbalancephasediagram}) shows that on the BEC side,
the Chandrasekhar-Clogston limit approaches 100\%. So even if
there are only a few minority atoms in a majority Fermi sea, they
can form bosonic molecules and Bose-Einstein condense.  The
observed deviation from 100\% is probably due to finite
temperature.   On resonance,  the critical population imbalance
converged towards $\approx 70\%$ when the temperature was
varied~\cite{zwie05imbalance}.

\begin{figure}[h]
\begin{center}
\includegraphics[width=3in]{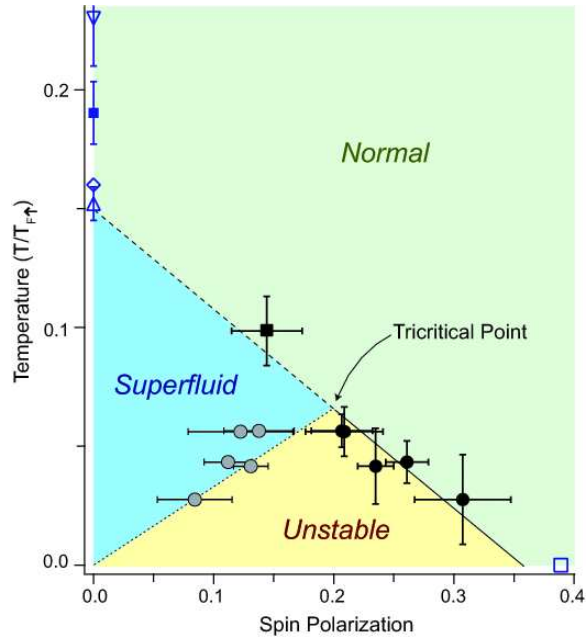}
\caption[ Phase diagram for a homogeneous spin-polarized Fermi
gas.]{Phase diagram for a homogeneous
spin-polarized Fermi gas with resonant interactions, as a function
of spin polarization and temperature.  Below the tricritical point
phase boundaries have been observed, characteristic for a first
order phase transition.  The polarization jumps from the value in
the superfluid (shown by the gray solid circles) to the higher
value in the normal phase (black solid circles). Above the
tricritical point, no abrupt change in spin polarization was
observed, and the phase transition in the center of the cloud was
determined from the onset of pair condensation (black square). The
blue open symbols show theoretical predictions for the critical
temperature of a homogeneous equal mixture (see
Ref.~\cite{shin07phasediagram} for a full
discussion).}\label{f:phasediagramtricrit}
\end{center}
\end{figure}

First hints for phase separation between the normal and superfluid
phase were seen in refs.~\cite{zwie05imbalance,part06phase}. Using
tomographic techniques, a sharp separation between a superfluid
core and a partially polarized normal phase was
found~\cite{shin06phase}. Finally, the phase diagram of a
spin-polarized Fermi gas at unitarity was obtained, by mapping out
the superfluid phase versus temperature and density
imbalance~\cite{shin07phasediagram}. Using tomographic techniques,
spatial discontinuities in the spin polarization were revealed,
the signature of a first-order phase transition that disappears at
a tricritical point (Fig.~\ref{f:phasediagramtricrit}). These
results are in excellent agreement with recent theoretical
predictions~\cite{lobo06,gubb07}. 
The Chandrasekhar-Clogston limit was not observed in the work at Rice, from which it was concluded that the critical imbalance on resonance was close to 100\%. The discrepancy to the MIT results is probably related to the
lower atom number and higher aspect ratio used in these
experiments~\cite{part06phase,part06deform}.

The phase diagram in Fig.~\ref{f:phasediagramtricrit} highlights
how far experimental studies on degenerate Fermi gases have
progressed since their first observation in 1999.  It is a rich
diagram including first and second order phase transitions, phase
separation and a tricritical point. It is expressed using only
local quantities (density, spin polarization, local Fermi
temperature) and therefore applies directly to the homogeneous
case. Experimentally, it required 3D spatial resolution using
tomographic techniques.  Finally, it is worth pointing out that the position of the
tricritical point could not be predicted and has even been
controversial until it had been experimentally determined.

\section{Conclusion}
\label{c:conclusion}

In this review paper, we have given a comprehensive description of
the concepts, methods and results involved in the exploration of the BEC-BCS
crossover with ultracold fermionic atoms.  Of course, many of the
techniques and concepts apply to other important areas where
experiments with ultracold fermions are conducted.

One area is atom optics and precision measurements.  The important
feature of fermions is the suppression of interactions in a
one-component gas due to Pauli blocking.  The Florence group
studied transport behavior of fermions in an optical lattice and
observed collisionally induced transport~\cite{ott04}, and conducting and insulating behavior of ideal
Fermi gases~\cite{pezz04}. The realization of
atom interferometry and observation of Bloch oscillation in a
Fermi gas~\cite{roat04} demonstrated the
potential of fermions for precision measurements unaffected by
atomic collisions~\cite{caru05}.

Another area is the study of mixtures of fermions with other species.
Bose-Fermi mixtures have been used to study how the addition of
fermionic atoms affects the bosonic superfluid-to-Mott-insulator
transition~\cite{ospe06,gunt06bosefermi}. Also intriguing is the study of
boson-mediated interactions between fermions~\cite{heis00,bijl00phonon}.  Interspecies
Feshbach resonances between fermionic and bosonic atoms have
already been identified~\cite{stan04,inou04}, and heteronuclear
molecules observed~\cite{ospe06hetero,zirb07hetero}. Mixtures between two fermionic
species (e.g. $^6$Li and $^{40}$K) may allow the study of
pairing and superfluidity where the pairs are made from atoms with
different masses~\cite{tagl08fermifermi,will07fermifermi}.

$p$-wave interactions provide a different way of pairing atoms, and
ultimately may connect the study of superfluidity in $^3$He
(which is based on $p$-wave pairing) with ultracold atoms. $p$-wave
Feshbach resonances have already been observed by various groups~\cite{rega03pwave,schu04fesh,zhan04pwave,gunt05pwave}, and $p$-wave molecules have been produced~\cite{gaeb07pwave}.

And finally, a whole new area is the study of fermions with
repulsive interactions in optical lattices.  At low temperature
and half filling, the system should be antiferromagnetic~\cite{hofs02hightc,wern05lattice}, and at lower filling, it may show d-wave superfluidity~\cite{hofs02hightc} and help to
elucidate the nature of pairing in high-temperature
superconductors. For a discussion of these effects, we refer to the contribution of A. Georges to these lecture notes.

With many new systems on the drawing board or already in
preparation, and with a flurry of theoretical papers predicting
new aspects of superfluidity and other novel phenomena, it seems
certain that we have exciting years ahead of us.

Work on ultracold fermions at MIT has been a tremendous team
effort, and we are grateful to the past and present collaborators
who have shared both the excitement and the hard work:  J.R.
Abo-Shaeer, J. K. Chin, K. Dieckmann, A. G\"orlitz, S. Gupta, Z.
Hadzibabic, A.J. Kerman, Y. Liu, D. E. Miller, S.M.F. Raupach, C.
Sanner, A. Schirotzek, C.H. Schunck, W. Setiawan, Y.-I. Shin, C. A.
Stan, and K. Xu.  We also acknowledge the fruitful interactions
with our colleagues in this rich and exciting
field, and we want to thank the organizers and participants of
the Varenna summer school for the stimulating atmosphere. We are
grateful to R. Gommers, A. Keshet, C. Sanner, A. Schirotzek, Y.-I. Shin, C.H.
Schunck and W. Zwerger for comments on the manuscript.

We want to thank the National Science Foundation, the Office of
Naval Research, NASA, the Army Research Office, DARPA,  and the
David and Lucile Packard Foundation for their encouragement and
financial support of this work during the past eight years.

\appendix
\bibliographystyle{varenna}
\bibliography{References} 

\begin{thebibliography}{100}

\bibitem{ande95}
{\sc Anderson, M.~H., Ensher, J.~R., Matthews, M.~R., Wieman, C.~E., and
  Cornell, E.~A.},
\newblock Observation of {Bose-Einstein} condensation in a dilute atomic vapor,
\newblock {\em Science, {\bf 269}\/} (1995) 198--201.

\bibitem{davi95bec}
{\sc Davis, K.~B., Mewes, M.-O., Andrews, M.~R., van Druten, N.~J., Durfee,
  D.~S., Kurn, D.~M., and Ketterle, W.},
\newblock {Bose-Einstein} condensation in a gas of sodium atoms,
\newblock {\em Phys. Rev. Lett., {\bf 75}\/} (1995) 3969--3973.

\bibitem{grei01mott}
{\sc Greiner, M., Mandel, O., Esslinger, T., Hänsch, T.~W., and Bloch, I.},
\newblock Quantum phase transition from a superfluid to a mott insulator in a
  gas of ultracold atoms,
\newblock {\em Nature, {\bf 415}\/} (2002) 39--44.

\bibitem{bloc05lattice}
{\sc Bloch, I.},
\newblock Ultracold quantum gases in optical lattices,
\newblock {\em Nature Physics, {\bf 1}\/} (2005) 23.

\bibitem{stwa76}
{\sc Stwalley, W.~C.},
\newblock Stability of spin-aligned hydrogen at low temperatures and high
  magnetic fields: New field-dependent scattering resonances and
  predissociations,
\newblock {\em Phys. Rev. Lett., {\bf 37}\/} (1976) 1628.

\bibitem{ties93}
{\sc Tiesinga, E., Verhaar, B.~J., and Stoof, H. T.~C.},
\newblock Threshold and resonance phenomena in ultracold ground-state
  collisions,
\newblock {\em Phys. Rev. A, {\bf 47}\/} (1993) 4114--4122.

\bibitem{inou98}
{\sc Inouye, S., Andrews, M.~R., Stenger, J., Miesner, H.-J., Stamper-Kurn,
  D.~M., and Ketterle, W.},
\newblock Observation of {F}eshbach resonances in a {Bose-Einstein} condensate,
\newblock {\em Nature, {\bf 392}\/} (1998) 151--154.

\bibitem{cour98fesh}
{\sc Courteille, P., Freeland, R.~S., Heinzen, D.~J., van Abeelen, F.~A., and
  Verhaar, B.~J.},
\newblock Observation of a {F}eshbach resonance in cold atom scattering,
\newblock {\em Phys. Rev. Lett., {\bf 81}\/} (1998) 69--72.

\bibitem{kett99var}
{\sc Ketterle, W., Durfee, D.~S., and Stamper-Kurn, D.~M.},
\newblock Making, probing and understanding {Bose-Einstein} condensates,
\newblock in {\em {Bose-Einstein} condensation in atomic gases, Proceedings of
  the International School of Physics Enrico {F}ermi, Course CXL, Varenna, 7-17
  July 1998}, edited by M.~Inguscio, S.~Stringari, and C.~Wieman, (IOS Press,
  Amsterdam) 1999, pp.~67--176.

\bibitem{sten98stro}
{\sc Stenger, J., Inouye, S., Andrews, M.~R., Miesner, H.-J., Stamper-Kurn,
  D.~M., and Ketterle, W.},
\newblock Strongly enhanced inelastic collisions in a {Bose-Einstein}
  condensate near {F}eshbach resonances,
\newblock {\em Phys. Rev. Lett., {\bf 82}\/} (1999) 2422--2425.

\bibitem{corn00JILA_collapse}
{\sc Cornish, S.~L., Claussen, N.~R., Roberts, J.~L., Cornell, E.~A., and
  Wieman, C.~E.},
\newblock Stable $^{85}${R}b {Bose-Einstein} condensates with widely tunable
  interactions,
\newblock {\em Phys. Rev. Lett., {\bf 85}\/} (2000) 1795--1798.

\bibitem{abra97}
{\sc Abraham, E. R.~I., McAlexander, W.~I., Gerton, J.~M., Hulet, R.~G., Côté,
  R., and Dalgarno, A.},
\newblock Triplet s-wave resonance in $^6${L}i collisions and scattering
  lengths of $^6${L}i and $^7${L}i,
\newblock {\em Phys. Rev. A, {\bf 55}\/} (1997) R3299.

\bibitem{stoo96superfluid}
{\sc Stoof, H. T.~C., Houbiers, M., Sackett, C.~A., and Hulet, R.~G.},
\newblock Superfluidity of spin-polarized $^{6}${L}i,
\newblock {\em Phys. Rev. Lett., {\bf 76}\/} (1996) 10.

\bibitem{houb98}
{\sc Houbiers, M., Stoof, H. T.~C., McAlexander, W.~I., and Hulet, R.~G.},
\newblock Elastic and inelastic collisions of $^6${L}i atoms in magnetic and
  optical traps,
\newblock {\em Phys. Rev. A, {\bf 57}\/} (1998) R1497.

\bibitem{cubi03}
{\sc Cubizolles, J., Bourdel, T., Kokkelmans, S. J. J. M.~F., Shlyapnikov,
  G.~V., and Salomon, C.},
\newblock Production of long-lived ultracold {Li$_2$} molecules from a {F}ermi
  gas,
\newblock {\em Phys. Rev. Lett., {\bf 91}\/} (2003) 240401.

\bibitem{stre03}
{\sc Strecker, K.~E., Partridge, G.~B., and Hulet, R.~G.},
\newblock Conversion of an atomic {F}ermi gas to a long-lived molecular {B}ose
  gas,
\newblock {\em Phys. Rev. Lett., {\bf 91}\/} (2003) 080406.

\bibitem{joch03lith}
{\sc Jochim, S., Bartenstein, M., Altmeyer, A., Hendl, G., Chin, C.,
  Hecker-Denschlag, J., and Grimm, R.},
\newblock Pure gas of optically trapped molecules created from fermionic atoms,
\newblock {\em Phys. Rev. Lett., {\bf 91}\/} (2003) 240402.

\bibitem{rega03lifetime}
{\sc Regal, C.~A., Greiner, M., and Jin, D.~S.},
\newblock Lifetime of molecule-atom mixtures near a {F}eshbach resonance in
  $^{40}${K},
\newblock {\em Phys. Rev. Lett., {\bf 92}\/} (2004) 083201.

\bibitem{petr03three_body}
{\sc Petrov, D.~S.},
\newblock Three-body problem in {F}ermi gases with short-range interparticle
  interaction,
\newblock {\em Phys. Rev. A, {\bf 67}\/} (2003) 010703.

\bibitem{stoo99var}
{\sc Stoof, H. T.~C., and Houbiers, M.},
\newblock Condensed matter physics with trapped atomic {F}ermi gases,
\newblock in {\em {Bose-Einstein} condensation in atomic gases, Proceedings of
  the International School of Physics Enrico {F}ermi, Course CXL, Varenna, 7-17
  July 1998}, edited by M.~Inguscio, S.~Stringari, and C.~Wieman, (IOS Press,
  Amsterdam) 1999, pp.~537--553.

\bibitem{houb97}
{\sc Houbiers, M., Ferwerda, R., Stoof, H. T.~C., McAlexander, W.~I., Sackett,
  C.~A., and Hulet, R.~G.},
\newblock Superfluid state of atomic $^6${L}i in a magnetic trap,
\newblock {\em Phys. Rev. A, {\bf 56}\/} (1997) 4864.

\bibitem{zhan99}
{\sc Zhang, W., Sackett, C.~A., and Hulet, R.~G.},
\newblock Optical detection of a {Bardeen-Cooper-Schrieffer} phase transition
  in a trapped gas of fermionic atoms,
\newblock {\em Phys. Rev. A, {\bf 60}\/} (1999) 504.

\bibitem{weig00}
{\sc Weig, F., and Zwerger, W.},
\newblock Optical detection of a {BCS} transition of lithium--6 in harmonic
  traps,
\newblock {\em Europhys. Lett., {\bf 49}\/} (2000) 282.

\bibitem{torm00}
{\sc Törmä, P., and Zoller, P.},
\newblock Laser probing of atomic {C}ooper pairs,
\newblock {\em Phys. Rev. Lett., {\bf 85}\/} (2000) 487.

\bibitem{bara00}
{\sc Baranov, M.~A., and Petrov, D.~S.},
\newblock Low-energy collective excitations in a superfluid trapped {F}ermi
  gas,
\newblock {\em Phys. Rev. A, {\bf 62}\/} (2000) 041601(R).

\bibitem{ming01}
{\sc Minguzzi, A., and Tosi, M.~P.},
\newblock Scissors mode in a superfluid {Fermi} gas,
\newblock {\em Phys. Rev. A, {\bf 63}\/} (2001) 023609.

\bibitem{chio02}
{\sc Chiofalo, M.~L., Kokkelmans, S. J. J. M.~F., Milstein, J.~N., and Holland,
  M.~J.},
\newblock Signatures of resonance superfluidity in a quantum {F}ermi gas,
\newblock {\em Phys. Rev. Lett., {\bf 88}\/} (2002) 090402.

\bibitem{onne13}
{\sc Onnes, K.},
\newblock Investigations into the properties of substances at low temperatures,
  which have led, amongst other things, to the preparation of liquid helium,
  nobel lecture, december 13, 1913,
\newblock in {\em Nobel Lectures, Physics 1901-1921}, (Elsevier Publishing
  Company, Amsterdam) 1967.

\bibitem{scha58}
{\sc Schafroth, M.~R.},
\newblock Remarks on the {M}eissner effect,
\newblock {\em Phys. Rev., {\bf 111}\/} (1958) 72.

\bibitem{reyn50}
{\sc Reynolds, C.~A., Serin, B., Wright, W.~H., and Nesbitt, L.~B.},
\newblock Superconductivity of isotopes of mercury,
\newblock {\em Phys. Rev., {\bf 78}\/} (1950) 487.

\bibitem{maxw50}
{\sc Maxwell, E.},
\newblock Isotope effect in the superconductivity of mercury,
\newblock {\em Phys. Rev., {\bf 78}\/} (1950) 477.

\bibitem{froh50}
{\sc Fröhlich, H.},
\newblock Theory of the superconducting state. {I}. the ground state at the
  absolute zero of temperature,
\newblock {\em Phys. Rev., {\bf 79}\/} (1950) 845.

\bibitem{coop56}
{\sc Cooper, L.~N.},
\newblock Bound electron pairs in a degenerate {F}ermi gas,
\newblock {\em Phys. Rev., {\bf 104}\/} (1956) {1189--1190}.

\bibitem{bard57}
{\sc Bardeen, J., {C}ooper, L.~N., and Schrieffer, J.~R.},
\newblock Theory of superconductivity,
\newblock {\em Phys. Rev., {\bf 108}\/} (1957) 1175.

\bibitem{lond64}
{\sc London, F.},
\newblock {\em Superfluids, Vol. {II}},
\newblock (Dover, New York) 1964.

\bibitem{schr99super}
{\sc Schrieffer, J.~R.},
\newblock {\em Theory of Superconductivity},
\newblock (Westview Press, Boulder, Colorado) 1999.

\bibitem{popo66}
{\sc Popov, V.~N.},
\newblock Theory of a {Bose} gas produced by bound states of {Fermi} particles,
\newblock {\em Zh. Eksp. Teor. Fiz., {\bf 50}\/} (1966) 1550,
\newblock [Sov. Phys. JETP 23, 1034 (1968)].

\bibitem{keld68}
{\sc Keldysh, L.~V., and Kozlov, A.~N.},
\newblock Collective properties of excitons in semiconductors,
\newblock {\em Zh. Eksp. Teor. Fiz., {\bf 54}\/} (1968) 978,
\newblock [Sov. Phys. JETP 27, 521 (1968)].

\bibitem{eagl69}
{\sc Eagles, D.~M.},
\newblock Possible pairing without superconductivity at low carrier
  concentrations in bulk and thin-film superconducting semiconductors,
\newblock {\em Phys. Rev., {\bf 186}\/} (1969) 456.

\bibitem{legg80}
{\sc Leggett, A.~J.},
\newblock Diatomic molecules and {C}ooper pairs,
\newblock in {\em Modern Trends in the Theory of Condensed Matter. Proceedings
  of the XVIth Karpacz Winter School of Theoretical Physics, Karpacz, Poland},
  (Springer-Verlag, Berlin) 1980, pp.~13--27.

\bibitem{nozi85}
{\sc Nozières, P., and Schmitt-Rink, S.},
\newblock {B}ose condensation in an attractive fermion gas: from weak to strong
  coupling superconductivity,
\newblock {\em J. Low Temp. Phys., {\bf 59}\/} (1985) 195--211.

\bibitem{ginz91}
{\sc Ginzburg, V.~L.},
\newblock High-temperature superconductivity (history and general review),
\newblock {\em Soviet Physics Uspekhi, {\bf 34}\/} (1991) 283--288.

\bibitem{bedn86hightc}
{\sc Bednorz, J.~G., and Müller, K.~A.},
\newblock Possible high-{$T_c$} superconductivity in the {Ba-La-Cu-O} system,
\newblock {\em Z. Phys. B, {\bf 64}\/} (1986) 189.

\bibitem{lee06hightc}
{\sc Lee, P.~A., Nagaosa, N., and Wen, X.-G.},
\newblock Doping a {Mott} insulator: Physics of high-temperature
  superconductivity,
\newblock {\em Rev. Mod. Phys., {\bf 78}\/} (2006) 17.

\bibitem{ho04uni}
{\sc Ho, T.-L.},
\newblock Universal thermodynamics of degenerate quantum gases in the unitarity
  limit,
\newblock {\em Phys. Rev. Lett., {\bf 92}\/} (2004) 090402.

\bibitem{myat97}
{\sc Myatt, C.~J., Burt, E.~A., Ghrist, R.~W., Cornell, E.~A., and Wieman,
  C.~E.},
\newblock Production of two overlapping {Bose-Einstein} condensates by
  sympathetic cooling,
\newblock {\em Phys. Rev. Lett., {\bf 78}\/} (1997) 586--589.

\bibitem{bloc01symp}
{\sc Bloch, I., Greiner, M., Mandel, O., Hänsch, T.~W., and Esslinger, T.},
\newblock Sympathetic cooling of $^{85}${R}b and $^{87}${R}b,
\newblock {\em Phys. Rev. A, {\bf 64}\/} (2001) 021402--4.

\bibitem{stam98odt}
{\sc Stamper-Kurn, D.~M., Andrews, M.~R., Chikkatur, A.~P., Inouye, S.,
  Miesner, H.-J., Stenger, J., and Ketterle, W.},
\newblock Optical confinement of a {Bose-Einstein} condensate,
\newblock {\em Phys. Rev. Lett., {\bf 80}\/} (1998) 2027--2030.

\bibitem{dema99}
{\sc DeMarco, B., and Jin, D.~S.},
\newblock Onset of {F}ermi degeneracy in a trapped atomic gas,
\newblock {\em Science, {\bf 285}\/} (1999) 1703--1706.

\bibitem{roat02}
{\sc Roati, G., Riboli, F., Modugno, G., and Inguscio, M.},
\newblock {F}ermi-{B}ose quantum degenerate $^{40}${K} - $^{87}${R}b mixture
  with attractive interaction,
\newblock {\em Phys. Rev. Lett., {\bf 89}\/} (2002) 150403.

\bibitem{trus01}
{\sc Truscott, A.~G., Strecker, K.~E., McAlexander, W.~I., Partridge, G.~B.,
  and Hulet, R.~G.},
\newblock Observation of {F}ermi pressure in a gas of trapped atoms,
\newblock {\em Science, {\bf 291}\/} (2001) 2570--2572.

\bibitem{schr01}
{\sc Schreck, F., Khaykovich, L., Corwin, K.~L., Ferrari, G., Bourdel, T.,
  Cubizolles, J., and Salomon, C.},
\newblock Quasipure {Bose-Einstein} condensate immersed in a {F}ermi sea,
\newblock {\em Phys. Rev. Lett., {\bf 87}\/} (2001) 080403.

\bibitem{gran02}
{\sc Granade, S.~R., Gehm, M.~E., O'Hara, K.~M., and Thomas, J.~E.},
\newblock All-optical production of a degenerate {F}ermi gas,
\newblock {\em Phys. Rev. Lett., {\bf 88}\/} (2002) 120405.

\bibitem{hadz02}
{\sc Hadzibabic, Z., Stan, C.~A., Dieckmann, K., Gupta, S., Zwierlein, M.~W.,
  Görlitz, A., and Ketterle, W.},
\newblock Two species mixture of quantum degenerate {B}ose and {F}ermi gases,
\newblock {\em Phys. Rev. Lett., {\bf 88}\/} (2002) 160401.

\bibitem{joch03bec}
{\sc Jochim, S., Bartenstein, M., Altmeyer, A., Hendl, G., Riedl, S., Chin, C.,
  Hecker-Denschlag, J., and Grimm, R.},
\newblock {Bose-Einstein} condensation of molecules,
\newblock {\em Science, {\bf 302}\/} (2003) 2101.

\bibitem{diec02fesh}
{\sc Dieckmann, K., Stan, C.~A., Gupta, S., Hadzibabic, Z., Schunck, C., and
  Ketterle, W.},
\newblock Decay of an ultracold fermionic lithium gas near a {F}eshbach
  resonance,
\newblock {\em Phys. Rev. Lett., {\bf 89}\/} (2002) 203201.

\bibitem{loft02}
{\sc Loftus, T., Regal, C.~A., Ticknor, C., Bohn, J.~L., and Jin, D.~S.},
\newblock Resonant control of elastic collisions in an optically trapped
  {F}ermi gas of atoms,
\newblock {\em Phys. Rev. Lett., {\bf 88}\/} (2002) 173201.

\bibitem{ohar02}
{\sc O'Hara, K.~M., Hemmer, S.~L., Granade, S.~R., Gehm, M.~E., Thomas, J.~E.,
  V.Venturi, Tiesinga, E., and Williams, C.~J.},
\newblock Measurement of the zero crossing in a {F}eshbach resonance of
  fermionic $^6${L}i,
\newblock {\em Phys. Rev. A, {\bf 66}\/} (2002) 041401(R).

\bibitem{joch02}
{\sc Jochim, S., Bartenstein, M., Hendl, G., Hecker-Denschlag, J., Grimm, R.,
  Mosk, A., and Weidemüller, M.},
\newblock Magnetic field control of elastic scattering in a cold gas of
  fermionic lithium atoms,
\newblock {\em Phys. Rev. Lett., {\bf 89}\/} (2002) 273202.

\bibitem{ohar02science}
{\sc O'Hara, K.~M., Hemmer, S.~L., Gehm, M.~E., Granade, S.~R., and Thomas,
  J.~E.},
\newblock Observation of a strongly interacting degenerate {F}ermi gas of
  atoms,
\newblock {\em Science, {\bf 298}\/} (2002) 2179.

\bibitem{rega03mol}
{\sc Regal, C.~A., Ticknor, C., Bohn, J.~L., and Jin, D.~S.},
\newblock Creation of ultracold molecules from a {F}ermi gas of atoms,
\newblock {\em Nature, {\bf 424}\/} (2003) 47.

\bibitem{herb03cs_mol}
{\sc Herbig, J., Kraemer, T., Mark, M., Weber, T., Chin, C., Nägerl, H.-C., and
  Grimm, R.},
\newblock Preparation of a pure molecular quantum gas,
\newblock {\em Science, {\bf 301}\/} (2003) 1510.

\bibitem{xu03na_mol}
{\sc Xu, K., Mukaiyama, T., Abo-Shaeer, J.~R., Chin, J.~K., Miller, D.~E., and
  Ketterle, W.},
\newblock Formation of quantum-degenerate sodium molecules,
\newblock {\em Phys. Rev. Lett., {\bf 91}\/} (2003) 210402.

\bibitem{durr04mol}
{\sc Dürr, S., Volz, T., Marte, A., and Rempe, G.},
\newblock Observation of molecules produced from a {Bose-Einstein} condensate,
\newblock {\em Phys. Rev. Lett., {\bf 92}\/} (2004) 020406.

\bibitem{grei03molbec}
{\sc Greiner, M., Regal, C.~A., and Jin, D.~S.},
\newblock Emergence of a molecular {Bose{-}Einstein} condensate from a {F}ermi
  gas,
\newblock {\em Nature, {\bf 426}\/} (2003) 537.

\bibitem{zwie03molBEC}
{\sc Zwierlein, M.~W., Stan, C.~A., Schunck, C.~H., Raupach, S. M.~F., Gupta,
  S., Hadzibabic, Z., and Ketterle, W.},
\newblock Observation of {Bose-Einstein} condensation of molecules,
\newblock {\em Phys. Rev. Lett., {\bf 91}\/} (2003) 250401.

\bibitem{bart04}
{\sc Bartenstein, M., Altmeyer, A., Riedl, S., Jochim, S., Chin, C.,
  {Hecker-Denschlag}, J., and Grimm, R.},
\newblock Crossover from a molecular {Bose-Einstein} condensate to a degenerate
  {F}ermi gas,
\newblock {\em Phys. Rev. Lett., {\bf 92}\/} (2004) 120401.

\bibitem{zwie05vort}
{\sc Zwierlein, M.~W., Abo-Shaeer, J.~R., Schirotzek, A., Schunck, C.~H., and
  Ketterle, W.},
\newblock Vortices and superfluidity in a strongly interacting {F}ermi gas,
\newblock {\em Nature, {\bf 435}\/} (2005) 1047--1051.

\bibitem{rega04}
{\sc Regal, C.~A., Greiner, M., and Jin, D.~S.},
\newblock Observation of resonance condensation of fermionic atom pairs,
\newblock {\em Phys. Rev. Lett., {\bf 92}\/} (2004) 040403.

\bibitem{zwie04rescond}
{\sc Zwierlein, M.~W., Stan, C.~A., Schunck, C.~H., Raupach, S. M.~F., Kerman,
  A.~J., and Ketterle, W.},
\newblock Condensation of pairs of fermionic atoms near a {F}eshbach resonance,
\newblock {\em Phys. Rev. Lett., {\bf 92}\/} (2004) 120403.

\bibitem{bour04coll}
{\sc Bourdel, T., Khaykovich, L., Cubizolles, J., Zhang, J., Chevy, F.,
  Teichmann, M., Tarruell, L., Kokkelmans, S. J. J. M.~F., and Salomon, C.},
\newblock Experimental study of the {BEC}-{BCS} crossover region in lithium-6,
\newblock {\em Phys. Rev. Lett., {\bf 93}\/} (2004) 050401.

\bibitem{kina05heat}
{\sc Kinast, J., Turlapov, A., Thomas, J.~E., Chen, Q., Stajic, J., and Levin,
  K.},
\newblock Heat capacity of a strongly-interacting {F}ermi gas,
\newblock {\em Science, {\bf 307}\/} (2005) 1296--1299.

\bibitem{kina04sfluid}
{\sc Kinast, J., Hemmer, S.~L., Gehm, M.~E., Turlapov, A., and Thomas, J.~E.},
\newblock Evidence for superfluidity in a resonantly interacting {F}ermi gas,
\newblock {\em Phys. Rev. Lett., {\bf 92}\/} (2004) 150402.

\bibitem{bart04coll}
{\sc Bartenstein, M., Altmeyer, A., Riedl, S., Jochim, S., Chin, C.,
  Hecker-Denschlag, J., and Grimm, R.},
\newblock Collective excitations of a degenerate gas at the {BEC}-{BCS}
  crossover,
\newblock {\em Phys. Rev. Lett., {\bf 92}\/} (2004) 203201.

\bibitem{chin04gap}
{\sc Chin, C., Bartenstein, M., Altmeyer, A., Riedl, S., Jochim, S.,
  Hecker-Denschlag, J., and Grimm, R.},
\newblock Observation of the pairing gap in a strongly interacting {F}ermi gas,
\newblock {\em Science, {\bf 305}\/} (2004) 1128.

\bibitem{part05}
{\sc Partridge, G.~B., Strecker, K.~E., Kamar, R.~I., Jack, M.~W., and Hulet,
  R.~G.},
\newblock Molecular probe of pairing in the {BEC}-{BCS} crossover,
\newblock {\em Phys. Rev. Lett., {\bf 95}\/} (2005) 020404.

\bibitem{schu07pair}
{\sc Schunck, C.~H., Shin, Y., Schirotzek, A., Zwierlein, M.~W., and Ketterle,
  W.},
\newblock Pairing without superfluidity: The ground state of an imbalanced
  {Fermi} mixture,
\newblock {\em Science, {\bf 316}\/} (2007) 867.

\bibitem{zwie05imbalance}
{\sc Zwierlein, M.~W., Schirotzek, A., Schunck, C.~H., and Ketterle, W.},
\newblock Fermionic superfluidity with imbalanced spin populations,
\newblock {\em Science, {\bf 311}\/} (2006) 492--496.

\bibitem{part06phase}
{\sc Partridge, G.~B., Li, W., Kamar, R.~I., Liao, Y., and Hulet, R.~G.},
\newblock Pairing and phase separation in a polarized {F}ermi gas,
\newblock {\em Science, {\bf 311}\/} (2006) 503.

\bibitem{shin06phase}
{\sc Shin, Y., Zwierlein, M.~W., Schunck, C.~H., Schirotzek, A., and Ketterle,
  W.},
\newblock Observation of phase separation in a strongly interacting imbalanced
  {F}ermi gas,
\newblock {\em Phys. Rev. Lett., {\bf 97}\/} (2006) 030401.

\bibitem{part06deform}
{\sc Partridge, G.~B., Li, W., Liao, Y., Hulet, R.~G., Haque, M., and Stoof, H.
  T.~C.},
\newblock Deformation of a trapped {Fermi} gas with unequal spin populations,
\newblock {\em Phys. Rev. Lett., {\bf 97}\/} (2006) 190407.

\bibitem{shin07phasediagram}
{\sc Shin, Y., Schunck, C.~H., Schirotzek, A., and Ketterle, W.},
\newblock Phase diagram of a two-component {Fermi} gas with resonant
  interactions,
\newblock Nature, in print; preprint arXiv:0709.3027.

\bibitem{chin06}
{\sc Chin, J.~K., Miller, D.~E., Liu, Y., Stan, C., Setiawan, W., Sanner, C.,
  Xu, K., and Ketterle, W.},
\newblock Evidence for superfluidity of ultracold fermions in an optical
  lattice,
\newblock {\em Nature, {\bf 443}\/} (2006) 961--964.

\bibitem{jose07sound}
{\sc Joseph, J., Clancy, B., Luo, L., Kinast, J., Turlapov, A., and Thomas,
  J.~E.},
\newblock Measurement of sound velocity in a {Fermi} gas near a {Feshbach}
  resonance,
\newblock {\em Phys. Rev. Lett., {\bf 98}\/} (2007) 170401.

\bibitem{mill07critical}
{\sc Miller, D.~E., Chin, J.~K., Stan, C.~A., Liu, Y., Setiawan, W., Sanner,
  C., and Ketterle, W.},
\newblock Critical velocity for superfluid flow across the {BEC-BCS} crossover,
\newblock {\em Phys. Rev. Lett., {\bf 99}\/} (2007) 070402.

\bibitem{gaeb07pwave}
{\sc Gaebler, J.~P., Stewart, J.~T., Bohn, J.~L., and Jin, D.~S.},
\newblock p-wave {Feshbach} molecules,
\newblock {\em Phys. Rev. Lett., {\bf 98}\/} (2007) 200403.

\bibitem{rom06}
{\sc Rom, T., Best, T., v.~Oosten, D., Schneider, U., Fölling, S., Paredes, B.,
  and Bloch, I.},
\newblock Free fermion antibunching in a degenerate atomic {Fermi} gas released
  from an optical lattice,
\newblock {\em Nature, {\bf 444}\/} (2006) 733--736.

\bibitem{tuor07lithiumambient}
{\sc Tuoriniemi, J., Juntunen-Nurmilaukas, K., Uusvuori, J., Pentti, E.,
  Salmela, A., and Sebedash, A.},
\newblock Superconductivity in lithium below 0.4 millikelvin at ambient
  pressure,
\newblock {\em Nature, {\bf 447}\/} (2007) 187--189.

\bibitem{scha07b}
{\sc Schäfer, T.},
\newblock What atomic liquids can teach us about quark liquids,
\newblock preprint hep-ph/0703141.

\bibitem{back05Phobos}
{\sc Back, B.~B.},
\newblock {\it et al.}, {The} {PHOBOS} perspective on discoveries at {RHIC},
\newblock {\em Nuclear Physics A, {\bf 757}\/} (2005) 28--101.

\bibitem{kina05damping}
{\sc Kinast, J., Turlapov, A., and Thomas, J.~E.},
\newblock Damping of a unitary {Fermi} gas,
\newblock {\em Phys. Rev. Lett., {\bf 94}\/} (2005) 170404.

\bibitem{kovt05visc}
{\sc Kovtun, P.~K., Son, D.~T., and Starinets, A.~O.},
\newblock Viscosity in strongly interacting quantum field theories from black
  hole physics,
\newblock {\em Phys. Rev. Lett., {\bf 94}\/} (2005) 111601.

\bibitem{scha07visc}
{\sc Schäfer, T.},
\newblock Ratio of shear viscosity to entropy density for trapped fermions in
  the unitarity limit,
\newblock {\em Phys. Rev. A, {\bf 76}\/} (2007) 063618--5.

\bibitem{casa04}
{\sc Casalbuoni, R., and Nardulli, G.},
\newblock Inhomogeneous superconductivity in condensed matter and {QCD},
\newblock {\em Rev. Mod. Phys., {\bf 76}\/} (2004) 263,
\newblock and references therein.

\bibitem{alfo01}
{\sc Alford, M.},
\newblock Color-superconducting quark matter,
\newblock {\em Annual Review of Nuclear and Particle Science, {\bf 51}\/}
  (2001) 131--160.

\bibitem{bowe02}
{\sc Bowers, J.~A., and Rajagopal, K.},
\newblock Crystallography of color superconductivity,
\newblock {\em Phys. Rev. D, {\bf 66}\/} (2002) 065002.

\bibitem{mcna06helium}
{\sc McNamara, J.~M., Jeltes, T., Tychkov, A.~S., Hogervorst, W., and Vassen,
  W.},
\newblock Degenerate {Bose-Fermi} mixture of metastable atoms,
\newblock {\em Phys. Rev. Lett., {\bf 97}\/} (2006) 080404.

\bibitem{fuku07ytterbium}
{\sc Fukuhara, T., Takasu, Y., Kumakura, M., and Takahashi, Y.},
\newblock Degenerate {Fermi} gases of {Ytterbium},
\newblock {\em Phys. Rev. Lett., {\bf 98}\/} (2007) 030401.

\bibitem{schu07tbp}
{\sc Schunck, C.~H., Schirotzek, A., Shin, Y.-I., and Ketterle, W.},
\newblock (to be published).

\bibitem{hadz03big_fermi}
{\sc Hadzibabic, Z., Gupta, S., Stan, C.~A., Schunck, C.~H., Zwierlein, M.~W.,
  Dieckmann, K., and Ketterle, W.},
\newblock Fifty-fold improvement in the number of quantum degenerate fermionic
  atoms,
\newblock {\em Phys. Rev. Lett., {\bf 91}\/} (2003) 160401.

\bibitem{inou04}
{\sc Inouye, S., Goldwin, J., Olsen, M.~L., Ticknor, C., Bohn, J.~L., and Jin,
  D.~S.},
\newblock Observation of heteronuclear {F}eshbach resonances in a mixture of
  bosons and fermions,
\newblock {\em Phys. Rev. Lett., {\bf 93}\/} (2004) 183201.

\bibitem{kohl05fermilattice}
{\sc Köhl, M., Moritz, H., Stöferle, T., Günter, K., and Esslinger, T.},
\newblock Fermionic atoms in a three dimensional optical lattice: Observing
  {F}ermi surfaces, dynamics, and interactions,
\newblock {\em Phys. Rev. Lett., {\bf 94}\/} (2005) 080403.

\bibitem{ospe06hetero}
{\sc Ospelkaus, C., Ospelkaus, S., Humbert, L., Ernst, P., Sengstock, K., and
  Bongs, K.},
\newblock Ultracold heteronuclear molecules in a {3D} optical lattice,
\newblock {\em Phys. Rev. Lett., {\bf 97}\/} (2006) 120402.

\bibitem{silb05}
{\sc Silber, C., Günther, S., Marzok, C., Deh, B., Courteille, P.~W., and
  Zimmermann, C.},
\newblock Quantum-degenerate mixture of fermionic lithium and bosonic rubidium
  gases,
\newblock {\em Phys. Rev. Lett., {\bf 95}\/} (2005) 170408.

\bibitem{tagl08fermifermi}
{\sc Taglieber, M., Voigt, A.-C., Aoki, T., Hänsch, T.~W., and Dieckmann, K.},
\newblock Quantum degenerate two-species {Fermi-Fermi} mixture coexisting with
  a {Bose-Einstein} condensate,
\newblock {\em Phys. Rev. Lett., {\bf 100}\/} (2008) 010401.

\bibitem{stan05oven}
{\sc Stan, C.~A., and Ketterle, W.},
\newblock Multiple species atom source for laser-cooling experiments,
\newblock {\em Rev. Sci. Instr., {\bf 76}\/} (2005) 063113.

\bibitem{barr01}
{\sc Barrett, M.~D., Sauer, J.~A., and Chapman, M.~S.},
\newblock All-optical formation of an atomic {Bose-Einstein} condensate,
\newblock {\em Phys. Rev. Lett., {\bf 87}\/} (2001) 010404--4.

\bibitem{webe03}
{\sc Weber, T., Herbig, J., Mark, M., Nägerl, H.-C., and Grimm, R.},
\newblock {Bose-Einstein} condensation of cesium,
\newblock {\em Science, {\bf 299}\/} (2003) 232--235.

\bibitem{dumk06na_all_optical}
{\sc Dumke, R., Johanning, M., Gomez, E., Weinstein, J.~D., Jones, K.~M., and
  Lett, P.~D.},
\newblock All-optical generation and photoassociative probing of sodium
  {Bose-Einstein} condensates,
\newblock {\em New Journal of Physics, {\bf 8}\/} (2006) 64.

\bibitem{taka03}
{\sc Takasu, Y., Maki, K., Komori, K., Takano, T., Honda, K., Kumakura, M.,
  Yabuzaki, T., and Takahashi, Y.},
\newblock Spin-singlet {Bose-Einstein} condensation of two-electron atoms,
\newblock {\em Phys. Rev. Lett., {\bf 91}\/} (2003) 040404.

\bibitem{holl00_evap_fermi}
{\sc Holland, M.~J., DeMarco, B., and Jin, D.~S.},
\newblock Evaporative cooling of a two-component degenerate {F}ermi gas,
\newblock {\em Phys. Rev. A, {\bf 61}\/} (2000) 053610.

\bibitem{geis02}
{\sc Geist, W., and Kennedy, T. A.~B.},
\newblock Evaporative cooling of mixed atomic fermions,
\newblock {\em Phys. Rev. A, {\bf 65}\/} (2002) 063617.

\bibitem{timm01fermi-heat}
{\sc Timmermans, E.},
\newblock Degenerate fermion gas heating by hole creation,
\newblock {\em Phys. Rev. Lett., {\bf 87}\/} (2001) 240403--4.

\bibitem{carr04}
{\sc Carr, L.~D., Bourdel, T., and Castin, Y.},
\newblock Limits of sympathetic cooling of fermions by zero-temperature bosons
  due to particle losses,
\newblock {\em Phys. Rev. A, {\bf 69}\/} (2004) 033603--14.

\bibitem{idzi05}
{\sc Idziaszek, Z., Santos, L., and Lewenstein, M.},
\newblock Sympathetic cooling of trapped fermions by bosons in the presence of
  particle losses,
\newblock {\em Europhys. Lett., {\bf 70}\/} (2005) 572--578.

\bibitem{wout02}
{\sc Wouters, M., Tempere, J., and Devreese, J.~T.},
\newblock Three-fluid description of the sympathetic cooling of a boson-fermion
  mixture,
\newblock {\em Phys. Rev. A, {\bf 66}\/} (2002) 043414.

\bibitem{onof02}
{\sc Onofrio, R., and Presilla, C.},
\newblock Reaching {Fermi} degeneracy in two-species optical dipole traps,
\newblock {\em Phys. Rev. Lett., {\bf 89}\/} (2002) 100401.

\bibitem{timm98super}
{\sc Timmermans, E., and Côté, R.},
\newblock Superfluidity in sympathetic cooling with atomic {Bose-Einstein}
  condensates,
\newblock {\em Phys. Rev. Lett., {\bf 80}\/} (1998) 3419--3423.

\bibitem{molm98}
{\sc Mølmer, K.},
\newblock {B}ose condensates and {F}ermi gases at zero temperature,
\newblock {\em Phys. Rev. Lett., {\bf 80}\/} (1998) 1804.

\bibitem{modu02}
{\sc Modugno, G., Roati, G., Riboli, F., Ferlaino, F., Brecha, R.~J., and
  Inguscio, M.},
\newblock Collapse of a degenerate {Fermi} gas,
\newblock {\em Science, {\bf 297}\/} (2002) 2240.

\bibitem{ospe05}
{\sc Ospelkaus, C., Ospelkaus, S., Sengstock, K., and Bongs, K.},
\newblock Interaction-driven dynamics of $^{40}${K} / $^{87}${R}b
  {F}ermi-{B}ose gas mixtures in the large particle number limit,
\newblock {\em Phys. Rev. Lett., {\bf 96}\/} (2006) 020401.

\bibitem{grim00}
{\sc Grimm, R., Weidemüller, M., and Ovchinnikov, Y.~B.},
\newblock Optical dipole traps for neutral atoms,
\newblock {\em Adv. At. Mol. Phys., {\bf 42}\/} (2000) 95--170.

\bibitem{take95cs}
{\sc Takekoshi, T., and Knize, R.~J.},
\newblock {CO}$_2$ laser trap for cesium atoms,
\newblock {\em Optics Lett., {\bf 21}\/} (1995) 77.

\bibitem{bour03}
{\sc Bourdel, T., Cubizolles, J., Khaykovich, L., Magalh{\~a}es, K. M.~F.,
  Kokkelmans, S. J. J. M.~F., Shlyapnikov, G.~V., and Salomon, C.},
\newblock Measurement of the interaction energy near a {F}eshbach resonance in
  a $^6${L}i {F}ermi gas,
\newblock {\em Phys. Rev. Lett., {\bf 91}\/} (2003) 020402.

\bibitem{mosk01res}
{\sc Mosk, A., Jochim, S., Moritz, H., Elsässer, T., Weidemüller, M., and
  Grimm, R.},
\newblock Resonator-enhanced optical dipole trap for fermionic lithium atoms,
\newblock {\em Optics Lett., {\bf 26}\/} (2001) 1837--1839.

\bibitem{sort00}
{\sc Sortais, Y., Bize, S., Nicolas, C., Clairon, A., Salomon, C., and
  Williams, C.},
\newblock Cold collision frequency shifts in a $^{87}${R}b atomic fountain,
\newblock {\em Phys. Rev. Lett., {\bf 85}\/} (2000) 3117.

\bibitem{gupt03rf}
{\sc Gupta, S., Hadzibabic, Z., Zwierlein, M.~W., Stan, C.~A., Dieckmann, K.,
  Schunck, C.~H., v.~Kempen, E. G.~M., Verhaar, B.~J., and Ketterle, W.},
\newblock {RF} spectroscopy of ultracold fermions,
\newblock {\em Science, {\bf 300}\/} (2003) 1723--1726.

\bibitem{zwie03}
{\sc Zwierlein, M.~W., Hadzibabic, Z., Gupta, S., and Ketterle, W.},
\newblock Spectroscopic insensitivity to cold collisions in a two-state mixture
  of fermions,
\newblock {\em Phys. Rev. Lett., {\bf 91}\/} (2003) 250404.

\bibitem{okte99}
{\sc Oktel, M.~{\"O}., and Levitov, L.~S.},
\newblock Optical excitations in a nonideal {Bose} gas,
\newblock {\em Phys. Rev. Lett., {\bf 83}\/} (1999) 6.

\bibitem{okte02}
{\sc Oktel, M.~{\"O}., Killian, T.~C., Kleppner, D., and Levitov, L.~S.},
\newblock Sum rule for the optical spectrum of a trapped gas,
\newblock {\em Phys. Rev. A, {\bf 65}\/} (2002) 033617.

\bibitem{rega03fesh}
{\sc Regal, C.~A., and Jin, D.~S.},
\newblock Measurement of positive and negative scattering lengths in a {F}ermi
  gas of atoms,
\newblock {\em Phys. Rev. Lett., {\bf 90}\/} (2003) 230404.

\bibitem{baym07}
{\sc Baym, G., Pethick, C.~J., Yu, Z., and Zwierlein, M.~W.},
\newblock Coherence and clock shifts in ultracold {Fermi} gases with resonant
  interactions,
\newblock {\em Phys. Rev. Lett., {\bf 99}\/} (2007) 190407.

\bibitem{punk07rf}
{\sc Punk, M., and Zwerger, W.},
\newblock Theory of rf-spectroscopy of strongly interacting fermions,
\newblock {\em Phys. Rev. Lett., {\bf 99}\/} (2007) 170404.

\bibitem{shin07rf}
{\sc Shin, Y., Schunck, C.~H., Schirotzek, A., and Ketterle, W.},
\newblock Tomographic {RF} spectroscopy of a trapped {Fermi} gas at unitarity,
\newblock {\em Phys. Rev. Lett., {\bf 99}\/} (2007) 090403.

\bibitem{bart04fesh}
{\sc Bartenstein, M., Altmeyer, A., Riedl, S., Geursen, R., Jochim, S., Chin,
  C., Hecker-Denschlag, J., Grimm, R., Simoni, A., Tiesinga, E., Williams,
  C.~J., and Julienne, P.~S.},
\newblock Precise determination of $^6${L}i cold collision parameters by
  radio-frequency spectroscopy on weakly bound molecules,
\newblock {\em Phys. Rev. Lett., {\bf 94}\/} (2004) 103201.

\bibitem{jone06}
{\sc Jones, K.~M., Tiesinga, E., Lett, P.~D., and Julienne, P.~S.},
\newblock Ultracold photoassociation spectroscopy: {Long}-range molecules and
  atomic scattering,
\newblock {\em Rev. Mod. Phys., {\bf 78}\/} (2006) 483--53.

\bibitem{huts06}
{\sc Hutson, J.~M., and Sold{\'a}n, P.},
\newblock Molecule formation in ultracold atomic gases,
\newblock {\em International Reviews in Physical Chemistry, {\bf 25}\/} (2006)
  497 -- 526.

\bibitem{lero69}
{\sc LeRoy, R.~J., and Bernstein, R.~B.},
\newblock Dissociation energy and long-range potential of diatomic molecules
  from vibrational spacings of higher levels,
\newblock {\em The Journal of Chemical Physics, {\bf 52}\/} (1969) 3869.

\bibitem{gao00}
{\sc Gao, B.},
\newblock Zero-energy bound or quasibound states and their implications for
  diatomic systems with an asymptotic {van der Waals} interaction,
\newblock {\em Phys. Rev. A, {\bf 62}\/} (2000) 050702(R).

\bibitem{mies00}
{\sc Mies, F.~H., Tiesinga, E., and Julienne, P.~S.},
\newblock Manipulation of {F}eshbach resonances in ultracold atomic collisions
  using time-dependent magnetic fields,
\newblock {\em Phys. Rev. A, {\bf 61}\/} (2000) 022721.

\bibitem{hodb05mol}
{\sc Hodby, E., Thompson, S.~T., Regal, C.~A., Greiner, M., Wilson, A.~C., Jin,
  D.~S., Cornell, E.~A., and Wieman, C.~E.},
\newblock Production efficiency of ultracold {F}eshbach molecules in bosonic
  and fermionic systems,
\newblock {\em Phys. Rev. Lett., {\bf 94}\/} (2005) 120402.

\bibitem{kohl06feshbachreview}
{\sc Köhler, T., G{\'o}ral, K., and Julienne, P.~S.},
\newblock Production of cold molecules via magnetically tunable {Feshbach}
  resonances,
\newblock {\em Rev. Mod. Phys., {\bf 78}\/} (2006) 1311.

\bibitem{thom05mol}
{\sc Thompson, S.~T., Hodby, E., and Wieman, C.~E.},
\newblock Ultracold molecule production via a resonant oscillating magnetic
  field,
\newblock {\em Phys. Rev. Lett., {\bf 95}\/} (2005) 190404.

\bibitem{schu04fesh}
{\sc Schunck, C.~H., Zwierlein, M.~W., Stan, C.~A., Raupach, S. M.~F.,
  Ketterle, W., Simoni, A., Tiesinga, E., Williams, C.~J., and Julienne,
  P.~S.},
\newblock {F}eshbach resonances in fermionic lithium-6,
\newblock {\em Phys. Rev. A, {\bf 71}\/} (2004) 045601.

\bibitem{rega03pwave}
{\sc Regal, C.~A., Ticknor, C., Bohn, J.~L., and Jin, D.~S.},
\newblock Tuning p-wave interactions in an ultracold {F}ermi gas of atoms,
\newblock {\em Phys. Rev. Lett., {\bf 90}\/} (2003) 053201.

\bibitem{zhan04pwave}
{\sc Zhang, J., Kempen, E. G. M.~V., Bourdel, T., Khaykovich, L., Cubizolles,
  J., Chevy, F., Teichmann, M., Tarruell, L., Kokkelmans, S. J. J. M.~F., and
  Salomon, C.},
\newblock P-wave {F}eshbach resonances of ultra-cold $^6${L}i,
\newblock {\em Phys. Rev. A, {\bf 70}\/} (2004) 030702.

\bibitem{stan04}
{\sc Stan, C.~A., Zwierlein, M.~W., Schunck, C.~H., Raupach, S. M.~F., and
  Ketterle, W.},
\newblock Observation of {F}eshbach resonances between two different atomic
  species,
\newblock {\em Phys. Rev. Lett., {\bf 93}\/} (2004) 143001.

\bibitem{robe98}
{\sc Roberts, J.~L., Claussen, N.~R., Jr., J.~B., Greene, C.~H., Cornell,
  E.~A., and Wieman, C.~E.},
\newblock Resonant magnetic field control of elastic scattering in cold
  $^{87}${Rb},
\newblock {\em Phys. Rev. Lett., {\bf 81}\/} (1998) 5109.

\bibitem{muka04}
{\sc Mukaiyama, T., Abo-Shaeer, J.~R., Xu, K., Chin, J.~K., and Ketterle, W.},
\newblock Dissociation and decay of ultracold sodium molecules,
\newblock {\em Phys. Rev. Lett., {\bf 92}\/} (2004) 180402.

\bibitem{durr04diss}
{\sc Dürr, S., Volz, T., and Rempe, G.},
\newblock Dissociation of ultracold molecules with {Feshbach} resonances,
\newblock {\em Phys. Rev. A, {\bf 70}\/} (2004) 031601.

\bibitem{chwe04mol}
{\sc Chwedenczuk, J., G{\'o}ral, K., Köhler, T., and Julienne, P.~S.},
\newblock Molecular production in two component atomic {Fermi} gases,
\newblock {\em Phys. Rev. Lett., {\bf 93}\/} (2004) 260403.

\bibitem{ho04proj}
{\sc Diener, R.~B., and Ho, T.-L.},
\newblock Projecting fermion pair condensates into molecular condensates,
\newblock {P}reprint cond-mat/0404517.

\bibitem{pera05}
{\sc Perali, A., Pieri, P., and Strinati, G.~C.},
\newblock Extracting the condensate density from projection experiments with
  {F}ermi gases,
\newblock {\em Phys. Rev. Lett., {\bf 95}\/} (2005) 010407.

\bibitem{altm05}
{\sc Altman, E., and Vishwanath, A.},
\newblock Dynamic projection on {F}eshbach molecules: A probe of pairing and
  phase fluctuations,
\newblock {\em Phys. Rev. Lett., {\bf 95}\/} (2005) 110404.

\bibitem{yuzb05noneq}
{\sc Yuzbashyan, E.~A., Altshuler, B.~L., Kuznetsov, V.~B., and Enolskii,
  V.~Z.},
\newblock Nonequilibrium {C}ooper pairing in the nonadiabatic regime,
\newblock {\em Phys. Rev. B, {\bf 72}\/} (2005) 220503.

\bibitem{gehm03coll}
{\sc Gehm, M.~E., Hemmer, S.~L., O'Hara, K.~M., and Thomas, J.~E.},
\newblock Unitarity-limited elastic collision rate in a harmonically trapped
  {F}ermi gas,
\newblock {\em Phys. Rev. A, {\bf 68}\/} (2003) 011603(R).

\bibitem{Bara04Coex}
{\sc Barankov, R.~A., and Levitov, L.~S.},
\newblock Atom-molecule coexistence and collective dynamics near a {F}eshbach
  resonance of cold fermions,
\newblock {\em Phys. Rev. Lett., {\bf 93}\/} (2004) 130403.

\bibitem{bara04Rabi}
{\sc Barankov, R.~A., Levitov, L.~S., and Spivak, B.~Z.},
\newblock Collective {Rabi} oscillations and solitons in a time-dependent {BCS}
  pairing problem,
\newblock {\em Phys. Rev. Lett., {\bf 93}\/} (2004) 160401.

\bibitem{sten98spin}
{\sc Stenger, J., Inouye, S., Stamper-Kurn, D.~M., Miesner, H.-J., Chikkatur,
  A.~P., and Ketterle, W.},
\newblock Spin domains in ground-state {Bose-Einstein} condensates,
\newblock {\em Nature, {\bf 396}\/} (1998) 345--348.

\bibitem{chan04}
{\sc Chang, M.-S., Hamley, C.~D., Barrett, M.~D., Sauer, J.~A., Fortier, K.~M.,
  Zhang, W., You, L., and Chapman, M.~S.},
\newblock Observation of spinor dynamics in optically trapped $^{87}${R}b
  {Bose-Einstein} condensates,
\newblock {\em Phys. Rev. Lett., {\bf 92}\/} (2004) 140403.

\bibitem{schm04spinor}
{\sc Schmaljohann, H., Erhard, M., Kronjager, J., Kottke, M., v.~Staa, S.,
  Cacciapuoti, L., Arlt, J.~J., Bongs, K., and Sengstock, K.},
\newblock Dynamics of {F=2} spinor {Bose-Einstein} condensates,
\newblock {\em Phys. Rev. Lett., {\bf 92}\/} (2004) 040402.

\bibitem{higb05spinBEC}
{\sc Higbie, J.~M., Sadler, L.~E., Inouye, S., Chikkatur, A.~P., Leslie, S.~R.,
  Moore, K.~L., Savalli, V., and Stamper-Kurn, D.~M.},
\newblock Direct nondestructive imaging of magnetization in a spin-1
  {Bose-Einstein} gas,
\newblock {\em Phys. Rev. Lett., {\bf 95}\/} (2005) 050401.

\bibitem{drib02abel}
{\sc Dribinski, V., Ossadtchi, A., Mandelshtam, V., and Reisler, H.},
\newblock Reconstruction of {Abel}-transformable images: The basis-set
  expansion {Abel} transform method,
\newblock {\em Rev. Sci. Inst., {\bf 73}\/} (2002) 2634.

\bibitem{ozer02}
{\sc Ozeri, R., Steinhauer, J., Katz, N., and Davidson, N.},
\newblock Direct observation of the phonon energy in a {Bose-Einstein}
  condensate by tomographic imaging,
\newblock {\em Phys. Rev. Lett., {\bf 88}\/} (2002) 220401.

\bibitem{bugg04scatt}
{\sc Buggle, C., Leonard, J., von Klitzing, W., and Walraven, J. T.~M.},
\newblock Interferometric determination of the s and d-wave scattering
  amplitudes in {Rb},
\newblock {\em Phys. Rev. Lett., {\bf 93}\/} (2004) 173202--4.

\bibitem{thom04scatt}
{\sc Thomas, N.~R., Kjaergaard, N., Julienne, P.~S., and Wilson, A.~C.},
\newblock Imaging of s and d partial-wave interference in quantum scattering of
  identical bosonic atoms,
\newblock {\em Phys. Rev. Lett., {\bf 93}\/} (2004) 173201--4.

\bibitem{huan87}
{\sc Huang, K.},
\newblock {\em Statistical Mechanics},
\newblock (Wiley, New York) 1987.

\bibitem{pera04temp}
{\sc Perali, A., Pieri, P., Pisani, L., and Strinati, G.~C.},
\newblock {BCS}-{BEC} crossover at finite temperature for superfluid trapped
  {F}ermi atoms,
\newblock {\em Phys. Rev. Lett., {\bf 92}\/} (2004) 220404.

\bibitem{peth02bec}
{\sc Pethick, C.~J., and Smith, H.},
\newblock {\em Bose-Einstein Condensation in Dilute Gases},
\newblock (Cambridge University Press, Cambridge) 2002.

\bibitem{holl01}
{\sc Holland, M., Kokkelmans, S. J. J. M.~F., Chiofalo, M.~L., and Walser, R.},
\newblock Resonance superfluidity in a quantum degenerate {F}ermi gas,
\newblock {\em Phys. Rev. Lett., {\bf 87}\/} (2001) 120406.

\bibitem{staj05dens}
{\sc Stajic, J., Chen, Q., and Levin, K.},
\newblock Density profiles of strongly interacting trapped {F}ermi gases,
\newblock {\em Phys. Rev. Lett., {\bf 94}\/} (2005) 060401.

\bibitem{grei05corr}
{\sc Greiner, M., Regal, C.~A., Stewart, J.~T., and Jin, D.~S.},
\newblock Probing pair-correlated fermionic atoms through correlations in atom
  shot noise,
\newblock {\em Phys. Rev. Lett., {\bf 94}\/} (2005) 110401.

\bibitem{stam98coll}
{\sc Stamper-Kurn, D.~M., Miesner, H.-J., Inouye, S., Andrews, M.~R., and
  Ketterle, W.},
\newblock Collisionless and hydrodynamic excitations of a {Bose-Einstein}
  condensate,
\newblock {\em Phys. Rev. Lett., {\bf 81}\/} (1998) 500--503.

\bibitem{shva03hydro}
{\sc Shvarchuck, I., Buggle, C., Petrov, D.~S., Kemmann, M., von Klitzing, W.,
  Shlyapnikov, G.~V., and Walraven, J. T.~M.},
\newblock Hydrodynamic behavior in expanding thermal clouds of $^{87}${Rb},
\newblock {\em Phys. Rev. A, {\bf 68}\/} (2003) 063603.

\bibitem{kaga97bose}
{\sc Kagan, Y., Surkov, E.~L., and Shlyapnikov, G.~V.},
\newblock Evolution of a {B}ose gas in anisotropic time-dependent traps,
\newblock {\em Phys. Rev. A, {\bf 55}\/} (1997) R18.

\bibitem{thom05virial}
{\sc Thomas, J.~E., Kinast, J., and Turlapov, A.},
\newblock Virial theorem and universality in a unitary {Fermi} gas,
\newblock {\em Phys. Rev. Lett., {\bf 95}\/} (2005) 120402.

\bibitem{guer99osc}
{\sc Guéry-Odelin, D., Zambelli, F., Dalibard, J., and Stringari, S.},
\newblock Collective oscillations of a classical gas confined in harmonic
  traps,
\newblock {\em Phys. Rev. A, {\bf 60}\/} (1999) 4851.

\bibitem{pedr03}
{\sc Pedri, P., Guéry-Odelin, D., and Stringari, S.},
\newblock Dynamics of a classical gas including dissipative and mean-field
  effects,
\newblock {\em Phys. Rev. A, {\bf 68}\/} (2003) 043608.

\bibitem{meno02}
{\sc Menotti, C., Pedri, P., and Stringari, S.},
\newblock Expansion of an interacting {F}ermi gas,
\newblock {\em Phys. Rev. Lett., {\bf 89}\/} (2002) 250402.

\bibitem{cast96}
{\sc Castin, Y., and Dum, R.},
\newblock {Bose-Einstein} condensation in time dependent traps,
\newblock {\em Phys. Rev. Lett., {\bf 77}\/} (1996) 5315--5319.

\bibitem{hu04coll}
{\sc Hu, H., Minguzzi, A., Liu, X.-J., and Tosi, M.~P.},
\newblock Collective modes and ballistic expansion of a {Fermi} gas in the
  {BCS-BEC} crossover,
\newblock {\em Phys. Rev. Lett., {\bf 93}\/} (2004) 190403.

\bibitem{cast04scal}
{\sc Castin, Y.},
\newblock Exact scaling transform for a unitary quantum gas in a time dependent
  harmonic potential,
\newblock {\em Comptes Rendus Physique, {\bf 5}\/} (2004) 407--410.

\bibitem{luo07entropy}
{\sc Luo, L., Clancy, B., Joseph, J., Kinast, J., and Thomas, J.~E.},
\newblock Measurement of the entropy and critical temperature of a strongly
  interacting {Fermi} gas,
\newblock {\em Phys. Rev. Lett., {\bf 98}\/} (2007) 080402.

\bibitem{bulg06TC}
{\sc Bulgac, A., Drut, J.~E., and Magierski, P.},
\newblock Spin 1/2 fermions in the unitary regime at finite temperature,
\newblock {\em Phys. Rev. Lett., {\bf 96}\/} (2006) 090404.

\bibitem{nara98semi}
{\sc Naraschewski, M., and Stamper-Kurn, D.~M.},
\newblock Analytical description of a trapped semi-ideal {B}ose gas at finite
  temperature,
\newblock {\em Phys. Rev. A, {\bf 58}\/} (1998) 2423.

\bibitem{land77qm}
{\sc Landau, L.~D., and Lifshitz, E.~M.},
\newblock {\em Quantum Mechanics: Non-Relativistic Theory},
\newblock (Pergamon Press, New York) 1987.

\bibitem{flam99scatt}
{\sc Flambaum, V.~V., Gribakin, G.~F., and Harabati, C.},
\newblock Analytical calculation of cold-atom scattering,
\newblock {\em Phys. Rev. A, {\bf 59}\/} (1999) 1998.

\bibitem{melo93}
{\sc {S{\'a} de Melo}, C. A.~R., Randeria, M., and Engelbrecht, J.~R.},
\newblock Crossover from {BCS} to {B}ose superconductivity: Transition
  temperature and time-dependent {Ginzburg-Landau} theory,
\newblock {\em Phys. Rev. Lett., {\bf 71}\/} (1993) 3202.

\bibitem{haus99}
{\sc Haussmann, R.},
\newblock {\em Self-consistent Quantum Field Theory and Bosonization for
  Strongly Correlated Electron Systems},
\newblock (Springer Verlag, Berlin) 1999.

\bibitem{bray71}
{\sc Brayshaw, D.~D.},
\newblock Off-shell t matrix for the boundary-condition model,
\newblock {\em Phys. Rev. C, {\bf 3}\/} (1971) 35--45.

\bibitem{kohn65}
{\sc Kohn, W., and Luttinger, J.~M.},
\newblock New mechanism for superconductivity,
\newblock {\em Phys. Rev. Lett., {\bf 15}\/} (1965) 524.

\bibitem{bara98}
{\sc Baranov, M.~A., and Petrov, D.~S.},
\newblock Critical temperature and {Ginzburg-Landau} equation for a trapped
  {F}ermi gas,
\newblock {\em Phys. Rev. A, {\bf 58}\/} (1998) R801.

\bibitem{rand89bound}
{\sc Randeria, M., Duan, J.-M., and Shieh, L.-Y.},
\newblock Bound states, {C}ooper pairing, and {B}ose condensation in two
  dimensions,
\newblock {\em Phys. Rev. Lett., {\bf 62}\/} (1989) 981.

\bibitem{orti05bcs}
{\sc Ortiz, G., and Dukelsky, J.},
\newblock {BCS}-to-{BEC} crossover from the exact {BCS} solution,
\newblock {\em Phys. Rev. A, {\bf 72}\/} (2005) 043611.

\bibitem{mari98becbcs}
{\sc Marini, M., Pistolesi, F., and Strinati, G.~C.},
\newblock Evolution from {BCS} superconductivity to {B}ose condensation:
  analytic results for the crossover in three dimensions,
\newblock {\em Eur. Phys. J. B, {\bf 1}\/} (1998) 151--159.

\bibitem{bogo58}
{\sc Bogoliubov, N.~N.},
\newblock On a new method in the theory of superconductivity,
\newblock {\em Nuovo Cimento, {\bf 7}\/} (1958) 794.

\bibitem{vala58}
{\sc Valatin, J.},
\newblock Comments on the theory of superconductivity,
\newblock {\em Nuovo Cimento, {\bf 7}\/} (1958) 843.

\bibitem{duke04review}
{\sc Dukelsky, J., Pittel, S., and Sierra, G.},
\newblock Colloquium: Exactly solvable {Richardson}-{Gaudin} models for
  many-body quantum systems,
\newblock {\em Rev. Mod. Phys., {\bf 76}\/} (2004) 643.

\bibitem{gork61}
{\sc Gor'kov, L.~P., and Melik-Barkhudarov, T.~K.},
\newblock Contribution to the theory of superfluidity in an imperfect {F}ermi
  gas,
\newblock {\em Zh. Eskp. Theor. Fiz., {\bf 40}\/} (1961) 1452,
\newblock [Sov. Phys. JETP 34, 61 (1961)].

\bibitem{petr04dimers}
{\sc Petrov, D.~S., Salomon, C., and Shlyapnikov, G.~V.},
\newblock Weakly bound dimers of fermionic atoms,
\newblock {\em Phys. Rev. Lett., {\bf 93}\/} (2004) 090404.

\bibitem{pier00becbcs}
{\sc Pieri, P., and Strinati, G.~C.},
\newblock Strong-coupling limit in the evolution from {BCS} superconductivity
  to {Bose-Einstein} condensation,
\newblock {\em Phys. Rev. B, {\bf 61}\/} (2000) 15370.

\bibitem{holl04bosefermi}
{\sc Holland, M., Menotti, C., and Viverit, L.},
\newblock The role of boson-fermion correlations in the resonance theory of
  superfluids,
\newblock preprint cond-mat/0404234.

\bibitem{hu06becbcs}
{\sc Hu, H., Liu, X.-J., and Drummond, P.~D.},
\newblock Equation of state of a superfluid {Fermi} gas in the {BCS}-{BEC}
  crossover,
\newblock {\em Europhys. Lett., {\bf 74}\/} (2006) 574--580.

\bibitem{pist94xi}
{\sc Pistolesi, F., and Strinati, G.~C.},
\newblock Evolution from {BCS} superconductivity to {B}ose condensation: Role
  of the parameter $k_{F} \xi$,
\newblock {\em Phys. Rev. B, {\bf 49}\/} (1994) 6356.

\bibitem{skor57abf}
{\sc Skorniakov, G.~V., and Ter-Martirosian, K.~A.},
\newblock Three-body problem for short-range forces, {I}, scattering of
  low-energy neutrons by deuterons,
\newblock {\em Zh. Eksp. Teor. Fiz., {\bf 31}\/} (1956) 775,
\newblock [JETP Letters 4, 648 (1957)].

\bibitem{chin05rf}
{\sc Chin, C., and Julienne, P.~S.},
\newblock Radio-frequency transitions on weakly bound ultracold molecules,
\newblock {\em Phys. Rev. A, {\bf 71}\/} (2005) 012713.

\bibitem{pera08rf}
{\sc Perali, A., Pieri, P., and Strinati, G.~C.},
\newblock Competition between final-state and pairing-gap effects in the
  radio-frequency spectra of ultracold {Fermi} atoms,
\newblock {\em Phys. Rev. Lett., {\bf 100}\/} (2008) 010402.

\bibitem{ohas03coll}
{\sc Ohashi, Y., and Griffin, A.},
\newblock Superfluidity and collective modes in a uniform gas of {F}ermi atoms
  with a {F}eshbach resonance,
\newblock {\em Phys. Rev. A, {\bf 67}\/} (2003) 063612--24.

\bibitem{comb06coll}
{\sc Combescot, R., Kagan, M.~Y., and Stringari, S.},
\newblock Collective mode of homogeneous superfluid {F}ermi gases in the
  {BEC}-{BCS} crossover,
\newblock {\em Phys. Rev. A, {\bf 74}\/} (2006) 042717--14.

\bibitem{gior07fermi}
{\sc Giorgini, S., Pitaevskii, L.~P., and Stringari, S.},
\newblock Theory of ultracold {Fermi} gases,
\newblock preprint arXiv:0706.3360.

\bibitem{altm07precision}
{\sc Altmeyer, A., Riedl, S., Kohstall, C., Wright, M.~J., Geursen, R.,
  Bartenstein, M., Chin, C., Hecker-Denschlag, J., and Grimm, R.},
\newblock Precision measurements of collective oscillations in the {BEC-BCS}
  crossover,
\newblock {\em Phys. Rev. Lett., {\bf 98}\/} (2007) 040401.

\bibitem{lifs80statphys2}
{\sc Lifshitz, E.~M., and Pitaevskii, L.~P.},
\newblock {\em Statistical Physics, Part 2},
\newblock (Elsevier, Amsterdam) 1980.

\bibitem{sens06vort}
{\sc Sensarma, R., Randeria, M., and Ho, T.-L.},
\newblock Vortices in superfluid {F}ermi gases through the {BEC} to {BCS}
  crossover,
\newblock {\em Phys. Rev. Lett., {\bf 96}\/} (2006) 090403.

\bibitem{drec92}
{\sc Drechsler, M., and Zwerger, W.},
\newblock Crossover from {BCS}-superconductivity to {B}ose-condensation,
\newblock {\em Annalen der Physik, {\bf 1}\/} (1992) 15--23.

\bibitem{haus07bcsbec}
{\sc Haussmann, R., Rantner, W., Cerrito, S., and Zwerger, W.},
\newblock Thermodynamics of the {BCS-BEC} crossover,
\newblock {\em Phys. Rev. A, {\bf 75}\/} (2007) 023610.

\bibitem{baym99tc}
{\sc Baym, G., Blaizot, J.-P., Holzmann, M., Laloë, F., and Vautherin, D.},
\newblock The transition temperature of the dilute interacting {Bose} gas,
\newblock {\em Phys. Rev. Lett., {\bf 83}\/} (1999) 1703.

\bibitem{baym00tc}
{\sc Baym, G., Blaizot, J.-P., and Zinn-Justin, J.},
\newblock The transition temperature of the dilute interacting {Bose} gas for
  {N} internal states,
\newblock {\em Europhys. Lett., {\bf 49}\/} (2000) 150--155.

\bibitem{baym01tc}
{\sc Baym, G., Blaizot, J.-P., Holzmann, M., Laloë, F., and Vautherin, D.},
\newblock {Bose-Einstein} transition in a dilute interacting gas,
\newblock {\em Eur. Phys. J. B, {\bf 24}\/} (2001) 107--124.

\bibitem{arno01tc}
{\sc Arnold, P., and Moore, G.},
\newblock {BEC} transition temperature of a dilute homogeneous imperfect {Bose}
  gas,
\newblock {\em Phys. Rev. Lett., {\bf 87}\/} (2001) 120401.

\bibitem{kash01tc}
{\sc Kashurnikov, V.~A., Prokof'ev, N.~V., and Svistunov, B.~V.},
\newblock Critical temperature shift in weakly interacting {Bose} gas,
\newblock {\em Phys. Rev. Lett., {\bf 87}\/} (2001) 120402.

\bibitem{nish06epsilon}
{\sc Nishida, Y., and Son, D.~T.},
\newblock Epsilon expansion for a {Fermi} gas at infinite scattering length,
\newblock {\em Phys. Rev. Lett., {\bf 97}\/} (2006) 050403.

\bibitem{buro06TC}
{\sc Burovski, E., Prokof'ev, N., Svistunov, B., and Troyer, M.},
\newblock Critical temperature and thermodynamics of attractive fermions at
  unitarity,
\newblock {\em Phys. Rev. Lett., {\bf 96}\/} (2006) 160402.

\bibitem{penr56}
{\sc Penrose, O., and Onsager, L.},
\newblock {Bose-Einstein} condensation and liquid helium,
\newblock {\em Phys. Rev., {\bf 104}\/} (1956) 576.

\bibitem{andr97int}
{\sc Andrews, M.~R., Townsend, C.~G., Miesner, H.-J., Durfee, D.~S., Kurn,
  D.~M., and Ketterle, W.},
\newblock Observation of interference between two {Bose-Einstein} condensates,
\newblock {\em Science, {\bf 275}\/} (1997) 637--641.

\bibitem{bloc00coh}
{\sc Bloch, I., Hänsch, T.~W., and Esslinger, T.},
\newblock Measurement of the spatial coherence of a trapped {B}ose gas at the
  phase transition,
\newblock {\em Nature, {\bf 403}\/} (2000) 166--170.

\bibitem{madi00}
{\sc Madison, K.~W., Chevy, F., Wohlleben, W., and Dalibard, J.},
\newblock Vortex formation in a stirred {Bose-Einstein} condensate,
\newblock {\em Phys. Rev. Lett., {\bf 84}\/} (2000) 806--809.

\bibitem{abos01latt}
{\sc Abo-Shaeer, J.~R., Raman, C., Vogels, J.~M., and Ketterle, W.},
\newblock Observation of vortex lattices in {Bose-Einstein} condensates,
\newblock {\em Science, {\bf 292}\/} (2001) 476--479.

\bibitem{hodb02vort}
{\sc Hodby, E., Hechenblaikner, G., Hopkins, S.~A., Maragò, O.~M., and Foot,
  C.~J.},
\newblock Vortex nucleation in {Bose-Einstein} condensates in an oblate, purely
  magnetic potential,
\newblock {\em Phys. Rev. Lett., {\bf 88}\/} (2002) 010405--4.

\bibitem{enge02}
{\sc Engels, P., Coddington, I., Haljan, P.~C., and Cornell, E.~A.},
\newblock Nonequilibrium effects of anisotropic compression applied to vortex
  lattices in {Bose-Einstein} condensates,
\newblock {\em Phys. Rev. Lett., {\bf 89}\/} (2002) 100403.

\bibitem{camp97ODLRO}
{\sc Campbell, C.~E.},
\newblock {Bose-Einstein} condensation, pairing and {ODLRO}: a view from
  coordinate space,
\newblock in {\em Condensed Matter Theories}, vol.~12, (Nova Science, New York)
  1997, p.~131.

\bibitem{sala05}
{\sc Salasnich, L., Manini, N., and Parola, A.},
\newblock Condensate fraction of a {F}ermi gas in the {BCS}-{BEC} crossover,
\newblock {\em Phys. Rev. A, {\bf 72}\/} (2005) 023621.

\bibitem{astr05cond}
{\sc Astrakharchik, G.~E., Boronat, J., Casulleras, J., and Giorgini, S.},
\newblock Momentum distribution and condensate fraction of a fermion gas in the
  {BCS}-{BEC} crossover,
\newblock {\em Phys. Rev. Lett., {\bf 95}\/} (2005) 230405.

\bibitem{gior97}
{\sc Giorgini, S., Pitaevskii, L.~P., and Stringari, S.},
\newblock Scaling and thermodynamics of a trapped {B}ose-condensed gas,
\newblock {\em Phys. Rev. Lett., {\bf 78}\/} (1997) 3987--3990.

\bibitem{tisz38}
{\sc Tisza, L.},
\newblock Transport phenomena in helium {II},
\newblock {\em Nature, {\bf 141}\/} (1938) 913.

\bibitem{land41}
{\sc Landau, L.},
\newblock Theory of the superfluidity of helium {II},
\newblock {\em Phys. Rev., {\bf 60}\/} (1941) 356.

\bibitem{legg75}
{\sc Leggett, A.~J.},
\newblock A theoretical description of the new phases of liquid $^3${H}e,
\newblock {\em Rev. Mod. Phys., {\bf 47}\/} (1975) 331.

\bibitem{enge97becbcs}
{\sc Engelbrecht, J.~R., Randeria, M., and {S{\'a} de Melo}, C. A.~R.},
\newblock {BCS} to {Bose} crossover: Broken-symmetry state,
\newblock {\em Phys. Rev. B, {\bf 55}\/} (1997) 15153.

\bibitem{andr03becbcs}
{\sc Andrenacci, N., Pieri, P., and Strinati, G.~C.},
\newblock Evolution from {BCS} superconductivity to {Bose-Einstein}
  condensation: Current correlation function in the broken-symmetry phase,
\newblock {\em Phys. Rev. B, {\bf 68}\/} (2003) 144507.

\bibitem{tayl06sf}
{\sc Taylor, E., Griffin, A., Fukushima, N., and Ohashi, Y.},
\newblock Pairing fluctuations and the superfluid density through the {BCS-BEC}
  crossover,
\newblock {\em Phys. Rev. A, {\bf 74}\/} (2006) 063626--15.

\bibitem{abri75}
{\sc Abrikosov, A.~A., Gor'kov, L.~P., and Dzyaloshinski, I.~E.},
\newblock {\em Methods of Quantum Field Theory in Statistical Physics},
\newblock (Dover Publications, New York) 1975.

\bibitem{abri60}
{\sc Abrikosov, A.~A., and Gor'kov, L.~P.},
\newblock Contribution to the theory of superconducting alloys with
  paramagnetic impurities,
\newblock {\em Zh. Eksp. Teor. Fiz., {\bf 39}\/} (1960) 1781,
\newblock [Sov. Phys. JETP 12, 1243 (1961)].

\bibitem{tink04sc}
{\sc Tinkham, M.},
\newblock {\em Introduction to Superconductivity},
\newblock (Dover, Mineola, New York) 2004.

\bibitem{stin97ginzburg}
{\sc Stintzing, S., and Zwerger, W.},
\newblock {Ginzburg-Landau} theory of superconductors with short coherence
  length,
\newblock {\em Phys. Rev. B, {\bf 56}\/} (1997) 9004.

\bibitem{wert63}
{\sc Werthamer, N.~R.},
\newblock Theory of a local superconductor in a magnetic field,
\newblock {\em Phys. Rev., {\bf 132}\/} (1963) 663.

\bibitem{jose66}
{\sc Josephson, B.~D.},
\newblock Relation between the superfluid density and order parameter for
  superfluid {He} near {Tc},
\newblock {\em Phys. Lett., {\bf 21}\/} (1966) 608--609.

\bibitem{klei99crit}
{\sc Kleinert, H.},
\newblock Critical exponents from seven-loop strong-coupling $\phi^{4}$ theory
  in three dimensions,
\newblock {\em Phys. Rev. D, {\bf 60}\/} (1999) 085001.

\bibitem{shee06phase}
{\sc Sheehy, D.~E., and Radzihovsky, L.},
\newblock {BEC}-{BCS} crossover, phase transitions and phase separation in
  polarized resonantly-paired superfluids,
\newblock {\em Annals of Physics, {\bf 322}\/} (2007) 1790.

\bibitem{Son05}
{\sc Son, D.~T., and Stephanov, M.~A.},
\newblock Phase diagram of cold polarized {F}ermi gas,
\newblock {\em Phys. Rev. A, {\bf 74}\/} (2006) 013614.

\bibitem{prok07polaron}
{\sc Prokof'ev, N., and Svistunov, B.},
\newblock {Fermi-Polaron}: Diagrammatic {Monte Carlo} for divergent
  sign-alternating series,
\newblock preprint arxiv.org:0707.4259.

\bibitem{fesh58}
{\sc Feshbach, H.},
\newblock A unified theory of nuclear reactions,
\newblock {\em Annals of Physics, {\bf 5}\/} (1958) 357--390.

\bibitem{fesh62}
{\sc Feshbach, H.},
\newblock A unified theory of nuclear reactions {II},
\newblock {\em Annals of Physics, {\bf 19}\/} (1962) 287--313.

\bibitem{fano61phaseshifts}
{\sc Fano, U.},
\newblock Effects of configuration interaction on intensities and phase shifts,
\newblock {\em Phys. Rev., {\bf 124}\/} (1961) 1866.

\bibitem{abel99}
{\sc van Abeelen, F.~A., and Verhaar, B.~J.},
\newblock Time-dependent {F}eshbach resonance scattering and anomalous decay of
  a {N}a {Bose-Einstein} condensate,
\newblock {\em Phys. Rev. Lett., {\bf 83}\/} (1999) 1550--1553.

\bibitem{yuro99}
{\sc Yurovsky, V.~A., Ben-Reuven, A., Julienne, P.~S., and Williams, C.~J.},
\newblock Atom loss from {Bose-Einstein} condensates due to {F}eshbach
  resonance,
\newblock {\em Phys. Rev. A, {\bf 60}\/} (1999) R765--R768.

\bibitem{donl02mol}
{\sc Donley, E.~A., Claussen, N.~R., Thompson, S.~T., and Wieman, C.~E.},
\newblock {Atom{-}molecule} coherence in a {B}ose–{E}instein condensate,
\newblock {\em Nature, {\bf 417}\/} (2002) 529.

\bibitem{fedi96reco}
{\sc Fedichev, P.~O., Reynolds, M.~W., and Shlyapnikov, G.~V.},
\newblock Three-body recombination of ultracold atoms to a weakly bound s
  level,
\newblock {\em Phys. Rev. Lett., {\bf 77}\/} (1996) 2921--2924.

\bibitem{niel99boserecomb}
{\sc Nielsen, E., and Macek, J.~H.},
\newblock Low-energy recombination of identical bosons by three-body
  collisions,
\newblock {\em Phys. Rev. Lett., {\bf 83}\/} (1999) 1566.

\bibitem{esry99recomb}
{\sc Esry, B.~D., Greene, C.~H., and Burke, J.~P.},
\newblock Recombination of three atoms in the ultracold limit,
\newblock {\em Phys. Rev. Lett., {\bf 83}\/} (1999) 1751.

\bibitem{beda00boserecomb}
{\sc Bedaque, P.~F., Braaten, E., and Hammer, H.-W.},
\newblock Three-body recombination in {Bose} gases with large scattering
  length,
\newblock {\em Phys. Rev. Lett., {\bf 85}\/} (2000) 908.

\bibitem{petr04bose}
{\sc Petrov, D.~S.},
\newblock Three-boson problem near a narrow {Feshbach} resonance,
\newblock {\em Phys. Rev. Lett., {\bf 93}\/} (2004) 143201.

\bibitem{moer95res}
{\sc Moerdijk, A.~J., Verhaar, B.~J., and Axelsson, A.},
\newblock Resonances in ultra-cold collisions of $^6${L}i, $^7${L}i and
  $^{23}${N}a,
\newblock {\em Phys. Rev. A, {\bf 51}\/} (1995) 4852--4861.

\bibitem{duin04feshbachreview}
{\sc Duine, R.~A., and Stoof, H. T.~C.},
\newblock Atom-molecule coherence in {Bose} gases,
\newblock {\em Phys. Rep., {\bf 396}\/} (2004) 115--195.

\bibitem{timm01}
{\sc Timmermans, E., Furuya, K., Milonni, P.~W., and Kerman, A.~K.},
\newblock Prospect of creating a composite {F}ermi {B}ose superfluid,
\newblock {\em Phys. Lett. A, {\bf 285}\/} (2001) 228.

\bibitem{comb03fesh}
{\sc Combescot, R.},
\newblock {Feshbach} resonance in dense ultracold {Fermi} gases,
\newblock {\em Phys. Rev. Lett., {\bf 91}\/} (2003) 120401.

\bibitem{bruu04eff}
{\sc Bruun, G.~M., and Pethick, C.~J.},
\newblock Effective theory of {Feshbach} resonances and many-body properties of
  {Fermi} gases,
\newblock {\em Phys. Rev. Lett., {\bf 92}\/} (2004) 140404.

\bibitem{bruu04univ}
{\sc Bruun, G.~M.},
\newblock Universality of a two-component {Fermi} gas with a resonant
  interaction,
\newblock {\em Phys. Rev. A, {\bf 70}\/} (2004) 053602.

\bibitem{palo04res}
{\sc {De Palo}, S., Chiofalo, M.~L., Holland, M.~J., and Kokkelmans, S. J. J.
  M.~F.},
\newblock Resonance effects on the crossover of bosonic to fermionic
  superfluidity,
\newblock {\em Phys. Lett. A, {\bf 327}\/} (2004) 490--499.

\bibitem{simo05fesh}
{\sc Simonucci, S., Pieri, P., and Strinati, G.~C.},
\newblock Broad versus narrow {F}ano-{F}eshbach resonances in the {BCS}-{BEC}
  crossover with trapped {F}ermi atom,
\newblock {\em Europhys. Lett., {\bf 69}\/} (2005) 713.

\bibitem{szym05}
{\sc Szymanska, M.~H., G{\'o}ral, K., Köhler, T., and Burnett, K.},
\newblock Conventional character of the {BCS-BEC} crossover in ultracold gases
  of $^{40}${K},
\newblock {\em Phys. Rev. A, {\bf 72}\/} (2005) 013610.

\bibitem{bour04phd}
{\sc Bourdel, T.},
\newblock {\em Gaz de {F}ermi en interaction forte : Du condensat de molécules
  aux paires de {C}ooper},
\newblock PhD thesis, Laboratoire Kastler Brossel, Ecole Normale Supérieure, 24
  rue Lhomond, 75231 Paris 05, France, 2004.

\bibitem{falc04}
{\sc Falco, G.~M., and Stoof, H. T.~C.},
\newblock Crossover temperature of {Bose-Einstein} condensation in an atomic
  {F}ermi gas,
\newblock {\em Phys. Rev. Lett., {\bf 92}\/} (2004) 130401.

\bibitem{marc04res}
{\sc Marcelis, B., v.~Kempen, E. G.~M., Verhaar, B.~J., and Kokkelmans, S. J.
  J. M.~F.},
\newblock {F}eshbach resonances with large background scattering length:
  Interplay with open-channel resonances,
\newblock {\em Phys. Rev. A, {\bf 70}\/} (2004) 012701.

\bibitem{kemp04}
{\sc v.~Kempen, E. G.~M., Marcelis, B., and Kokkelmans, S. J. J. M.~F.},
\newblock Formation of fermionic molecules via interisotope {F}eshbach
  resonances,
\newblock {\em Phys. Rev. A, {\bf 70}\/} (2004) 050701(R).

\bibitem{huan95}
{\sc Huang, K.},
\newblock {Bose-Einstein} condensation and superfluidity,
\newblock in {\em {Bose-Einstein} Condensation}, edited by A.~Griffin,
  D.~Snoke, and S.~Stringari, (Cambridge University Press, Cambridge) 1995,
  pp.~31--50.

\bibitem{lieb631dbose}
{\sc Lieb, E.~H., and Liniger, W.},
\newblock Exact analysis of an interacting {Bose} gas. {I}. {The} general
  solution and the ground state,
\newblock {\em Phys. Rev., {\bf 130}\/} (1963) 1605.

\bibitem{lieb631dboseb}
{\sc Lieb, E.~H.},
\newblock Exact analysis of an interacting {Bose} gas. {II}. {The} excitation
  spectrum,
\newblock {\em Phys. Rev., {\bf 130}\/} (1963) 1616.

\bibitem{bish782dSF}
{\sc Bishop, D.~J., and Reppy, J.~D.},
\newblock Study of the superfluid transition in two-dimensional $^{4}${He}
  films,
\newblock {\em Phys. Rev. Lett., {\bf 40}\/} (1978) 1727.

\bibitem{hohe67longrange}
{\sc Hohenberg, P.~C.},
\newblock Existence of long-range order in one and two dimensions,
\newblock {\em Phys. Rev., {\bf 158}\/} (1967) 383.

\bibitem{bere72BKT}
{\sc Berezinskii, V.~L.},
\newblock Destruction of long-range order in one-dimensional and
  two-dimensional systems possessing a continuous symmetry group. {II}.
  {Quantum} systems.,
\newblock {\em Sov. Phys. JETP, {\bf 34}\/} (1972) 610.

\bibitem{kost73BKT}
{\sc Kosterlitz, J.~M., and Thouless, D.~J.},
\newblock Ordering, metastability and phase transitions in two-dimensional
  systems,
\newblock {\em Journal of Physics C: Solid State Physics, {\bf 6}\/} (1973)
  1181.

\bibitem{schi712dbec}
{\sc Schick, M.},
\newblock Two-dimensional system of hard-core bosons,
\newblock {\em Phys. Rev. A, {\bf 3}\/} (1971) 1067.

\bibitem{huan92}
{\sc Huang, K., and Meng, H.-F.},
\newblock Hard-sphere {Bose} gas in random external potentials,
\newblock {\em Phys. Rev. Lett., {\bf 69}\/} (1992) 644.

\bibitem{gior94disorder}
{\sc Giorgini, S., Pitaevskii, L., and Stringari, S.},
\newblock Effects of disorder in a dilute {Bose} gas,
\newblock {\em Phys. Rev. B, {\bf 49}\/} (1994) 12938.

\bibitem{astr02disorder}
{\sc Astrakharchik, G.~E., Boronat, J., Casulleras, J., and Giorgini, S.},
\newblock Superfluidity versus {Bose-Einstein} condensation in a {Bose} gas
  with disorder,
\newblock {\em Phys. Rev. A, {\bf 66}\/} (2002) 023603.

\bibitem{yuka07disorder}
{\sc Yukalov, V.~I., Yukalova, E.~P., Krutitsky, K.~V., and Graham, R.},
\newblock {Bose-Einstein}-condensed gases in arbitrarily strong random
  potentials,
\newblock {\em Phys. Rev. A, {\bf 76}\/} (2007) 053623--11.

\bibitem{bloc07review}
{\sc Bloch, I., Dalibard, J., and Zwerger, W.},
\newblock Many-body physics with ultracold gases,
\newblock Rev. Mod. Phys., to be published, preprint arXiv:0704.3011.

\bibitem{ande59dirty}
{\sc Anderson, P.~W.},
\newblock Theory of dirty superconductors,
\newblock {\em Journal of Physics and Chemistry of Solids, {\bf 11}\/} (1959)
  26--30.

\bibitem{orso07disorder}
{\sc Orso, G.},
\newblock {BCS-BEC} crossover in a random external potential,
\newblock {\em Phys. Rev. Lett., {\bf 99}\/} (2007) 250402.

\bibitem{legg73helium}
{\sc Leggett, A.~J.},
\newblock Topics in the theory of helium,
\newblock {\em Physica Fennica, {\bf 8}\/} (1973) 125--170.

\bibitem{fish73super}
{\sc Fisher, M.~E., Barber, M.~N., and Jasnow, D.},
\newblock Helicity modulus, superfluidity, and scaling in isotropic systems,
\newblock {\em Phys. Rev. A, {\bf 8}\/} (1973) 1111.

\bibitem{lieb02bose}
{\sc Lieb, E.~H., Seiringer, R., and Yngvason, J.},
\newblock Superfluidity in dilute trapped {Bose} gases,
\newblock {\em Phys. Rev. B, {\bf 66}\/} (2002) 134529.

\bibitem{matt99vort}
{\sc Matthews, M.~R., Anderson, B.~P., Haljan, P.~C., Hall, D.~S., Wieman,
  C.~E., and Cornell, E.~A.},
\newblock Vortices in a {Bose-Einstein} condensate,
\newblock {\em Phys. Rev. Lett., {\bf 83}\/} (1999) 2498--2501.

\bibitem{rama99}
{\sc Raman, C., Köhl, M., Onofrio, R., Durfee, D.~S., Kuklewicz, C.~E.,
  Hadzibabic, Z., and Ketterle, W.},
\newblock Evidence for a critical velocity in a {Bose-Einstein} condensed gas,
\newblock {\em Phys. Rev. Lett., {\bf 83}\/} (1999) 2502--2505.

\bibitem{onof00sup}
{\sc Onofrio, R., Raman, C., Vogels, J.~M., Abo-Shaeer, J.~R., Chikkatur,
  A.~P., and Ketterle, W.},
\newblock Observation of superfluid flow in a {Bose-Einstein} condensed gas,
\newblock {\em Phys. Rev. Lett., {\bf 85}\/} (2000) 2228--2231.

\bibitem{sche99}
{\sc Schecter, D.~A., Dubin, D. H.~E., Fine, K.~S., and Driscoll, C.~F.},
\newblock Vortex crystals from {2D} {Euler} flow: Experiment and simulation,
\newblock {\em Phys. Fluids, {\bf 11}\/} (1999) 905.

\bibitem{onof00}
{\sc Onofrio, R., Durfee, D.~S., Raman, C., Köhl, M., Kuklewicz, C.~E., and
  Ketterle, W.},
\newblock Surface excitations in a {Bose-Einstein} condensate,
\newblock {\em Phys. Rev. Lett., {\bf 84}\/} (2000) 810--813.

\bibitem{guer99scissors}
{\sc Guéry-Odelin, D., and Stringari, S.},
\newblock Scissors mode and superfluidity of a trapped {Bose-Einstein}
  condensed gas,
\newblock {\em Phys. Rev. Lett., {\bf 83}\/} (1999) 4452--4455.

\bibitem{mara00scis}
{\sc Maragò, O.~M., Hopkins, S.~A., Arlt, J., Hodby, E., Hechenblaikner, G.,
  and Foot, C.~J.},
\newblock Observation of the scissors mode and evidence for superfluidity of a
  trapped {Bose-Einstein} condensed gas,
\newblock {\em Phys. Rev. Lett., {\bf 84}\/} (2000) 2056--2059.

\bibitem{edwa02}
{\sc Edwards, M., Clark, C.~W., Pedri, P., Pitaevskii, L., and Stringari, S.},
\newblock Consequence of superfluidity on the expansion of a rotating
  {Bose-Einstein} condensate,
\newblock {\em Phys. Rev. Lett., {\bf 88}\/} (2002) 070405.

\bibitem{hech02}
{\sc Hechenblaikner, G., Hodby, E., Hopkins, S.~A., Maragò, O.~M., and Foot,
  C.~J.},
\newblock Direct observation of irrotational flow and evidence of superfluidity
  in a rotating {Bose-Einstein} condensate,
\newblock {\em Phys. Rev. Lett., {\bf 88}\/} (2002) 070406.

\bibitem{modu03}
{\sc Modugno, M., Modugno, G., Roati, G., Fort, C., and Inguscio, M.},
\newblock Scissors mode of an expanding {Bose-Einstein} condensate,
\newblock {\em Phys. Rev. A, {\bf 67}\/} (2003) 023608.

\bibitem{clan07irrot}
{\sc Clancy, B., Luo, L., and Thomas, J.~E.},
\newblock Observation of nearly perfect irrotational flow in normal and
  superfluid strongly interacting {Fermi} gases,
\newblock {\em Phys. Rev. Lett., {\bf 99}\/} (2007) 140401.

\bibitem{wrig07finite}
{\sc Wright, M.~J., Riedl, S., Altmeyer, A., Kohstall, C., Guajardo, E. R.~S.,
  Hecker-Denschlag, J., and Grimm, R.},
\newblock Finite-temperature collective dynamics of a {Fermi} gas in the
  {BEC-BCS} crossover,
\newblock {\em Phys. Rev. Lett., {\bf 99}\/} (2007) 150403.

\bibitem{cozz03}
{\sc Cozzini, M., and Stringari, S.},
\newblock {F}ermi gases in slowly rotating traps: Superfluid versus collisional
  hydrodynamics,
\newblock {\em Phys. Rev. Lett., {\bf 91}\/} (2002) 070401.

\bibitem{gupt04coll}
{\sc Gupta, S., Hadzibabic, Z., Anglin, J.~R., and Ketterle, W.},
\newblock Collisions in zero temperature {F}ermi gases,
\newblock {\em Phys. Rev. Lett., {\bf 92}\/} (2004) 100401.

\bibitem{kinn04}
{\sc Kinnunen, J., Rodriguez, M., and Törmä, P.},
\newblock Signatures of superfluidity for {F}eshbach-resonant {F}ermi gases,
\newblock {\em Phys. Rev. Lett., {\bf 92}\/} (2004) 230403--4.

\bibitem{kina04hydr}
{\sc Kinast, J., Turlapov, A., and Thomas, J.~E.},
\newblock Breakdown of hydrodynamics in the radial breathing mode of a
  strongly-interacting {F}ermi gas,
\newblock {\em Phys. Rev. A, {\bf 70}\/} (2004) 051401.

\bibitem{grei04corr}
{\sc Greiner, M., Regal, C.~A., Ticknor, C., Bohn, J.~L., and Jin, D.~S.},
\newblock Detection of spatial correlations in an ultracold gas of fermions,
\newblock {\em Phys. Rev. Lett., {\bf 92}\/} (2004) 150405.

\bibitem{avde05}
{\sc Avdeenkov, A.~V., and Bohn, J.~L.},
\newblock Pair wave functions in atomic {F}ermi condensates,
\newblock {\em Phys. Rev. A, {\bf 71}\/} (2005) 023609.

\bibitem{chen06phase}
{\sc Chen, Q., Regal, C.~A., Greiner, M., Jin, D.~S., and Levin, K.},
\newblock Understanding the superfluid phase diagram in trapped {F}ermi gases,
\newblock {\em Phys. Rev. A, {\bf 73}\/} (2006) 041601.

\bibitem{zwie04form}
{\sc Zwierlein, M.~W., Schunck, C.~H., Stan, C.~A., Raupach, S. M.~F., and
  Ketterle, W.},
\newblock Formation dynamics of a fermion pair condensate,
\newblock {\em Phys. Rev. Lett., {\bf 94}\/} (2005) 180401.

\bibitem{zwie06direct}
{\sc Zwierlein, M.~W., Schunck, C.~H., Schirotzek, A., and Ketterle, W.},
\newblock Direct observation of the superfluid phase transition in ultracold
  {F}ermi gases,
\newblock {\em Nature, {\bf 442}\/} (2006) 54--58.

\bibitem{onsa49quant}
{\sc Onsager, L.},
\newblock (discussion on a paper by {C.J. Gorter}),
\newblock {\em Nuovo Cimento Suppl., {\bf 6}\/} (1949) 249--50.

\bibitem{feyn53}
{\sc Feynman, R.~P.},
\newblock Atomic theory of liquid helium near absolute zero,
\newblock {\em Phys. Rev., {\bf 91}\/} (1953) 1301.

\bibitem{feyn54}
{\sc Feynman, R.~P.},
\newblock Atomic theory of the two-fluid model of liquid helium,
\newblock {\em Phys. Rev., {\bf 94}\/} (1954) 262.

\bibitem{shin04doublevortex}
{\sc Shin, Y., Saba, M., Vengalattore, M., Pasquini, T.~A., Sanner, C.,
  Leanhardt, A.~E., Prentiss, M., Pritchard, D.~E., and Ketterle, W.},
\newblock Dynamical instability of a doubly quantized vortex in a
  {Bose-Einstein} condensate,
\newblock {\em Phys. Rev. Lett., {\bf 93}\/} (2004) 160406.

\bibitem{abri57}
{\sc Abrikosov, A.~A.},
\newblock On the magnetic properties of superconductors of the second group,
\newblock {\em Zh. Eksp. Teor. Fiz., {\bf 32}\/} (1957) 1442,
\newblock [Sov. Phys. JETP 5, 1174 (1957)].

\bibitem{angl01vort}
{\sc Anglin, J.~R.},
\newblock Local vortex generation and the surface mode spectrum of large
  {Bose-Einstein} condensates,
\newblock {\em Phys. Rev. Lett., {\bf 87}\/} (2001) 240401--4.

\bibitem{dalf01crit}
{\sc Dalfovo, F., and Stringari, S.},
\newblock Shape deformations and angular-momentum transfer in trapped
  {Bose-Einstein} condensates,
\newblock {\em Phys. Rev. A, {\bf 63}\/} (2001) 011601--4.

\bibitem{rama01nuc}
{\sc Raman, C., Abo-Shaeer, J.~R., Vogels, J.~M., Xu, K., and Ketterle, W.},
\newblock Vortex nucleation in a stirred {Bose-Einstein} condensate,
\newblock {\em Phys. Rev. Lett., {\bf 87}\/} (2001) 210402.

\bibitem{vine58}
{\sc Vinen, W.~F.},
\newblock Detection of single quanta of circulation in rotating helium {II},
\newblock {\em Nature, {\bf 181}\/} (1958) 1524.

\bibitem{deav61}
{\sc Deaver, B.~S., and Fairbank, W.~M.},
\newblock Experimental evidence for quantized flux in superconducting
  cylinders,
\newblock {\em Phys. Rev. Lett., {\bf 7}\/} (1961) 43.

\bibitem{doll61}
{\sc Doll, R., and Näbauer, M.},
\newblock Experimental proof of magnetic flux quantization in a superconducting
  ring,
\newblock {\em Phys. Rev. Lett., {\bf 7}\/} (1961) 51.

\bibitem{trau67}
{\sc Träuble, H., and Essmann, U.},
\newblock The direct observation of individual flux lines in type {II}
  superconductors,
\newblock {\em Phys. Lett., {\bf 24A}\/} (1967) 526.

\bibitem{sarm67}
{\sc Sarma, N.~V.},
\newblock Direct evidence for the laminar and flux line models of mixed state
  in type {II} superconductors,
\newblock {\em Phys. Lett. A, {\bf 25}\/} (1967) 315.

\bibitem{yarm79}
{\sc Yarmchuk, E.~J., Gordon, M. J.~V., and Packard, R.~E.},
\newblock Observation of stationary vortex arrays in rotating superfluid
  helium,
\newblock {\em Phys. Rev. Lett., {\bf 43}\/} (1979) 214.

\bibitem{davi91he3}
{\sc Davis, J.~C., Close, J.~D., Zieve, R., and Packard, R.~E.},
\newblock Observation of quantized circulation in superfluid $^3${H}e-{B},
\newblock {\em Phys. Rev. Lett., {\bf 66}\/} (1991) 329.

\bibitem{alpa95neutron}
{\sc Alpar, M., and Pines, D.},
\newblock in {\em The Lives of the Neutron Star: Conference Proceedings},
  edited by M.~Alpar and J.~Paradijs, (Kluwer Academic, Dordrecht) 1995.

\bibitem{donn91}
{\sc Donnelly, R.~J.},
\newblock {\em Quantized vortices in Helium {II}},
\newblock (Cambridge University Press, Cambridge) 1991.

\bibitem{halj01vort}
{\sc Haljan, P.~C., Coddington, I., Engels, P., and Cornell, E.~A.},
\newblock Driving {Bose-Einstein}-condensate vorticity with a rotating normal
  cloud,
\newblock {\em Phys. Rev. Lett., {\bf 87}\/} (2001) 210403--4.

\bibitem{stri96coll}
{\sc Stringari, S.},
\newblock Collective excitations of a trapped {B}ose-condensed gas,
\newblock {\em Phys. Rev. Lett., {\bf 77}\/} (1996) 2360--2363.

\bibitem{madi01}
{\sc Madison, K.~W., Chevy, F., Bretin, V., and Dalibard, J.},
\newblock Stationary states of a rotating {Bose-Einstein} condensate: Routes to
  vortex nucleation,
\newblock {\em Phys. Rev. Lett., {\bf 86}\/} (2001) 4443--4446.

\bibitem{abos02_form}
{\sc Abo-Shaeer, J.~R., Raman, C., and Ketterle, W.},
\newblock Formation and decay of vortex lattices in {Bose-Einstein} condensates
  at finite temperatures,
\newblock {\em Phys. Rev. Lett., {\bf 88}\/} (2002) 070409.

\bibitem{bulg03}
{\sc Bulgac, A., and Yu, Y.},
\newblock The vortex state in a strongly coupled dilute atomic fermionic
  superfluid,
\newblock {\em Phys. Rev. Lett., {\bf 91}\/} (2003) 190404.

\bibitem{chie06vortex}
{\sc Chien, C.-C., He, Y., Chen, Q., and Levin, K.},
\newblock Ground-state description of a single vortex in an atomic {F}ermi gas:
  From {BCS} to {B}ose--{E}instein condensation,
\newblock {\em Phys. Rev. A, {\bf 73}\/} (2006) 041603.

\bibitem{toni05}
{\sc Tonini, G., Werner, F., and Castin, Y.},
\newblock Formation of a vortex lattice in a rotating {BCS} {F}ermi gas,
\newblock {\em Eur. Phys. J. D, {\bf 39}\/} (2006) 283.

\bibitem{angl99}
{\sc Anglin, J.~R., and Zurek, W.~H.},
\newblock Vortices in the wake of rapid {Bose-Einstein} condensation,
\newblock {\em Phys. Rev. Lett., {\bf 83}\/} (1999) 1707.

\bibitem{shva02non_eq_bec}
{\sc Shvarchuck, I., Buggle, C., Petrov, D.~S., Dieckmann, K., Zielonkowski,
  M., Kemmann, M., Tiecke, T.~G., v.~Klitzing, W., Shlyapnikov, G.~V., and
  Walraven, J. T.~M.},
\newblock {Bose-Einstein} condensation into nonequilibrium states studied by
  condensate focusing,
\newblock {\em Phys. Rev. Lett., {\bf 89}\/} (2002) 270404.

\bibitem{zhur01diss}
{\sc Zhuravlev, O.~N., Muryshev, A.~E., and Fedichev, P.~O.},
\newblock Dissipative dynamics of vortex arrays in anisotropic traps,
\newblock {\em Phys. Rev. A, {\bf 64}\/} (2001) 053601.

\bibitem{fedi02vortex}
{\sc Fedichev, P.~O., and Muryshev, A.~E.},
\newblock Equilibrium properties and dissipative dynamics of vortex arrays in
  trapped {B}ose-condensed gases,
\newblock {\em Phys. Rev. A, {\bf 65}\/} (2002) 061601.

\bibitem{schu06pair}
{\sc Schunck, C.~H., Zwierlein, M.~W., Schirotzek, A., and Ketterle, W.},
\newblock Superfluid expansion of a rotating {F}ermi gas,
\newblock {\em Phys. Rev. Lett., {\bf 98}\/} (2007) 050404.

\bibitem{ante07}
{\sc Antezza, M., Cozzini, M., and Stringari, S.},
\newblock Breathing modes of a fast rotating {Fermi} gas,
\newblock {\em Phys. Rev. A, {\bf 75}\/} (2007) 053609.

\bibitem{houb97crit}
{\sc Houbiers, M., Stoof, H. T.~C., and Cornell, E.~A.},
\newblock Critical temperature of a trapped {B}ose gas: Mean-field theory and
  fluctuations,
\newblock {\em Phys. Rev. A, {\bf 56}\/} (1997) 2041.

\bibitem{gehm03stab}
{\sc Gehm, M.~E., Hemmer, S.~L., Granade, S.~R., O'Hara, K.~M., and Thomas,
  J.~E.},
\newblock Mechanical stability of a strongly interacting {Fermi} gas of atoms,
\newblock {\em Phys. Rev. A, {\bf 68}\/} (2003) 011401.

\bibitem{stew06pot}
{\sc Stewart, J.~T., Gaebler, J.~P., Regal, C.~A., and Jin, D.~S.},
\newblock Potential energy of a $^{40}${K} {Fermi} gas in the {BCS-BEC}
  crossover,
\newblock {\em Phys. Rev. Lett., {\bf 97}\/} (2006) 220406.

\bibitem{bake99neutron}
{\sc Baker, G.~A.},
\newblock Neutron matter model,
\newblock {\em Phys. Rev. C, {\bf 60}\/} (1999) 054311.

\bibitem{stee00power}
{\sc Steele, J.~V.},
\newblock Effective field theory power counting at finite density,
\newblock preprint nucl-th/0010066.

\bibitem{heis01}
{\sc Heiselberg, H.},
\newblock {F}ermi systems with long scattering lengths,
\newblock {\em Phys. Rev. A, {\bf 63}\/} (2001) 043606.

\bibitem{pera04bcsbec}
{\sc Perali, A., Pieri, P., and Strinati, G.~C.},
\newblock Quantitative comparison between theoretical predictions and
  experimental results for the {BCS-BEC} crossover,
\newblock {\em Phys. Rev. Lett., {\bf 93}\/} (2004) 100404.

\bibitem{hu07universal}
{\sc Hu, H., Drummond, P.~D., and Liu, X.-J.},
\newblock Universal thermodynamics of strongly interacting {Fermi} gases,
\newblock {\em Nature Physics, {\bf 3}\/} (2007) 469--472.

\bibitem{carl03}
{\sc Carlson, J., Chang, S.-Y., Pandharipande, V.~R., and Schmidt, K.~E.},
\newblock Superfluid {F}ermi gases with large scattering length,
\newblock {\em Phys. Rev. Lett., {\bf 91}\/} (2003) 050401.

\bibitem{astr04}
{\sc Astrakharchik, G.~E., Boronat, J., Casulleras, J., and Giorgini, S.},
\newblock Equation of state of a {F}ermi gas in the {BEC}-{BCS} crossover: A
  {Q}uantum {M}onte-{C}arlo study,
\newblock {\em Phys. Rev. Lett., {\bf 93}\/} (2004) 200404.

\bibitem{carl05}
{\sc Carlson, J., and Reddy, S.},
\newblock Asymmetric two-component fermion systems in strong coupling,
\newblock {\em Phys. Rev. Lett., {\bf 95}\/} (2005) 060401.

\bibitem{rega05}
{\sc Regal, C.~A., Greiner, M., Giorgini, S., Holland, M., and Jin, D.~S.},
\newblock Momentum distribution of a {Fermi} gas of atoms in the {BCS-BEC}
  crossover,
\newblock {\em Phys. Rev. Lett., {\bf 95}\/} (2005) 250404.

\bibitem{chen06momentum}
{\sc Chen, Q., Regal, C.~A., Jin, D.~S., and Levin, K.},
\newblock Finite-temperature momentum distribution of a trapped {F}ermi gas,
\newblock {\em Phys. Rev. A, {\bf 74}\/} (2006) 011601.

\bibitem{jin96coll}
{\sc Jin, D.~S., Ensher, J.~R., Matthews, M.~R., Wieman, C.~E., and Cornell,
  E.~A.},
\newblock Collective excitations of a {Bose-Einstein} condensate in a dilute
  gas,
\newblock {\em Phys. Rev. Lett., {\bf 77}\/} (1996) 420.

\bibitem{mewe96coll}
{\sc Mewes, M.-O., Andrews, M.~R., van Druten, N.~J., Kurn, D.~M., Durfee,
  D.~S., Townsend, C.~G., and Ketterle, W.},
\newblock Collective excitations of a {Bose-Einstein} condensate in a magnetic
  trap,
\newblock {\em Phys. Rev. Lett., {\bf 77}\/} (1996) 988--991.

\bibitem{jin97}
{\sc Jin, D.~S., Matthews, M.~R., Ensher, J.~R., Wieman, C.~E., and Cornell,
  E.~A.},
\newblock Temperature-dependent damping and frequency shifts in collective
  excitations of a dilute {Bose-Einstein} condensate,
\newblock {\em Phys. Rev. Lett., {\bf 78}\/} (1997) 764.

\bibitem{dalf99rmp}
{\sc Dalfovo, F., Giorgini, S., Pitaevskii, L.~P., and Stringari, S.},
\newblock Theory of {B}ose--{E}instein condensation in trapped gases,
\newblock {\em Rev. Mod. Phys., {\bf 71}\/} (1999) 463--512.

\bibitem{gens01coll}
{\sc Gensemer, S.~D., and Jin, D.~S.},
\newblock Transition from collisionless to hydrodynamic behavior in an
  ultracold {Fermi} gas,
\newblock {\em Phys. Rev. Lett., {\bf 87}\/} (2001) 173201.

\bibitem{andr97prop}
{\sc Andrews, M.~R., Kurn, D.~M., Miesner, H.-J., Durfee, D.~S., Townsend,
  C.~G., Inouye, S., and Ketterle, W.},
\newblock Propagation of sound in a {Bose-Einstein} condensate,
\newblock {\em Phys. Rev. Lett., {\bf 79}\/} (1997) 553--556.

\bibitem{astr07private}
{\sc Astrakharchik, G.},
\newblock quoted for private communication in J. Joseph et al., {\it Phys. Rev.
  Lett.} {\bf 98} (2007), 170401.

\bibitem{kinn04pairing}
{\sc Kinnunen, J., Rodriguez, M., and Torma, P.},
\newblock Pairing gap and in-gap excitations in trapped fermionic superfluids,
\newblock {\em Science, {\bf 305}\/} (2004) 1131--1133.

\bibitem{duan05lattice}
{\sc Duan, L.-M.},
\newblock Effective {Hamiltonian} for fermions in an optical lattice across a
  {Feshbach} resonance,
\newblock {\em Phys. Rev. Lett., {\bf 95}\/} (2005) 243202.

\bibitem{pari07finitetemp}
{\sc Parish, M.~M., Marchetti, F.~M., Lamacraft, A., and Simons, B.~D.},
\newblock Finite-temperature phase diagram of a polarized {Fermi} condensate,
\newblock {\em Nature Physics, {\bf 3}\/} (2007) 124--128.

\bibitem{chan62}
{\sc Chandrasekhar, B.~S.},
\newblock A note on the maximum critical field of high-field superconductors,
\newblock {\em Appl. Phys. Lett., {\bf 1}\/} (1962) 7.

\bibitem{clog62}
{\sc Clogston, A.~M.},
\newblock Upper limit for the critical field in hard superconductors,
\newblock {\em Phys. Rev. Lett., {\bf 9}\/} (1962) 266.

\bibitem{lobo06}
{\sc Lobo, C., Recati, A., Giorgini, S., and Stringari, S.},
\newblock Normal state of a polarized {Fermi} gas at unitarity,
\newblock {\em Phys. Rev. Lett., {\bf 97}\/} (2006) 200403--4.

\bibitem{gubb07}
{\sc Gubbels, K.~B., and Stoof, H. T.~C.},
\newblock Renormalization group theory for the imbalanced {Fermi} gas,
\newblock preprint arXiv:0711.2963.

\bibitem{ott04}
{\sc Ott, H., d.~Mirandes, E., Ferlaino, F., Roati, G., Modugno, G., and
  Inguscio, M.},
\newblock Collisionally induced transport in periodic potentials,
\newblock {\em Phys. Rev. Lett., {\bf 92}\/} (2004) 160601.

\bibitem{pezz04}
{\sc Pezze, L., Pitaevskii, L., Smerzi, A., Stringari, S., Modugno, G.,
  d.~Mirandes, E., Ferlaino, F., Ott, H., Roati, G., and Inguscio, M.},
\newblock Insulating behavior of a trapped ideal {Fermi} gas,
\newblock {\em Phys. Rev. Lett., {\bf 93}\/} (2004) 120401.

\bibitem{roat04}
{\sc Roati, G., d.~Mirandes, E., Ferlaino, F., Ott, H., Modugno, G., and
  Inguscio, M.},
\newblock Atom interferometry with trapped {Fermi} gases,
\newblock {\em Phys. Rev. Lett., {\bf 92}\/} (2004) 230402.

\bibitem{caru05}
{\sc Carusotto, I., Pitaevskii, L., Stringari, S., Modugno, G., and Inguscio,
  M.},
\newblock Sensitive measurement of forces at the micron scale using {Bloch}
  oscillations of ultracold atoms,
\newblock {\em Phys. Rev. Lett., {\bf 95}\/} (2005) 093202.

\bibitem{ospe06}
{\sc Ospelkaus, S., Ospelkaus, C., Wille, O., Succo, M., Ernst, P., Sengstock,
  K., and Bongs, K.},
\newblock Localization of bosonic atoms by fermionic impurities in a
  three-dimensional optical lattice,
\newblock {\em Phys. Rev. Lett., {\bf 96}\/} (2006) 180403.

\bibitem{gunt06bosefermi}
{\sc Günter, K., Stöferle, T., Moritz, H., Köhl, M., and Esslinger, T.},
\newblock {Bose-Fermi} mixtures in a three-dimensional optical lattice,
\newblock {\em Phys. Rev. Lett., {\bf 96}\/} (2006) 180402.

\bibitem{heis00}
{\sc Heiselberg, H., Pethick, C.~J., Smith, H., and Viverit, L.},
\newblock Influence of induced interactions on the superfluid transition in
  dilute {F}ermi gases,
\newblock {\em Phys. Rev. Lett., {\bf 85}\/} (2000) 2418.

\bibitem{bijl00phonon}
{\sc Bijlsma, M.~J., Heringa, B.~A., and Stoof, H. T.~C.},
\newblock Phonon exchange in dilute {Fermi-Bose} mixtures: Tailoring the
  {Fermi-Fermi} interaction,
\newblock {\em Phys. Rev. A, {\bf 61}\/} (2000) 053601.

\bibitem{zirb07hetero}
{\sc Zirbel, J.~J., Ni, K.-K., Ospelkaus, S., D'Incao, J.~P., Wieman, C.~E.,
  Ye, J., and Jin, D.~S.},
\newblock Collisional stability of fermionic {Feshbach} molecules,
\newblock preprint arxiv:0710.2479.

\bibitem{will07fermifermi}
{\sc Wille, E., Spiegelhalder, F.~M., Kerner, G., Naik, D., Trenkwalder, A.,
  Hendl, G., Schreck, F., Grimm, R., Tiecke, T.~G., Walraven, J. T.~M.,
  Kokkelmans, S. J. J. M.~F., Tiesinga, E., and Julienne, P.~S.},
\newblock Exploring an ultracold {Fermi-Fermi} mixture: {Interspecies}
  {Feshbach} resonances and scattering properties of $^6${L}i and $^{40}${K},
\newblock preprint arXiv:0711.2916.

\bibitem{gunt05pwave}
{\sc Günter, K., Stöferle, T., Moritz, H., Köhl, M., and Esslinger, T.},
\newblock p-wave interactions in low-dimensional fermionic gases,
\newblock {\em Phys. Rev. Lett., {\bf 95}\/} (2005) 230401.

\bibitem{hofs02hightc}
{\sc Hofstetter, W., Cirac, J.~I., Zoller, P., Demler, E., and Lukin, M.~D.},
\newblock High-temperature superfluidity of fermionic atoms in optical
  lattices,
\newblock {\em Phys. Rev. Lett., {\bf 89}\/} (2002) 220407.

\bibitem{wern05lattice}
{\sc Werner, F., Parcollet, O., Georges, A., and Hassan, S.~R.},
\newblock Interaction-induced adiabatic cooling and antiferromagnetism of cold
  fermions in optical lattices,
\newblock {\em Phys. Rev. Lett., {\bf 95}\/} (2005) 056401--4.

\end{thebibliography}

\end{document}